\documentclass[a4paper,11pt]{article}
\pdfoutput=1
\usepackage{jheppub}
\usepackage[utf8]{inputenc}
\usepackage{url}
\usepackage{mathrsfs}  
\usepackage{pdflscape}
\usepackage{tikz}
\usetikzlibrary{positioning}
\usetikzlibrary{arrows}
\usetikzlibrary{decorations.pathreplacing}
\usetikzlibrary{shapes.geometric}
\usetikzlibrary{calc}
\usepackage{refcount}
\usepackage{booktabs}
\usepackage{todonotes}
\usepackage{enumitem}
\usepackage{adjustbox}
\usepackage{afterpage}

\newcommand{\specialcell}[2][c]{%
  \begin{tabular}[#1]{@{}c@{}}#2\end{tabular}}
  
\newcommand{\aONEquiv}[2]{\scalebox{.678}{\begin{tikzpicture}
    \node[gauge,label=below:{#1}] (1) at (0,0) {};
    \node[flavour,label=above:{#2}] (1a) at (0,1) {};
    \draw (1a)--(1);
\end{tikzpicture}}}
\newcommand{\aquiv}[4]{\scalebox{.678}{\begin{tikzpicture}
    \node[gauge,label=below:{#3}] (2) at (-1,0) {};
    \node[gauge,label=below:{#4}] (1) at (0,0) {};
    \node[flavour,label=above:{#2}] (1a) at (0,1) {};
    \node[flavour,label=above:{#1}] (2a) at (-1,1) {};
    \draw (1a)--(1)--(2)--(2a);
\end{tikzpicture}}}
\newcommand{\bquiv}[4]{\scalebox{.678}{\begin{tikzpicture}
    \node[gauge,label=below:{#3}] (2) at (1,0) {};
    \node[gauge,label=below:{#4}] (1) at (0,0) {};
    \node[flavour,label=above:{#1}] (1a) at (0,1) {};
    \node[flavour,label=above:{#2}] (2a) at (1,1) {};
    \draw (1a)--(1) (2)--(2a) (0.4,0.2)--(0.6,0)--(0.4,-0.2);
    \draw[transform canvas={yshift=-1pt}] (1)--(2);
    \draw[transform canvas={yshift=1pt}] (1)--(2);
\end{tikzpicture}}} 
\newcommand{\cquiv}[4]{\scalebox{.678}{\begin{tikzpicture}
    \node[gauge,label=below:{#3}] (2) at (-1,0) {};
    \node[gauge,label=below:{#4}] (1) at (0,0) {};
    \node[flavour,label=above:{#2}] (1a) at (0,1) {};
    \node[flavour,label=above:{#1}] (2a) at (-1,1) {};
    \draw (1a)--(1) (2)--(2a) (-0.4,0.2)--(-0.6,0)--(-0.4,-0.2);
    \draw[transform canvas={yshift=-1pt}] (1)--(2);
    \draw[transform canvas={yshift=1pt}] (1)--(2);
\end{tikzpicture}}}

\newcommand{\gr}[3]{\left[\mathrm{Gr}_{#1} \right]^{#2}_{#3}}
\newcommand{\grb}[3]{\overline{\left[\mathrm{Gr}_{#1} \right]}^{#2}_{#3}}
\newcommand{\w}[3]{\left[\mathcal{W}_{#1} \right]^{#2}_{#3}}
\newcommand{\wb}[3]{\overline{\left[\mathcal{W}_{#1} \right]}^{#2}_{#3}}

\tikzset{skeleton/.style={draw=red,minimum size=0.35mm,fill=white,circle, draw}}
\tikzset{flavour2/.style={draw=none,minimum size=0.4cm,fill=white,regular polygon sides=4,draw}}
\tikzset{gauge/.style={inner sep=1mm,draw=none,fill=white,minimum size=2mm,circle, draw}}
\tikzset{flavour/.style={draw=none,minimum size=0.3mm,fill=white, regular polygon,regular polygon sides=4,draw}}
\tikzset{bd/.style={circle, draw=black, inner sep=0pt, fill=black, minimum size=2mm}}
\tikzset{bd2/.style={circle, draw=black!50, inner sep=0pt, fill=black!50, minimum size=2mm}}
\tikzset{rd/.style={circle, draw=red!50, inner sep=0pt, fill=red!50, minimum size=2mm}}
\tikzset{gd/.style={circle, draw=green!50, inner sep=0pt, fill=green!50, minimum size=2mm}}
\tikzset{bl/.style={circle, draw=blue!30, inner sep=0pt, fill=blue!30, minimum size=2mm}}
\tikzset{gauge3/.style={draw=none,minimum size=0.35mm,fill=white,circle, draw}}
\tikzset{miniBlue/.style={draw=none,minimum size=0.35mm,fill=blue,circle, draw}}
\tikzset{oliveX/.style={draw=none,minimum size=0.35mm,fill=none,circle, draw}}
\tikzset{olivesquare/.style={draw=none,minimum size=0.35mm,fill=white, regular polygon,regular polygon sides=4,draw}}
\tikzset{none/.style={draw=none}}
\tikzset{flavor1/.style={draw=none,minimum size=0.35mm,fill=white, regular polygon,regular polygon sides=4,draw}}
\tikzset{flavour2/.style={draw=none,minimum size=0.35mm,fill=white, regular polygon,regular polygon sides=4,draw}}
\tikzset{Reddotted/.style={dashed,color=red}}
\tikzset{Bluedotted/.style={dashed,color=blue}}
\pgfdeclarelayer{edgelayer}
\pgfdeclarelayer{nodelayer}
\pgfsetlayers{edgelayer,nodelayer,main}
\tikzset{brace/.style={decorate,decoration={brace,amplitude=10pt}}}
\DeclareFontFamily{U}{rcjhbltx}{}
\DeclareFontShape{U}{rcjhbltx}{m}{n}{<->rcjhbltx}{}
\DeclareSymbolFont{hebrewletters}{U}{rcjhbltx}{m}{n}

\DeclareMathSymbol{\beth}{\mathord}{hebrewletters}{98}\let\bet\beth
\DeclareMathSymbol{\lamed}{\mathord}{hebrewletters}{108}

\usetikzlibrary{decorations.pathmorphing}
\tikzset{D5/.style={circle, fill=blue,inner sep=3pt}}
\tikzset{D5o/.style={circle, fill=olive,inner sep=3pt}}
\tikzset{wiggler/.style={decorate, decoration={snake}}}

\preprint{Imperial/TP/21/AH/01}
\title{Branes, Quivers, and the Affine Grassmannian}
\author[\lamed]{Antoine Bourget,}
\author[\lamed]{Julius F. Grimminger,}
\author[\lamed]{Amihay Hanany,}
\author[\beth]{Marcus Sperling,}
\author[\lamed]{and Zhenghao Zhong\,}
\affiliation[\lamed]{Theoretical Physics Group, The Blackett Laboratory, Imperial College London, Prince Consort Road
London, SW7 2AZ, UK}
\affiliation[\beth]{Yau Mathematical Sciences Center, Tsinghua University, Haidian District, Beijing, 100084, China}
\emailAdd{a.bourget@imperial.ac.uk}
\emailAdd{julius.grimminger17@imperial.ac.uk}
\emailAdd{a.hanany@imperial.ac.uk}
\emailAdd{marcus.sperling@univie.ac.at}
\emailAdd{zhenghao.zhong14@imperial.ac.uk}

\abstract{Brane systems provide a large class of gauge theories that arise in string theory. This paper demonstrates how such brane systems fit with a somewhat exotic geometric object, called the affine Grassmannian.
This gives a strong motivation to study physical aspects of the affine Grassmannian. Explicit quivers are presented throughout the paper, and a quiver addition algorithm to generate the affine Grassmannian is introduced. An important outcome of this study is a set of quivers for new elementary slices. 
}

\begin{document}
\maketitle

\section{Introduction}

Branes play a crucial role in our current understanding of string theory.
Their world-volume low energy effective action has been occupying numerous papers and is still used as an essential tool in studying a whole variety of string backgrounds.
Furthermore, the existence of brane systems opens up a window to the study of theories when an effective Lagrangian either does not exist or is beyond reach.
This again increases the level of understanding in a significant way.

This paper is devoted to yet another conceptual advancement in the study of brane systems in string theory. It has to do with a somewhat exotic object -- and so far less studied within the string theory community -- the affine Grassmannian \cite{beauville1994conformal,kumar1994infinite,beauville1998picard,pappas2008twisted,sorgerlectures,beilinson1991quantization}. This object plays a prominent role in the so-called geometric Satake correspondence in the geometric Langlands program \cite{lusztig1983singularities,ginzburg1995perverse,mirkovic2007geometric}, which is partly responsible for the considerable interest it has attracted in mathematics in the past two decades. 
A close inspection reveals that the affine Grassmannian is also precisely the type of object one should study when trying to understand brane systems, hence making it an important element in physical systems.  

As is argued below, brane systems of the type of Hanany-Witten setups \cite{Hanany:1996ie} fit into the affine Grassmannian, and they do so in a host of space-time dimensions ranging from 6 to 3, and possibly even lower.
A definition of the affine Grassmannian, suitable to a physics audience is presented below, together with examples and explanations. Special attention is paid to slices in the affine Grassmannian \cite{2003math:5095M,kamnitzer2014yangians,Braverman:2016pwk} which turn out to fit with the physical setting, namely each slice is a moduli space of vacua of an interacting theory, with or without a Lagrangian; or viewed from the brane perspective, a Kraft-Procesi transition \cite{kraft1980minimal,Kraft1982} between two phases of the brane system \cite{Cabrera:2016vvv,Cabrera:2017njm}. The connection between Coulomb branches viewed as moduli spaces of singular monopoles \cite{Chalmers:1996xh,Hanany:1996ie,deBoer:1996mp,deBoer:1996ck,Cherkis:1997aa,Tong:1998fa} and slices in the affine Grassmannians was made in \cite{Bullimore:2015lsa,Braverman:2016pwk,finkelberg2017double,Nakajima:2019olw}, using previous developments \cite{braverman2010pursuing,Cremonesi:2013lqa,Nakajima:2015txa,Nakajima2015,Braverman:2016wma,Nakajima:2017bdt}. The mathematical properties of the slices in connection with symplectic geometry and the relation with quiver theories have been further studied in \cite{kamnitzer2014yangians,kamnitzer2018reducedness,muthiah2019symplectic,weekes2019generators,kamnitzer2019category}. 

We note that so far we do not have a brane construction for the affine Grassmannian of exceptional groups. Slices in these affine Grassmannians can nevertheless be studied using quivers. This involves a new algorithm of quiver addition, which is explained in detail below. 

Let us turn to Hasse diagrams \cite{Bourget:2019aer}. Hasse diagrams are a depiction of a partial order. They are a useful tool for characterizing the phase structure of a given theory. In such diagrams, consisting of nodes connected by edges, each node represents a fixed set of massless states of the physical system. Tuning moduli without moving to a different phase corresponds to changing the effective masses of massive particles, while keeping the set of massless states fixed. When the set of massless states of one phase, call it phase $a$, is contained in the set of massless states of another phase, call it phase $b$, we assign the order $b<a$. Suppose there exists no phase $c$ such that $b<c<a$ then $a$ and $b$ are connected by an edge due to the partial order $b<a$. The edge connecting node $a$ to node $b$ represents the minimal set of moduli one needs to tune in order to move from phase $a$ phase $b$. The tuned moduli form a conical singularity and supersymmetry dictates the type of such a singularity. For $3d$ $\mathcal{N}=4$ Coulomb branches, and Higgs branches of theories with 8 supercharges in dimensions $3-6$, this singularity is a symplectic singularity \cite{beauville2000symplectic}, or a hyper K\"ahler cone -- as more commonly referred to in the physics literature.

As Hasse diagrams for moduli spaces are composed of nodes and edges, 
it is natural to put some additional structure by embedding such a diagram into a lattice. Indeed, it is one of the features of the affine Grassmannian of a Lie group $G$, that the Hasse diagram for any slice originates from such a lattice, making the conceptual understanding of such Hasse diagram much easier. The leaves of the affine Grassmannian (called Schubert cells)\footnote{We note that these Schubert cells may be of odd complex dimension, and their closure hence not a symplectic singularity. The transverse slice between two Schubert cells, however, is a symplectic singularity.} are in one-to-one correspondence with dominant coweights. Transverse slices between two leaves are in one-to-one correspondence with lattice vectors that connect one node to another. These lattice vectors, in turn, are spanned by positive coroots, and if the linear combination is non-negative then there exists a directed path between one point and another; hence, inducing a natural partial order in the lattice \cite{stembridge1998partial}. Therefore, slices in the affine Grassmannian are particularly nice, as their Hasse diagrams can be understood as all directed paths between two points in the principal Weyl chamber. The fact that this structure fits beautifully with brane systems is not surprising, as brane systems can be viewed as algebraic objects. This just emphasizes the point that slices in the affine Grassmannian form very interesting physical systems. The stratification of the slices in the affine Grassmannian into symplectic leaves can be related to monopole bubbling \cite{Ito:2011ea,Gomis:2011pf,Dedushenko:2017avn,Dedushenko:2018icp,Assel:2019iae,Assel:2019yzd}. Brane constructions in relation to monopole bubbling were studied in \cite{Assel:2019yzd}. 

The simplest affine Grassmannian is that of $\mathrm{SL}(2,\mathbb{C})$, where the Hasse diagram is a semi infinite linear diagram. Each slice in the affine Grassmannian of $\mathrm{SL}(2,\mathbb{C})$ is an SQCD type moduli space (specifically its Coulomb branch) with a unitary gauge group as opposed to special unitary. Hasse diagrams of SQCD theories were already shown to be linear in \cite[Table 1]{Bourget:2019aer}. One important point made in the present paper is that this line is actually connecting two points on the principal Weyl chamber of $\mathrm{SL}(2,\mathbb{C})$. It is reassuring to see that for the simplest and most studied gauge theories -- SQCD -- there is a correspondence with the simplest possible affine Grassmannian -- the affine Grassmannian of $\mathrm{SL}(2,\mathbb{C})$. 
If one identifies SQCD theories with the more mathematical name, framed $A_1$ quiver theories, then it is easy to see the generalization. A slice in the affine Grassmannian of the group $G$ is given by the Coulomb branch of a framed Dynkin diagram of type $G$. 

It is useful to introduce a simple characterisation of a moduli space by an integer number, for example the dimension or the number of leaves. A new measure of the simplicity of a moduli space is given by the following integer number. 
For a generic moduli space $\mathcal{M}$ with a Hasse diagram $\mathfrak{H}$ one can define the \emph{disposition}\footnote{We propose the name disposition to mean the tendency of the theory to gain more massless states in several inequivalent ways.} $\mathfrak{D}(n)$ of a node $n\in\mathfrak{H}$ as the number of minimal degenerations of the corresponding leaf,
\begin{equation}
\label{eq:disposition}
    \mathfrak{D}(n)=\textnormal{ number of minimal degenerations of n.}
\end{equation}
One can further define the disposition $\mathfrak{D}(\mathcal{M})$ of a moduli space as the maximal disposition over all nodes in its Hasse diagram,
\begin{equation}
\label{eq:disposition1}
    \mathfrak{D}(\mathcal{M})=\max_{n\in\mathfrak{H}}\left(\mathfrak{D}(n)\right)\,.
\end{equation}
One would say that a moduli space with a smaller disposition is simpler than a moduli space with a larger disposition. The disposition of the affine Grassmannian is the rank of the group. The disposition of a slice in the affine Grassmannian is hence bounded from above by the rank of $G$. The disposition of the Coulomb branch or Higgs branch of SQCD is $1$; the lowest possible. The disposition of the entire moduli space of SQCD is $2$. In general it is much easier to obtain and generalise Hasse diagrams with a small disposition. However, Hasse diagrams for slices in the affine Grassmannian can be well understood due to their embedding into the coweight lattice, even when the disposition is very large.

Another important integer number is the minimal number of generators of the chiral ring, call it $g$. A striking feature of any slice in the affine Grassmannian of a group $G$ is the fact that $g/\mathrm{dim}(G)$ is again an integer number.

\afterpage{
\begin{landscape}
\begin{table}[]
    \centering
           \vspace*{-2cm}\begin{tabular}{c|c|c} \toprule 
Slice & Framed quiver & Unframed quiver  \\  \midrule            
               $a_n$ & \raisebox{-.5\height}{\begin{tikzpicture}[xscale=.7,yscale=.7]
    \node[gauge,label=below:{1}] (0) at (0,0) {};
    \node[gauge,label=below:{1}] (1) at (1,0) {};
    \node (2) at (2,0) {$\cdots$};
    \node[gauge,label=below:{1}] (3) at (3,0) {};
    \node[gauge,label=below:{1}] (4) at (4,0) {};
    \node[flavour,label=above:{1}] (0a) at (0,1) {};
    \node[flavour,label=above:{1}] (4a) at (4,1) {};
    \draw (0a)--(0)--(1)--(2)--(3)--(4)--(4a);
\end{tikzpicture}}   & \raisebox{-.5\height}{\begin{tikzpicture}[xscale=.7,yscale=.7]
    \node[gauge,label=below:{1}] (0) at (0,0) {};
    \node[gauge,label=below:{1}] (1) at (1,0) {};
    \node (2) at (2,0) {$\cdots$};
    \node[gauge,label=below:{1}] (3) at (3,0) {};
    \node[gauge,label=below:{1}] (4) at (4,0) {};
    \node[gauge,label=above:{1}] (0a) at (2,1) {};
    \draw (0a)--(0)--(1)--(2)--(3)--(4)--(0a);
\end{tikzpicture}} \\
        $b_n$ &  \raisebox{-.5\height}{\begin{tikzpicture}[xscale=.7,yscale=.7]
    \node[gauge,label=below:{1}] (0) at (0,0) {};
    \node[gauge,label=below:{2}] (1) at (1,0) {};
    \node (2) at (2,0) {$\cdots$};
    \node[gauge,label=below:{2}] (3) at (3,0) {};
    \node[gauge,label=below:{1}] (4) at (4,0) {};
    \node[flavour,label=above:{1}] (1a) at (1,1) {};
    \draw (1)--(1a);
    \draw (0)--(1)--(2)--(3) (3.4,0.2)--(3.6,0)--(3.4,-0.2);
    \draw[transform canvas={yshift=-1pt}] (3)--(4);
    \draw[transform canvas={yshift=1pt}] (3)--(4);
\end{tikzpicture}} & \raisebox{-.5\height}{\begin{tikzpicture}[xscale=.7,yscale=.7]
    \node[gauge,label=below:{1}] (0) at (0,0) {};
    \node[gauge,label=below:{2}] (1) at (1,0) {};
    \node (2) at (2,0) {$\cdots$};
    \node[gauge,label=below:{2}] (3) at (3,0) {};
    \node[gauge,label=below:{1}] (4) at (4,0) {};
    \node[gauge,label=above:{1}] (1a) at (1,1) {};
    \draw (1)--(1a);
    \draw (0)--(1)--(2)--(3) (3.4,0.2)--(3.6,0)--(3.4,-0.2);
    \draw[transform canvas={yshift=-1pt}] (3)--(4);
    \draw[transform canvas={yshift=1pt}] (3)--(4);
\end{tikzpicture}} \\
        $c_n$ & \raisebox{-.5\height}{\begin{tikzpicture}[xscale=.7,yscale=.7]
    \node[gauge,label=below:{1}] (0) at (0,0) {};
    \node[gauge,label=below:{1}] (1) at (1,0) {};
    \node (2) at (2,0) {$\cdots$};
    \node[gauge,label=below:{1}] (3) at (3,0) {};
    \node[gauge,label=below:{1}] (4) at (4,0) {};
    \node[flavour,label=above:{1}] (0a) at (0,1) {};
    \draw (0)--(0a);
    \draw (0)--(1)--(2)--(3) (3.6,0.2)--(3.4,0)--(3.6,-0.2);
    \draw[transform canvas={yshift=-1pt}] (3)--(4);
    \draw[transform canvas={yshift=1pt}] (3)--(4);
\end{tikzpicture}} & \raisebox{-.5\height}{\begin{tikzpicture}[xscale=.7,yscale=.7]
    \node[gauge,label=below:{1}] (0) at (0,0) {};
    \node[gauge,label=below:{1}] (1) at (1,0) {};
    \node (2) at (2,0) {$\cdots$};
    \node[gauge,label=below:{1}] (3) at (3,0) {};
    \node[gauge,label=below:{1}] (4) at (4,0) {};
    \node[gauge,label=above:{1}] (0a) at (0,1) {};
    \draw (0)--(1)--(2)--(3) (3.6,0.2)--(3.4,0)--(3.6,-0.2);
    \draw[transform canvas={yshift=-1pt}] (3)--(4);
    \draw[transform canvas={yshift=1pt}] (3)--(4);
    \draw[transform canvas={xshift=-1pt}] (0)--(0a);
    \draw[transform canvas={xshift=1pt}] (0)--(0a);
    \draw (-.2,.6)--(0,0.4)--(.2,.6);
\end{tikzpicture}} \\
        $d_n$ &  \raisebox{-.5\height}{\begin{tikzpicture}[xscale=.7,yscale=.7]
    \node[gauge,label=below:{1}] (0) at (0,0) {};
    \node[gauge,label=below:{2}] (1) at (1,0) {};
    \node (2) at (2,0) {$\cdots$};
    \node[gauge,label=below:{2}] (3) at (3,0) {};
    \node[gauge,label=below:{1}] (4) at (4,0) {};
    \node[gauge,label=above:{1}] (5) at (3,1) {};
    \node[flavour,label=above:{1}] (1a) at (1,1) {};
    \draw (1)--(1a) (3)--(5);
    \draw (0)--(1)--(2)--(3)--(4);
\end{tikzpicture}} &  \raisebox{-.5\height}{\begin{tikzpicture}[xscale=.7,yscale=.7]
    \node[gauge,label=below:{1}] (0) at (0,0) {};
    \node[gauge,label=below:{2}] (1) at (1,0) {};
    \node (2) at (2,0) {$\cdots$};
    \node[gauge,label=below:{2}] (3) at (3,0) {};
    \node[gauge,label=below:{1}] (4) at (4,0) {};
    \node[gauge,label=above:{1}] (5) at (3,1) {};
    \node[gauge,label=above:{1}] (1a) at (1,1) {};
    \draw (1)--(1a) (3)--(5);
    \draw (0)--(1)--(2)--(3)--(4);
\end{tikzpicture}}\\ 
        $e_{6}$ & \raisebox{-.5\height}{\begin{tikzpicture}[xscale=.7,yscale=.7]
    \node[gauge,label=below:{1}] (0) at (0,0) {};
    \node[gauge,label=below:{2}] (1) at (1,0) {};
    \node[gauge,label=below:{3}] (2) at (2,0) {};
    \node[gauge,label=below:{2}] (3) at (3,0) {};
    \node[gauge,label=below:{1}] (4) at (4,0) {};
    \node[gauge,label=right:{2}] (5) at (2,1) {};
    \node[flavour,label=right:{1}] (1a) at (2,2) {};
    \draw (0)--(1)--(2)--(3)--(4) (2)--(5)--(1a);
\end{tikzpicture}}  & \raisebox{-.5\height}{\begin{tikzpicture}[xscale=.7,yscale=.7]
    \node[gauge,label=below:{1}] (0) at (0,0) {};
    \node[gauge,label=below:{2}] (1) at (1,0) {};
    \node[gauge,label=below:{3}] (2) at (2,0) {};
    \node[gauge,label=below:{2}] (3) at (3,0) {};
    \node[gauge,label=below:{1}] (4) at (4,0) {};
    \node[gauge,label=right:{2}] (5) at (2,1) {};
    \node[gauge,label=right:{1}] (1a) at (2,2) {};
    \draw (0)--(1)--(2)--(3)--(4) (2)--(5)--(1a);
\end{tikzpicture}} \\
        $e_{7}$ & \raisebox{-.5\height}{\begin{tikzpicture}[xscale=.7,yscale=.7]
    \node[gauge,label=below:{1}] (0) at (0,0) {};
    \node[gauge,label=below:{2}] (1) at (1,0) {};
    \node[gauge,label=below:{3}] (2) at (2,0) {};
    \node[gauge,label=below:{4}] (3) at (3,0) {};
    \node[gauge,label=above:{2}] (7) at (3,1) {};
    \node[gauge,label=below:{3}] (4) at (4,0) {};
    \node[gauge,label=below:{2}] (5) at (5,0) {};
    \node[flavour,label=above:{1}] (6) at (5,1) {};
    \draw (0)--(1)--(2)--(3)--(4)--(5)--(6) (3)--(7);
\end{tikzpicture}}  &  \raisebox{-.5\height}{\begin{tikzpicture}[xscale=.7,yscale=.7]
    \node[gauge,label=below:{1}] (0) at (0,0) {};
    \node[gauge,label=below:{2}] (1) at (1,0) {};
    \node[gauge,label=below:{3}] (2) at (2,0) {};
    \node[gauge,label=below:{4}] (3) at (3,0) {};
    \node[gauge,label=above:{2}] (7) at (3,1) {};
    \node[gauge,label=below:{3}] (4) at (4,0) {};
    \node[gauge,label=below:{2}] (5) at (5,0) {};
    \node[gauge,label=above:{1}] (6) at (5,1) {};
    \draw (0)--(1)--(2)--(3)--(4)--(5)--(6) (3)--(7);
\end{tikzpicture}}  \\
        $e_{8}$ & \raisebox{-.5\height}{\begin{tikzpicture}[xscale=.7,yscale=.7]
    \node[flavour,label=above:{1}] (0) at (1,1) {};
    \node[gauge,label=below:{2}] (1) at (1,0) {};
    \node[gauge,label=below:{3}] (2) at (2,0) {};
    \node[gauge,label=below:{4}] (3) at (3,0) {};
    \node[gauge,label=below:{5}] (4) at (4,0) {};
    \node[gauge,label=below:{6}] (5) at (5,0) {};
    \node[gauge,label=below:{4}] (6) at (6,0) {};
    \node[gauge,label=below:{2}] (7) at (7,0) {};
    \node[gauge,label=above:{3}] (8) at (5,1) {};
    \draw (0)--(1)--(2)--(3)--(4)--(5)--(6)--(7) (5)--(8);
\end{tikzpicture}}  &  \raisebox{-.5\height}{\begin{tikzpicture}[xscale=.7,yscale=.7]
    \node[gauge,label=above:{1}] (0) at (1,1) {};
    \node[gauge,label=below:{2}] (1) at (1,0) {};
    \node[gauge,label=below:{3}] (2) at (2,0) {};
    \node[gauge,label=below:{4}] (3) at (3,0) {};
    \node[gauge,label=below:{5}] (4) at (4,0) {};
    \node[gauge,label=below:{6}] (5) at (5,0) {};
    \node[gauge,label=below:{4}] (6) at (6,0) {};
    \node[gauge,label=below:{2}] (7) at (7,0) {};
    \node[gauge,label=above:{3}] (8) at (5,1) {};
    \draw (0)--(1)--(2)--(3)--(4)--(5)--(6)--(7) (5)--(8);
\end{tikzpicture}}  \\ \bottomrule 
    \end{tabular} \hspace{2cm}
    \begin{tabular}{c|c|c} \toprule 
Slice & Framed quiver & Unframed quiver  \\  \midrule  
   $f_4$ & \raisebox{-.5\height}{\begin{tikzpicture}[xscale=.7,yscale=.7]
    \node[gauge,label=below:{1}] (2) at (2,0) {};
    \node[gauge,label=below:{2}] (3) at (3,0) {};
    \node[gauge,label=below:{3}] (4) at (4,0) {};
    \node[gauge,label=below:{2}] (5) at (5,0) {};
    \node[flavour,label=above:{1}] (6) at (5,1) {};
    \draw (2)--(3) (3.6,0.2)--(3.4,0)--(3.6,-0.2) (4)--(5)--(6);
    \draw[transform canvas={yshift=-1pt}] (3)--(4);
    \draw[transform canvas={yshift=1pt}] (3)--(4);
\end{tikzpicture}} & 
 \raisebox{-.5\height}{\begin{tikzpicture}[xscale=.7,yscale=.7]
    \node[gauge,label=below:{1}] (2) at (2,0) {};
    \node[gauge,label=below:{2}] (3) at (3,0) {};
    \node[gauge,label=below:{3}] (4) at (4,0) {};
    \node[gauge,label=below:{2}] (5) at (5,0) {};
    \node[gauge,label=above:{1}] (6) at (5,1) {};
    \draw (2)--(3) (3.6,0.2)--(3.4,0)--(3.6,-0.2) (4)--(5)--(6);
    \draw[transform canvas={yshift=-1pt}] (3)--(4);
    \draw[transform canvas={yshift=1pt}] (3)--(4);
\end{tikzpicture}} \\ 
 $g_2$ & \raisebox{-.5\height}{\begin{tikzpicture}[xscale=.7,yscale=.7]
    \node[gauge,label=below:{1}] (3) at (3,0) {};
    \node[gauge,label=below:{2}] (4) at (4,0) {};
    \node[flavour,label=above:{1}] (5) at (4,1) {};
    \draw[transform canvas={yshift=-2pt}] (3)--(4);
    \draw[transform canvas={yshift=0pt}] (3)--(4);
    \draw[transform canvas={yshift=2pt}] (3)--(4);
    \draw (3.6,0.2)--(3.4,0)--(3.6,-0.2) (4)--(5);
\end{tikzpicture}} & 
 \raisebox{-.5\height}{\begin{tikzpicture}[xscale=.7,yscale=.7]
    \node[gauge,label=below:{1}] (3) at (3,0) {};
    \node[gauge,label=below:{2}] (4) at (4,0) {};
    \node[gauge,label=above:{1}] (5) at (4,1) {};
    \draw[transform canvas={yshift=-2pt}] (3)--(4);
    \draw[transform canvas={yshift=0pt}] (3)--(4);
    \draw[transform canvas={yshift=2pt}] (3)--(4);
    \draw (3.6,0.2)--(3.4,0)--(3.6,-0.2) (4)--(5);
\end{tikzpicture}}  \\
 $ac_n$ & \raisebox{-.5\height}{\begin{tikzpicture}[xscale=.7,yscale=.7]
    \node[flavour,label=above:{1}] (10) at (4,1) {}; 
    \node[gauge,label=below:{1}] (0) at (0,0) {};
    \node[gauge,label=below:{1}] (1) at (1,0) {};
    \node (2) at (2,0) {$\cdots$};
    \node[gauge,label=below:{1}] (3) at (3,0) {};
    \node[gauge,label=below:{1}] (4) at (4,0) {};
    \node[flavour,label=above:{1}] (0a) at (0,1) {};
    \draw (0)--(0a);
    \draw (0)--(1)--(2)--(3) (3.6,0.2)--(3.4,0)--(3.6,-0.2) (4)--(10);
    \draw[transform canvas={yshift=-1pt}] (3)--(4);
    \draw[transform canvas={yshift=1pt}] (3)--(4);
\end{tikzpicture}} & \raisebox{-.5\height}{\begin{tikzpicture}[xscale=.7,yscale=.7]
    \node[gauge,label=below:{1}] (0) at (0,0) {};
    \node[gauge,label=below:{1}] (1) at (1,0) {};
    \node (2) at (2,0) {$\cdots$};
    \node[gauge,label=below:{1}] (3) at (3,0) {};
    \node[gauge,label=below:{1}] (4) at (4,0) {};
    \node[gauge,label=above:{1}] (0a) at (0,1) {};
    \draw (0)--(1)--(2)--(3) (3.6,0.2)--(3.4,0)--(3.6,-0.2);
    \draw (0a) .. controls (4,1) .. (4);
    \draw[transform canvas={yshift=-1pt}] (3)--(4);
    \draw[transform canvas={yshift=1pt}] (3)--(4);
    \draw[transform canvas={xshift=-1pt}] (0)--(0a);
    \draw[transform canvas={xshift=1pt}] (0)--(0a);
    \draw (-.2,.6)--(0,0.4)--(.2,.6);
\end{tikzpicture}} \\
        $ag_2$ & \raisebox{-.5\height}{\begin{tikzpicture}[xscale=.7,yscale=.7]
    \node[gauge,label=below:{1}] (3) at (3,0) {};
    \node[gauge,label=below:{1}] (4) at (4,0) {};
    \node[flavour,label=above:{1}] (5) at (4,1) {};
    \node[flavour,label=above:{1}] (2) at (3,1) {};
    \draw[transform canvas={yshift=-2pt}] (3)--(4);
    \draw[transform canvas={yshift=0pt}] (3)--(4);
    \draw[transform canvas={yshift=2pt}] (3)--(4);
    \draw (3.6,0.2)--(3.4,0)--(3.6,-0.2) (4)--(5) (2)--(3);
\end{tikzpicture}} & 
 \raisebox{-.5\height}{\begin{tikzpicture}[xscale=.7,yscale=.7]
    \node[gauge,label=below:{1}] (3) at (3,0) {};
    \node[gauge,label=below:{1}] (4) at (4,0) {};
    \node[gauge,label=above:{1}] (5) at (3,1) {};
    \draw[transform canvas={yshift=-2pt}] (3)--(4);
    \draw[transform canvas={yshift=0pt}] (3)--(4);
    \draw[transform canvas={yshift=2pt}] (3)--(4);
    \draw[transform canvas={xshift=-2pt}] (3)--(5);
    \draw[transform canvas={xshift=0pt}] (3)--(5);
    \draw[transform canvas={xshift=2pt}] (3)--(5);
    \draw (3.6,0.2)--(3.4,0)--(3.6,-0.2) (4)--(5);
    \draw (2.8,.6)--(3,0.4)--(3.2,.6);
\end{tikzpicture}} \\
        $cg_2$ &\raisebox{-.5\height}{\begin{tikzpicture}[xscale=.7,yscale=.7]
    \node[gauge,label=below:{1}] (3) at (3,0) {};
    \node[gauge,label=below:{1}] (4) at (4,0) {};
    \node[flavour,label=above:{1}] (2) at (3,1) {};
    \draw[transform canvas={yshift=-2pt}] (3)--(4);
    \draw[transform canvas={yshift=0pt}] (3)--(4);
    \draw[transform canvas={yshift=2pt}] (3)--(4);
    \draw (3.6,0.2)--(3.4,0)--(3.6,-0.2) (2)--(3);
\end{tikzpicture}} & 
 \raisebox{-.5\height}{\begin{tikzpicture}[xscale=.7,yscale=.7]
    \node[gauge,label=below:{1}] (3) at (3,0) {};
    \node[gauge,label=below:{1}] (4) at (4,0) {};
    \node[gauge,label=above:{1}] (5) at (3,1) {};
    \draw[transform canvas={yshift=-2pt}] (3)--(4);
    \draw[transform canvas={yshift=0pt}] (3)--(4);
    \draw[transform canvas={yshift=2pt}] (3)--(4);
    \draw[transform canvas={xshift=-2pt}] (3)--(5);
    \draw[transform canvas={xshift=0pt}] (3)--(5);
    \draw[transform canvas={xshift=2pt}] (3)--(5);
    \draw (3.6,0.2)--(3.4,0)--(3.6,-0.2);
    \draw (2.8,.6)--(3,0.4)--(3.2,.6);
\end{tikzpicture}} \\
        $h_{n,k}$ & \raisebox{-.5\height}{\begin{tikzpicture}[xscale=.7,yscale=.7]
    \node[gauge,label=below:{1}] (0) at (0,0) {};
    \node[gauge,label=below:{1}] (1) at (1,0) {};
    \node (2) at (2,0) {$\cdots$};
    \node[gauge,label=below:{1}] (3) at (3,0) {};
    \node[gauge,label=below:{1}] (4) at (4,0) {};
    \node[flavour,label=above:{1}] (0a) at (0,1) {};
    \draw (0)--(0a);
    \draw (0)--(1)--(2)--(3) (3.6,0.2)--(3.4,0)--(3.6,-0.2);
    \draw[transform canvas={yshift=-2pt}] (3)--(4);
    \draw[transform canvas={yshift=0pt}] (3)--(4);
    \draw[transform canvas={yshift=2pt}] (3)--(4);
    \node at (3.5,.5) {$k$};
\end{tikzpicture}} & \raisebox{-.5\height}{\begin{tikzpicture}[xscale=.7,yscale=.7]
    \node[gauge,label=below:{1}] (0) at (0,0) {};
    \node[gauge,label=below:{1}] (1) at (1,0) {};
    \node (2) at (2,0) {$\cdots$};
    \node[gauge,label=below:{1}] (3) at (3,0) {};
    \node[gauge,label=below:{1}] (4) at (4,0) {};
    \node[gauge,label=above:{1}] (0a) at (0,1) {};
    \draw (0)--(1)--(2)--(3) (3.6,0.2)--(3.4,0)--(3.6,-0.2);
    \draw[transform canvas={yshift=-2pt}] (3)--(4);
    \draw[transform canvas={yshift=0pt}] (3)--(4);
    \draw[transform canvas={yshift=2pt}] (3)--(4);
    \draw[transform canvas={xshift=-2pt}] (0)--(0a);
    \draw[transform canvas={xshift=0pt}] (0)--(0a);
    \draw[transform canvas={xshift=2pt}] (0)--(0a);
    \draw (-.2,.6)--(0,0.4)--(.2,.6);
     \node at (3.5,.5) {$k$};
     \node at (-.5,.5) {$k$};
\end{tikzpicture}} \\
        $\overline{h}_{n,k}$ & \raisebox{-.5\height}{\begin{tikzpicture}[xscale=.7,yscale=.7]
    \node[flavour,label=above:{1}] (10) at (4,1) {}; 
    \node[gauge,label=below:{1}] (0) at (0,0) {};
    \node[gauge,label=below:{1}] (1) at (1,0) {};
    \node (2) at (2,0) {$\cdots$};
    \node[gauge,label=below:{1}] (3) at (3,0) {};
    \node[gauge,label=below:{1}] (4) at (4,0) {};
    \node[flavour,label=above:{1}] (0a) at (0,1) {};
    \draw (0)--(0a);
    \draw (0)--(1)--(2)--(3) (3.6,0.2)--(3.4,0)--(3.6,-0.2) (4)--(10);
    \draw[transform canvas={yshift=-2pt}] (3)--(4);
    \draw[transform canvas={yshift=0pt}] (3)--(4);
    \draw[transform canvas={yshift=2pt}] (3)--(4);
     \node at (3.5,.5) {$k$};
\end{tikzpicture}} & \raisebox{-.5\height}{\begin{tikzpicture}[xscale=.7,yscale=.7]
    \node[gauge,label=below:{1}] (0) at (0,0) {};
    \node[gauge,label=below:{1}] (1) at (1,0) {};
    \node (2) at (2,0) {$\cdots$};
    \node[gauge,label=below:{1}] (3) at (3,0) {};
    \node[gauge,label=below:{1}] (4) at (4,0) {};
    \node[gauge,label=above:{1}] (0a) at (0,1) {};
    \draw (0)--(1)--(2)--(3) (3.6,0.2)--(3.4,0)--(3.6,-0.2);
    \draw (0a) .. controls (4,1) .. (4);
    \draw[transform canvas={yshift=-2pt}] (3)--(4);
    \draw[transform canvas={yshift=0pt}] (3)--(4);
    \draw[transform canvas={yshift=2pt}] (3)--(4);
    \draw[transform canvas={xshift=-2pt}] (0)--(0a);
    \draw[transform canvas={xshift=0pt}] (0)--(0a);
    \draw[transform canvas={xshift=2pt}] (0)--(0a);
    \draw (-.2,.6)--(0,0.4)--(.2,.6);
     \node at (3.5,.5) {$k$};
     \node at (-.5,.5) {$k$};
\end{tikzpicture}}  \\      
   $A_n$ & \raisebox{-.5\height}{\begin{tikzpicture}[xscale=.7,yscale=.7]
    \node[flavour,label=right:{$n+1$}] (1) at (0,1) {}; 
    \node[gauge,label=right:{1}] (0) at (0,0) {};
    \draw (0)--(1);
\end{tikzpicture}} & \raisebox{-.5\height}{\begin{tikzpicture}[xscale=.7,yscale=.7]
    \node[gauge,label=right:{1}] (1) at (0,1) {}; 
    \node[gauge,label=right:{1}] (0) at (0,0) {};
    \draw[transform canvas={xshift=2pt}] (0)--(1);
    \draw[transform canvas={xshift=0pt}] (0)--(1);
    \draw[transform canvas={xshift=-2pt}] (0)--(1);
     \node at (-.8,.5) {$n+1$};
\end{tikzpicture}}  \\ \bottomrule   
    \end{tabular}
    \caption{Most up-to-date, but incomplete list of unitary quivers without loops for elementary slices usable in the quiver subtraction algorithm. In each case we provide two quivers, a framed version and an equivalent unframed version, where a $\mathrm{U}(1)$ should be ungauged on the long node. For $a_n$, $b_n$, $c_n$, $d_n$, $ac_n$, $h_{n,k}$ and $\bar{h}_{n,k}$ there are $n$ gauge nodes in the framed quiver and $n+1$ gauge nodes in the unframed quiver. Notice that $h_{n,1}=\mathbb{H}^n$, $h_{n,2}=c_n$, $h_{2,3}=cg_2$, $\overline{h}_{n,1}=a_n$, $\overline{h}_{n,2}=ac_n$, and $\overline{h}_{2,3}=ag_2$.}
    \label{tab:elem}
\end{table}
\end{landscape}}

One should further make a distinction between finite dimensional Lie groups and infinite dimensional generalizations. By analogy with the finite dimensional case, one would identify (the Coulomb branch of) a framed affine quiver with a slice in the affine Grassmannian of the affine Lie group. Naturally, the structure of the moduli space, and consequently its Hasse diagram, become significantly more complex. The disposition for a generic slice in the affine Grassmannian of an affine group appears not to be bounded from above. Consequently, if the group is infinite dimensional one expects a much more complicated moduli space than that of a finite dimensional group. Prominent examples of slices in the affine Grassmannian of affine groups are moduli spaces of instantons. While they are well studied spaces in physics, we see that they are much more challenging objects than slices in the affine Grassmannian of finite groups.

The present paper focuses on slices in the affine Grassmannian of a finite dimensional Lie group, which in this sense are simpler than moduli spaces of instantons. This is reflected in several ways. Many of the slices are complete intersections. Many of these are closures of nilpotent orbits. The generators of the chiral ring are simple and, as argued above, the Hasse diagram for slices of algebras of finite type is simpler than the Hasse diagram for the moduli space of instantons. These features make these moduli spaces more tractable objects and, hence, simpler to study.

In the long run, the lessons learned from the affine Grassmannian of finite dimensional groups will hopefully help us to tackle more difficult moduli spaces.

A nice by-product of studying the affine Grassmannian is the encounter of the so-called \emph{quasi-minimal singularities} introduced in \cite{2003math:5095M}, which are elementary slices that do not appear in the study of nilpotent orbits. This is an important update to our arsenal of quivers to be used in quiver subtraction. We display all unitary quivers without loops which we know to be elementary slices in Table \ref{tab:elem}. The classification of these slices is still an open problem, and every new addition is hence very exciting.

\paragraph{Plan of the paper.}
In Section \ref{sec:affGr}, we provide a short review of the construction of the affine Grassmannian, and put forward the important concepts of orbits, stratification and transverse slices. In Section \ref{sec:quivers}, we construct quivers for the transverse slices and compute them explicitly for groups of rank 1 and 2, along with the Hasse diagrams. This allows to identify new quivers for the quasi-minimal singularities. We show how this construction can be reproduced using brane setups with orientifolds for the classical groups in Section \ref{sec:branes}, and for any group using the new algorithm of quiver addition in Section \ref{sec:quiveradd}. Finally, we end with an analysis of the generators of the infinite dimensional transverse slices using Hilbert series in Section \ref{sec:Hilbert}. An appendix reviews allowed configurations of branes in the presence of ON planes. 

\section{The affine Grassmannian}
\label{sec:affGr}

In this section, we give the definitions that apply to the discussions in this paper. Our notations are summarized at the end of this section in Table \ref{tab:notation}. 

Our description tries to avoid excessive technicalities as it is aimed at physicists primarily. For instance, we describe the affine Grassmannian as an infinite dimensional variety and not as an ind-scheme; we also identify schemes with their underlying topological space. For a more formal treatment, we refer to  \cite{segal1985loop,beauville1994conformal,gortz2010affine,2016arXiv160305593Z,baumann2018notes,RicharzLec}. A summary of the main properties of the affine Grassmannian is presented in \cite[Section 2]{braverman2010pursuing}. A very explicit treatment with examples is available in  \cite{AcharLec,brunson2014matrix}. For an introduction in the context of gauge theory, see \cite{Kapustin:2006pk,Witten:2009mh}, and in the context of conformal field theories see \cite{Frenkel:2005pa}.

\subsection{Formal power series}

We define three important structures that underlie all the construction of the affine Grassmannian, reviewed below. We use throughout a formal variable $t$. 
\begin{itemize}
    \item First, we have the ring of formal power series in $t$, denoted $\mathbb{C}[[t]]$.  
Here formal means that we do not worry about convergence issues. For examples of elements of $\mathbb{C}[[t]]$, let us mention the polynomials in $t$, the rational functions like $\frac{1}{1-t} = \sum_{i=0}^{\infty} t^i$, transcendental functions like $e^t = \sum_{i=0}^{\infty} t^i / i!$, or non-convergent series like $\sum_{i=0}^{\infty} t^i  i!$. Note that the important condition is that an element of $\mathbb{C}[[t]]$ is a formal series in $t$ with only non-negative powers of $t$. Geometrically, $\mathbb{C}[[t]]$ can be seen as the ring of functions on the unit complex disk $D = \{ z \in \mathbb{C} \mid |z| < 1 \}$. Equivalently, the unit disk $D$ is the spectrum of $\mathbb{C}[[t]]$. 
\item The second object we consider is the ring of polynomials in $t^{-1}$, denoted $\mathbb{C}[t^{-1}]$. 
We draw the attention of the reader on two differences between $\mathbb{C}[t^{-1}]$ and $\mathbb{C}[[t]]$: in $\mathbb{C}[t^{-1}]$ the powers of $t$ are non-positive, and the series expansion needs to terminate (a polynomial has a finite degree). 
\item Finally, we need the field of fractions of $\mathbb{C}[[t]]$, denoted $\mathbb{C}((t))$. 
This can be defined as the set of all fractions $f(t)/g(t)$ for $f(t) , g(t) \in \mathbb{C}[[t]]$ with $g(t) \neq 0$. Equivalently, this is the set of formal series that can be written in the form $\sum_{i=N}^{\infty} a_i t^i$ for some $N \in \mathbb{Z}$ and $a_i \in \mathbb{C}$. The spectrum of $\mathbb{C}((t))$ is the punctured disk $D^{\ast} = \{ z \in \mathbb{C} \mid  0 < |z| < 1 \}$. 
\end{itemize}
The ring $\mathbb{C}[[t]]$ is a \emph{discrete valuation ring}, where the valuation is given by the degree of the lowest non-zero monomial. More generally for 
\begin{equation}
\label{valuation}
   f = \sum\limits_{i=N}^{\infty} f_i t^i \in \mathbb{C}((t)) 
\end{equation}
with $N \in \mathbb{Z}$ and $f_N \neq 0$, the valuation is $\nu (f) = N$. 
As the name indicates, $\mathbb{C}((t))$ is a field, so every non-zero element is invertible. 
We remark that $\mathbb{C}[[t]]$ is the ring of integers of $\mathbb{C}((t))$, and that the multiplicative group $\mathbb{C}[[t]]^{\ast}$ of invertible elements in $\mathbb{C}[[t]]$ is 
\begin{equation}
    \mathbb{C}[[t]]^{\ast} = \left\{ \sum\limits_{i=0}^{\infty} a_i t^i \mid a_i \in \mathbb{C} \textrm{  and  } a_0 \neq 0 \right\} \, . 
\end{equation}
Finally, the only invertible elements in $\mathbb{C}[t^{-1}]$ are the non-zero constants.

\subsection{Groups valued in formal power series}

\subsubsection*{The groups $G((t))$}
For any characteristic 0 field $K$, $\mathrm{GL}(n,K)$ is the group of invertible $n \times n$ matrices with entries in $K$. As mentioned in the previous subsection, $\mathbb{C}((t))$ is a field and we can use it to consider the group $\mathrm{GL}(n,\mathbb{C}((t)))$ of invertible $n \times n$ matrices with entries in $\mathbb{C}((t))$. Such a matrix is invertible if and only if its determinant is not identically zero. More generally, let $G$ be an algebraic subgroup of $\mathrm{GL}(n,\mathbb{C})$, which means that 
\begin{equation}
    G = \{ M \in \mathrm{GL}(n , \mathbb{C}) \mid P_j (M) = 0 \}
\end{equation}
for a certain collection of polynomials $P_j$, $j =1,...,J$. This includes the classical groups $\mathrm{SL}(n,\mathbb{C})$, $\mathrm{SO}(n,\mathbb{C})$, $\mathrm{Sp}(2n,\mathbb{C})$,\footnote{Note that $\mathrm{Sp}(2n,\mathbb{C})$ denotes the symplectic group of rank $n$ over the complex numbers.} and also the exceptional groups. We define 
\begin{equation}
    G((t)) := \{ M \in \mathrm{GL}(n , \mathbb{C}((t))) \mid P_j (M) = 0 \} \, . 
\end{equation}

\paragraph{Example.} 
For $G = \mathrm{SL}(2,\mathbb{C})$, we have  
\begin{equation}
\label{SL2((t))}
    G((t)) := \left\{ \left( \begin{array}{cc}
        a(t) & b(t) \\
        c(t) & d(t)
    \end{array} \right) \mid a(t),b(t),c(t),d(t) \in \mathbb{C}((t)) \textrm{  and  } a(t)d(t)-b(t)c(t) = 1 \right\} \, . 
\end{equation}

\subsubsection*{The groups $G[[t]]$ and $G[t^{-1}]$}
We can mimic this construction with $\mathbb{C}((t))$ replaced with $\mathbb{C}[[t]]$. One can consider the group $\mathrm{GL}(n,\mathbb{C}[[t]])$ of matrices with coefficients in $\mathbb{C}[[t]]$ and which are invertible in $\mathbb{C}[[t]]$. Note that a necessary and sufficient condition for this to be the case is that the determinant should belong to $\mathbb{C}[[t]]^{\ast}$, i.e. have a series expansion with non-zero coefficient of degree 0. Up to this subtlety, the definition of $G[[t]]$ is exactly as before: 
\begin{equation}
    G[[t]] := \{ M \in \mathrm{GL}(n , \mathbb{C}[[t]]) \mid P_j (M) = 0 \} \, . 
\end{equation}
Similarly, one can construct $G[t^{-1}]$, bearing in mind that by the remark above $\mathrm{GL}(n,\mathbb{C}[t^{-1}])$ is characterized by the property that the determinant is a non-vanishing constant.

\subsubsection*{The groups $G_1[t^{-1}]$}
Finally, we define 
\begin{equation}
    G_1[t^{-1}] := \{ M \in  G[t^{-1}] \mid M|_{t \rightarrow \infty} = \mathbf{1} \} \, . 
\end{equation}
In other words, $G_1[t^{-1}]$ is the subgroup of $G[t^{-1}]$ where the degree 0 term in the series expansion of the elements are the identity matrix.

\subsection{\texorpdfstring{Lattices and the affine Grassmannian}{Lattices and the affine Grassmannian}}

In this section, we finally give the definition of the affine Grassmannian for the group $G$.  This involves spaces of lattices, so we begin as a warm-up with a brief reminder about lattices. 

Consider a vector space $V$ of dimension $n$. A \emph{lattice} in $V$ is a discrete subgroup of $V$ which is isomorphic to $\mathbb{Z}^n$. One way to construct a lattice explicitly is as follows: pick a basis $(v_1 , \dots , v_n)$ of $V$, and consider the set of all linear combinations with integer coefficients: 
\begin{equation}
    \mathcal{L} = \left\{ \sum\limits_{i=1}^n a_i v_i \mid a_i \in \mathbb{Z} \right\} \, . 
\end{equation}
So a basis of $V$ defines a unique lattice $\mathcal{L} \subset V$, but the converse is not true: many different bases give the same lattice. Two bases correspond to the same lattice if and only if one is obtained from the other by multiplication by a matrix in $ \mathrm{GL}(n,\mathbb{Z})$, the set of invertible matrices with integer coefficients and whose inverse also has integer coefficients. On the other hand, the set of all bases of $V$ can be identified with the set $ \mathrm{GL}(n,\mathbb{R})$, once a reference basis is picked: one just writes the components of the basis vectors as \emph{columns} of the matrix. We now turn to the crucial point of the argument: we have chosen to write the basis vectors as columns of the matrix $M \in  \mathrm{GL}(n,\mathbb{R})$; then for any $P \in  \mathrm{GL}(n,\mathbb{Z})$, the matrix $MP$ corresponds to the same lattice, as the columns of $MP$ are linear combinations with integer coefficients of the columns of $M$. However $PM$ does not a priori give the same lattice. Therefore the set of inequivalent lattices is
\begin{equation}
    \textrm{Set of inequivalent lattices in } V =  \mathrm{GL}(n,\mathbb{R}) /  \mathrm{GL}(n,\mathbb{Z}) \, , 
\end{equation}
where the quotient is taken on the \emph{right}. Multiplication on the \emph{left} gives instead an action of $ \mathrm{GL}(n,\mathbb{Z})$ on the set of lattices. In the context of the affine Grassmannian, it is this action which generates the orbits we are interested in. 

We now repeat the above discussion, but replace the base field $\mathbb{R}$ by $\mathbb{C}((t))$ and the integers $\mathbb{Z}$ by $\mathbb{C}[[t]]$. A \emph{lattice} $\mathcal{L}$ in $V = \mathbb{C}((t))^n$ is a free $\mathbb{C}[[t]]$-submodule of rank $n$. 
A lattice is fully specified by a basis, in other words an $n \times n$ matrix of elements of $\mathbb{C}((t))$ with non-zero determinant, or equivalently a matrix in $ \mathrm{GL}(n,\mathbb{C}((t)))$. We identify the basis vectors with the \emph{columns} of the matrix. Multiplying such a matrix on the \emph{right} by a matrix in $ \mathrm{GL}(n,\mathbb{C}[[t]])$ gives a matrix representing the same lattice. On the other hand, multiplication on the \emph{left} by a matrices in $ \mathrm{GL}(n,\mathbb{C}[[t]])$ gives a family of possibly inequivalent lattices that we call the \emph{orbit} of the initial lattice. 
We can identify the set of lattices as the set of equivalence classes of elements of $\mathrm{GL}(n,\mathbb{C}((t)))$ with equivalence relation given by multiplication on the right by $\mathrm{GL}(n,\mathbb{C}[[t]])$.

\paragraph{Definition.}
The affine Grassmannian $\mathrm{Gr}_{G}$ for the group $G=\mathrm{GL}(n,\mathbb{C})$ is, as a set, the set of all lattices in $\mathbb{C}((t))^n$. By the remark above, we have 
\begin{equation}
    \mathrm{Gr}_{\mathrm{GL}(n,\mathbb{C})} \simeq  \mathrm{GL}(n,\mathbb{C}((t))) / \mathrm{GL}(n,\mathbb{C}[[t]]) \, . 
\end{equation}
More generally, for a group $G$ we define 
\begin{equation}
    \mathrm{Gr}_{G} \simeq  G((t)) / G[[t]] \, . 
\end{equation}
This can be seen as an infinite dimensional variety. 
The group $G[[t]]$ acts on the left on the points of the affine Grassmannian $\mathrm{Gr}_{G}$, thus generating \emph{orbits}, which are Zariski open algebraic sets. The closure of the orbits are finite dimensional (generically singular) projective varieties.   

\subsection{\texorpdfstring{Explicit description of the $\mathrm{SL}(2,\mathbb{C})$ and $\mathrm{PSL}(2,\mathbb{C})$ affine Grassmannians}{Explicit description of the A1 affine Grassmannians}}

In this subsection, we digress from the general discussion to illustrate it with explicit computations at rank 1. Some of these computations can be found in \cite[Proposition 2.6]{beauville1994conformal}.  

\subsubsection*{Points of $\mathrm{Gr}_{\mathrm{SL}(2,\mathbb{C})}$}

In this paragraph, we set $G = \mathrm{SL}(2,\mathbb{C})$. 
The group $G((t))$ has been explicitly written in (\ref{SL2((t))}). A matrix in this set represents a lattice as: 
\begin{equation}
  \Lambda \left[ \left( \begin{array}{cc}
        a  & b  \\
        c & d 
    \end{array} \right) \right] = \left\{u (t)   \left( \begin{array}{c}
        a(t) \\
        c(t) 
    \end{array} \right) +   v (t) \left( \begin{array}{c}
        b(t) \\
        d(t) 
    \end{array} \right) \bigm\vert  u ,  v \in G[[t]] \right\} \, . 
\end{equation}
For any matrix $R \in G[[t]]$, it is clear with this description that 
\begin{equation}
  \Lambda \left[ \left( \begin{array}{cc}
        a  & b  \\
        c & d 
    \end{array} \right) \right] =  \Lambda \left[ \left( \begin{array}{cc}
        a  & b  \\
        c & d 
    \end{array} \right) R \right] \, . 
\end{equation}
The points in $\mathrm{Gr}_{G}$ are equivalence classes of  matrices 
\begin{equation}
  M =  \left( \begin{array}{cc}
        a  & b  \\
        c & d 
    \end{array} \right) \in  G((t))  \qquad \textrm{under}  \qquad M  \sim M R \, , \quad R =  \left( \begin{array}{cc}
        \alpha  & \beta  \\
        \gamma & \delta 
    \end{array} \right) \in G[[t]] \, . 
\end{equation}
This relation is non-trivial because $R$ is restricted to lie in $G[[t]]$, so $\alpha(t)$, $\beta(t)$, $\gamma(t)$, and $\delta(t)$ cannot have negative powers of $t$, while $a(t)$, $b(t)$, $c(t)$, and $d(t)$ may have negative powers. We can see this in action explicitly by trying to implement Gaussian elimination. Let us assume for definiteness that the most singular coefficient of $M$ (the coefficient with lowest valuation) is $d$, and set $\lambda = - 2\nu(d) \geq 0$ (recall that the valuation was defined below equation (\ref{valuation})). We can write $d = t^{- \lambda /2} d_0$ where $\nu (d_0) = 0$. The first step of Gaussian elimination uses 
\begin{equation}
    R_1 = \left( \begin{array}{cc}
        d_0  & 0  \\
        - c t^{\lambda /2} & d_0^{-1} 
    \end{array} \right) \in G[[t]] \quad \textrm{giving} \quad  MR_1 = \left( \begin{array}{cc}
        t^{\lambda /2 }  & b d_0^{-1}    \\
        0 & t^{-\lambda /2} 
    \end{array} \right) \, . 
\end{equation}
For the second step one can write 
\begin{equation}
    b d_0^{-1}  = \left( \sum\limits_{i=\nu (b)}^{\lambda/2 -1} x_i t^i  \right) + t^{\lambda/2}  b'
\end{equation}
where $\nu(b') \geq 0$, so that we obtain
\begin{equation}
\label{explicitMRR}
    R_2 = \left( \begin{array}{cc}
        1  & -b'  \\
        0 & 1 
    \end{array} \right) \in G[[t]] \quad \textrm{giving} \quad  MR_1R_2 = \left( \begin{array}{cc}
        t^{\lambda/2}  & \sum\limits_{i=- \lambda/2}^{\lambda/2 -1} x_i t^i    \\
        0 & t^{-\lambda/2} 
    \end{array} \right) \, . 
\end{equation}
The points in $\mathrm{Gr}_G$ are parametrized by $\lambda \in 2 \mathbb{N}$, and then for a given $\lambda$ by $\lambda$ complex numbers $ (x_{- \lambda /2 } , x_{- \lambda /2  +1} , \dots , x_{\lambda/2  -1} )$. 

\subsubsection*{Orbits}
We now describe the orbits under the right action of $G[[t]]$. From the point of view of Gaussian elimination, this means we can now combine rows of the matrix $M$ together. It is clear from the explicit representative (\ref{explicitMRR}) that all the points of $\mathrm{Gr}_G$ corresponding to the same $\lambda \in 2 \mathbb{N}$ form a single orbit, and conversely points corresponding to $\lambda \neq \lambda '$ belong to different orbits. Therefore, the orbits, which we denote by $\gr{G}{\lambda}{}$, are labeled by $\lambda \in 2 \mathbb{N}$, and can be defined by a distinguished element $M_{\lambda}$: 
\begin{equation}
\label{Mlambda}
    \gr{G}{\lambda}{} = G[[t]] \cdot M_{\lambda} \, , \qquad M_{\lambda} = \left( \begin{array}{cc}
        t^{\lambda/2}  & 0  \\
        0 &   t^{-\lambda/2}
    \end{array} \right) \, . 
\end{equation}
Elements of $2 \mathbb{N}$ can be seen as the positive coweights of $G$. 

Consider an orbit $\gr{G}{\lambda}{}$. For every $u \in \mathbb{C}^{\ast}$, we have 
\begin{equation}
\label{eqLimitMat}
     \left( \begin{array}{cc}
        0  & u  \\
        -u^{-1} &  t u^{-1}  
    \end{array} \right) M_{\lambda}\left( \begin{array}{cc}
        u^{-1}  & 0 \\
          u^{-1} t^{\lambda -1}  &   u
    \end{array} \right) = \left( \begin{array}{cc}
        t^{\lambda/2 -1}  & u^2 t^{-\lambda/2}  \\
        0 &   t^{-\lambda/2 +1 }
    \end{array} \right) 
\end{equation}
which when $u \rightarrow 0$ reaches the orbit $\gr{G}{\lambda -2}{}$. This shows that the orbit $\gr{G}{\lambda}{}$ is open and contains  $\gr{G}{\lambda -2}{}$ in its closure. Therefore there is a partial order in the orbits defined by closure inclusions. The difference between the coweights labeling two adjacent orbits, $\lambda - (\lambda -2) = 2$, can be interpreted as the positive coroot in $G$.

\subsubsection*{Points and orbits of $\mathrm{Gr}_{\mathrm{PSL}(2,\mathbb{C})}$}
\begin{figure}[t]
    \centering
\begin{tikzpicture} 
\node (00) [bd] at (4,0) {};
\node (02) [bd] at (4,2) {};
\node (04) [bd] at (4,4) {};
\node (06) [bd] at (4,6) {};
\draw (00) edge (02);
\draw (02) edge (04);
\draw (04) edge (06);
\draw (06) edge [dotted] (4,7.5);
\node at (5,0) {$\lambda = 0$};
\node at (5,2) {$\lambda = 2$};
\node at (5,4) {$\lambda = 4$};
\node at (5,6) {$\lambda = 6$};
\node at (4,-2) {Component $I = 0 \in \mathbb{Z}_2$};
\node (01) [bd] at (10,1) {};
\node (03) [bd] at (10,3) {};
\node (05) [bd] at (10,5) {};
\node (07) [bd] at (10,7) {};
\draw (01) edge (03);
\draw (03) edge (05);
\draw (05) edge (07);
\draw (07) edge [dotted] (10,7.5);
\node at (11,1) {$\lambda = 1$};
\node at (11,3) {$\lambda = 3$};
\node at (11,5) {$\lambda = 5$};
\node at (11,7) {$\lambda = 7$};
\node at (10,-2) {Component $I = 1 \in \mathbb{Z}_2$};
\end{tikzpicture}
    \caption{Hasse diagram for the orbits in the affine Grassmannian of $\mathrm{PSL}(2,\mathbb{C})$. }
    \label{fig:a1HasseSimple}
\end{figure}
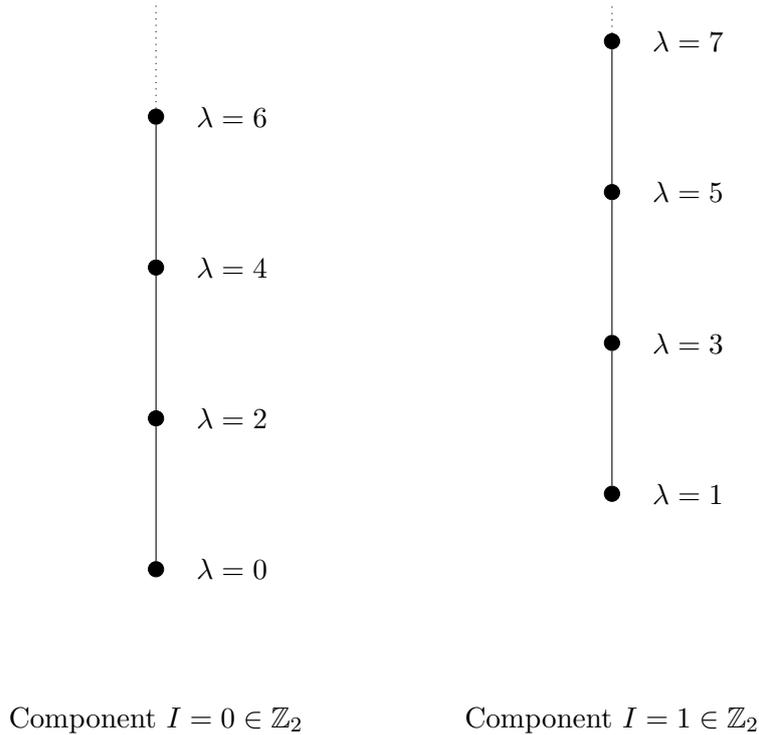

In this paragraph we discuss $\mathrm{PSL}(2,\mathbb{C})$, seen as $\mathrm{GL}(2,\mathbb{C})$ quotiented by scalar matrices. In $\mathrm{GL}(2,\mathbb{C})$, the difference with the previous analysis is that the determinant condition is lifted, so the orbits can be labeled by two integers $a$ and $b$, with $a \geq b$. In $\mathrm{PSL}(2,\mathbb{C})$ though we have
\begin{equation}
    \left( \begin{array}{cc}
        t^a  & 0 \\
         0  &   t^b
    \end{array} \right) \sim  \left( \begin{array}{cc}
        t^{a-b}  & 0 \\
         0  &   1
    \end{array} \right)  \, , 
\end{equation}
so the orbits can be labeled by $\lambda = a-b \in \mathbb{N}$. A computation analog to (\ref{eqLimitMat}) shows that again the orbit $\gr{\mathrm{PSL}(2,\mathbb{C})}{\lambda -2}{}$ lies in the closure of $\gr{\mathrm{PSL}(2,\mathbb{C})}{\lambda}{}$. There are two connected components in $\mathrm{Gr}_{\mathrm{PSL}(2,\mathbb{C})}$: one is the union of all the orbits $\gr{\mathrm{PSL}(2,\mathbb{C})}{\lambda}{}$ for $\lambda$ even, which gives $\mathrm{Gr}_{\mathrm{SL}(2,\mathbb{C})}$, and the other is the union of all the orbits $\gr{\mathrm{PSL}(2,\mathbb{C})}{\lambda}{}$ for $\lambda$ odd. The complex dimension of $\gr{\mathrm{PSL}(2,\mathbb{C})}{\lambda}{}$ is $\lambda$ in all cases, so it is even in one component and odd in the other. 
This is summarized in Figure \ref{fig:a1HasseSimple}. 

\subsection{Orbits in the affine Grassmannian}

We now resume the general discussion and detail how the observations of the previous subsections generalize to an arbitrary (semisimple complex algebraic) group $G$. An important role is played by the coweights and the coroots of $G$, so we begin with a reminder of a few definitions. 

Let us pick a maximal torus and a Borel subgroup $T \subset B \subset G$, with Lie algebras $\mathfrak{h} \subset \mathfrak{b} \subset \mathfrak{g}$. A character of $G$ is a homomorphism $T \rightarrow \mathbb{C}^{\ast}$, while a cocharacter is a homomorphism $\mathbb{C}^{\ast} \rightarrow T$. The differentials of characters are weights, while the differentials of cocharacters are coweights. We call $\Lambda$ the coweight lattice of $G$ and $\Lambda_0$ the coroot lattice of $G$. The goal of this subsection is to define a collection of elements $M_{\lambda} \in G((t))$ generalizing (\ref{Mlambda}) from which we build orbits in the affine Grassmannian, so it is natural to require that $\lambda$ be a \emph{coweight}, and not a weight. The coweights of $G$ can be used to label the (highest-weight) irreducible representations of the Langlands dual group $G^{\vee}$.
We use the Killing form to identify the space of weights and the space of coweights. The Borel subalgebra $\mathfrak{b}$ fixes a set of simple roots, which in turn defines a set of fundamental weights in the usual fashion. We call $\Lambda^+$ (respectively $\Lambda_0^+$) the set of linear combinations with non-negative integer coefficients of the fundamental coweights (resp. simple coroots).  Any weight or coweight can be expressed in the basis of the fundamental weights $(\varpi_i)_{i =1 , \dots , r}$, and we denote a (co)weight by its coordinates in this basis 
\begin{equation}
    \lambda  = \sum\limits_{i=1}^r  \lambda_i \varpi_i \, . 
\end{equation}
Note that coweights are expressed in the basis of fundamental weights, as opposed to fundamental coweights, because of the identification above. 
The coweights are partially ordered as follows. For $\lambda$ and $\mu$ two coweights, 
\begin{equation}
\label{partialOrder}
    \lambda \leq \mu \quad \Longleftrightarrow \quad  \begin{array}{c}
         \mu - \lambda \textrm{ is a linear combination of  }    \\
           \textrm{ simple coroots with coefficients in } \mathbb{N} 
    \end{array} \quad \Longleftrightarrow \quad  \mu - \lambda \in \Lambda_0^+  \, . 
\end{equation}
This partial order defines a Hasse diagram for the dominant coweights. This Hasse diagram has connected components labeled by classes $\Lambda / \Lambda_0$. 

\paragraph{Definition.}
Given a coweight $\lambda$ we have a corresponding homomorphism $\mathbb{C}^{\ast} \rightarrow T$. Denoting by $t \in \mathbb{C}^{\ast}$ the formal variable, the image of this homomorphism is called $M_{\lambda}(t)  := M_{\lambda}$.  

\vspace{1em}

We have identified the affine Grassmannian of $G$ with the space 
\begin{equation}
    \mathrm{Gr}_G = G((t))/G[[t]] \, , 
\end{equation}
where a matrix in $G((t))$ specifies a lattice by giving an explicit basis, and the quotient by $G[[t]]$ eliminating the arbitrariness in the choice of such a basis. We also recall that given a matrix $M \in G((t))$, multiplying on the right by an element of $G[[t]]$ gives the same lattice, as we just said, but multiplying on the left gives a different lattice. We can denote points and orbits in the affine Grassmannian as follows: 
\begin{itemize}
    \item $M \cdot G[[t]]$ is the point in $\mathrm{Gr}_G$ corresponding to $M$.
    \item $G[[t]] \cdot M \cdot G[[t]]$ is the orbit of that point in $\mathrm{Gr}_G$. 
\end{itemize}
Given a coweight $\lambda$ of $G$, we can build an orbit in the affine Grassmannian, which is denoted: 
\begin{equation}
\label{orbits}
    \gr{G}{\lambda}{} = G[[t]] \cdot M_{\lambda}  \cdot G[[t]] \, .   
\end{equation}

Two coweight $\lambda$ and $\lambda '$ which belong to the same Weyl group orbit give the same orbit in the affine Grassmannian, $\gr{G}{\lambda}{} = \gr{G}{\lambda '}{}$, so we can restrict our attention to the orbits $\gr{G}{\lambda}{} $ for $\lambda$ dominant. It turns out that these orbits provide a partition of the affine Grassmannian, which can be written as a disjoint union 
\begin{equation}
    \mathrm{Gr}_{G} = \bigcup\limits_{\lambda \in \Lambda^+} \gr{G}{\lambda}{}  \, . 
\end{equation}
The (Zariski) closure of $\gr{G}{\lambda}{}$ is 
\begin{equation}
   \grb{G}{\lambda}{} =  \gr{G}{\leq \lambda}{} = \bigcup\limits_{\mu \in \Lambda^+, \mu \leq \lambda} \gr{G}{\mu}{}  \, . 
\end{equation}
Therefore, the partial order on the dominant coweights $\Lambda^+$ provides the Hasse diagram for the $G[[t]]$ orbits in the affine Grassmannian $\mathrm{Gr}_{G}$. There is one Hasse diagram for each connected component of $\mathrm{Gr}_{G}$, and by construction we have 
\begin{equation}
\label{fundamentalGroup}
    \pi_0 (\mathrm{Gr}_{G}) = \Lambda / \Lambda_0 = \pi_1 (G) \, . 
\end{equation}
Elements of this group will be denoted by an index $I$. 
We now introduce the notion of lowest dominant coweights. A coweight $\Omega$ is a lowest dominant coweight if it is dominant, and if in addition there is no other dominant coweight $\omega$ such that $\omega<\Omega$. There are as many components in the affine Grassmannian of $G$ as there are lowest dominant coweights of $G$; we call them $\Omega_I$, where the index $I \in \pi_1 (G)$. We list the lowest dominant coweights for simple groups in Table \ref{tab:LDC}.

\begin{table}[ht]
    \centering
    \begin{tabular}{c|c|c}
        \toprule
        $G$ & $\pi_1(G)=Z(G^\vee)$ & $\Omega_I$\\
        \midrule[0.1em]
        $\mathrm{PSL}(n+1,\mathbb{C})$ & $\mathbb{Z}_{n+1}$ & $\Omega_0=[0,\dots,0]$ \\
        & & $\Omega_l=[\underbrace{0,\dots,0}_{l-1},1,0\dots,0]$\\
        \midrule
        $\mathrm{SO}(2n+1,\mathbb{C})$ & $\mathbb{Z}_{2}$ & $\Omega_0=[0,0,\dots,0]$ \\
        & & $\Omega_1=[1,0,\dots,0]$\\
        \midrule
        $ \mathrm{PSp}(2n,\mathbb{C})$ & $\mathbb{Z}_{2}$ & $\Omega_0=[0,\dots,0,0]$ \\
        & & $\Omega_1=[0,\dots,0,1]$ \\
        \midrule
        $\mathrm{PSO}(4n,\mathbb{C})$ & $\mathbb{Z}_{2}\times \mathbb{Z}_{2}$ & $\Omega_{(0,0)}=[0,0,\dots,0,0,0]$ \\
        & & $\Omega_{(1,1)}=[1,0,\dots,0,0,0]$\\
        & & $\Omega_{(1,0)}=[0,0,\dots,0,1,0]$\\
        & & $\Omega_{(0,1)}=[0,0,\dots,0,0,1]$\\
        \midrule
        $\mathrm{PSO}(4n+2,\mathbb{C})$ & $\mathbb{Z}_{4}$ & $\Omega_0=[0,0,\dots,0,0,0]$ \\
        & & $\Omega_{1}=[0,0,\dots,0,1,0]$\\
        & & $\Omega_{2}=[1,0,\dots,0,0,0]$\\
        & & $\Omega_{3}=[0,0,\dots,0,0,1]$\\
        \midrule
        $\mathrm{E}_6/\mathbb{Z}_3$ & $\mathbb{Z}_3$ & $\Omega_0=[0,0,0,0,0,0]$ \\
        & & $\Omega_1=[1,0,0,0,0,0]$ \\
        & & $\Omega_2=[0,0,0,0,1,0]$ \\
        \midrule
        $\mathrm{E}_7/\mathbb{Z}_2$& $\mathbb{Z}_2$ & $\Omega_0=[0,0,0,0,0,0,0]$ \\
        & & $\Omega_1=[1,0,0,0,0,0,0]$ \\
        \midrule
        $\mathrm{E}_8$ & trivial & $\Omega_0=[0,0,0,0,0,0,0,0]$ \\
        \midrule
        $\mathrm{F}_4$ & trivial & $\Omega_0=[0,0,0,0]$ \\
        \midrule
        $\mathrm{G}_2$ & trivial & $\Omega_0=[0,0]$ \\
        \bottomrule
    \end{tabular}
    \caption{Lowest dominant coweights $\Omega_I$ (given in the basis of fundamental coweights) for centerless simple groups $G$. We denote elements of $\pi_1(G)=Z(G^\vee)$ by the letter $I$. To each $\Omega_I$ there is a component labelled $I$ in the affine Grassmannian of $G$.}
    \label{tab:LDC}
\end{table}

The $\gr{G}{\lambda}{}$ are called \emph{Schubert cells} and the $\gr{G}{\leq \lambda}{}$ are called \emph{Schubert varieties}. 
The Schubert cell $\gr{G}{\lambda}{}$ is a smooth variety of dimension 
\begin{equation}
    \mathrm{dim}_{\mathbb{C}} \gr{G}{\lambda}{} = \langle 2 \rho , \lambda \rangle \, , 
\end{equation}
where $\rho$ is the Weyl vector \cite[Prop 2.1.5]{2016arXiv160305593Z}. 
The Schubert varieties can be singular spaces, but are well behaved: the singularities are always normal, Gorenstein and Cohen-Macaulay \cite[Theorem 2.1.21]{2016arXiv160305593Z}. Their complex dimension can be even or odd, but the parity of the complex dimension remains the same in a given connected component of $\mathrm{Gr}_{G}$.

\begin{table}
    \centering
        \begin{tabular}{c|c} 
        \toprule 
         Notation & Explanation \\  \midrule [.08em]
          $\mathbb{C}[t]$ & Ring of polynomials in $t$  \\  \midrule [.03em]
          $\mathbb{C}[[t]]$ & Ring of formal power series in $t$  \\  \midrule [.03em]
          $\mathbb{C}((t))$ & \specialcell{Field of formal Laurent series in $t$, \\ also field of fractions of $\mathbb{C}[[t]]$} \\  \midrule [.03em]
          $G$ & Simple algebraic subgroup of $\mathrm{GL}(n,\mathbb{C})$  \\  \midrule [.03em]
          $\mathfrak{g}$ & Lie algebra of $G$  \\  \midrule [.03em]
          $C_{\mathfrak{g}}$ & Cartan matrix of  $\mathfrak{g}$  \\  \midrule [.03em]
          $\Lambda$ & Coweight lattice of $G$  \\  \midrule [.03em]
          $\Lambda_0$ & Coroot lattice of $G$  \\  \midrule [.03em]
          $(\alpha_i)_{i=1 , \dots , r}$, $(\alpha^{\vee}_i)_{i=1 , \dots , r}$ & Simple roots and simple coroots of $\mathfrak{g}$  \\  \midrule [.03em]
          $(\varpi_i)_{i=1 , \dots , r}$, $(\varpi^{\vee}_i)_{i=1 , \dots , r}$ & Fundamental weights and fundamental coweights of $G$  \\  \midrule [.03em]
          $\rho$ & Weyl vector, half-sum of the positive roots  \\  \midrule [.03em]
          $G[t]$  & Group $G$ ``with coefficients in $\mathbb{C}[t]$"  \\  \midrule [.03em]
          $G[[t]]$ & Group $G$ ``with coefficients in $\mathbb{C}[[t]]$" \\  \midrule [.03em]
          $G((t))$  & Group $G$ ``with coefficients in $\mathbb{C}((t))$"  \\  \midrule [.03em]
          $G_1[t^{-1}]$ & \specialcell{Subgroup of $G[t^{-1}]$ where the $t^0$ \\ part is the identity of $G$} \\  \midrule [.03em]
          $\mathrm{Gr}_G=G((t))/G[[t]]$ & Affine Grassmannian of $G$  \\  \midrule [.03em]
          $M_{\lambda}$ & Image in $G((t))$ of the coweight $\lambda \in \Lambda$  \\  \midrule [.03em]
          $\gr{G}{\lambda}{}=G[[t]] \cdot   M_{\lambda} \cdot G[[t]]$ & \specialcell{Schubert cell associated to dominant coweight $\lambda$, \\ a smooth variety of complex dimension $\langle 2 \rho , \lambda \rangle$}  \\  \midrule [.03em]
          $\grb{G}{\lambda}{}=\bigsqcup_{\sigma\leq\lambda}\gr{G}{\lambda}{}$ & \specialcell{Schubert variety, the Zariski closure of the \\ Schubert cell $\gr{G}{\lambda}{}$, of dimension $\langle 2 \rho , \lambda \rangle$}  \\   \midrule [.03em]
        $\gr{G}{}{\lambda}=G[t^{-1}] \cdot M_{\lambda}  \cdot G[[t]]$ & Opposite Schubert Cell (infinite dimensional) \\   \midrule [.03em]
          $\w{G}{}{\lambda}=G_1[t^{-1}] \cdot  M_{\lambda}  \cdot G[[t]]$ & \specialcell{Transverse slice to Schubert cell \\ in $\mathrm{Gr}_G$ (infinite dimensional)}  \\  \midrule [.03em]
          $\wb{G}{\mu}{\lambda}=\w{G}{}{\lambda}\cap \grb{G}{\mu}{}$  & \specialcell{Transverse slice from $\mu$ to $\lambda$, a symplectic \\ singularity of complex dimension  $\langle 2 \rho ,\mu - \lambda \rangle$ }\\  \midrule [.03em]
          $\Omega_I$  &  \specialcell{Lowest dominant coweights (Table \ref{tab:LDC}), \\ with index $I$ in the fundamental group (\ref{fundamentalGroup})} \\  \midrule [.03em]
          $\w{G}{}{\Omega_I}$  &  \specialcell{Transverse slices to the lowest Schubert cells} \\  \bottomrule 
        \end{tabular}
    \caption{Summary of the notations. }
    \label{tab:notation}
\end{table}

\subsection{Slices in the affine Grassmannian}
\label{subsectionSlicesAG}

One can use other groups than $G[[t]]$ to generate orbits. In particular, we can use $G[t^{-1}]$ to generate the orbit 
\begin{equation}
    \gr{G}{}{\lambda} := G[t^{-1}] \cdot M_{\lambda} \cdot G[[t]] \, , 
\end{equation}
which is called an \emph{opposite Schubert cell}. From this one can build an (infinite dimensional, but finite codimensional) \emph{opposite Schubert variety}
\begin{equation}
    \gr{G}{}{\geq \lambda} :=  \bigcup\limits_{\mu \in \Lambda^+, \mu \geq \lambda} \gr{G}{}{\mu}  \, .  
\end{equation}
Note that $\gr{G}{}{\lambda} \cap \gr{G}{\lambda}{} = G \cdot M_{\lambda} \cdot G[[t]]$ is not reduced to a point. This motivates the introduction of a last type of orbits, in which $G$ does not act at degree 0, namely 
\begin{equation}
    \w{G}{}{\lambda} := G_1 [t^{-1}] \cdot M_{\lambda} \cdot G[[t]] \, . 
\end{equation} 
The intersection of these orbits with Schubert cells and Schubert varieties give the spaces which are the main focus of this paper: 
\begin{equation}
\w{G}{\mu}{\lambda}=\w{G}{}{\lambda}\cap \gr{G}{\mu}{}  \qquad \textrm{and} \qquad  \wb{G}{\mu}{\lambda}=\w{G}{}{\lambda}\cap \grb{G}{\mu}{} \, . 
\end{equation}
The spaces $\wb{G}{\mu}{\lambda}$ intersect $\gr{G}{\nu}{}$ transversely for any $\lambda \leq \nu \leq \mu$, and are called the \emph{slices} in the affine Grassmannian. They are varieties with symplectic singularities \cite{beauville2000symplectic,kamnitzer2014yangians}, with dimension
\begin{equation}
    \mathrm{dim}_{\mathbb{C}} \wb{G}{\mu}{\lambda} = \langle 2 \rho , \mu - \lambda \rangle \,  . 
\end{equation}
In particular, when $\mu - \lambda$ is a simple coroot $\alpha^{\vee}$, the two orbits $\gr{G}{\mu}{}$ and $\gr{G}{\lambda}{}$ are adjacent in the Hasse diagram, the elementary transverse slice has complex dimension $2$, and one can show that \cite[Example 2.2]{kamnitzer2014yangians}
\begin{equation}
\label{Aslice}
    \wb{G}{\lambda + \alpha^{\vee}}{\lambda} =  \mathbb{C}^2 / \mathbb{Z}_{2+ \langle  \lambda , \alpha \rangle} = A_{1+ \langle  \lambda , \alpha \rangle} \, . 
\end{equation}
This shows that the ``generic" elementary slice in the affine Grassmannian is a Kleinian singularity of type $A$. 
More generally, elementary slices correspond to pairs of adjacent coweights according to the partial order (\ref{partialOrder}). This problem has been studied in detail in \cite{stembridge1998partial,2003math:5095M,juteau2008modular}, and the conclusion is that the elementary transverse slices fall into three categories \cite{2003math:5095M}: 
\begin{itemize}
    \item Kleinian singularities of type $A$, (\ref{Aslice}), when $\mu - \lambda$ is a simple coroot. These are denoted $A_n$. 
    \item Closure of minimal nilpotent orbits, when $\mu - \lambda$ is the short dominant coroot of the subalgebra defined by the nonvanishing of $\mu - \lambda$. These are denoted by $a_n$, $b_n$, $c_n$, $d_n$, $e_6$, $e_7$, $e_8$, $f_4$ and $g_2$. 
    \item \emph{Quasi-minimal} singularities,\footnote{The terminology comes from \cite{2003math:5095M}. } which can be of three types: 
    \begin{enumerate}
        \item An infinite family called $ac_n$, for $n \geq 2$;
        \item The $ag_2$ singularity;
        \item The $cg_2$ singularity. 
    \end{enumerate}
\end{itemize}
The existence of the elementary slices other than Kleinian of type $A$ is caused by geometric constraints in the Weyl chambers when approaching walls. The disposition  $\mathfrak{D}$\footnote{The disposition $\mathfrak{D}$ of a leaf and a  moduli space is defined in \eqref{eq:disposition} and \eqref{eq:disposition1} respectively in the Introduction.} of the affine Grassmannian $\mathrm{Gr}_G$ is $r = \mathrm{rank}(G)$. Close to a codimension $d$ wall of the Weyl chamber, the disposition of a generic coweight decreases to $r-d$. 

In the next section we see this at work in the affine Grassmannian at rank 2. 
The quasi-minimal singularities are studied in detail in Section \ref{newslices}.

\section{Quivers and Hasse diagrams}
\label{sec:quivers}
In this section we explore the affine Grassmannian, or rather the slices in the affine Grassmannian, of simple groups. The two coweights which define a slice can be used to produce a quiver, whose Coulomb branch is the slice.

\subsection{General formula for slices}
\label{subsectionGeneralFormulas}

From now on we assume that $G$ is simple with rank $r$, so that its Lie algebra can be characterized by a connected Dynkin diagram. 
In \cite{Braverman:2016pwk,finkelberg2017double}, following the series of works \cite{braverman2010pursuing,Nakajima:2015txa,Braverman:2016wma,Nakajima:2017bdt} the slices in the affine Grassmannian $\mathrm{Gr}_G$ have been identified with Coulomb branches of certain good 3d $\mathcal{N}=4$ quiver gauge theories. In this subsection, we summarize this connection and give explicit formulas.

\paragraph{From slice to quiver.}
Let $\lambda$ and $\mu$ be two coweights. We assume that
\begin{itemize}
    \item $\mu$ is dominant. 
    \item $\mu - \lambda$ is in the positive coroot lattice. 
\end{itemize}
Then we can construct a quiver $\mathsf{Q}_{\lambda }^{\mu}$ as follows: 
\begin{itemize}
    \item There are $r$ gauge nodes, connected as in the Dynkin diagram of $\mathfrak{g}$, with node ranks 
    \begin{equation}
        k_i = \langle \varpi_i , \mu - \lambda \rangle \, , 
    \end{equation}
    with $i = 1 , \dots , r$. When $\lambda$ and $\mu$ are identified with their components in the basis of fundamental weights, the vector of gauge nodes ranks is $\mathbf{k} = C_\mathfrak{g}^{-1} \cdot (\mu - \lambda)$.  
    \item There are $r$ flavor nodes, connected to the $r$ gauge nodes, with ranks given by 
    \begin{equation}
        N_i = \langle \alpha_i^{\vee} , \mu \rangle \, ,
    \end{equation}
    with $i = 1 , \dots , r$. When $\mu$ is identified with its components in the basis of fundamental weights, $\mathbf{N} = \mu$. 
\end{itemize}
Note that the condition that $\mu - \lambda$ be in the positive coroot lattice guarantees precisely that $\mathbf{k}$ is an element of $\mathbb{N}^r$.

\paragraph{From good quiver to slice.}
Consider a \emph{good} quiver such that the gauge nodes form the Dynkin diagram of a simple Lie algebra $\mathfrak{g}$. We denote by $\mathbf{k} = (k_i)_{i=1 , \dots , r}$ the ranks of the gauge nodes and $\mathbf{N} = (N_i)_{i=1 , \dots , r}$ the ranks of the flavor nodes, with $r$ the rank of $\mathfrak{g}$. To this quiver we associate two coweights: 
\begin{itemize}
    \item The coweight $\mu$ is given by the flavor nodes, $\mu = \sum\limits_{i=1}^r n_i \varpi_i$; 
    \item The coweight $\lambda$ is given by\footnote{Note that the components of the coweight $\lambda$ gives the \emph{balance} of the gauge nodes of the quiver. The assumption that the quiver is good implies that $\lambda$ is dominant. } $\lambda = \mathbf{N} - C_\mathfrak{g} \cdot \mathbf{k}$. 
\end{itemize}

\paragraph{Example.}
For instance if one considers the algebra $C_2$ and coweights expressed in the basis of fundamental weights\footnote{For conventions regarding the $C_2$ algebra, see Figure \ref{fig:c2rootSystem}. } $\lambda = [2,0]$ and $\mu = [2,2]$, with the matrix $C_{\mathfrak{g}} = \left( \begin{array}{cc}
    2 & -1 \\
   -2  & 2
\end{array} \right)$ one gets $\mu - \lambda = [0,2]$. This is in the positive coroot lattice, and $C_{\mathfrak{g}}^{-1} \cdot (\mu - \lambda) = [1,2]$ so the quiver is 
\begin{equation}
   \mathsf{Q}^{\mu}_{\lambda} = \scalebox{1.3}{\raisebox{-.5\height}{\cquiv{2}{2}{1}{2}}} \, . 
\end{equation}
Note that the gauge node on the right is balanced, corresponding to the vanishing component of $\lambda$. 

If instead one considers $\lambda=[0,1]$ and $\mu = [2,2]$, then $\mu - \lambda = [2,1]$ is not in the positive coroot lattice. Accordingly, $C_{\mathfrak{g}}^{-1} \cdot (\mu - \lambda) = [\frac{5}{2},3]$ is not a vector of integers and no quiver can be defined. 

Note that it is not necessary that the coweight $\lambda$ be dominant. For instance consider the following pair: 
\begin{equation}
    \lambda = [-1,2] \qquad \mu = [4,6] \qquad  \mathsf{Q}^{\mu}_{\lambda} = \scalebox{1.3
}{\raisebox{-.5\height}{\cquiv{4}{6}{7}{9}}} \, . 
\end{equation}
Note that the node on the left is under-balanced.

\paragraph{Generalized Slices.}
When $\lambda$ is not dominant, the construction of the quiver $\mathsf{Q}^{\mu}_{\lambda}$ is still perfectly well defined. The quiver is not good in the sense of \cite{Gaiotto:2008ak}, and its Coulomb branch does not correspond to a slice in $\mathrm{Gr}_G$. However the Coulomb branch of $\mathsf{Q}^{\mu}_{\lambda}$ can be identified with the generalized slices of \cite{Braverman:2016pwk}, see also \cite{Zhou:2020bwa}. 

\paragraph{Explicit quivers.}
The quivers $\mathsf{Q}^{\mu}_{\lambda}$ for algebras of classical types are gathered in Table \ref{tab:explicitQuivers}. We use the following explicit form for the inverse of the Cartan matrix for $\mathfrak{g} = A_n$: 
\begin{equation}
    C_{\mathfrak{g}}^{-1} = \left( \mathrm{min}(i,j) - \frac{ij}{n+1} \right)_{ 1 \leq i,j \leq n} \, . 
\end{equation} 

\subsection{Hasse diagrams for affine Grassmannians at ranks 1 and 2}

\begin{figure}[t]
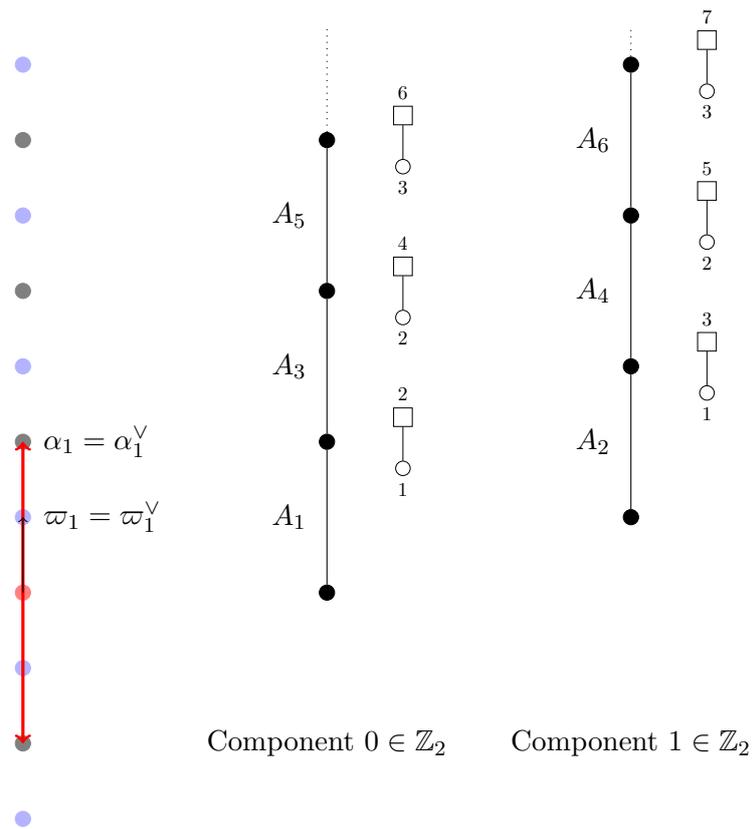

    \centering
%
    \caption{
    On the left: weight and coweight lattice of $A_1$. On the right: Bottom of the Hasse diagram for $\mathrm{PSL}(2,\mathbb{C})$. See detailed caption on page \pageref{captions}. 
    }
    \label{fig:a1rootSystem}
\end{figure}

\begin{figure}[t]
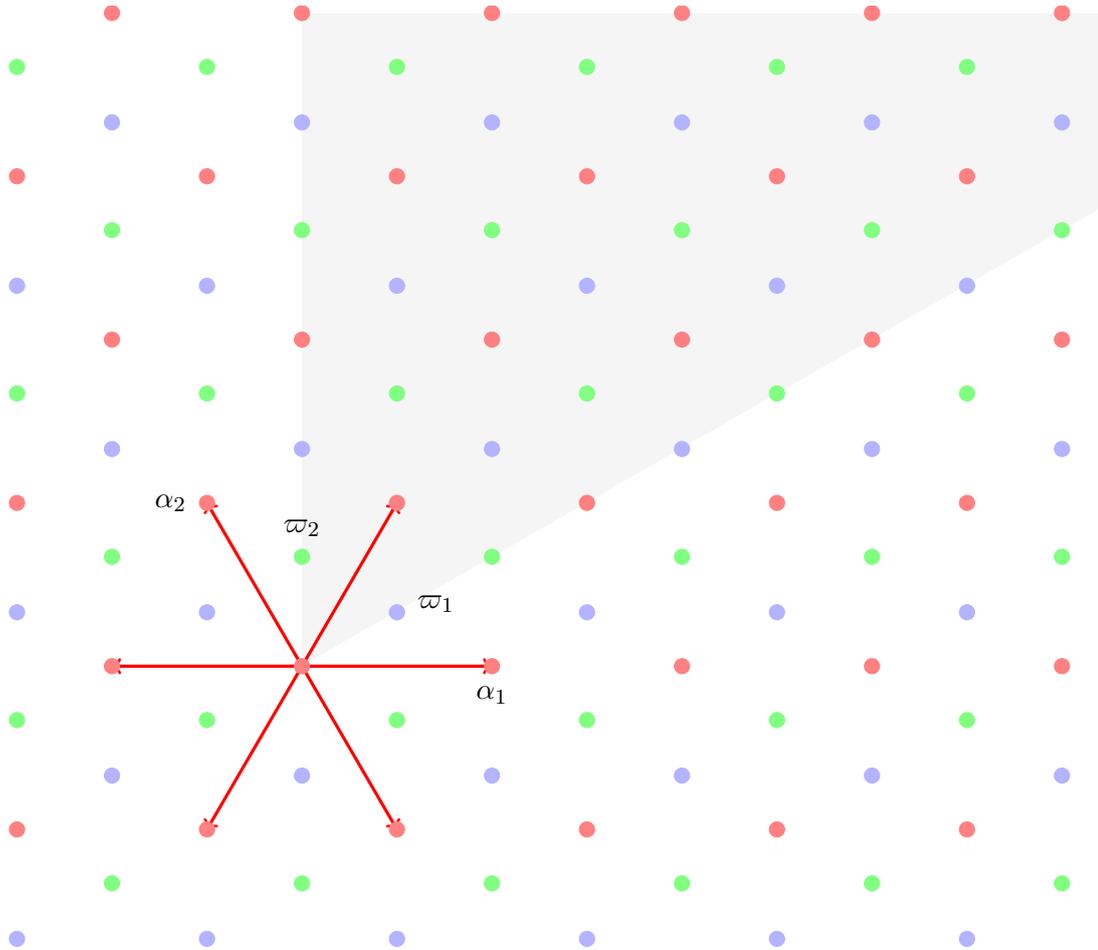

    \centering
\scalebox{1.00}{%
}
    \caption{Summary of $A_2$ data. See detailed caption on page \pageref{captions}.  The red dots correspond to the coroot lattice $\Lambda_0$, while the blue dots are the lattice $\Lambda_0+\varpi_1$ and the green dots are the lattice $\Lambda_0+\varpi_2$. }
    \label{fig:a2rootSystem2}
\end{figure}

\begin{figure}[t]
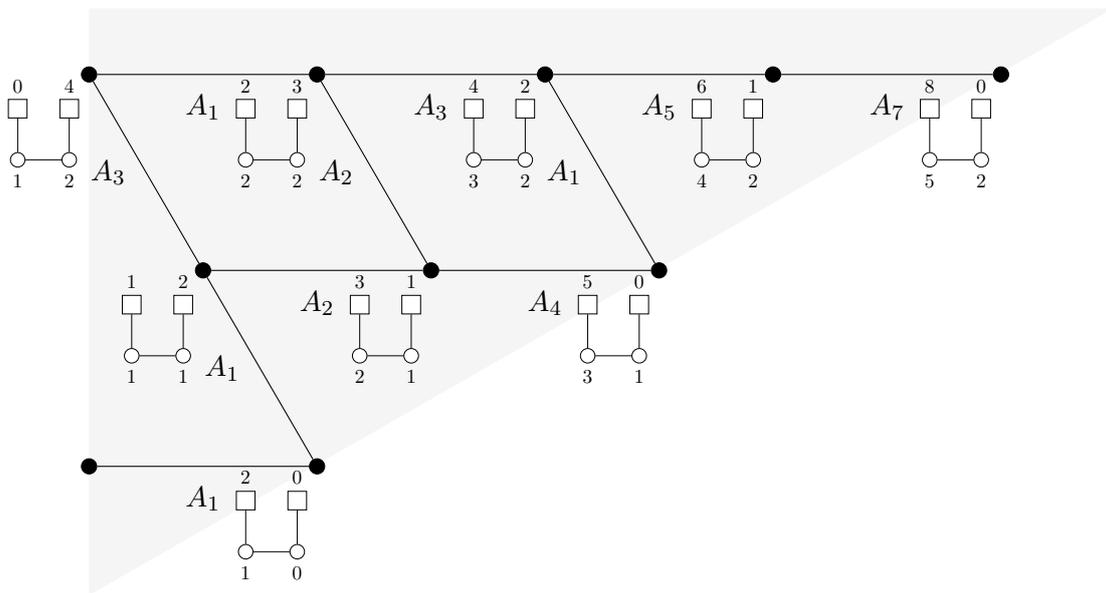

    \centering
%
    \caption{Bottom of the Hasse diagram for $\mathrm{PSL}(3,\mathbb{C})$, component $-1 \in \mathbb{Z}_3$. See detailed caption on page \pageref{captions}. The third component $1 \in \mathbb{Z}_3$ is obtained by symmetry.  }
    \label{fig:a2AGComponent2}
\end{figure}

\begin{figure}[t]
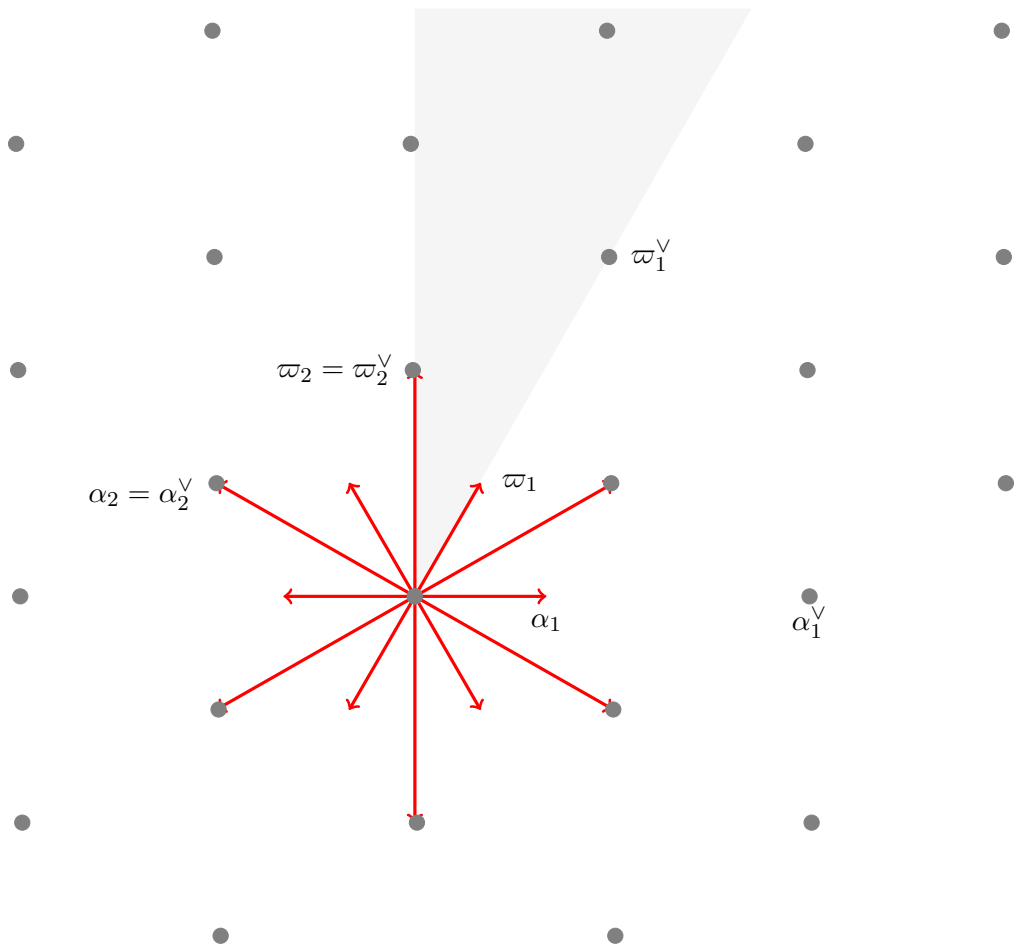

    \centering
\scalebox{1.00}{%
}
    \caption{Summary of $\mathrm{G}_2$ data. See detailed caption on page \pageref{captions}. }
    \label{fig:g2rootSystem}
\end{figure}

\begin{figure}[t]
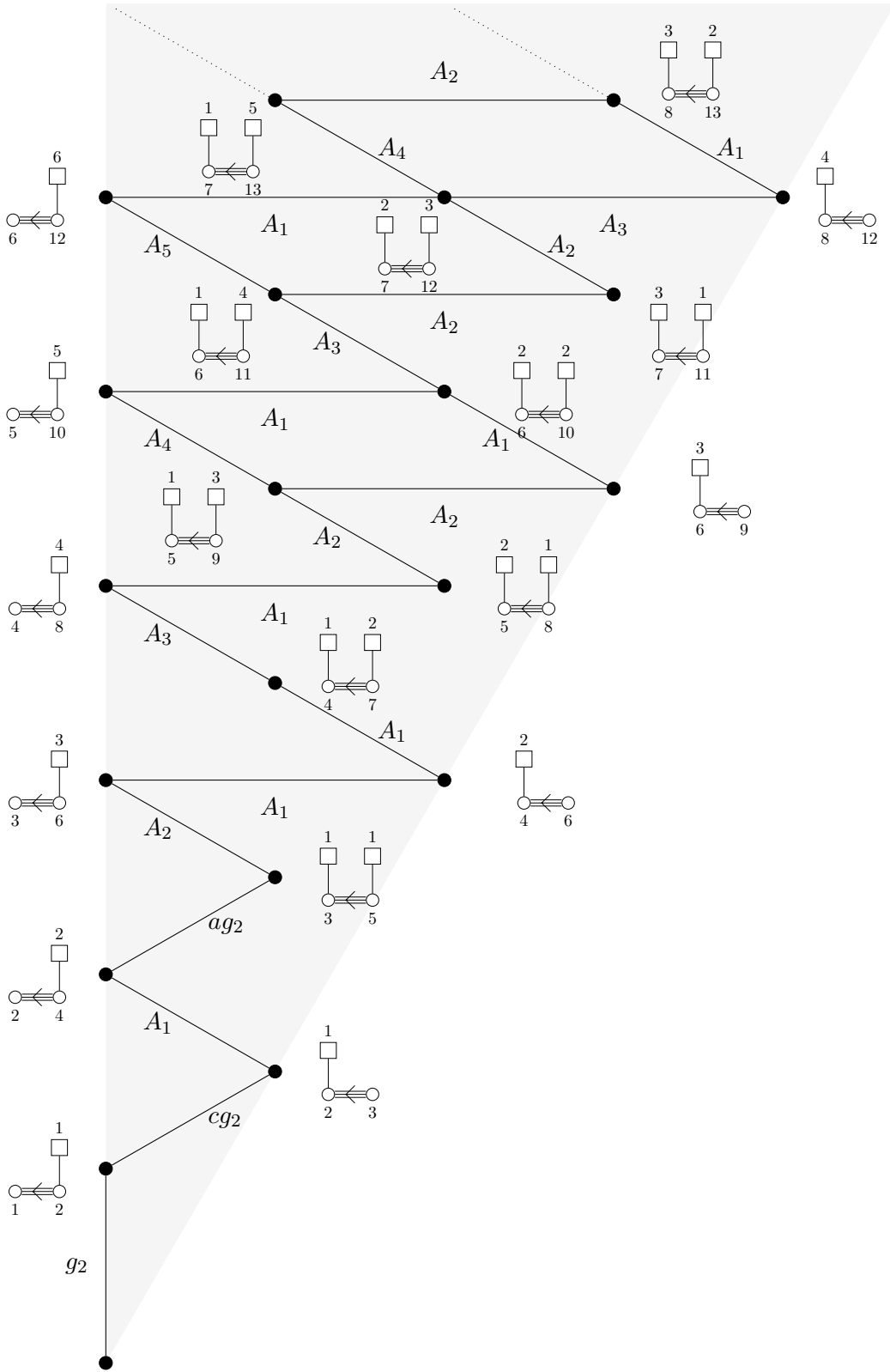

    \centering
%
}};
\end{tikzpicture}
    \caption{Bottom of the Hasse diagram for $\mathrm{G}_2$. See detailed caption on page \pageref{captions}.}
\label{fig:g2AGComponent1}
\end{figure}

In this subsection, we use the description reviewed above to construct the bottom part of the Hasse diagrams for complex simple groups of ranks 1 and 2. This can be achieved by focusing only on the centerless groups associated to the Lie algebra (the affine Grassmannian for groups with non-trivial center are obtained by taking only the relevant connected components). This is presented in Figures \ref{fig:a1rootSystem} to \ref{fig:g2AGComponent1}. 

\paragraph{Caption for Figures \ref{fig:a1rootSystem}-\ref{fig:g2AGComponent1}.} \label{captions}
For each Lie algebra, we first draw a diagram with our Lie algebra conventions, which can be found in the tables of \cite{gorbatsevich1994lie}. This diagram represents both the Cartan subalgebra and its dual, identified via the Killing form. The red arrows represent the roots, the black arrow represent the fundamental weights. The simple roots and coroots, and the fundamental weights and coweight are explicitly labeled. The dots denote the coweight lattice, where different colors are used for different elements of the group (\ref{fundamentalGroup}), and therefore correspond to different connected components in (the Hasse diagram of) the affine Grassmannian. The Weyl chamber, defined by the fundamental coweights, is shaded in gray. 

Then for each element of the group (\ref{fundamentalGroup}) we draw a Hasse diagram for the few lowest orbits in the corresponding component of the affine Grassmannian. The dots in the Hasse diagram are the coweights and correspond to orbits (\ref{orbits}), while the lines connect adjacent coweights according to the partial order (\ref{partialOrder}). Dotted lines indicate that the Hasse diagram is infinite. 

A connected component can be labeled by a lowest coweight $\Omega_I$. In that component, next to each dot corresponding to a coweight $\mu$, we draw the quiver $\mathsf{Q}^{\mu}_{\Omega_I}$. The Coulomb branch of this quiver is the transverse slice to the orbit at the origin of the Hasse diagram, $\wb{G}{\mu}{\Omega_I}$. 
Finally, next to each line we indicate the nature of the elementary transverse slice, using the notations specified at the end of Section \ref{subsectionSlicesAG}. 

\paragraph{Disposition and quasi-minimal slices.}
From the Hasse diagrams presented in this section, one can observe that the appearance of slices beyond Kleinian singularities of type $A$ is related to space constraints close to the walls of the Weyl chamber. In the $A_2$ case, a non-zero coweight $\lambda$ always has at least one coweight $\mu$ such that $\mu - \lambda$ is a simple coroot. This is related to the basic geometric fact that the Weyl chamber has an angle of $\frac{\pi}{3}$ at the origin, which is the angle for equilateral triangles. 

In the $B_2$ case, the Weyl chamber has an angle $\frac{\pi}{4}$ at the origin. Adding any simple coroot to the coweight $\varpi^{\vee}_1$ in Figure \ref{fig:b2rootSystem} gives a coweight outside the Weyl chamber. In order to stay in the Weyl chamber, one has to add the non simple coroot $\alpha^{\vee}_1 + \alpha^{\vee}_2$. This leads to an elementary slice with quasi-minimal singularity in Figure \ref{fig:b2AGComponent2}. The same obviously applies to the $C_2$ case. 

Finally in the $\mathrm{G}_2$ case, there are two non-zero coweights which are such that adding any simple coroot gives a coweight outside the Weyl chamber, namely $\varpi^{\vee}_2$ and $2\varpi^{\vee}_2$. This gives rise to two elementary slices with quasi-minimal singularities in Figure \ref{fig:g2AGComponent1}. 

\afterpage{
\begin{landscape}
\begin{table}[ht]
    \centering
\begin{tabular}{c|c} \toprule
        Algebra & Quiver \\ \midrule
        $A_n$ &    \raisebox{-.5\height}{\begin{tikzpicture}
   \node[gauge,label=below:{$\sum\limits_{j=1}^n \frac{j\nu_{n+1-j}}{n+1} $}] (1) at (0,0) {};
    \node (2) at (3,0) {$\cdots$};
    \node[gauge,label=below:{$\sum\limits_{j=1}^n \left( \mathrm{min}(i,j) - \frac{ij}{n+1} \right) \nu_j $}] (3) at (6,0) {};
    \node (4) at (9,0) {$\cdots$};
    \node[gauge,label=below:{$\sum\limits_{j=1}^n \left(  \frac{n(n+1)-nj}{n+1} \right) \nu_j $}] (5) at (12,0) {};
    \node[flavour,label=above:{$\mu_1$}] (1a) at (0,1) {};
    \node[flavour,label=above:{$\mu_i$}] (3a) at (6,1) {};
    \node[flavour,label=above:{$\mu_n$}] (5a) at (12,1) {};
    \draw (1a)--(1)--(2)--(3)--(4)--(5)--(5a);
    \draw (3a)--(3);
\end{tikzpicture}} \\ \midrule 
        $B_n$ &    \raisebox{-.5\height}{\begin{tikzpicture}
    \node[gauge,label=below:{\scalebox{.8}{$\sum\limits_{j=1}^n \nu_j$}}] (1) at (4,0) {};
    \node[] (2) at (6,0) {$\cdots$};
    \node[gauge,label=below:{\scalebox{.8}{$\sum\limits_{k=1}^i \sum\limits_{j=k}^n \nu_j$}}] (3) at (8,0) {};
    \node[] (4) at (10,0) {$\cdots$};
    \node[gauge,label=below:{\scalebox{.8}{$\sum\limits_{k=1}^{n-1} \sum\limits_{j=k}^n \nu_j$}}] (5) at (12,0) {};
    \node[gauge,label=below:{\scalebox{.8}{$\frac{1}{2}\sum\limits_{k=1}^{n} \sum\limits_{j=k}^n \nu_j$}}] (6) at (15,0) {};
    \node[flavour,label=above:{$\mu_1$}] (1a) at (4,1) {};
    \node[flavour,label=above:{$\mu_i$}] (3a) at (8,1) {};
    \node[flavour,label=above:{$\mu_{n-1}$}] (5a) at (12,1) {};
    \node[flavour,label=above:{$\mu_{n}$}] (6a) at (15,1) {};
    \draw (1a)--(1) (3)--(3a) (5)--(5a) (6)--(6a) (13.4,0.2)--(13.6,0)--(13.4,-0.2);
    \draw (1)--(2)--(3)--(4)--(5);
    \draw[transform canvas={yshift=-1pt}] (5)--(6);
    \draw[transform canvas={yshift=1pt}] (5)--(6);
\end{tikzpicture}} \\
\bottomrule 
    \end{tabular}
    \caption{For algebras of classical types, we depict the quivers $\mathsf{Q}^{\mu}_{\lambda}$ where $\mu = \sum\limits_{j=1}^n \mu_i \varpi_i$ and $\lambda = \sum\limits_{j=1}^n \lambda_i \varpi_i$ are coweights with $\mu$ dominant and $\nu := \mu - \lambda$ in the positive coroot lattice. If $\lambda$ is dominant, their 3d $\mathcal{N}=4$ Coulomb branch is the transverse slice $\wb{G}{\mu}{\lambda}$.   }
    \label{tab:explicitQuivers}
\end{table}
\end{landscape}}

\begin{figure}[t]
    \centering
%
    \caption{Bottom of the Hasse diagram for $\mathrm{PSL}(3,\mathbb{C})$, component $0 \in \mathbb{Z}_3$. See detailed caption on page \pageref{captions}.  }
    \label{fig:a2AGComponent1}
\end{figure}

\begin{figure}[t]
    \centering
\scalebox{1.00}{%
}
    \caption{Summary of $B_2$ data. See detailed caption on page \pageref{captions}. The black dots correspond to the coroot lattice $\Lambda_0$, while the blue dots are the lattice $\Lambda_0+\varpi_1^{\vee}$. }
    \label{fig:b2rootSystem}
\end{figure}

\begin{figure}[t]
    \centering
%
    \caption{Bottom of the Hasse diagram for $\mathrm{PSO}(5,\mathbb{C})$, component $0 \in \mathbb{Z}_2$. See detailed caption on page \pageref{captions}. }
    \label{fig:b2AGComponent1}
\end{figure}

\begin{figure}[t]
    \centering
%
    \caption{Bottom of the Hasse diagram for $\mathrm{PSO}(5,\mathbb{C})$, component $1 \in \mathbb{Z}_2$. See detailed caption on page \pageref{captions}. }
    \label{fig:b2AGComponent2}
\end{figure}

\begin{figure}[t]
    \centering
\scalebox{1.00}{%
}
    \caption{Summary of $C_2$ data. See detailed caption on page \pageref{captions}. The black dots correspond to the coroot lattice $\Lambda_0$, while the blue dots are the lattice $\Lambda_0+\varpi_2^{\vee}$. }
    \label{fig:c2rootSystem}
\end{figure}

\begin{figure}[t]
    \centering
%
    \caption{Bottom of the Hasse diagram for $\mathrm{Sp}(4,\mathbb{C})$, component $0 \in \mathbb{Z}_2$. See detailed caption on page \pageref{captions}.}
    \label{fig:c2AGComponent1}
\end{figure}

\begin{figure}[t]
    \centering
 %
    \caption{Bottom of the Hasse diagram for $\mathrm{Sp}(4,\mathbb{C})$, component $1 \in \mathbb{Z}_2$. See detailed caption on page \pageref{captions}.}
    \label{fig:c2AGComponent2}
\end{figure}

\clearpage 

\subsection{Quivers for quasi-minimal singularities}\label{newslices}

In this section we focus on the elementary slices in the affine Grassmannian that did not make an appearance in \cite{Bourget:2019aer}, corresponding to the quasi-minimal singularities reviewed in Section \ref{subsectionSlicesAG}: 
\begin{itemize}
    \item The $ac_n$ slices are the lowest slices in the component of $\mathrm{Gr}_{\mathrm{Sp}(n)}$ not connected to the identity. 
    \item The slices $cg_2$ and $ag_2$, which appear in $\mathrm{Gr}_{\mathrm{G}_2}$, and were identified in Figure \ref{fig:g2AGComponent1}. 
\end{itemize}
In this section, we study magnetic quivers for these elementary slices, which are crucial in the algorithm of quiver subtraction and quiver addition, see Section \ref{sec:quiveradd}. They are deduced from the coweights that are used to define them, using the formulas of Section \ref{subsectionGeneralFormulas}. The quivers are gathered in Table \ref{tab:elem}. 

\subsubsection*{The $ac_n$ slices}
We start with the $ac_n$ singularity. It is the Coulomb branch of the quiver
\begin{equation}
    \raisebox{-.5\height}{
    \begin{tikzpicture}
        \node at (-2,0.5) {$ac_n$:};
    \node[gauge,label=below:{$1$}] (1) at (0,0) {};
    \node[gauge,label=below:{$1$}] (2) at (1,0) {};
    \node (3) at (2,0) {$\dots$};
    \node[gauge,label=below:{$1$}] (4) at (3,0) {};
    \node[gauge,label=below:{$1$}] (5) at (4,0) {};
    \node[gauge,label=below:{$1$}] (6) at (5,0) {};
    \node[flavour,label=left:{$1$}] (1a) at (0,1) {};
    \node[flavour,label=right:{$1$}] (6a) at (5,1) {};
    \draw (1a)--(1)--(2)--(3)--(4)--(5) (6)--(6a) (4.6,0.2)--(4.4,0)--(4.6,-0.2);
    \draw[transform canvas={yshift=-1.5pt}] (5)--(6);
    \draw[transform canvas={yshift=1.5pt}] (5)--(6);
    \draw [decorate,decoration={brace,amplitude=5pt},xshift=0pt,yshift=-20pt]
    (5,0)--(0,0) node [black,midway,xshift=0pt,yshift=-15pt] {$n$};
    \end{tikzpicture}
    }
\end{equation}
The global symmetry is $\mathfrak{su}(n)\times \mathfrak{u}(1)$, which is consistent with the rightmost node being overbalanced, while the other nodes are balanced and form the $A_{n-1}$ Dynkin diagram.  The Coulomb branch Hilbert series can be encapsulated in the form of a highest weight generating function (HWG) \cite{Hanany:2014dia}:
\begin{eqnarray}
    \mathrm{HWG}[ac_n] &=& \frac{1-\mu_1^2\mu_{n-1}^2t^6}{(1-t^2)(1-\mu_1\mu_{n-1}t^2)(1- q^{-1} \mu_{1}^2t^3 )(1-q\mu_{n-1}^2t^3)} \\ 
    &=& \mathrm{PE}\left[ t^2 + \mu_1\mu_{n-1}t^2 + (q^{-1} \mu_{1}^2 +q \mu_{n-1}^2) t^{3} - \mu_{1}^2  \mu_{n-1}^2 t^{6}\right] \nonumber
\end{eqnarray}
where $\mu_i$ and $q$ are highest weight fugacities of $\mathfrak{su}(n)$ and $\mathfrak{u}(1)$ respectively. From the HWG we can also obtain the unrefined Hilbert series. For $ac_2$, the unrefined Hilbert series is:
\begin{equation}
    \mathrm{HS}[ac_2]=\frac{1+2t^2+4t^3+2t^4+t^6}{(1-t^2)^2(1-t^3)^2} \, . 
\end{equation}
In addition, in Table \ref{acnPL} we provide the refined plethystic logarithm (PL) of the Hilbert series which encodes the representation content of the generators and relations of the Coulomb branch. 

\begin{table}[t]
    \centering
    \begin{tabular}{c|l}\toprule
   $n$     & PL coefficient  \\ \midrule
2         &\begin{tabular}{l}$ t^2$: \; $[2]+[0]$ \\ $t^3$: \; $(q+q^{-1})[2]$\\ $t^4$: \; $-[0]$\\ $t^5$: \;  $-(q+q^{-1})([0]+[2])$  \\ $t^6$: \;  $-q^2 [0] - q^{-2}[0] - [4]-[2]-[0]$  \end{tabular}
\\ \midrule
3 & \begin{tabular}{l}$t^2$:\; $[11]+[00]$\\ $t^3$:\; $q^{-1}[20]+q [02]$\\ $t^4:$\; $-[00]-[11]$\\ $t^5:$ \; $-  q^{-1}([01]+[12]+[20])-q([02]+[10]+[21])$  \\ $t^6$: \;  $-q^2 [20] - q^{-2}[02] - [22]+[11]-[00]$  \end{tabular} \\ \midrule
4& \begin{tabular}{l}$t^2$:\; $[101]+[000]$ \\ $t^3$:\; $q^{-1} [200]+q[002]$ \\ $t^4$: \; $-[000]-[020]-[101]$ \\ $t^5$: \; $-q^{-1}([010]+[111]+[200]) - q ([002]+[010]+[111])$  \\ $t^6$: \;  $(2-q^2-q^{-2})[020]-[202]+[210]+[012]+[101]-[000]$  \end{tabular}\\ \midrule
5& \begin{tabular}{l}$t^2$:\; $[1001]+[0000]$ \\ $t^3$:\; $q^{-1} [2000]+q[0002]$ \\ $t^4$: \; $-[0110]-[1001]-[0000]$ \\ $t^5$: \; $-q^{-1}([1101]+[2000]+[0100]) - q ([1011]+[0002]+[0010])$  \\ $t^6$: \;  $-q^2[0020]-q^{-2}[0200] -[2002]+[1200]+[0021]+[0102]$ \\ $ \qquad +[2010]+2[0110]+[1001]-[0000]$  \end{tabular} \\ \bottomrule
    \end{tabular}
    \caption{$ac_n$ refined PL up to order $t^6$. $[\cdots]$ are the Dynkin labels for $\mathfrak{su}(n)$. Note that the number of generators is equal to the dimension of the adjoint representation of $C_n$, and they form a natural embedding of $\mathrm{U}(n)$ in $\mathrm{Sp}(n)$.
    }
    \label{acnPL}
\end{table}

\begin{table}[t]
    \centering
    \begin{tabular}{c|l}\toprule
        & PL coefficient \\ \midrule
$ag_2$        &\begin{tabular}{l}$ t^2$: \; $[2]+[0]$ \\ $t^3$: \; $(q + q^{-1})[3]$ \\   $t^4$: \; $-[0]$ \\ $t^5$: \; $-(q + q^{-1})([3]+[1])$  \\ $t^6$: \; $-(q^2 + q^{-2})[2]-[6]-[4]-[2]-[0]$ \end{tabular} \\ \midrule 
$cg_2$         &\begin{tabular}{l}$ t^2$: \; $[2]+[0]$ \\  $t^4$: \; $(q + q^{-1})[3]-[0]$ \\$t^6$ \; $-(q+q^{-1})([3]+[1])$\end{tabular}
\\ \bottomrule
    \end{tabular}
    \caption{$ag_2$ and $cg_2$ refined PL up to order $t^6$. $[\dots]$ are the Dynkin labels for $\mathfrak{su}(2)$. Note that the number of generators is equal to the dimension of the adjoint representation of $\mathrm{G}_2$, and they form a natural embedding of $\mathrm{SU}(2)$ in $\mathrm{G}_2$.  }
    \label{ag2PL}
\end{table}

\subsubsection*{The $ag_2$ and $cg_2$ slices}
For the $\mathrm{G}_2$ affine Grassmannian, we have two new elementary slices which are the $ag_2$ and $cg_2$. For $ag_2$, the quiver is
\begin{equation}
    \raisebox{-.5\height}{
\label{acnquiver}
    \begin{tikzpicture}
    \node at (-2,0.5) {$ag_2$:};
    \node[gauge,label=below:{$1$}] (1) at (0,0) {};
    \node[gauge,label=below:{$1$}] (2) at (1,0) {};
    \node[flavour,label=left:{$1$}] (1a) at (0,1) {};
    \node[flavour,label=right:{$1$}] (2a) at (1,1) {};
    \draw (1a)--(1)--(2)--(2a) (0.6,0.2)--(0.4,0)--(0.6,-0.2);
    \draw[transform canvas={yshift=-2pt}] (1)--(2);
    \draw[transform canvas={yshift=2pt}] (1)--(2);
    \end{tikzpicture}
    }
\end{equation}
with global symmetry $\mathfrak{su}(2) \times \mathfrak{u}(1)$. The HWG takes the form:
\begin{eqnarray}
    \mathrm{HWG}[cg_2] &=& \frac{(1-\mu^6 t^8)}{(1-t^2) (1-\mu^2t^2)\left(1-  q^{-1} \mu^3 t^4 \right)(1-\mu^3 q t^4)} \\  &=& \mathrm{PE}[t^2 + \mu^2 t^2 + (q + q^{-1})\mu^3 t^4 - \mu^6 t^8] \, \nonumber 
\end{eqnarray}
where $\mu$ and $q$ are the highest weight fugacities of $\mathfrak{su}(2)$ and $\mathfrak{u}(1)$ respectively. The unrefined Hilbert series is 
\begin{equation}
     \mathrm{HS}[cg_2]=\frac{1+2 t^2+8 t^4+2 t^6+t^8}{(1-t^2)^2(1-t^4)^2}
\end{equation}
and the refined PL is given in Table \ref{ag2PL}. 
For $cg_2$ the quiver is
\begin{equation}
    \begin{tikzpicture}
    \node at (-2,0.5) {$cg_2$:};
    \node[gauge,label=below:{$1$}] (1) at (0,0) {};
    \node[gauge,label=below:{$1$}] (2) at (1,0) {};
    \node[flavour,label=right:{$1$}] (2a) at (0,1) {};
    \draw (2)--(1)--(2a) (0.6,0.2)--(0.4,0)--(0.6,-0.2);
    \draw[transform canvas={yshift=-2pt}] (1)--(2);
    \draw[transform canvas={yshift=2pt}] (1)--(2);
    \end{tikzpicture}
\end{equation}
where the global symmetry is $\mathfrak{su}(2)\times \mathfrak{u}(1)$. The HWG is:
\begin{eqnarray}
    \mathrm{HWG}[ag_2] &=& \frac{1-\mu^6 t^6}{\left(1-t^2\right) (1-\mu^2t^2) \left(1-  q^{-1} \mu^3 t^3 \right) \left(1-q \mu^3 t^3\right)} \\  &=&\mathrm{PE}[t^2 + \mu^2 t^2 + (q + q^{-1})\mu^3 t^3 - \mu^6 t^6]  \nonumber 
\end{eqnarray}
where $\mu$ and $q$ are the highest weight fugacities of $\mathfrak{su}(2)$ and $\mathfrak{u}(1)$ respectively. The unrefined Hilbert series takes the form:
\begin{equation}
     \mathrm{HS}[ag_2]=\frac{1+2t^2+6t^3+2t^4+t^6}{(1-t^2)^2(1-t^3)^2}
\end{equation}
with the refined PL given in Table \ref{acnPL}. We note that the $cg_2$ singularity is the orbifold $\mathbb{C}^4/\mathbb{Z}_3$ where $\mathbb{Z}_3$ acts on $\mathbb{C}^4$ with charges $(+1,+1,-1,-1)$. In the next paragraph we use this observation to define two families of singularities which contain $ac_n$, $ag_2$ and $cg_2$.

\subsubsection*{The $h_{n,k}$ and $\bar{h}_{n,k}$ slices}
The $ac_n$, $ag_2$ and $cg_2$ singularities, along with Kleinian singularities and closures of minimal nilpotent orbits are all the singularities that occur as elementary slices in affine Grassmannians. However, this does not mean that we have a complete list of \emph{all possible} elementary slices for symplectic singularities. In \cite{Bourget:2020asf}, we introduced a two parameter family $h_{n,k} \cong \mathbb{H}^n / \mathbb{Z}_k$ of singularities which generalizes the $c_n$ singularities ($c_n = h_{n,2}$). They were used to characterize certain elementary slices in Higgs branches of 4d $\mathcal{N}=2$ rank 1 SCFTs. 
The quiver for $h_{n,k}$ is  
\begin{equation}
\label{hquiver}
  h_{n,k}:   \raisebox{-.5\height}{
\begin{tikzpicture}
	\begin{pgfonlayer}{nodelayer}
		\node [style=gauge3] (0) at (0, 0) {};
		\node [style=gauge3] (1) at (1, 0) {};
		\node [style=gauge3] (3) at (-1, 0) {};
		\node [style=none] (4) at (-2, 0) {$\dots$};
		\node [style=gauge3] (5) at (-3, 0) {};
		\node [style=gauge3] (6) at (-4, 0) {};
		\node [style=none] (9) at (-2.5, 0) {};
		\node [style=none] (10) at (-1.5, 0) {};
		\node [style=none] (11) at (0, 0.1) {};
		\node [style=none] (12) at (1, 0.1) {};
		\node [style=none] (13) at (1, -0.1) {};
		\node [style=none] (14) at (0, -0.1) {};
		\node [style=none] (15) at (0.7, 0.25) {};
		\node [style=none] (16) at (0.7, -0.25) {};
		\node [style=none] (17) at (0.45, 0) {};
		\node [style=none] (18) at (-4, 1.5) {1};
		\node [style=none] (20) at (-4, -0.5) {1};
		\node [style=none] (21) at (-3, -0.5) {1};
		\node [style=none] (22) at (-1, -0.5) {1};
		\node [style=none] (23) at (0, -0.5) {1};
		\node [style=none] (24) at (1, -0.5) {1};
		\node [style=none] (25) at (-4, -0.75) {};
		\node [style=none] (27) at (1, -0.75) {};
		\node [style=flavor1] (28) at (-4, 1) {};
		\node [style=none] (29) at (-1.5, -1.25) {$n$};
		\node [style=none] (30) at (0.5, 0.6) {$k$};
	\end{pgfonlayer}
	\begin{pgfonlayer}{edgelayer}
		\draw (6) to (5);
		\draw (5) to (9.center);
		\draw (10.center) to (3);
		\draw (3) to (0);
		\draw (0) to (1);
		\draw (11.center) to (12.center);
		\draw (13.center) to (14.center);
		\draw (15.center) to (17.center);
		\draw (17.center) to (16.center);
		\draw (28) to (6);
		\draw [style=brace] (27.center) to (25.center);
	\end{pgfonlayer}
\end{tikzpicture}
}
\end{equation}
with $\mathfrak{u}(n)$ global symmetry. 

Along a similar line of logic, we expect the generalization of $ac_n$ and $ag_2$ a two parameter family of singularities that we call $\bar{h}_{n,k}$, with $a_2=\bar{h}_{n,1}$, $ac_n = \bar{h}_{n,2}$ and $ag_2 = \bar{h}_{2,3}$. The quiver for $\bar{h}_{n,k}$ is
\begin{equation}
\label{hbarquiver}
  \bar{h}_{n,k}:   \raisebox{-.5\height}{
\begin{tikzpicture}
	\begin{pgfonlayer}{nodelayer}
		\node [style=gauge3] (0) at (0, 0) {};
		\node [style=gauge3] (1) at (1, 0) {};
		\node [style=gauge3] (3) at (-1, 0) {};
		\node [style=none] (4) at (-2, 0) {$\dots$};
		\node [style=gauge3] (5) at (-3, 0) {};
		\node [style=gauge3] (6) at (-4, 0) {};
		\node [style=none] (9) at (-2.5, 0) {};
		\node [style=none] (10) at (-1.5, 0) {};
		\node [style=none] (11) at (0, 0.1) {};
		\node [style=none] (12) at (1, 0.1) {};
		\node [style=none] (13) at (1, -0.1) {};
		\node [style=none] (14) at (0, -0.1) {};
		\node [style=none] (15) at (0.7, 0.25) {};
		\node [style=none] (16) at (0.7, -0.25) {};
		\node [style=none] (17) at (0.45, 0) {};
		\node [style=none] (18) at (-4, 1.5) {1};
		\node [style=none] (20) at (-4, -0.5) {1};
		\node [style=none] (21) at (-3, -0.5) {1};
		\node [style=none] (22) at (-1, -0.5) {1};
		\node [style=none] (23) at (0, -0.5) {1};
		\node [style=none] (24) at (1, -0.5) {1};
		\node [style=none] (25) at (-4, -0.75) {};
		\node [style=none] (27) at (1, -0.75) {};
		\node [style=flavor1] (28) at (-4, 1) {};
		\node [style=none] (29) at (-1.5, -1.25) {$n$};
		\node [style=flavor1] (30) at (1, 1) {};
		\node [style=none] (31) at (1, 1.5) {1};
		\node [style=none] (32) at (0.5, 0.5) {$k$};
	\end{pgfonlayer}
	\begin{pgfonlayer}{edgelayer}
		\draw (6) to (5);
		\draw (5) to (9.center);
		\draw (10.center) to (3);
		\draw (3) to (0);
		\draw (0) to (1);
		\draw (11.center) to (12.center);
		\draw (13.center) to (14.center);
		\draw (15.center) to (17.center);
		\draw (17.center) to (16.center);
		\draw (28) to (6);
		\draw [style=brace] (27.center) to (25.center);
		\draw (30) to (13.center);
	\end{pgfonlayer}
\end{tikzpicture}
}
\end{equation}
where the global symmetry is also $\mathfrak{u}(n)$. The Coulomb branch is no longer an orbifold. 

The HWGs for the singularities $h_{n,k}$ and $\bar{h}_{n,k}$ are 
\begin{eqnarray}
\mathrm{HWG}[h_{n,k}] &=& \mathrm{PE}\left[ t^2 + \mu_1\mu_{n-1}t^2 + (q^{-1} \mu_{1}^k +q \mu_{n-1}^k) t^{k} - \mu_{1}^k  \mu_{n-1}^k t^{2k}\right] \\ 
\mathrm{HWG}[\bar{h}_{n,k}] &=& \mathrm{PE}\left[ t^2 + \mu_1\mu_{n-1}t^2 + (q^{-1} \mu_{1}^k +q \mu_{n-1}^k) t^{k+1} - \mu_{1}^k  \mu_{n-1}^k t^{2k+2}\right]
\end{eqnarray}
where $\mu_i$ and $q$ are the highest weight fugacities of $\mathfrak{u}(n)\cong \mathfrak{su}(n)\times \mathfrak{u}(1)$. The terms in the HWGs have a natural interpretation: $t^2 + \mu_1\mu_{n-1}t^2$ corresponds to the adjoint of the global symmetry $\mathfrak{u}(n)$. The term $q^{-1} \mu_{1}^k  t^{k}$ (respectively $q^{-1} \mu_{1}^k  t^{k+1}$) corresponds to the rightmost gauge node in the quiver (\ref{hquiver}) (resp. (\ref{hbarquiver})), which have imbalance $k-2$ (resp. $k-1$). The order of the symmetric product follows from the order of the non simply laced edge. This representation is complex, so charge conjugation requires the addition of the term $q \mu_{n-1}^k  t^{k}$ (resp. $q \mu_{n-1}^k  t^{k+1}$), and of a relation $-  \mu_{1}^k \mu_{n-1}^k  t^{2k}$ (resp. $-  \mu_{1}^k \mu_{n-1}^k  t^{2k+2}$). 

A highest weight variety for $h_{n,k}$ is $\mathbb{C} \times \mathbb{C}^2/\mathbb{Z}_k$, with equation $xy=z^k$. A highest weight variety for $\bar{h}_{n,k}$ is the threefold defined by the equation $xy=z^k w$ for $(x,y,z,w) \in \mathbb{C}^4$. 

\subsubsection*{Unframed quivers.}
So far we have considered quivers with an explicit framing (flavor nodes).  
Magnetic quivers for higher dimensional supersymmetric gauge theories ($4d$ $\mathcal{N}=2$, $5d$ $\mathcal{N}=1$ and $6d$ $\mathcal{N}=(1,0)$ theories) usually arise from brane configurations, and are more conveniently written without framing. For unframed quivers, it is understood that a $\mathrm{U}(1)$ node has to be ungauged on a long node \cite{Bourget:2019aer,Bourget:2020mez}. The quivers discussed in this section are summarized in both framed and unframed form in Table \ref{tab:elem}.

\section{Branes}
\label{sec:branes}
In this section we construct brane systems for the affine Grassmannian of classical groups consisting of NS5, D5, and D3 branes in Type IIB String Theory as first developed in \cite{Hanany:1996ie}, possibly in presence of an ON plane. The method of reading quivers from such brane systems, including quivers for transverse slices, is reviewed in Appendix \ref{app:orientifolds}.

\paragraph{Affine Grassmannian.} As discussed in Section \ref{sec:affGr}, the affine Grassmannian $\mathrm{Gr}_G$ of a group $G$ consists of several connected components, one for 
each lowest dominant coweight $\Omega_I$, see Table \ref{tab:LDC}. Each connected component of the affine Grassmannian is infinite dimensional, as are the transverse slices $\w{G}{}{\lambda}$. The transverse slices $\wb{G}{\mu}{\lambda}$ however are finite dimensional and can be constructed as a Coulomb branch\footnote{Higgs branch constructions are possible, but are not discussed in this paper.} of a quiver theory as discussed in Section \ref{sec:quivers}. Every such quiver describes the low energy theory of a brane system, which can be obtained using simple rules (see Appendix \ref{app:orientifolds}). The goal of this section however, is to turn this around and to generate the affine Grassmannian, or rather any $\w{G}{}{\lambda}$, from a brane system. For every component $I$ in the affine Grassmannian a brane system is proposed whose phases correspond to the symplectic leaves of $\w{G}{}{\Omega_I}$, the transverse slices to the lowest Schubert cells. From this brane system all of $\w{G}{}{\Omega_I}$ can be constructed. Since the objective is to construct an infinite dimensional space, an infinite number of moduli (D3 branes) must be present in the brane system. To study the symplectic leaves of $\w{G}{}{\Omega_I}$ from bottom-up, we restrict the movements of almost every D3 brane and only turn on few moduli at a time, moving from one symplectic leaf of $\w{G}{}{\Omega_I}$ to another. Starting at the lowest leaf, i.e.\ all D3 branes rest at the origin, this process generates the transverse slices to the lowest leaf in the component at hand. This bottom-up construction solely uses Kraft-Procesi transitions \cite{kraft1980minimal,Kraft1982,Cabrera:2016vvv} in brane systems, without the need to rely on the results of Section \ref{sec:quivers}.  As a result, the affine Grassmannian arises naturally in brane systems.

All other finite slices $\wb{G}{\mu}{\lambda}$ in the affine Grassmannian are contained in the brane systems as (generically non-elementary) Kraft-Procesi transitions. Furthermore any $\w{G}{}{\lambda}$ may be constructed bottom-up from the proposed brane systems by first moving to the phase in the brane system which corresponds to $\w{G}{\lambda}{\Omega_I}$, ignoring the already turned on D3 moduli, and proceeding with Kraft-Procesi transitions.

\subsection{\texorpdfstring{The $A$-type affine Grassmannian}{The A-type affine Grassmannian}}

The affine Grassmannian of $\mathrm{PSL}(n,\mathbb{C})$ consists of $n$ connected components. We propose a brane system for each such component. More precisely, we propose a brane system for $\w{\mathrm{PSL}(n,\mathbb{C})}{}{\Omega_I}$ for every lowest dominant coweight $\Omega_I$. 

\paragraph{Component $0\in\mathbb{Z}_n$.} The component connected to the origin (component $0$) is in fact the affine Grassmannian of $\mathrm{SL}(n,\mathbb{C})$.
The proposed brane system for 
$\w{\mathrm{PSL}(n,\mathbb{C})}{}{[0,\dots,0]}$ consists of $n$ NS5 branes with an infinite number of D5 branes on both sides, and an infinite number of D3 branes extended between every neighbouring pair of D5 branes. In detail, 
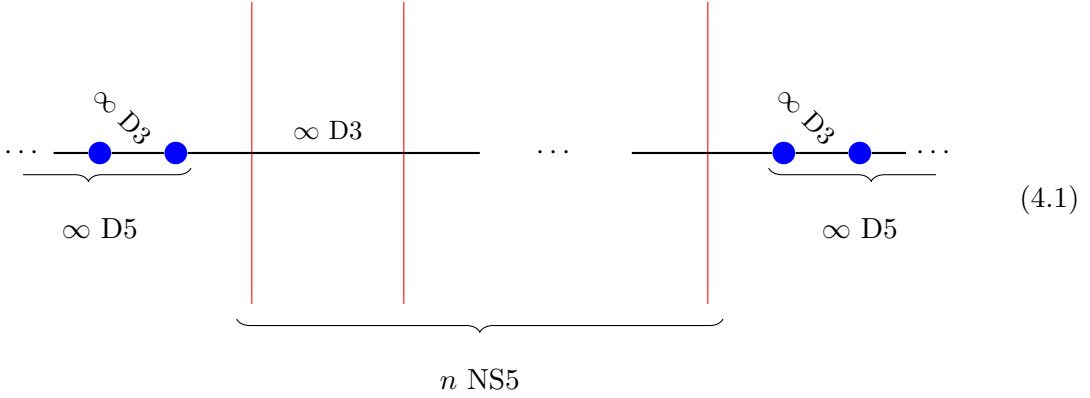
\begin{equation}
\label{braneInfinity}
    \vcenter{\hbox{
	\begin{tikzpicture}
		\draw[red] (0,0)--(0,4) (2,0)--(2,4) (6,0)--(6,4);
		\node at (4,2) {$\cdots$};
		\node (1) at (-3,2) {$\cdots$};
		\node[D5] (2) at (-2,2) {};
		\node[D5] (3) at (-1,2) {};
		\node[D5] (4) at (7,2) {};
		\node[D5] (5) at (8,2) {};
		\node (6) at (9,2) {$\cdots$};
		\draw[thick] (1)--(2)--(3)--(3,2) (5,2)--(4)--(5)--(6);
		\node[rotate=-45] at (-1.7,2.5) {\small $\infty$ D3};
		\node at (1,2.3) {\small $\infty$ D3};
		\node[rotate=-45] at (7.3,2.5) {\small $\infty$ D3};
		\draw [decorate,decoration={brace,amplitude=5pt}] (6.2,-0.2)--(-0.2,-0.2);
		\node at (3,-1) {$n$ NS5};
		\begin{scope}
		\clip (-3,1) rectangle(3,2);
		\draw [decorate,decoration={brace,amplitude=5pt}] (-0.8,1.8)--(-3.5,1.8);
		\end{scope}
		\node at (-2,1) {$\infty$ D5};
		\begin{scope}
		\clip (6,1) rectangle(9,2);
		\draw [decorate,decoration={brace,amplitude=5pt}] (9.5,1.8)--(6.8,1.8);
		\end{scope}
		\node at (8,1) {$\infty$ D5};
	\end{tikzpicture}}}
\end{equation}
When drawing the brane system, we generally do not depict branes which are not essential to study the symplectic leaf at hand, i.e.\ we suppress those branes resting at the origin. Depicting the origin of $\w{\mathrm{PSL}(n,\mathbb{C})}{}{[0,\dots,0]}$ and suppressing all irrelevant branes leads to the following:
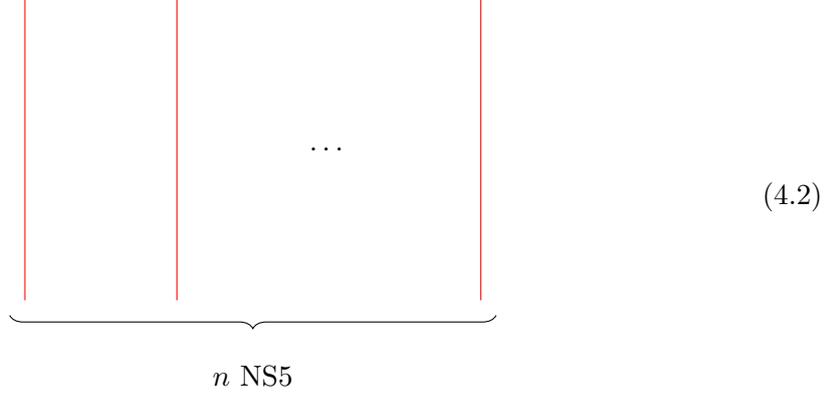
\begin{equation}
    \vcenter{\hbox{
	\begin{tikzpicture}
		\draw[red] (0,0)--(0,4) (2,0)--(2,4) (6,0)--(6,4);
		\node at (4,2) {$\cdots$};
		\draw [decorate,decoration={brace,amplitude=5pt}] (6.2,-0.2)--(-0.2,-0.2);
		\node at (3,-1) {$n$ NS5};
	\end{tikzpicture}}}
	\label{eq:AFirstCompGeneral}
\end{equation}
For this brane system we read the empty quiver:
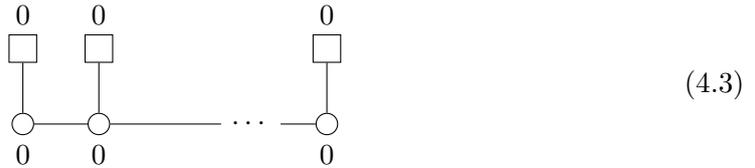
\begin{equation}
    \vcenter{\hbox{
	\begin{tikzpicture}
		\node[gauge,label=below:{$0$}] (g1) at (0,0) {};
		\node[flavour,label=above:{$0$}] (f1) at (0,1) {};
		\node[gauge,label=below:{$0$}] (g2) at (1,0) {};
		\node[flavour,label=above:{$0$}] (f2) at (1,1) {};
		\node[gauge,label=below:{$0$}] (g3) at (4,0) {};
		\node[flavour,label=above:{$0$}] (f3) at (4,1) {};
		\node (g) at (3,0) {$\cdots$};
		\draw (f1)--(g1)--(g2)--(g)--(g3)--(f3) (g2)--(f2);
	\end{tikzpicture}}}
\end{equation}
i.e.\ the origin of $\w{\mathrm{PSL}(n,\mathbb{C})}{}{[0,\dots,0]}$ corresponding to the lowest dominant coweight $\Omega_0=[0,0,\dots,0]$.

\paragraph{Other components.} The transverse slices to the lowest leaf in the remaining connected components of the affine Grassmannian of $\mathrm{PSL}(n,\mathbb{C})$ can be constructed from a brane system where a single D5 is in one of the intervals between two neighbouring NS5 branes. For example, placing a D5 brane in the first interval we obtain the following brane system:
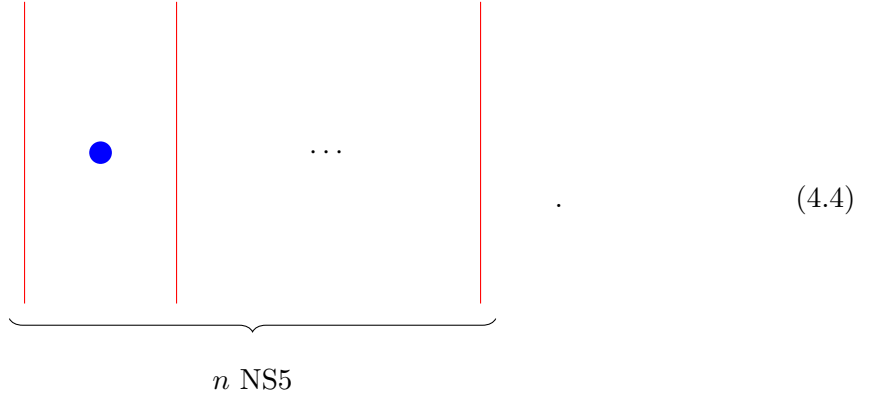
\begin{equation}
    \vcenter{\hbox{
	\begin{tikzpicture}
		\draw[red] (0,0)--(0,4) (2,0)--(2,4) (6,0)--(6,4);
		\node at (4,2) {$\cdots$};
		\draw [decorate,decoration={brace,amplitude=5pt}] (6.2,-0.2)--(-0.2,-0.2);
		\node at (3,-1) {$n$ NS5};
		\node[D5] at (1,2) {};
	\end{tikzpicture}}}\qquad .
	\label{eq:ASecondCompGeneral}
\end{equation}
For this brane system we read the quiver:
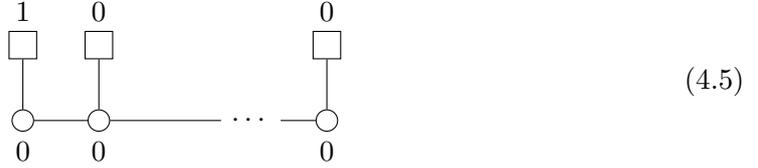
\begin{equation}
    \vcenter{\hbox{
	\begin{tikzpicture}
		\node[gauge,label=below:{$0$}] (g1) at (0,0) {};
		\node[flavour,label=above:{$1$}] (f1) at (0,1) {};
		\node[gauge,label=below:{$0$}] (g2) at (1,0) {};
		\node[flavour,label=above:{$0$}] (f2) at (1,1) {};
		\node[gauge,label=below:{$0$}] (g3) at (4,0) {};
		\node[flavour,label=above:{$0$}] (f3) at (4,1) {};
		\node (g) at (3,0) {$\cdots$};
		\draw (f1)--(g1)--(g2)--(g)--(g3)--(f3) (g2)--(f2);
	\end{tikzpicture}}}
\end{equation}
which represents the origin of $\wb{\mathrm{PSL}_n}{}{[1,0,\dots,0]}$, corresponding to the lowest dominant coweight $\Omega_0=[1,0,\dots,0]$.

The two brane systems in \eqref{eq:AFirstCompGeneral} and \eqref{eq:ASecondCompGeneral} contain an infinite number of D5 branes; however, they differ as the D5 brane depicted in \eqref{eq:ASecondCompGeneral} has a different linking number\footnote{All prescriptions for linking numbers are equivalent and can be used.} than any of the D5 branes in \eqref{eq:AFirstCompGeneral}.

\paragraph{Quotients of $\mathrm{SL}(n,\mathbb{C})$.}
For quotients $\mathrm{SL}(n,\mathbb{C})/\mathbb{Z}_k$, where $k$ divides $n$, $k$ connected components make up its affine Grassmannian. The starting points for the respective components are those with a single D5 placed in the $I$-th interval satisfying the condition 
\begin{equation}
    I\textnormal{ mod }\frac{n}{k}=0\;,
\end{equation}
which has $k$ solutions for $I\in\{0,\dots,n-1\}$.

For example, the possible starting intervals for a D5 (depicted by an $\times$) to be placed for a respective component of the affine Grassmannian of $\mathrm{SL}(6,\mathbb{C})/\mathbb{Z}_k$ are as follows: 
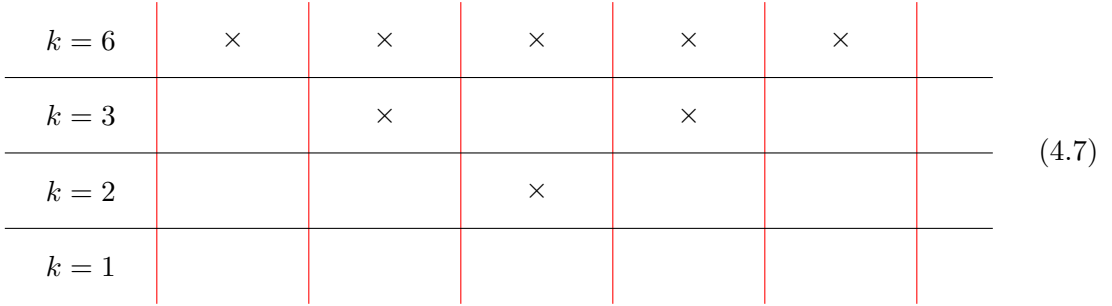
\begin{equation}
    \vcenter{\hbox{
	\begin{tikzpicture}
		\draw[red] (0,0)--(0,4) (2,0)--(2,4) (4,0)--(4,4) (6,0)--(6,4) (8,0)--(8,4) (10,0)--(10,4);
		\draw (-2,3)--(11,3) (-2,2)--(11,2) (-2,1)--(11,1);
		\node at (-1,3.5) {$k=6$};
		\node at (1,3.5) {$\times$};
		\node at (3,3.5) {$\times$};
		\node at (5,3.5) {$\times$};
		\node at (7,3.5) {$\times$};
		\node at (9,3.5) {$\times$};
		\node at (-1,2.5) {$k=3$};
		\node at (3,2.5) {$\times$};
		\node at (7,2.5) {$\times$};
		\node at (-1,1.5) {$k=2$};
		\node at (5,1.5) {$\times$};
		\node at (-1,0.5) {$k=1$};
	\end{tikzpicture}}}
	\label{eq:ComponentsSL6}
\end{equation}
Note that a brane system without a D5 in any interval corresponds to the component connected to the origin, the $0$-th component, which is always present for any $\mathrm{SL}(n,\mathbb{C})/\mathbb{Z}_k$.

\paragraph{Identifying the component.} Given a brane system for any slice in the affine Grassmannian of $\mathrm{PSL}(n,\mathbb{C})$, with $n_l$ D5 branes in the $l$-th interval, there is a simple formula to compute which component the corresponding slice belongs to:
\begin{equation}
    \sum_{l=1}^{n-1}ln_l\textnormal{ mod }n=I\qquad,
    \label{eq:ComponentLabelling}
\end{equation}
where $I$ labels the component. Now that we addressed the various components of the A-type affine Grassmannian, we can start exploring the stratification of each component.

\paragraph{Symplectic leaves, minimal transitions.} In order to study the different leaves of a given transverse slice in the affine Grassmannian one can start turning on Coulomb branch moduli of the brane system. However, this has to be done in a systematic way in order to identify every leaf. One has to carefully turn on a `minimal' amount of moduli in the following two ways.

In order to open up Coulomb branch directions, one needs to consider an interval of D5 branes that contains at least 2 NS5 branes. In order to open up a minimal direction, the D5 branes need to be the closest two D5 branes separated by at least two NS5 branes. This leaves two options:
\begin{enumerate}
    \item Activating a D3 brane between two neighbouring D5 branes (i.e.\ there is no D5 brane between them) separated by $p$ NS5 branes, where $p\geq 2$. This corresponds to the transverse slice $a_{p-1}$.
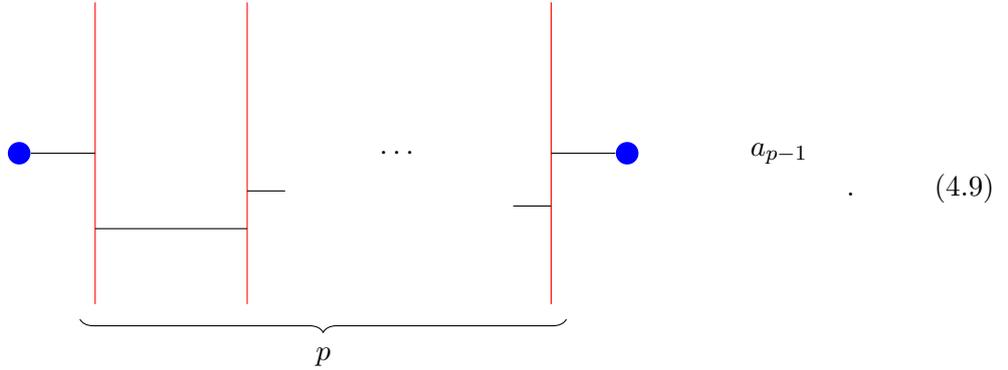
\begin{equation}
    \vcenter{\hbox{
	\begin{tikzpicture}
		\draw[red] (0,0)--(0,4) (2,0)--(2,4) (6,0)--(6,4);
		\node at (4,2) {$\cdots$};
		\draw [decorate,decoration={brace,amplitude=5pt}] (6.2,-0.2)--(-0.2,-0.2);
		\node[D5] (1) at (-1,2) {};
		\node[D5] (2) at (7,2) {};
		\draw (1)--(0,2) (0,1)--(2,1) (6,2)--(2) (2,1.5)--(2.5,1.5) (5.5,1.3)--(6,1.3);
		\node at (3,-0.7) {$p$};
		\node at (9,2) {$a_{p-1}$};
	\end{tikzpicture}}}\quad.
	\label{eq:antrans}
\end{equation}
    \item Activating a D3 brane between two D5 branes which have exactly 2 NS5 branes between them, with possibly $q\geq0$ D5 branes between the two NS5 branes. This corresponds to a transverse slice $A_{q+1}$.
\begin{equation}
    \vcenter{\hbox{
    \begin{tikzpicture}
        \draw[red] (0,0)--(0,4) (2,0)--(2,4);
        \node[D5] (1) at (-1,2) {};
        \node[D5] at (0.5,2) {};
        \draw [decorate,decoration={brace,amplitude=5pt}] (0.25,2.25)--(1.75,2.25);
        \node at (1,2.75) {$q$};
        \node at (1,2) {$\cdots$};
        \node[D5] at (1.5,2) {};
        \node[D5] (2) at (3,2) {};
        \draw (1)--(0,2) (0,1)--(2,1) (2,2)--(2);
        \node at (5,2) {$A_{q+1}$};
    \end{tikzpicture}}}\quad.
    \label{eq:Antrans}
\end{equation}
Clearly \eqref{eq:antrans} for $p=2$ and \eqref{eq:Antrans} for $q=0$ are the same transition, as $a_1=A_1$.
\end{enumerate}
This allows a transition from one leaf to another. The quiver can be read from the brane system. Its Coulomb branch contains the moduli which are turned on, and the moduli space is the closure of the corresponding leaf. This is analysed in detail for rank $1$ and $2$ in the following two sections.

\subsubsection{\texorpdfstring{The $A_1$ affine Grassmannian}{The A1 affine Grassmannian}}
\label{sec:A1branes}

The affine Grassmannian for $\mathrm{PSL}(2,\mathbb{C})$ has two connected components, corresponding to the two lowest dominant coweights, $[0]$ and $[1]$.

\paragraph{Component $0\in\mathbb{Z}_2$.} $\w{\mathrm{PSL}(2,\mathbb{C})}{}{[0]}$, or the $0$-th component of the affine Grassmannian of $\mathrm{PSL}(2,\mathbb{C})$, which is the affine Grassmannian of $\mathrm{SL}(2,\mathbb{C})$, is given by a brane system with infinitely many D3 branes which are free to move between two NS5 branes in the presence of D5 branes. The origin of $\w{\mathrm{PSL}(2,\mathbb{C})}{}{[0]}$ is where all these infinitely many branes are coincident. Let us draw the two NS5 branes and suppress the infinitely many D3 branes at the origin, as well as infinitely many D5 branes to the left and right.
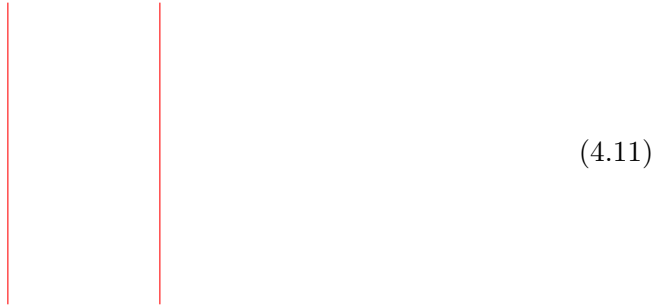
\begin{equation}
    \vcenter{\hbox{
    \begin{tikzpicture}
        \draw[red] (0,0)--(0,4) (2,0)--(2,4);
    \end{tikzpicture}}}
\end{equation}
We move to the lowest non-trivial leaf by pulling a D3 brane from the stack of infinitely many branes at the origin and placing it between two D5 branes and letting it slide along the NS5 branes, corresponding to an $A_1$ transition:
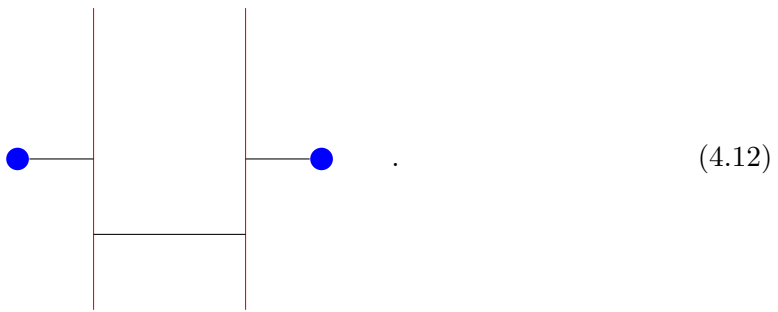
\begin{equation}
    \vcenter{\hbox{
    \begin{tikzpicture}
        \draw[red] (0,0)--(0,4) (2,0)--(2,4);
        \node[D5] (1) at (-1,2) {};
        \node[D5] (2) at (3,2) {};
        \draw (1)--(0,2) (2,2)--(2) (0,1)--(2,1);
    \end{tikzpicture}}}\qquad.
\end{equation}
The stuck D3 branes between the two pairs of D5 and NS5 can be annihilated by a Hanany-Witten transition and we can read the quiver whose Coulomb branch is the closure of the leaf:
\begin{equation}
    \vcenter{\hbox{
    \begin{tikzpicture}
        \draw[red] (0,0)--(0,4) (2,0)--(2,4);
        \node[D5] (1) at (0.5,2) {};
        \node[D5] (2) at (1.5,2) {};
        \draw (0,1)--(2,1);
    \end{tikzpicture}}}
    \qquad
    \vcenter{\hbox{
    \begin{tikzpicture}
        \node[gauge,label=below:{$1$}] (1) at (0,0) {};
        \node[flavour,label=above:{$2$}] (2) at (0,1) {};
        \draw (1)--(2);
    \end{tikzpicture}
    }}\qquad.
\end{equation}
The only way to keep moving onto higher leaves in the affine Grassmannian is by repeating the same process. However, the D5 branes already placed between the NS5 branes make the transverse slices $A_{2k-1}$ for the $k$-th transition of this type. A general leaf and the corresponding quiver are:
\begin{equation}
    \vcenter{\hbox{
    \begin{tikzpicture}
        \draw[red] (0,0)--(0,4) (2,0)--(2,4);
        \node[D5] (1) at (0.5,2) {};
        \node at (1,2.75) {$2k$};
        \draw [decorate,decoration={brace,amplitude=5pt}] (0.25,2.25)--(1.75,2.25);
        \node at (1,2) {$\cdots$};
        \node[D5] (2) at (1.5,2) {};
        \draw (0,1)--(2,1);
        \node at (1,0.75) {$\vdots$ $k$};
        \draw(0,0.3)--(2,0.3);
    \end{tikzpicture}}}
    \qquad
    \vcenter{\hbox{
    \begin{tikzpicture}
        \node[gauge,label=below:{$k$}] (1) at (0,0) {};
        \node[flavour,label=above:{$2k$}] (2) at (0,1) {};
        \draw (1)--(2);
    \end{tikzpicture}
    }}\qquad.
    \label{eq:A1_1st}
\end{equation}
Which is indeed what we find in Section \ref{sec:quivers}.

\paragraph{Component $1\in\mathbb{Z}_2$.} For the first connected component of $\mathrm{PSL}(2,\mathbb{C})$ we start with the following brane system:
\begin{equation}
    \vcenter{\hbox{
    \begin{tikzpicture}
        \draw[red] (0,0)--(0,4) (2,0)--(2,4);
        \node[D5] at (1,2) {};
    \end{tikzpicture}}}\qquad , 
\end{equation}
which depicts the origin of $\w{\mathrm{PSL}(2,\mathbb{C})}{}{[1]}$. The transverse slice for the $k$-th transition is $A_{2k}$, while the brane system and quiver for the leaf are:
\begin{equation}
    \vcenter{\hbox{
    \begin{tikzpicture}
        \draw[red] (0,0)--(0,4) (2,0)--(2,4);
        \node[D5] (1) at (0.5,2) {};
        \node at (1,2.75) {$2k+1$};
        \draw [decorate,decoration={brace,amplitude=5pt}] (0.25,2.25)--(1.75,2.25);
        \node at (1,2) {$\cdots$};
        \node[D5] (2) at (1.5,2) {};
        \draw (0,1)--(2,1);
        \node at (1,0.75) {$\vdots$ $k$};
        \draw(0,0.3)--(2,0.3);
    \end{tikzpicture}}}
    \qquad
    \vcenter{\hbox{
    \begin{tikzpicture}
        \node[gauge,label=below:{$k$}] (1) at (0,0) {};
        \node[flavour,label=above:{$2k+1$}] (2) at (0,1) {};
        \draw (1)--(2);
    \end{tikzpicture}
    }}\qquad.
    \label{eq:A1_2nd}
\end{equation}
The quivers in \eqref{eq:A1_1st} correspond to slices to the lowest leaf (origin) in the $0$-th component of the affine Grassmannian, while the quivers in \eqref{eq:A1_2nd} correspond to slices to the lowest leaf in the $1$-st component. The other slices, between any two leaves in a connected component, are realized as (generically non-minimal) Kraft-Procesi transitions \cite{Cabrera:2016vvv} in the brane system. The form of the quiver is a framed $A_1$ quiver, or an SQCD theory with $\mathrm{U}(k)$ gauge group and $N$ flavors, satisfying $N\ge2k$. $k$ is the number of minimal slices between the two leaves, and $N$ is even for slices in the $0$-th component, while $N$ is odd for slices in the $1$-st component. In this way every good quiver of $A_1$ type is realized in one of the two brane systems corresponding to the two connected components of the affine Grassmannian of $\mathrm{SL}(2,\mathbb{C})$.

\subsubsection{\texorpdfstring{The $A_2$ affine Grassmannian}{The A2 affine Grassmannian}}

The affine Grassmannian of $\mathrm{PSL}(3,\mathbb{C})$ has 3 disconnected components.

\paragraph{Component $0 \in \mathbb{Z}_3$.} The brane systems for the lower symplectic leaves and elementary slices between them for the component connected to the origin are depicted in Figure \ref{fig:A2branes}.

\paragraph{Component $1,2 \in \mathbb{Z}_3$.} The brane systems for the lower symplectic leaves and elementary slices between them for one of the components not connected to the origin are depicted in Figure \ref{fig:A2branes2}. The brane system for the other component is related to the one depicted in Figure \ref{fig:A2branes2} by a $\mathbb{Z}_2$ action $x^6\mapsto-x^6$, and we do not draw it.

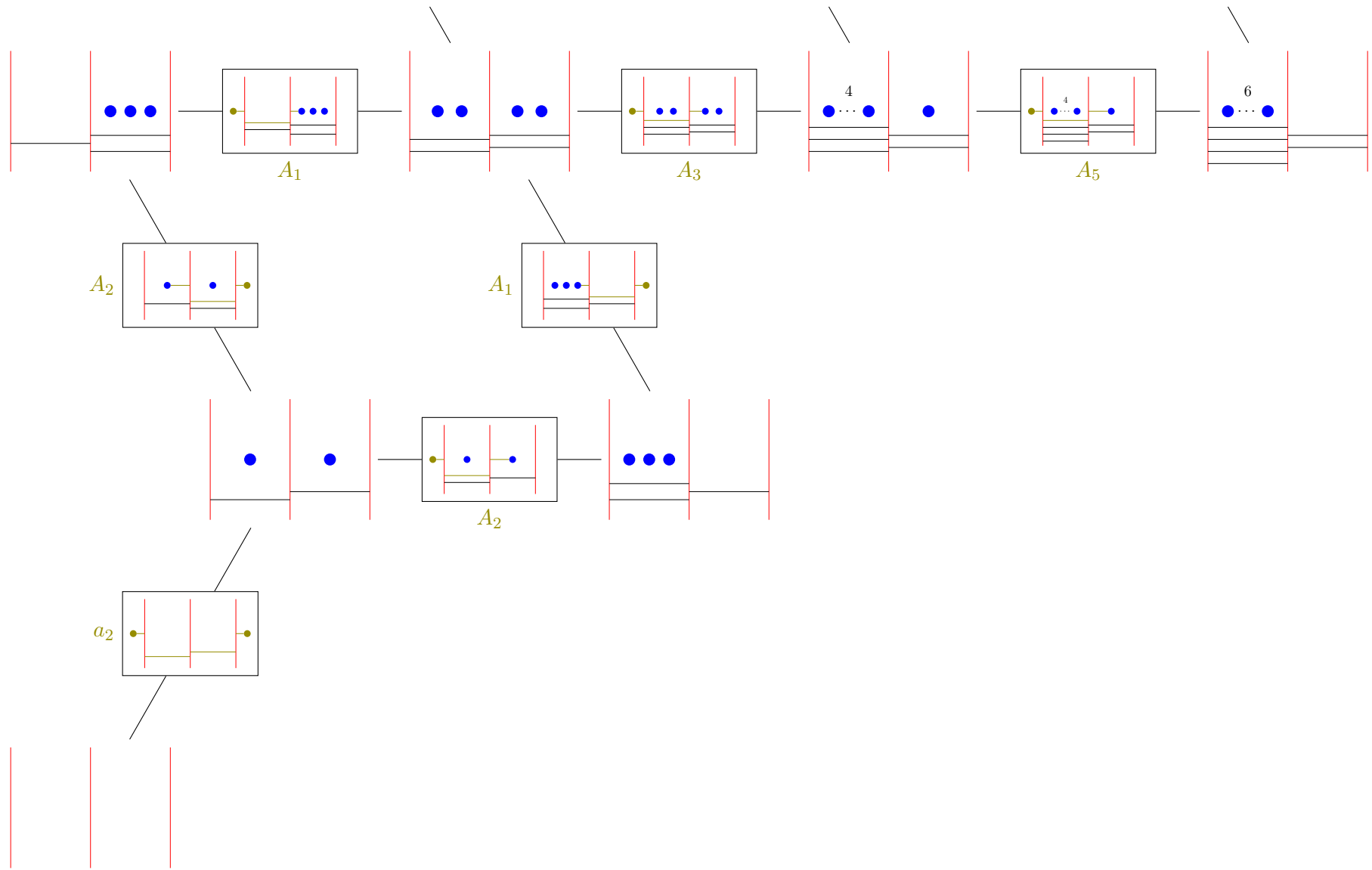
\begin{landscape}

\begin{figure}[t]
    \centering
    \begin{adjustbox}{center}
    \begin{tikzpicture}
    \def\x{7*1cm};
    \def\y{0.866*\x};
        \node (00) at (0*\x,0*\y) {$\scalebox{0.7}{\begin{tikzpicture}
            \draw[red] (0,0)--(0,3) (2,0)--(2,3) (4,0)--(4,3);
        \end{tikzpicture}}$};
        \node (11) at (.5*\x,1*\y) {$\scalebox{0.7}{\begin{tikzpicture}
            \draw[red] (0,0)--(0,3) (2,0)--(2,3) (4,0)--(4,3);
            \node[D5] at (1,1.5) {};
            \node[D5] at (3,1.5) {};
            \draw (0,0.5)--(2,0.5) (2,0.7)--(4,0.7);
        \end{tikzpicture}}$};
        \node (30) at (1.5*\x,1*\y) {$\scalebox{0.7}{\begin{tikzpicture}
            \draw[red] (0,0)--(0,3) (2,0)--(2,3) (4,0)--(4,3);
            \node[D5] at (0.5,1.5) {};
            \node[D5] at (1,1.5) {};
            \node[D5] at (1.5,1.5) {};
            \draw (0,0.5)--(2,0.5) (0,0.9)--(2,0.9) (2,0.7)--(4,0.7);
        \end{tikzpicture}}$};
        \node (03) at (0*\x,2*\y) {$\scalebox{0.7}{\begin{tikzpicture}
            \begin{scope}[xscale=-1]
            \draw[red] (0,0)--(0,3) (2,0)--(2,3) (4,0)--(4,3);
            \node[D5] at (0.5,1.5) {};
            \node[D5] at (1,1.5) {};
            \node[D5] at (1.5,1.5) {};
            \draw (0,0.5)--(2,0.5) (0,0.9)--(2,0.9) (2,0.7)--(4,0.7);
            \end{scope}
        \end{tikzpicture}}$};
        \node (22) at (1*\x,2*\y) {$\scalebox{0.7}{\begin{tikzpicture}
            \draw[red] (0,0)--(0,3) (2,0)--(2,3) (4,0)--(4,3);
            \node[D5] at (0.7,1.5) {};
            \node[D5] at (1.3,1.5) {};
            \node[D5] at (2.7,1.5) {};
            \node[D5] at (3.3,1.5) {};
            \draw (0,0.5)--(2,0.5) (0,0.8)--(2,0.8) (2,0.6)--(4,0.6) (2,0.9)--(4,0.9);
        \end{tikzpicture}}$};
        \node (41) at (2*\x,2*\y) {$\scalebox{0.7}{\begin{tikzpicture}
            \draw[red] (0,0)--(0,3) (2,0)--(2,3) (4,0)--(4,3);
            \node[D5] at (0.5,1.5) {};
            \node at (1,1.5) {$\cdots$};
            \node at (1,2) {$4$};
            \node[D5] at (1.5,1.5) {};
            \node[D5] at (3,1.5) {};
            \draw (0,0.5)--(2,0.5) (0,0.8)--(2,0.8) (0,1.1)--(2,1.1) (2,0.6)--(4,0.6) (2,0.9)--(4,0.9);
        \end{tikzpicture}}$};
        \node (60) at (3*\x,2*\y) {$\scalebox{0.7}{\begin{tikzpicture}
            \draw[red] (0,0)--(0,3) (2,0)--(2,3) (4,0)--(4,3);
            \node[D5] at (0.5,1.5) {};
            \node at (1,1.5) {$\cdots$};
            \node at (1,2) {$6$};
            \node[D5] at (1.5,1.5) {};
            \draw (0,0.5)--(2,0.5) (0,0.8)--(2,0.8) (0,1.1)--(2,1.1) (0,0.2)--(2,0.2) (2,0.6)--(4,0.6) (2,0.9)--(4,0.9);
        \end{tikzpicture}}$};
        \node[draw,label=left:{\color{olive}$a_2$}] (0011) at ($(00)!0.5!(11)$) {$\scalebox{0.4}{\begin{tikzpicture}
            \draw[red] (0,0)--(0,3) (2,0)--(2,3) (4,0)--(4,3);
            \node[D5o] (1) at (-0.5,1.5) {};
            \node[D5o] (2) at (4.5,1.5) {};
            \draw[olive,thick] (1)--(0,1.5) (0,0.5)--(2,0.5) (2,0.7)--(4,0.7) (4,1.5)--(2);
        \end{tikzpicture}}$};
        \node[draw,label=below:{\color{olive}$A_2$}] (1130) at ($(11)!0.5!(30)$) {$\scalebox{0.4}{\begin{tikzpicture}
            \draw[red] (0,0)--(0,3) (2,0)--(2,3) (4,0)--(4,3);
            \node[D5o] (1) at (-0.5,1.5) {};
            \node[D5] at (1,1.5) {};
            \node[D5] (2) at (3,1.5) {};
            \draw[olive,thick] (1)--(0,1.5) (0,0.8)--(2,0.8) (2,1.5)--(2);
            \draw (0,0.5)--(2,0.5) (2,0.7)--(4,0.7);
            \node at (4.5,1.5) {};
        \end{tikzpicture}}$};
        \node[draw,label=left:{\color{olive}$A_2$}] (1103) at ($(11)!0.5!(03)$) {$\scalebox{0.4}{\begin{tikzpicture}
            \begin{scope}[xscale=-1]
            \draw[red] (0,0)--(0,3) (2,0)--(2,3) (4,0)--(4,3);
            \node[D5o] (1) at (-0.5,1.5) {};
            \node[D5] at (1,1.5) {};
            \node[D5] (2) at (3,1.5) {};
            \draw[olive,thick] (1)--(0,1.5) (0,0.8)--(2,0.8) (2,1.5)--(2);
            \draw (0,0.5)--(2,0.5) (2,0.7)--(4,0.7);
            \node at (4.5,1.5) {};
            \end{scope}
        \end{tikzpicture}}$};
        \node[draw,label=left:{\color{olive}$A_1$}] (3022) at ($(30)!0.5!(22)$) {$\scalebox{0.4}{\begin{tikzpicture}
            \draw[red] (0,0)--(0,3) (2,0)--(2,3) (4,0)--(4,3);
            \node[D5] at (0.5,1.5) {};
            \node[D5] at (1,1.5) {};
            \node[D5] (1) at (1.5,1.5) {};
            \draw (0,0.5)--(2,0.5) (0,0.9)--(2,0.9) (2,0.7)--(4,0.7);
            \node[D5o] (2) at (4.5,1.5) {};
            \draw[olive,thick] (1)--(2,1.5) (4,1.5)--(2) (2,1)--(4,1);
            \node at (-0.5,1.5) {};
        \end{tikzpicture}}$};
        \node[draw,label=below:{\color{olive}$A_1$}] (0322) at ($(03)!0.5!(22)$) {$\scalebox{0.4}{\begin{tikzpicture}
            \begin{scope}[xscale=-1]
            \draw[red] (0,0)--(0,3) (2,0)--(2,3) (4,0)--(4,3);
            \node[D5] at (0.5,1.5) {};
            \node[D5] at (1,1.5) {};
            \node[D5] (1) at (1.5,1.5) {};
            \draw (0,0.5)--(2,0.5) (0,0.9)--(2,0.9) (2,0.7)--(4,0.7);
            \node[D5o] (2) at (4.5,1.5) {};
            \draw[olive,thick] (1)--(2,1.5) (4,1.5)--(2) (2,1)--(4,1);
            \node at (-0.5,1.5) {};
            \end{scope}
        \end{tikzpicture}}$};
        \node[draw,label=below:{\color{olive}$A_3$}] (2241) at ($(22)!0.5!(41)$) {$\scalebox{0.4}{\begin{tikzpicture}
            \draw[red] (0,0)--(0,3) (2,0)--(2,3) (4,0)--(4,3);
            \node[D5] at (0.7,1.5) {};
            \node[D5] at (1.3,1.5) {};
            \node[D5] (2) at (2.7,1.5) {};
            \node[D5] at (3.3,1.5) {};
            \draw (0,0.5)--(2,0.5) (0,0.8)--(2,0.8) (2,0.6)--(4,0.6) (2,0.9)--(4,0.9);
            \node[D5o] (1) at (-0.5,1.5) {};
            \draw[olive,thick] (1)--(0,1.5) (0,1.1)--(2,1.1) (2,1.5)--(2);
            \node at (4.5,1.5) {};
        \end{tikzpicture}}$};
        \node[draw,label=below:{\color{olive}$A_5$}] (4160) at ($(41)!0.5!(60)$) {$\scalebox{0.4}{\begin{tikzpicture}
            \draw[red] (0,0)--(0,3) (2,0)--(2,3) (4,0)--(4,3);
            \node[D5] at (0.5,1.5) {};
            \node at (1,1.5) {$\cdots$};
            \node at (1,2) {$4$};
            \node[D5] at (1.5,1.5) {};
            \node[D5] (2) at (3,1.5) {};
            \draw (0,0.2)--(2,0.2) (0,0.5)--(2,0.5) (0,0.8)--(2,0.8) (2,0.6)--(4,0.6) (2,0.9)--(4,0.9);
            \node[D5o] (1) at (-0.5,1.5) {};
            \draw[olive,thick] (1)--(0,1.5) (0,1.1)--(2,1.1) (2,1.5)--(2);
            \node at (4.5,1.5) {};
        \end{tikzpicture}}$};
        \draw (00)--(0011)--(11)--(1130)--(30)--(3022)--(22)--(2241)--(41)--(4160)--(60) (11)--(1103)--(03)--(0322)--(22);
        \draw (22)--(1*\x-0.3*0.5*\x,2*\y+0.3*1*\y) (41)--(2*\x-0.3*0.5*\x,2*\y+0.3*1*\y) (60)--(3*\x-0.3*0.5*\x,2*\y+0.3*1*\y);
    \end{tikzpicture}
    \end{adjustbox}
    \caption{Brane depiction of the low dimensional leaves [and elementary slices between them] of $\w{\mathrm{PSL}(3,\mathbb{C})}{}{[0,0]}$.}
    \label{fig:A2branes}
\end{figure}

\begin{figure}[t]
    \centering
    \begin{adjustbox}{center}
    \begin{tikzpicture}
    \def\x{7*1cm};
    \def\y{0.866*\x};
        \node (01) at (0*\x,0.66*\y) {$\scalebox{0.7}{\begin{tikzpicture}
            \draw[red] (0,0)--(0,3) (2,0)--(2,3) (4,0)--(4,3);
            \node[D5] at (3,1.5) {};
        \end{tikzpicture}}$};
        \node (20) at (1*\x,0.66*\y) {$\scalebox{0.7}{\begin{tikzpicture}
            \draw[red] (0,0)--(0,3) (2,0)--(2,3) (4,0)--(4,3);
            \node[D5] at (0.5,1.5) {};
            \node[D5] at (1.5,1.5) {};
            \draw (0,0.5)--(2,0.5);
        \end{tikzpicture}}$};
        \node (12) at (.5*\x,1.66*\y) {$\scalebox{0.7}{\begin{tikzpicture}
            \draw[red] (0,0)--(0,3) (2,0)--(2,3) (4,0)--(4,3);
            \node[D5] at (1,1.5) {};
            \node[D5] at (2.5,1.5) {};
            \node[D5] at (3.5,1.5) {};
            \draw (0,0.5)--(2,0.5) (2,0.6)--(4,0.6);
        \end{tikzpicture}}$};
        \node (31) at (1.5*\x,1.66*\y) {$\scalebox{0.7}{\begin{tikzpicture}
            \draw[red] (0,0)--(0,3) (2,0)--(2,3) (4,0)--(4,3);
            \node[D5] at (0.5,1.5) {};
            \node[D5] at (1,1.5) {};
            \node[D5] at (1.5,1.5) {};
            \node[D5] at (3,1.5) {};
            \draw (0,0.5)--(2,0.5) (0,0.7)--(2,0.7) (2,0.6)--(4,0.6);
        \end{tikzpicture}}$};
        \node (50) at (2.5*\x,1.66*\y) {$\scalebox{0.7}{\begin{tikzpicture}
            \draw[red] (0,0)--(0,3) (2,0)--(2,3) (4,0)--(4,3);
            \node[D5] at (0.5,1.5) {};
            \node at (1,1.5) {$\cdots$};
            \node at (1,2) {$5$};
            \node[D5] at (1.5,1.5) {};
            \draw (0,0.5)--(2,0.5) (0,0.7)--(2,0.7) (0,0.9)--(2,0.9) (2,0.6)--(4,0.6);
        \end{tikzpicture}}$};
        \node (04) at (0*\x,2.66*\y) {$\scalebox{0.7}{\begin{tikzpicture}
            \draw[red] (0,0)--(0,3) (2,0)--(2,3) (4,0)--(4,3);
            \node[D5] at (2.5,1.5) {};
            \node at (3,1.5) {$\cdots$};
            \node at (3,2) {$4$};
            \node[D5] at (3.5,1.5) {};
            \draw (0,0.5)--(2,0.5) (2,0.6)--(4,0.6) (2,0.8)--(4,0.8);
        \end{tikzpicture}}$};
        \node (23) at (1*\x,2.66*\y) {$\scalebox{0.7}{\begin{tikzpicture}
            \draw[red] (0,0)--(0,3) (2,0)--(2,3) (4,0)--(4,3);
            \node[D5] at (0.5,1.5) {};
            \node[D5] at (1.5,1.5) {};
            \node[D5] at (2.5,1.5) {};
            \node[D5] at (3,1.5) {};
            \node[D5] at (3.5,1.5) {};
            \draw (0,0.5)--(2,0.5) (0,0.7)--(2,0.7) (2,0.6)--(4,0.6) (2,0.8)--(4,0.8);
        \end{tikzpicture}}$};
        \node (42) at (2*\x,2.66*\y) {$\scalebox{0.7}{\begin{tikzpicture}
            \draw[red] (0,0)--(0,3) (2,0)--(2,3) (4,0)--(4,3);
            \node[D5] at (0.5,1.5) {};
            \node at (1,1.5) {$\cdots$};
            \node at (1,2) {$4$};
            \node[D5] at (1.5,1.5) {};
            \node[D5] at (2.5,1.5) {};
            \node[D5] at (3.5,1.5) {};
            \draw (0,0.5)--(2,0.5) (0,0.7)--(2,0.7) (0,0.9)--(2,0.9) (2,0.6)--(4,0.6) (2,0.8)--(4,0.8);
        \end{tikzpicture}}$};
        \node (61) at (3*\x,2.66*\y) {$\scalebox{0.7}{\begin{tikzpicture}
            \draw[red] (0,0)--(0,3) (2,0)--(2,3) (4,0)--(4,3);
            \node[D5] at (0.5,1.5) {};
            \node at (1,1.5) {$\cdots$};
            \node at (1,2) {$6$};
            \node[D5] at (1.5,1.5) {};
            \node[D5] at (3,1.5) {};
            \draw (0,0.5)--(2,0.5) (0,0.7)--(2,0.7) (0,0.9)--(2,0.9) (0,1.1)--(2,1.1) (2,0.6)--(4,0.6) (2,0.8)--(4,0.8);
        \end{tikzpicture}}$};
        \node[draw,label=below:{\color{olive}$A_1$}] (0120) at ($(01)!0.5!(20)$) {$\scalebox{0.4}{\begin{tikzpicture}
            \draw[red] (0,0)--(0,3) (2,0)--(2,3) (4,0)--(4,3);
            \node[D5o] (1) at (-0.5,1.5) {};
            \node[D5] (2) at (3.5,1.5) {};
            \draw[olive,thick] (1)--(0,1.5) (0,0.5)--(2,0.5) (2,1.5)--(2);
            \node at (4.5,1.5) {};
        \end{tikzpicture}}$};
        \node[draw,label=left:{\color{olive}$A_1$}] (2012) at ($(20)!0.5!(12)$) {$\scalebox{0.4}{\begin{tikzpicture}
            \draw[red] (0,0)--(0,3) (2,0)--(2,3) (4,0)--(4,3);
            \node[D5] at (0.5,1.5) {};
            \node[D5] (1) at (1.5,1.5) {};
            \node[D5o] (2) at (4.5,1.5) {};
            \draw (0,0.5)--(2,0.5);
            \draw[olive,thick] (1)--(2,1.5) (2,0.6)--(4,0.6) (4,1.5)--(2);
            \node at (-0.5,1.5) {};
        \end{tikzpicture}}$};
        \node[draw,label=below:{\color{olive}$A_2$}] (1231) at ($(12)!0.5!(31)$) {$\scalebox{0.4}{\begin{tikzpicture}
            \draw[red] (0,0)--(0,3) (2,0)--(2,3) (4,0)--(4,3);
            \node[D5o] (1) at (-0.5,1.5) {};
            \node[D5] at (1,1.5) {};
            \node[D5] (2) at (2.5,1.5) {};
            \node[D5] at (3.5,1.5) {};
            \draw (0,0.5)--(2,0.5) (2,0.6)--(4,0.6);
            \draw[olive,thick] (1)--(0,1.5) (0,0.7)--(2,0.7) (2,1.5)--(2);
            \node at (4.5,1.5) {};
        \end{tikzpicture}}$};
        \node[draw,label=below:{\color{olive}$A_4$}] (3150) at ($(31)!0.5!(50)$) {$\scalebox{0.4}{\begin{tikzpicture}
            \draw[red] (0,0)--(0,3) (2,0)--(2,3) (4,0)--(4,3);
            \node[D5o] (1) at (-0.5,1.5) {};
            \node[D5] at (0.5,1.5) {};
            \node[D5] at (1,1.5) {};
            \node[D5] at (1.5,1.5) {};
            \node[D5] (2) at (3,1.5) {};
            \draw (0,0.5)--(2,0.5) (0,0.7)--(2,0.7) (2,0.6)--(4,0.6);
            \draw[olive,thick] (1)--(0,1.5) (0,0.9)--(2,0.9) (2,1.5)--(2);
            \node at (4.5,1.5) {};
        \end{tikzpicture}}$};
        \node[draw,label=left:{\color{olive}$A_3$}] (1204) at ($(12)!0.5!(04)$) {$\scalebox{0.4}{\begin{tikzpicture}
            \draw[red] (0,0)--(0,3) (2,0)--(2,3) (4,0)--(4,3);
            \node[D5] (1) at (1,1.5) {};
            \node[D5] at (2.5,1.5) {};
            \node[D5] at (3.5,1.5) {};
            \node[D5o] (2) at (4.5,1.5) {};
            \draw (0,0.5)--(2,0.5) (2,0.6)--(4,0.6);
            \draw[olive,thick] (1)--(2,1.5) (2,0.8)--(4,0.8) (4,1.5)--(2);
            \node at (-0.5,1.5) {};
        \end{tikzpicture}}$};
        \node[draw,label=left:{\color{olive}$A_2$}] (3123) at ($(31)!0.5!(23)$) {$\scalebox{0.4}{\begin{tikzpicture}
            \draw[red] (0,0)--(0,3) (2,0)--(2,3) (4,0)--(4,3);
            \node[D5] at (0.5,1.5) {};
            \node[D5] at (1,1.5) {};
            \node[D5] (1) at (1.5,1.5) {};
            \node[D5] at (3,1.5) {};
            \node[D5o] (2) at (4.5,1.5) {};
            \draw (0,0.5)--(2,0.5) (0,0.7)--(2,0.7) (2,0.6)--(4,0.6);
            \draw[olive,thick] (1)--(2,1.5) (2,0.8)--(4,0.8) (4,1.5)--(2);
            \node at (-0.5,1.5) {};
        \end{tikzpicture}}$};
        \node[draw,label=left:{\color{olive}$A_1$}] (5042) at ($(50)!0.5!(42)$) {$\scalebox{0.4}{\begin{tikzpicture}
            \draw[red] (0,0)--(0,3) (2,0)--(2,3) (4,0)--(4,3);
            \node[D5] at (0.5,1.5) {};
            \node at (1,1.5) {$\cdots$};
            \node at (1,2) {$5$};
            \node[D5] (1) at (1.5,1.5) {};
            \node[D5o] (2) at (4.5,1.5) {};
            \draw (0,0.5)--(2,0.5) (0,0.7)--(2,0.7) (0,0.9)--(2,0.9) (2,0.6)--(4,0.6);
            \draw[olive,thick] (1)--(2,1.5) (2,0.8)--(4,0.8) (4,1.5)--(2);
            \node at (-0.5,1.5) {};
        \end{tikzpicture}}$};
        \node[draw,label=below:{\color{olive}$A_1$}] (0423) at ($(04)!0.5!(23)$) {$\scalebox{0.4}{\begin{tikzpicture}
            \draw[red] (0,0)--(0,3) (2,0)--(2,3) (4,0)--(4,3);
            \node[D5o] (1) at (-0.5,1.5) {};
            \node[D5] (2) at (2.5,1.5) {};
            \node at (3,1.5) {$\cdots$};
            \node at (3,2) {$4$};
            \node[D5] at (3.5,1.5) {};
            \draw (0,0.5)--(2,0.5) (2,0.6)--(4,0.6) (2,0.8)--(4,0.8);
            \draw[olive,thick] (1)--(0,1.5) (0,0.7)--(2,0.7) (2,1.5)--(2);
            \node at (4.5,1.5) {};
        \end{tikzpicture}}$};
        \node[draw,label=below:{\color{olive}$A_3$}] (2342) at ($(23)!0.5!(42)$) {$\scalebox{0.4}{\begin{tikzpicture}
            \draw[red] (0,0)--(0,3) (2,0)--(2,3) (4,0)--(4,3);
            \node[D5o] (1) at (-0.5,1.5) {};
            \node[D5] at (0.5,1.5) {};
            \node[D5] at (1.5,1.5) {};
            \node[D5] (2) at (2.5,1.5) {};
            \node[D5] at (3,1.5) {};
            \node[D5] at (3.5,1.5) {};
            \draw (0,0.5)--(2,0.5) (0,0.7)--(2,0.7) (2,0.6)--(4,0.6) (2,0.8)--(4,0.8);
            \draw[olive,thick] (1)--(0,1.5) (0,0.9)--(2,0.9) (2,1.5)--(2);
            \node at (4.5,1.5) {};
        \end{tikzpicture}}$};
        \node[draw,label=below:{\color{olive}$A_5$}] (4261) at ($(42)!0.5!(61)$) {$\scalebox{0.4}{\begin{tikzpicture}
            \draw[red] (0,0)--(0,3) (2,0)--(2,3) (4,0)--(4,3);
            \node[D5o] (1) at (-0.5,1.5) {};
            \node[D5] at (0.5,1.5) {};
            \node at (1,1.5) {$\cdots$};
            \node at (1,2) {$4$};
            \node[D5] at (1.5,1.5) {};
            \node[D5] (2) at (2.5,1.5) {};
            \node[D5] at (3.5,1.5) {};
            \draw (0,0.5)--(2,0.5) (0,0.7)--(2,0.7) (0,0.9)--(2,0.9) (2,0.6)--(4,0.6) (2,0.8)--(4,0.8);
            \draw[olive,thick] (1)--(0,1.5) (0,1.1)--(2,1.1) (2,1.5)--(2);
            \node at (4.5,1.5) {};
        \end{tikzpicture}}$};
        \draw (01)--(0120)--(20) (20)--(2012)--(12) (12)--(1231)--(31) (31)--(3150)--(50) (12)--(1204)--(04) (31)--(3123)--(23) (50)--(5042)--(42) (04)--(0423)--(23) (23)--(2342)--(42) (42)--(4261)--(61);
        \draw (23)--(1*\x-0.3*0.5*\x,2.66*\y+0.3*1*\y) (42)--(2*\x-0.3*0.5*\x,2.66*\y+0.3*1*\y) (61)--(3*\x-0.3*0.5*\x,2.66*\y+0.3*1*\y) (61)--(3*\x+0.4*\x,2.66*\y);
    \end{tikzpicture}
    \end{adjustbox}
    \caption{Brane depiction of the low dimensional leaves [and elementary slices between them] of $\w{\mathrm{PSL}(3,\mathbb{C})}{}{[0,1]}$.}
    \label{fig:A2branes2}
\end{figure}
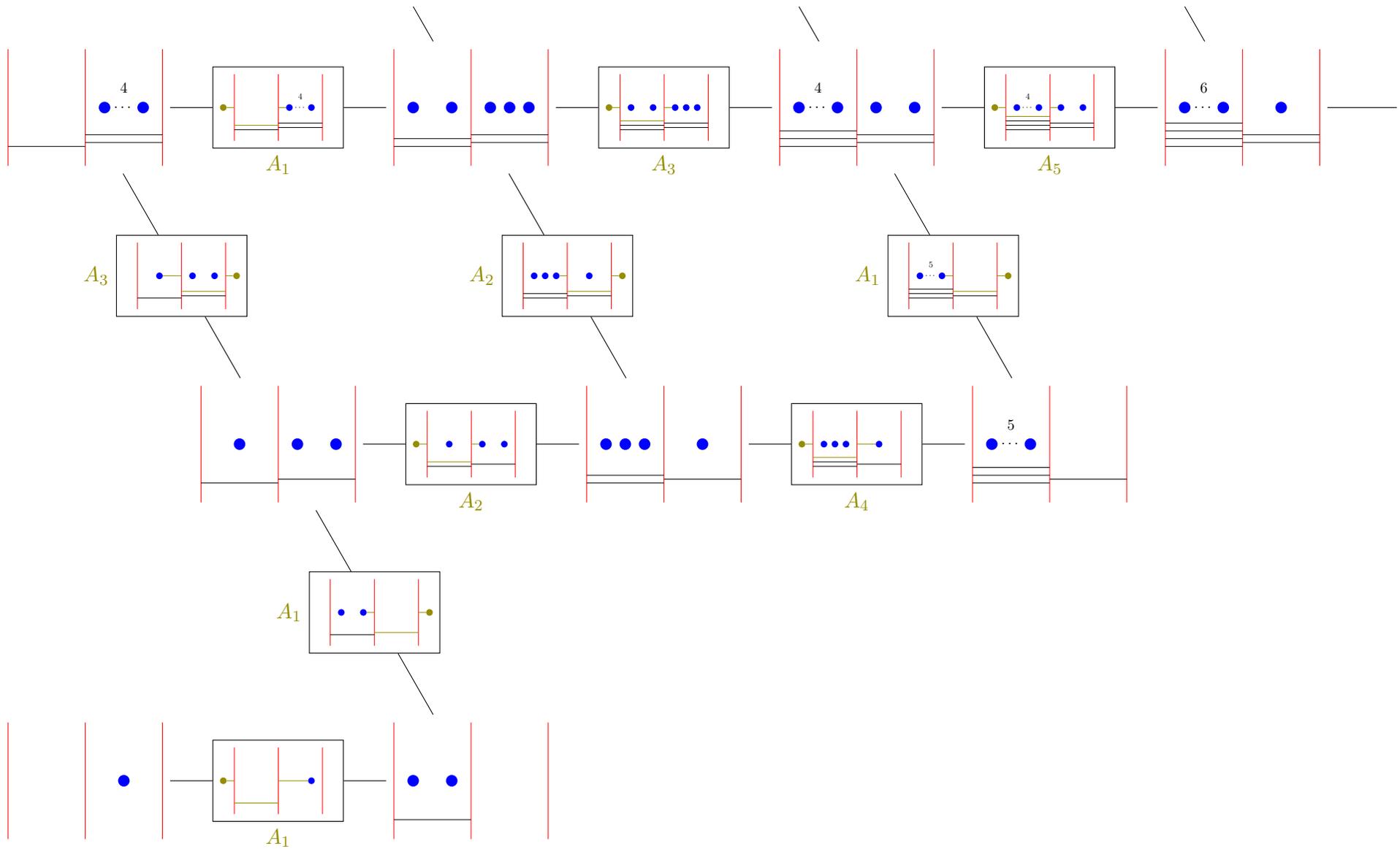

\end{landscape}

\subsection{\texorpdfstring{The $B$-type affine Grassmannian}{The B-type affine Grassmannian}}

For this setup, we require a ${\color{orange}\widetilde{\textnormal{ON}}^-}$ plane. In the main text we always draw the full covering space and refer to half branes simply as branes.

\paragraph{Component $0\in\mathbb{Z}_2$.} The affine Grassmannian of $\mathrm{Spin}(2n+1,\mathbb{C})$ is the $0$-th component of the affine Grassmannian of $\mathrm{SO}(2n+1,\mathbb{C})$. The brane system for $\w{\mathrm{SO}(2n+1,\mathbb{C})}{}{[0,0\dots,0]}$ is proposed to be
\begin{equation}
    \vcenter{\hbox{
    \begin{tikzpicture}
        \draw[red] (-3,0)--(-3,4) (-2,0)--(-2,4) (1,0)--(1,4) (3,0)--(3,4) (6,0)--(6,4) (7,0)--(7,4);
        \node at (-0.5,2) {$\cdots$};
        \node at (-1,-0.6) {$n$};
        \draw [decorate,decoration={brace,amplitude=5pt}] (1.1,-0.25)--(-3.1,-0.25);
        \node at (4.5,2) {$\cdots$};
        \node at (5,-0.6) {$n$};
        \draw [decorate,decoration={brace,amplitude=5pt}] (7.1,-0.25)--(2.9,-0.25);
        \draw[orange] (2,0)--(2,4);
        \node at (2,-0.5) {$\widetilde{\textnormal{ON}}^-$};
    \end{tikzpicture}
    }} \, , 
\end{equation}
which is depicted at the origin, with all D3 branes and D5 branes suppressed. The quiver representing the origin is:
\begin{equation}
    \raisebox{-.5\height}{\begin{tikzpicture}
        \node[gauge,label=below:{$0$}] (g1) at (0,0) {};
        \node[gauge,label=below:{$0$}] (g2) at (1,0) {};
        \node (g) at (2,0) {$\cdots$};
        \node[gauge,label=below:{$0$}] (g3) at (3,0) {};
        \node[gauge,label=below:{$0$}] (g4) at (4,0) {};
        \node[flavour,label=above:{$0$}] (f1) at (0,1) {};
        \node[flavour,label=above:{$0$}] (f2) at (1,1) {};
        \node[flavour,label=above:{$0$}] (f3) at (3,1) {};
        \node[flavour,label=above:{$0$}] (f4) at (4,1) {};
        \draw (f1)--(g1) (f2)--(g2) (f3)--(g3) (f4)--(g4);
        \draw (g1)--(g2)--(g)--(g3);
        \draw[transform canvas={yshift=-2pt}] (g3)--(g4);
        \draw[transform canvas={yshift=2pt}] (g3)--(g4);
        \draw (3.45,0.2)--(3.55,0)--(3.45,-0.2);
    \end{tikzpicture}}
\end{equation}

\paragraph{Component $1\in\mathbb{Z}_2$.} The brane system for the $1$-st component of the affine Grassmannian of $\mathrm{SO}(2n+1,\mathbb{C})$, is proposed to be\begin{equation}
    \vcenter{\hbox{
    \begin{tikzpicture}
        \draw[red] (-3,0)--(-3,4) (-2,0)--(-2,4) (1,0)--(1,4) (3,0)--(3,4) (6,0)--(6,4) (7,0)--(7,4);
        \node at (-0.5,2) {$\cdots$};
        \node at (-1,-0.6) {$n$};
        \draw [decorate,decoration={brace,amplitude=5pt}] (1.1,-0.25)--(-3.1,-0.25);
        \node at (4.5,2) {$\cdots$};
        \node at (5,-0.6) {$n$};
        \draw [decorate,decoration={brace,amplitude=5pt}] (7.1,-0.25)--(2.9,-0.25);
        \draw[orange] (2,0)--(2,4);
        \node at (2,-0.5) {$\widetilde{\textnormal{ON}}^-$};
        \node[D5] at (-2.5,2) {};
        \node[D5] at (6.5,2) {};
    \end{tikzpicture}
    }} \, , 
\end{equation}
which is depicted at the origin, with all D3 branes and D5 branes suppressed. The quiver representing the origin is 
\begin{equation}
    \raisebox{-.5\height}{\begin{tikzpicture}
        \node[gauge,label=below:{$0$}] (g1) at (0,0) {};
        \node[gauge,label=below:{$0$}] (g2) at (1,0) {};
        \node (g) at (2,0) {$\cdots$};
        \node[gauge,label=below:{$0$}] (g3) at (3,0) {};
        \node[gauge,label=below:{$0$}] (g4) at (4,0) {};
        \node[flavour,label=above:{$1$}] (f1) at (0,1) {};
        \node[flavour,label=above:{$0$}] (f2) at (1,1) {};
        \node[flavour,label=above:{$0$}] (f3) at (3,1) {};
        \node[flavour,label=above:{$0$}] (f4) at (4,1) {};
        \draw (f1)--(g1) (f2)--(g2) (f3)--(g3) (f4)--(g4);
        \draw (g1)--(g2)--(g)--(g3);
        \draw[transform canvas={yshift=-2pt}] (g3)--(g4);
        \draw[transform canvas={yshift=2pt}] (g3)--(g4);
        \draw (3.45,0.2)--(3.55,0)--(3.45,-0.2);
    \end{tikzpicture}} \, . 
\end{equation}
This is indeed the quiver representing the origin of $\w{\mathrm{SO}(2n+1,\mathbb{C})}{}{[1,0,\dots,0]}$.

\paragraph{Minimal transitions.} When activating D3 branes between D5 branes, one can follow the same procedure as for the A-type affine Grassmannian, when considering branes away from the ON (keeping in mind that in the covering space one has to move branes together with their mirror images). If however one wants to activate a D3 brane between a D5 brane and its image, in order to move the D3 brane along an NS5 brane and its image, one has to keep in mind the boundary conditions reviewed in Appendix \ref{app:orientifolds}. The D3 brane ending on a D5 brane and its image needs to be accompanied by its own mirror image. The associated transition is $b_n$ if the D5 branes are separated by $n$ NS5 branes (and their mirror images).
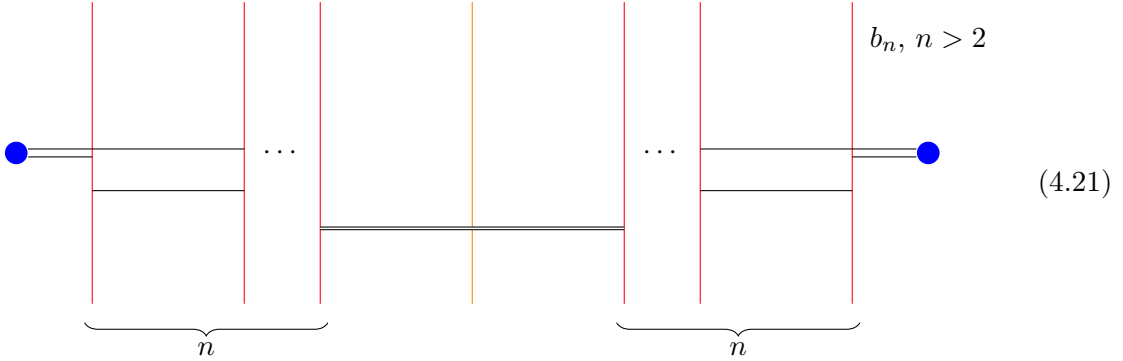
\begin{equation}
    \vcenter{\hbox{
    \begin{tikzpicture}
        \draw[red] (-3,0)--(-3,4) (-1,0)--(-1,4) (0,0)--(0,4) (4,0)--(4,4) (5,0)--(5,4) (7,0)--(7,4);
        \node at (-0.5,2) {$\cdots$};
        \node at (-1.5,-0.6) {$n$};
        \draw [decorate,decoration={brace,amplitude=5pt}] (0.1,-0.25)--(-3.1,-0.25);
        \node at (4.5,2) {$\cdots$};
        \node at (5.5,-0.6) {$n$};
        \draw [decorate,decoration={brace,amplitude=5pt}] (7.1,-0.25)--(3.9,-0.25);
        \draw[orange] (2,0)--(2,4);
        \node[D5] (1) at (-4,2) {};
        \node[D5] (2) at (8,2) {};
        \draw[transform canvas={yshift=-1.5pt}] (1)--(-3,2) (7,2)--(2);
        \draw[transform canvas={yshift=1.5pt}] (1)--(-1,2) (5,2)--(2);
        \draw (-3,1.5)--(-1,1.5) (5,1.5)--(7,1.5);
        \draw[double] (0,1)--(4,1);
        \node at (8,3.5) {$b_n$, $n>2$};
    \end{tikzpicture}
    }}
\end{equation}
In the case of $n=1$ one has to be slightly more careful. If we attempt to move the D3 brane and its mirror along the NS5 brane we would obtain a set-up breaking the S-rule:
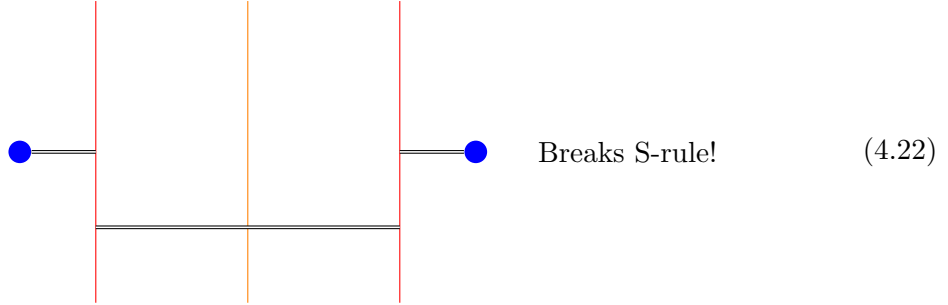
\begin{equation}
    \vcenter{\hbox{
    \begin{tikzpicture}
        \draw[red] (0,0)--(0,4) (4,0)--(4,4);
        \draw[orange] (2,0)--(2,4);
        \node[D5] (1) at (-1,2) {};
        \node[D5] (2) at (5,2) {};
        \draw[double] (1)--(0,2) (4,2)--(2) (0,1)--(4,1);
        \node at (7,2) {Breaks S-rule!};
    \end{tikzpicture}}}
\end{equation}
We therefore need to include another D5 brane and its mirror image and can now perform a $b_1=A_1$ transition:
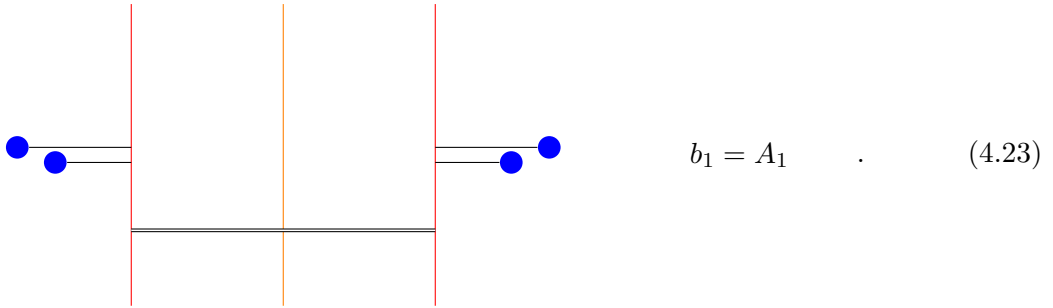
\begin{equation}
    \vcenter{\hbox{
    \begin{tikzpicture}
        \draw[red] (0,0)--(0,4) (4,0)--(4,4);
        \draw[orange] (2,0)--(2,4);
        \node[D5] (1) at (-1.5,2.1) {};
        \node[D5] (2) at (-1,1.9) {};
        \node[D5] (3) at (5,1.9) {};
        \node[D5] (4) at (5.5,2.1) {};
        \draw (1)--(0,2.1) (4,2.1)--(4) (2)--(0,1.9) (4,1.9)--(3);
        \draw[double] (0,1)--(4,1);
        \node at (8,2) {$b_1=A_1$};
    \end{tikzpicture}}}\qquad.
    \label{eq:b1trans}
\end{equation}
Drawing the D5 branes at different vertical positions is purely for convenience. In \eqref{eq:b1trans} more D5 branes could be present between the NS5 and the $\widetilde{\textnormal{ON}}^-$ plane leading to $A_k$ transitions.

When dealing with the $1$-st component for $\mathrm{SO}(5,\mathbb{C})$, i.e.\ $n=2$, one also has to be careful with the first transition. Activating a minimal number of moduli without breaking supersymmetry gives the configuration
\begin{equation}
    \vcenter{\hbox{
    \begin{tikzpicture}
        \draw[red] (0,0)--(0,4) (2,0)--(2,4) (6,0)--(6,4) (8,0)--(8,4);
        \draw[orange] (4,0)--(4,4);
        \node[D5] (1) at (-1,2) {};
        \node[D5] (2) at (9,2) {};
        \node[D5] at (1,2.5) {};
        \node[D5] at (7,2.5) {};
        \draw[double] (1)--(0,2) (8,2)--(2) (2,1)--(6,1);
        \draw (0,2)--(2,2) (6,2)--(8,2);
        \draw (0,1.5)--(2,1.5) (6,1.5)--(8,1.5);
        \node at (10.5,2) {$ac_2$};
    \end{tikzpicture}}}\quad.
    \label{eq:ab2trans}
\end{equation}
This transition is called $ac_2$ as it appears also in the type $C$ affine Grassmannians, and we use the isomorphism $C_2 = B_2$. For $n>2$ no new slices arise.

\subsubsection{\texorpdfstring{The $B_1$ affine Grassmannian}{The B1 affine Grassmannian}}
\label{sec:B1branes}

\paragraph{Component $0 \in \mathbb{Z}_2$.} The affine Grassmannian of $\mathrm{Spin}(3,\mathbb{C})$, i.e.\ component 0 of the affine Grassmannian of $\mathrm{SO}(3,\mathbb{C})$, is given by the brane system:
\begin{equation}
    \vcenter{\hbox{
    \begin{tikzpicture}
        \draw[red] (0,0)--(0,4) (4,0)--(4,4);
        \draw[orange] (2,0)--(2,4);
    \end{tikzpicture}}}\qquad,
\end{equation}
depicted at the origin of $\w{\mathrm{SO}(3,\mathbb{C})}{}{[0]}$. We can now move onto the lowest non-trivial leaf through a $b_1=A_1$ transition as in \eqref{eq:b1trans}. After a Hanany-Witten transition we can read off the quiver:
\begin{equation}
    \vcenter{\hbox{
    \begin{tikzpicture}
        \draw[red] (0,0)--(0,4) (4,0)--(4,4);
        \draw[orange] (2,0)--(2,4);
        \node[D5] (1) at (0.7,2) {};
        \node[D5] (2) at (1.3,2) {};
        \node[D5] (3) at (2.7,2) {};
        \node[D5] (4) at (3.3,2) {};
        \draw[double] (0,1)--(4,1);
    \end{tikzpicture}}}
    \qquad
    \vcenter{\hbox{
    \begin{tikzpicture}
        \node[gauge,label=below:{$1$}] (1) at (0,0) {};
        \node[flavour,label=above:{$2$}] (2) at (0,1) {};
        \draw (1)--(2);
    \end{tikzpicture}
    }} \, .  
\end{equation}
The only way to keep moving onto higher leaves in $\w{\mathrm{SO}(3,\mathbb{C})}{}{[0]}$ is by repeating the same operation, however the D5 branes already placed between the NS5 branes make the transverse slices $A_{2k-1}$ for the $k$-th transition of this type. A general leaf and the corresponding quiver are
\begin{equation}
    \vcenter{\hbox{
    \begin{tikzpicture}
        \draw[red] (0,0)--(0,4) (4,0)--(4,4);
        \draw[orange] (2,0)--(2,4);
        \node[D5] (1) at (0.5,2) {};
        \node at (1,2.3) {$2k$};
        \node at (1,2) {$\cdots$};
        \node[D5] (2) at (1.5,2) {};
        \node[D5] (3) at (2.5,2) {};
        \node at (3,2.3) {$2k$};
        \node at (3,2) {$\cdots$};
        \node[D5] (4) at (3.5,2) {};
        \draw[double] (0,1)--(4,1);
        \node at (1,0.75) {$\vdots$ $k$};
        \node at (3,0.75) {$\vdots$ $k$};
        \draw[double] (0,0.3)--(4,0.3);
    \end{tikzpicture}}}
    \qquad
    \vcenter{\hbox{
    \begin{tikzpicture}
        \node[gauge,label=below:{$k$}] (1) at (0,0) {};
        \node[flavour,label=above:{$2k$}] (2) at (0,1) {};
        \draw (1)--(2);
    \end{tikzpicture}
    }} \, , 
    \label{eq:B1_1st}
\end{equation}
which completely agrees with the construction in Section \ref{sec:A1branes}.

\paragraph{Component $1\in\mathbb{Z}_2$.} For the second connected component of $\mathrm{SO}(3,\mathbb{C})$ we start with:
\begin{equation}
    \vcenter{\hbox{
    \begin{tikzpicture}
        \draw[red] (0,0)--(0,4) (4,0)--(4,4);
        \draw[orange] (2,0)--(2,4);
        \node[D5] at (1,2) {};
        \node[D5] at (3,2) {};
    \end{tikzpicture}}}\qquad,
\end{equation}
depicting the origin of $\w{\mathrm{SO}(3,\mathbb{C})}{}{[1]}$
The transverse slice for the $k$-th transition is $A_{2k}$ and the brane system and quiver for the leaf are:
\begin{equation}
    \vcenter{\hbox{
    \begin{tikzpicture}
        \draw[red] (0,0)--(0,4) (4,0)--(4,4);
        \draw[orange] (2,0)--(2,4);
        \node[D5] (1) at (0.5,2) {};
        \node at (1,2.3) {$2k+1$};
        \node at (1,2) {$\cdots$};
        \node[D5] (2) at (1.5,2) {};
        \node[D5] (3) at (2.5,2) {};
        \node at (3,2.3) {$2k+1$};
        \node at (3,2) {$\cdots$};
        \node[D5] (4) at (3.5,2) {};
        \draw[double] (0,1)--(4,1);
        \node at (1,0.75) {$\vdots$ $k$};
        \node at (3,0.75) {$\vdots$ $k$};
        \draw[double] (0,0.3)--(4,0.3);
    \end{tikzpicture}}}
    \qquad
    \vcenter{\hbox{
    \begin{tikzpicture}
        \node[gauge,label=below:{$k$}] (1) at (0,0) {};
        \node[flavour,label=above:{$2k+1$}] (2) at (0,1) {};
        \draw (1)--(2);
    \end{tikzpicture}
    }} \, , 
    \label{eq:B1_2nd}
\end{equation}
which is again in agreement with Section \ref{sec:A1branes}.

\subsubsection{\texorpdfstring{The $B_2$ affine Grassmannian}{The B2 affine Grassmannian}}

\paragraph{Component $0\in\mathbb{Z}_2$.} This is depicted in Figure \ref{fig:B2Branes}.

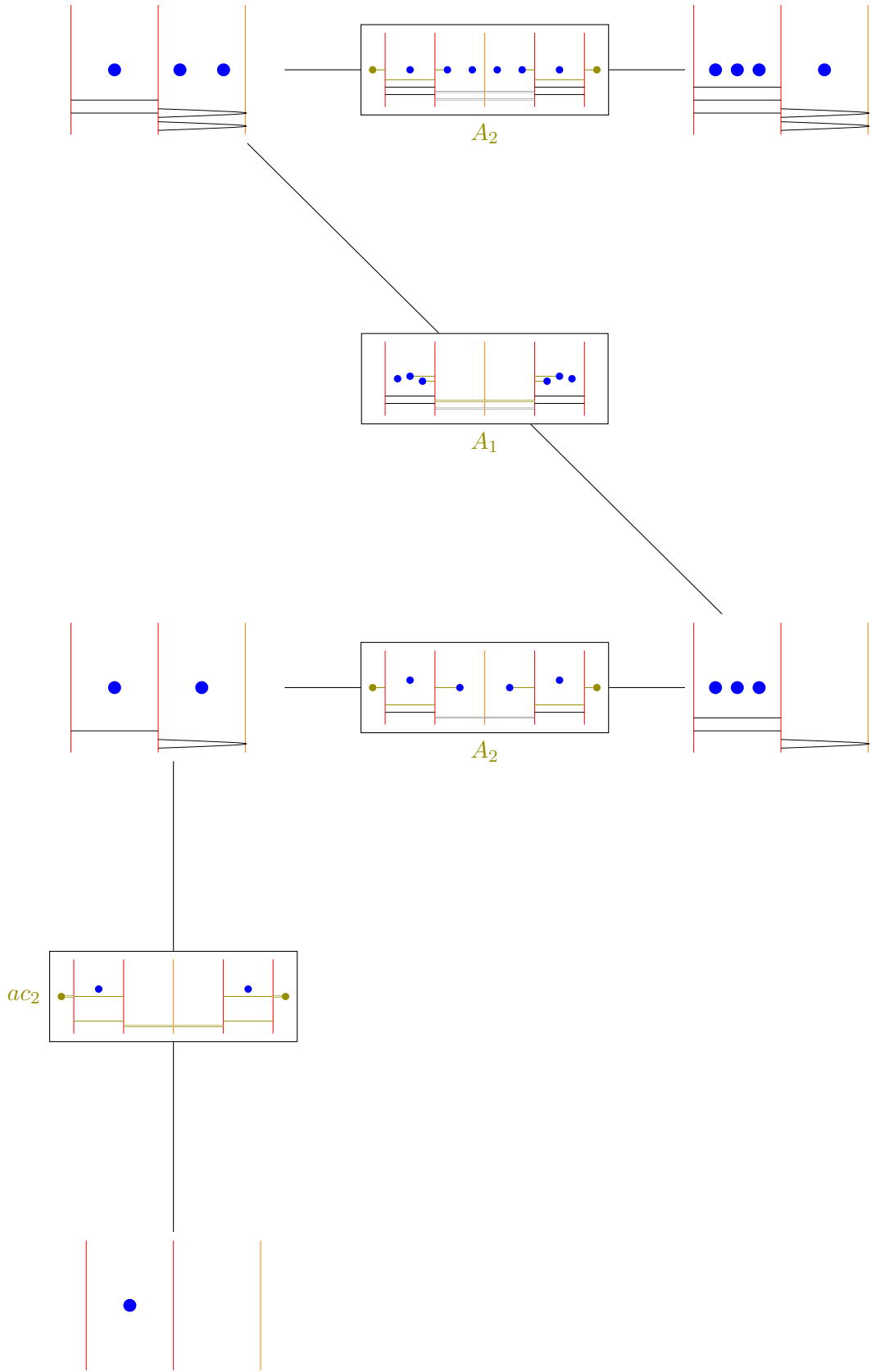
\begin{figure}[t]
    \centering
    \begin{adjustbox}{center}
    \begin{tikzpicture}
    \def\x{10*1cm};
        \node (00) at (0*\x,0*\x) {$\scalebox{0.7}{\begin{tikzpicture}
            \draw[red] (0,0)--(0,3) (2,0)--(2,3);
            \draw[orange] (4,0)--(4,3);
            \node[D5] at (1,1.5) {};
        \end{tikzpicture}}$};
        \node (01) at (0*\x,1*\x) {$\scalebox{0.7}{\begin{tikzpicture}
            \draw[red] (0,0)--(0,3) (2,0)--(2,3);
            \draw[orange] (4,0)--(4,3);
            \node[D5] at (3,1.5) {};
            \node[D5] at (1,1.5) {};
            \draw (0,0.5)--(2,0.5);
            \draw (2,0.3) .. controls (4.7,0.2) .. (2,0.1);
        \end{tikzpicture}}$};
        \node (20) at (1*\x,1*\x) {$\scalebox{0.7}{\begin{tikzpicture}
            \draw[red] (0,0)--(0,3) (2,0)--(2,3);
            \draw[orange] (4,0)--(4,3);
            \node[D5] at (0.5,1.5) {};
            \node[D5] at (1,1.5) {};
            \node[D5] at (1.5,1.5) {};
            \draw (0,0.5)--(2,0.5) (0,0.8)--(2,0.8);
            \draw (2,0.3) .. controls (4.7,0.2) .. (2,0.1);
        \end{tikzpicture}}$};
        \node (02) at (0*\x,2*\x) {$\scalebox{0.7}{\begin{tikzpicture}
            \draw[red] (0,0)--(0,3) (2,0)--(2,3);
            \draw[orange] (4,0)--(4,3);
            \node[D5] at (1,1.5) {};
            \node[D5] at (2.5,1.5) {};
            \node[D5] at (3.5,1.5) {};
            \draw (0,0.5)--(2,0.5) (0,0.8)--(2,0.8);
            \draw (2,0.3) .. controls (4.7,0.2) .. (2,0.1);
            \draw (2,0.6) .. controls (4.7,0.5) .. (2,0.4);
        \end{tikzpicture}}$};
        \node (21) at (1*\x,2*\x) {$\scalebox{0.7}{\begin{tikzpicture}
            \draw[red] (0,0)--(0,3) (2,0)--(2,3);
            \draw[orange] (4,0)--(4,3);
            \node[D5] at (0.5,1.5) {};
            \node[D5] at (1,1.5) {};
            \node[D5] at (1.5,1.5) {};
            \node[D5] at (3,1.5) {};
            \draw (0,0.5)--(2,0.5) (0,0.8)--(2,0.8) (0,1.1)--(2,1.1);
            \draw (2,0.3) .. controls (4.7,0.2) .. (2,0.1);
            \draw (2,0.6) .. controls (4.7,0.5) .. (2,0.4);
        \end{tikzpicture}}$};
       
        \node[draw,label=left:{\color{olive}$ac_2$}] (0001) at ($(00)!0.5!(01)$) {$\scalebox{0.4}{\begin{tikzpicture}
            \draw[red] (0,0)--(0,3) (2,0)--(2,3) (6,0)--(6,3) (8,0)--(8,3);
            \draw[orange] (4,0)--(4,3);
            \node[D5] at (1,1.8) {};
            \node[D5] at (7,1.8) {};
            \node[D5o] (1) at (-0.5,1.5) {};
            \node[D5o] (2) at (8.5,1.5) {};
            \draw[olive,double,thick] (1)--(0,1.5) (2,0.3)--(6,0.3) (8,1.5)--(2);
            \draw[olive,thick] (0,1.5)--(2,1.5) (0,0.5)--(2,0.5) (6,0.5)--(8,0.5) (8,1.5)--(6,1.5);
        \end{tikzpicture}}$};
        \node[draw,label=below:{\color{olive}$A_2$}] (0120) at ($(01)!0.5!(20)$) {$\scalebox{0.4}{\begin{tikzpicture}
            \draw[red] (0,0)--(0,3) (2,0)--(2,3) (6,0)--(6,3) (8,0)--(8,3);
            \draw[orange] (4,0)--(4,3);
            \node[D5] at (1,1.8) {};
            \node[D5] at (7,1.8) {};
            \node[D5o] (1) at (-0.5,1.5) {};
            \node[D5] (2) at (3,1.5) {};
            \node[D5] (3) at (5,1.5) {};
            \node[D5o] (4) at (8.5,1.5) {};
            \draw[olive,thick] (1)--(0,1.5) (0,0.8)--(2,0.8) (2,1.5)--(2) (3)--(6,1.5) (6,0.8)--(8,0.8) (8,1.5)--(4);
            \draw (0,0.5)--(2,0.5) (6,0.5)--(8,0.5);
            \draw[double] (2,0.3)--(6,0.3); 
        \end{tikzpicture}}$};
        \node[draw,label=below:{\color{olive}$A_1$}] (2002) at ($(20)!0.5!(02)$) {$\scalebox{0.4}{\begin{tikzpicture}
            \node at (-0.5,1.5) {};
            \node at (8.5,1.5) {};
            \draw[red] (0,0)--(0,3) (2,0)--(2,3) (6,0)--(6,3) (8,0)--(8,3);
            \draw[orange] (4,0)--(4,3);
            \node[D5] (1) at (1,1.6) {};
            \node[D5] (2) at (1.5,1.4) {};
            \node[D5] at (0.5,1.5) {};
            \node[D5] at (7.5,1.5) {};
            \node[D5] (3) at (6.5,1.4) {};
            \node[D5] (4) at (7,1.6) {};
            \draw[olive,thick] (1)--(2,1.6) (2)--(2,1.4) (6,1.4)--(3) (6,1.6)--(4);
            \draw[olive,thick,double] (2,0.6)--(6,0.6);
            \draw (0,0.5)--(2,0.5) (6,0.5)--(8,0.5) (0,0.8)--(2,0.8) (6,0.8)--(8,0.8);
            \draw[double] (2,0.3)--(6,0.3); 
        \end{tikzpicture}}$};
        \node[draw,label=below:{\color{olive}$A_2$}] (0221) at ($(02)!0.5!(21)$) {$\scalebox{0.4}{\begin{tikzpicture}
            \draw[red] (0,0)--(0,3) (2,0)--(2,3) (6,0)--(6,3) (8,0)--(8,3);
            \draw[orange] (4,0)--(4,3);
            \node[D5o] (1) at (-0.5,1.5) {};
            \node[D5] (2) at (2.5,1.5) {};
            \node[D5] (3) at (5.5,1.5) {};
            \node[D5o] (4) at (8.5,1.5) {};
            \node[D5] at (1,1.5) {};
            \node[D5] at (7,1.5) {};
            \node[D5] at (3.5,1.5) {};
            \node[D5] at (4.5,1.5) {};
            \draw[olive,thick] (1)--(0,1.5) (2)--(2,1.5) (6,1.5)--(3) (8,1.5)--(4) (0,1.1)--(2,1.1) (6,1.1)--(8,1.1);
            \draw (0,0.5)--(2,0.5) (6,0.5)--(8,0.5) (0,0.8)--(2,0.8) (6,0.8)--(8,0.8);
            \draw[double] (2,0.3)--(6,0.3) (2,0.6)--(6,0.6); 
        \end{tikzpicture}}$};
        \draw (00)--(0001)--(01)--(0120)--(20)--(2002)--(02)--(0221)--(21);
    \end{tikzpicture}
    \end{adjustbox}
    \caption{Brane depiction of the low dimensional leaves [and elementary slices between them] of $\w{\mathrm{SO}(5,\mathbb{C})}{}{[1,0]}$.}
    \label{fig:B2Branes2}
\end{figure}

\paragraph{Component $1\in\mathbb{Z}_2$.} This is depicted in Figure \ref{fig:B2Branes2}.

\begin{figure}[t]
    \centering
    \begin{adjustbox}{center}
    \begin{tikzpicture}
    \def\x{10*1cm};
        \node (00) at (0*\x,0*\x) {$\scalebox{0.7}{\begin{tikzpicture}
            \draw[red] (0,0)--(0,3) (2,0)--(2,3);
            \draw[orange] (4,0)--(4,3);
        \end{tikzpicture}}$};
        \node (01) at (0*\x,1*\x) {$\scalebox{0.7}{\begin{tikzpicture}
            \draw[red] (0,0)--(0,3) (2,0)--(2,3);
            \draw[orange] (4,0)--(4,3);
            \node[D5] at (3,1.5) {};
            \draw (0,0.5)--(2,0.5);
            \draw (2,0.3) .. controls (4.7,0.2) .. (2,0.1);
        \end{tikzpicture}}$};
        \node (20) at (1*\x,1*\x) {$\scalebox{0.7}{\begin{tikzpicture}
            \draw[red] (0,0)--(0,3) (2,0)--(2,3);
            \draw[orange] (4,0)--(4,3);
            \node[D5] at (0.5,1.5) {};
            \node[D5] at (1.5,1.5) {};
            \draw (0,0.5)--(2,0.5) (0,0.8)--(2,0.8);
            \draw (2,0.3) .. controls (4.7,0.2) .. (2,0.1);
        \end{tikzpicture}}$};
        \node (02) at (0*\x,2*\x) {$\scalebox{0.7}{\begin{tikzpicture}
            \draw[red] (0,0)--(0,3) (2,0)--(2,3);
            \draw[orange] (4,0)--(4,3);
            \node[D5] at (2.5,1.5) {};
            \node[D5] at (3.5,1.5) {};
            \draw (0,0.5)--(2,0.5) (0,0.8)--(2,0.8);
            \draw (2,0.3) .. controls (4.7,0.2) .. (2,0.1);
            \draw (2,0.6) .. controls (4.7,0.5) .. (2,0.4);
        \end{tikzpicture}}$};
        \node (21) at (1*\x,2*\x) {$\scalebox{0.7}{\begin{tikzpicture}
            \draw[red] (0,0)--(0,3) (2,0)--(2,3);
            \draw[orange] (4,0)--(4,3);
            \node[D5] at (0.5,1.5) {};
            \node[D5] at (1.5,1.5) {};
            \node[D5] at (3,1.5) {};
            \draw (0,0.5)--(2,0.5) (0,0.8)--(2,0.8) (0,1.1)--(2,1.1);
            \draw (2,0.3) .. controls (4.7,0.2) .. (2,0.1);
            \draw (2,0.6) .. controls (4.7,0.5) .. (2,0.4);
        \end{tikzpicture}}$};
       
        \node[draw,label=left:{\color{olive}$b_2$}] (0001) at ($(00)!0.5!(01)$) {$\scalebox{0.4}{\begin{tikzpicture}
            \draw[red] (0,0)--(0,3) (2,0)--(2,3) (6,0)--(6,3) (8,0)--(8,3);
            \draw[orange] (4,0)--(4,3);
            \node[D5o] (1) at (-0.5,1.5) {};
            \node[D5o] (2) at (8.5,1.5) {};
            \draw[olive,double,thick] (1)--(0,1.5) (2,0.3)--(6,0.3) (8,1.5)--(2);
            \draw[olive,thick] (0,1.5)--(2,1.5) (0,0.5)--(2,0.5) (6,0.5)--(8,0.5) (8,1.5)--(6,1.5);
        \end{tikzpicture}}$};
        \node[draw,label=below:{\color{olive}$A_1$}] (0120) at ($(01)!0.5!(20)$) {$\scalebox{0.4}{\begin{tikzpicture}
            \draw[red] (0,0)--(0,3) (2,0)--(2,3) (6,0)--(6,3) (8,0)--(8,3);
            \draw[orange] (4,0)--(4,3);
            \node[D5o] (1) at (-0.5,1.5) {};
            \node[D5] (2) at (3,1.5) {};
            \node[D5] (3) at (5,1.5) {};
            \node[D5o] (4) at (8.5,1.5) {};
            \draw[olive,thick] (1)--(0,1.5) (0,0.8)--(2,0.8) (2,1.5)--(2) (3)--(6,1.5) (6,0.8)--(8,0.8) (8,1.5)--(4);
            \draw (0,0.5)--(2,0.5) (6,0.5)--(8,0.5);
            \draw[double] (2,0.3)--(6,0.3); 
        \end{tikzpicture}}$};
        \node[draw,label=below:{\color{olive}$A_1$}] (2002) at ($(20)!0.5!(02)$) {$\scalebox{0.4}{\begin{tikzpicture}
            \node at (-0.5,1.5) {};
            \node at (8.5,1.5) {};
            \draw[red] (0,0)--(0,3) (2,0)--(2,3) (6,0)--(6,3) (8,0)--(8,3);
            \draw[orange] (4,0)--(4,3);
            \node[D5] (1) at (0.5,1.6) {};
            \node[D5] (2) at (1.5,1.4) {};
            \node[D5] (3) at (6.5,1.4) {};
            \node[D5] (4) at (7.5,1.6) {};
            \draw[olive,thick] (1)--(2,1.6) (2)--(2,1.4) (6,1.4)--(3) (6,1.6)--(4);
            \draw[olive,thick,double] (2,0.6)--(6,0.6);
            \draw (0,0.5)--(2,0.5) (6,0.5)--(8,0.5) (0,0.8)--(2,0.8) (6,0.8)--(8,0.8);
            \draw[double] (2,0.3)--(6,0.3); 
        \end{tikzpicture}}$};
        \node[draw,label=below:{\color{olive}$A_1$}] (0221) at ($(02)!0.5!(21)$) {$\scalebox{0.4}{\begin{tikzpicture}
            \draw[red] (0,0)--(0,3) (2,0)--(2,3) (6,0)--(6,3) (8,0)--(8,3);
            \draw[orange] (4,0)--(4,3);
            \node[D5o] (1) at (-0.5,1.5) {};
            \node[D5] (2) at (2.5,1.5) {};
            \node[D5] (3) at (5.5,1.5) {};
            \node[D5o] (4) at (8.5,1.5) {};
            \node[D5] at (3.5,1.5) {};
            \node[D5] at (4.5,1.5) {};
            \draw[olive,thick] (1)--(0,1.5) (2)--(2,1.5) (6,1.5)--(3) (8,1.5)--(4) (0,1.1)--(2,1.1) (6,1.1)--(8,1.1);
            \draw (0,0.5)--(2,0.5) (6,0.5)--(8,0.5) (0,0.8)--(2,0.8) (6,0.8)--(8,0.8);
            \draw[double] (2,0.3)--(6,0.3) (2,0.6)--(6,0.6); 
        \end{tikzpicture}}$};
        \draw (00)--(0001)--(01)--(0120)--(20)--(2002)--(02)--(0221)--(21);
    \end{tikzpicture}
    \end{adjustbox}
    \caption{Brane depiction of the low dimensional leaves [and elementary slices between them] of $\w{\mathrm{SO}(5,\mathbb{C})}{}{[0,0]}$.}
    \label{fig:B2Branes}
\end{figure}

\clearpage

\subsection{\texorpdfstring{The $C$-type affine Grassmannian}{The C-type affine Grassmannian}}

For this setup, we require a ${\color{green}\textnormal{ON}^+}$ plane. In the main text we draw the full covering space and refer to half branes simply as branes.

\paragraph{Component $0\in\mathbb{Z}_2$.} The brane system for the affine Grassmannian of $\mathrm{Sp}(2n,\mathbb{C})$, i.e.\ the $0$-th component of the affine Grassmannian of $ \mathrm{PSp}(2n,\mathbb{C})$, is proposed to be
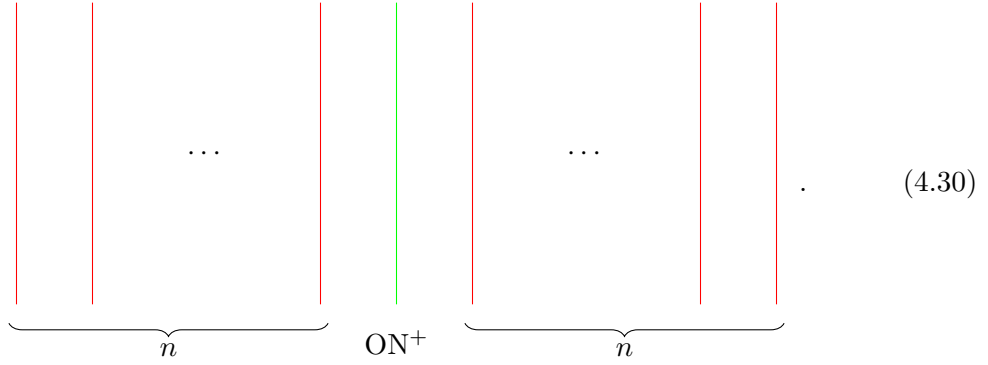
\begin{equation}
    \vcenter{\hbox{

    }} \, . 
\end{equation}
Depicted is the origin of $\w{\mathrm{PSp}(2n,\mathbb{C})}{}{[0,\dots,0,0]}$, with all D3 branes and D5 branes suppressed. The quiver representing the origin is:
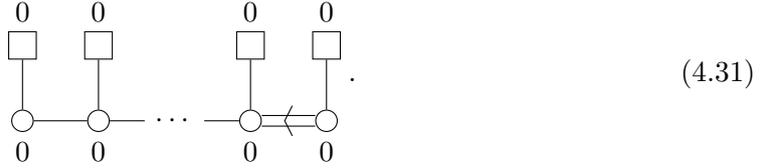
\begin{equation}
    \raisebox{-.5\height}{%
} \, . 
\end{equation}

\paragraph{Component $1\in\mathbb{Z}_2$.} The brane system for the $1$-st component of the affine Grassmannian of $ \mathrm{PSp}(2n,\mathbb{C})$, is proposed to be\begin{equation}
    \vcenter{\hbox{
    %
    }}  \, . 
\end{equation}
Depicted is the origin of $\w{\mathrm{PSp}(2n,\mathbb{C})}{}{[0,\dots,0,1]}$, with all D3 branes and D5 branes suppressed. The quiver representing the origin is:
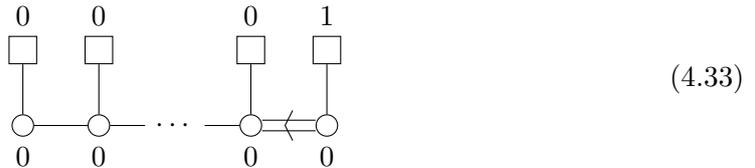
\begin{equation}
   \raisebox{-.5\height}{ %
}
\end{equation}

\paragraph{Minimal transitions.} In the $C$-type brane systems, we may perform the same transitions as in the $A$-type case. Additionally there are two more minimal transitions:
\begin{equation}
    \vcenter{\hbox{
    %
    }}
\end{equation}
For $n=1$ there are transitions corresponding to Kleinian singularities studied in detail in the next section.

\subsubsection{\texorpdfstring{The $C_1$ affine Grassmannian}{The C1 affine Grassmannian}}

\paragraph{Component $0 \in \mathbb{Z}_2$.} The affine Grassmannian of $ \mathrm{PSp}(2,\mathbb{C})$ is given by the brane system:
\begin{equation}
    \vcenter{\hbox{
    %
}}\qquad,
\end{equation}
depicted at the origin of $\w{\mathrm{PSp}(2,\mathbb{C})}{}{[0]}$. We can now move onto the lowest non-trivial leaf by activating a D3 brane between two D5 branes, which are mirror images, and moving the D3 brane along the NS5 branes, corresponding to a $A_1$ transition:
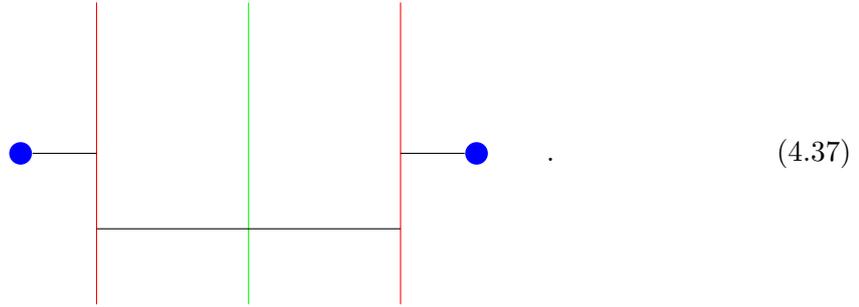
\begin{equation}
    \vcenter{\hbox{
    %
}}\qquad.
\end{equation}
After a Hanany-Witten transition we can read off the quiver:
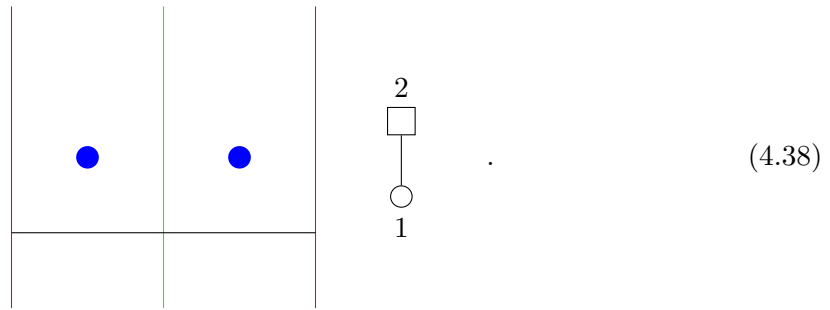
\begin{equation}
    \vcenter{\hbox{
    %
    }}\qquad.
\end{equation}
The only way to keep moving onto higher leaves in the affine Grassmannian is by repeating the same operation, however the D5 branes already placed between the NS5 branes make the transverse slices $A_{2k-1}$ for the $k$-th transition of this type. A general leaf and the corresponding quiver are:
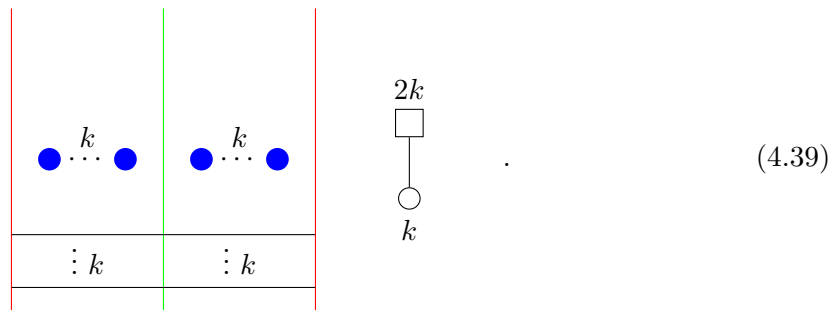
\begin{equation}
    \vcenter{\hbox{
    %
    }}\qquad.
    \label{eq:C1_1st}
\end{equation}
This is in agreement with the construction in both Section \ref{sec:A1branes} and \ref{sec:B1branes}.

\paragraph{Component $1 \in \mathbb{Z}_2$.} For the second connected component for $\mathrm{PSp}(2,\mathbb{C})$ we start with:
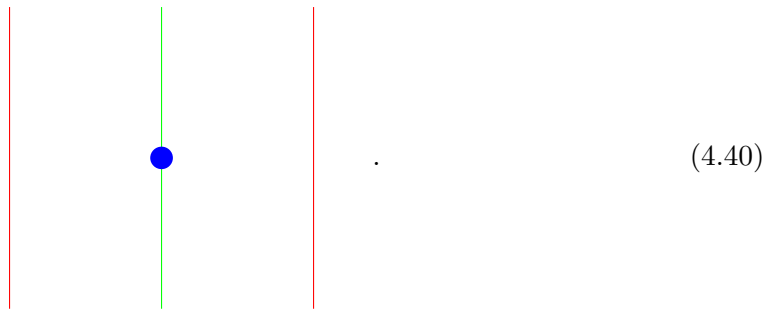
\begin{equation}
    \vcenter{\hbox{
    %
}}\qquad.
\end{equation}
Depicting the origin of $\w{\mathrm{PSp}(2,\mathbb{C})}{}{[1]}$. The transverse slice for the $k$-th transition is $A_{2k}$ and the brane system and quiver for the leaf are:
\begin{equation}
    \vcenter{\hbox{
    %
    }}\qquad.
    \label{eq:C1_2nd}
\end{equation}
This is in agreement with the construction in both Section \ref{sec:A1branes} and \ref{sec:B1branes}.

\subsubsection{\texorpdfstring{The $C_2$ affine Grassmannian}{The C2 affine Grassmannian}}

The two components are represented in Figures \ref{fig:C2Branes} and \ref{fig:C2Branes2}. 

\begin{figure}[t]
    \centering
    \begin{adjustbox}{center}
    %
    \end{adjustbox}
    \caption{Brane depiction of the low dimensional leaves [and elementary slices between them] of $\w{ \mathrm{PSp}(4,\mathbb{C})}{}{[0,0]}$.}
    \label{fig:C2Branes}
\end{figure}

\begin{figure}[t]
    \centering
    \begin{adjustbox}{center}
    %
}$};
        \draw (00)--(0010)--(10)--(1002)--(02)--(0220)--(20)--(2012)--(12)--(1230)--(30)--(3022)--(22)--(0422)--(04)--(1204)--(12);
        \draw (22)--(2*\x-0.3*\x,4*\x+0.3*\x) (22)--(2*\x+0.7*\x,4*\x);
    \end{tikzpicture}
    \end{adjustbox}
    \caption{Brane depiction of the low dimensional leaves [and elementary slices between them] of $\w{ \mathrm{PSp}(4,\mathbb{C})}{}{[0,1]}$.}
    \label{fig:C2Branes2}
\end{figure}


\subsection{\texorpdfstring{The $D$-type affine Grassmannian}{The D-type affine Grassmannian}}

For this setup, we require a ${\color{cyan}\textnormal{ON}^-}$ plane. In the main text we draw the full covering space and refer to half branes simply as branes.

\paragraph{Component $(0,0)\in\mathbb{Z}_2\times\mathbb{Z}_2$ or $0\in\mathbb{Z}_4$.} The brane system for the affine Grassmannian of $\mathrm{Spin}(2n,\mathbb{C})$, i.e.\ the $(0,0)$-th (for $n$ even) or $0$-th component (for $n$ odd) of the affine Grassmannian of $ \mathrm{PSO}(2n,\mathbb{C})$, is proposed to be
\begin{equation}
    \vcenter{\hbox{

    }}
\end{equation}
Depicted is at the origin of $\w{\mathrm{Spin}(2n,\mathbb{C})}{}{[0,0,\dots,0,0,0]}$, with all D3 branes and D5 branes suppressed. The quiver representing the origin is:
\begin{equation}
   \raisebox{-.5\height}{ %
} \, . 
\end{equation}
There are three more components.
\paragraph{Component $(1,1)\in\mathbb{Z}_2\times\mathbb{Z}_2$ or $2\in\mathbb{Z}_4$.}
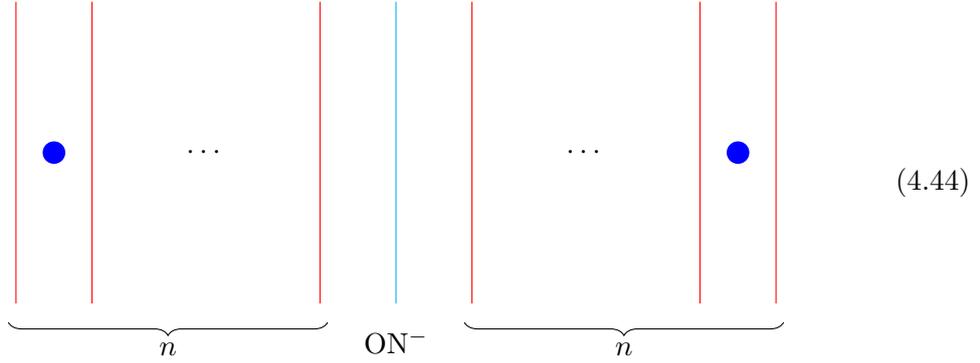
\begin{equation}
    \vcenter{\hbox{
    %
    }}
\end{equation}
Depicted at the origin of $\w{\mathrm{Spin}(2n,\mathbb{C})}{}{[1,0,\dots,0,0,0]}$, with all D3 branes and D5 branes suppressed. The quiver representing the origin is:
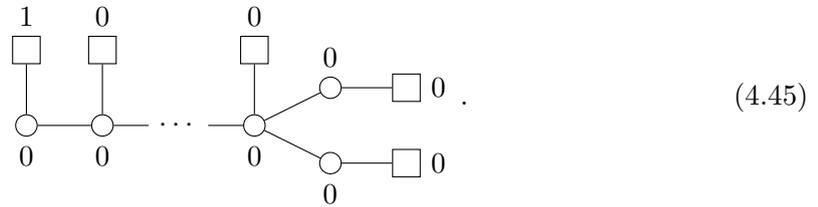
\begin{equation}
     \raisebox{-.5\height}{ %
} \, . 
\end{equation}

\paragraph{Component $(1,0)\in\mathbb{Z}_2\times\mathbb{Z}_2$ or $1\in\mathbb{Z}_4$.}
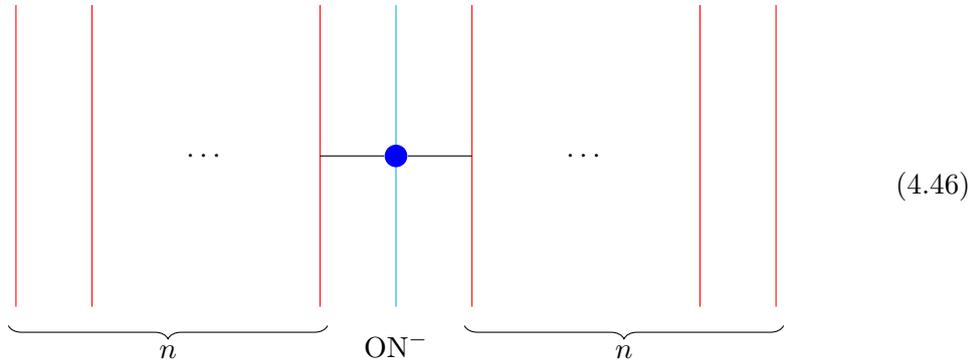
\begin{equation}
    \vcenter{\hbox{
    %
    }}
\end{equation}
Depicted at the origin of $\w{\mathrm{Spin}(2n,\mathbb{C})}{}{[0,0,\dots,0,1,0]}$, with all D3 branes and D5 branes suppressed. The quiver representing the origin is:
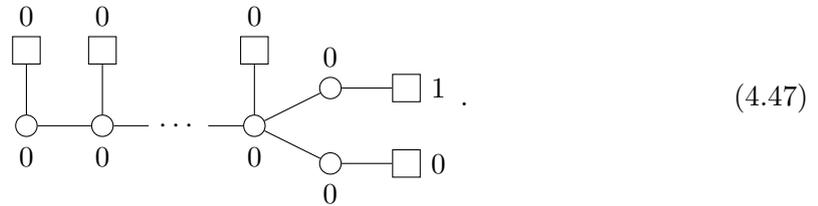
\begin{equation}
     \raisebox{-.5\height}{ %
} \, . 
\end{equation}

\paragraph{Component $(0,1)\in\mathbb{Z}_2\times\mathbb{Z}_2$ or $3\in\mathbb{Z}_4$.}
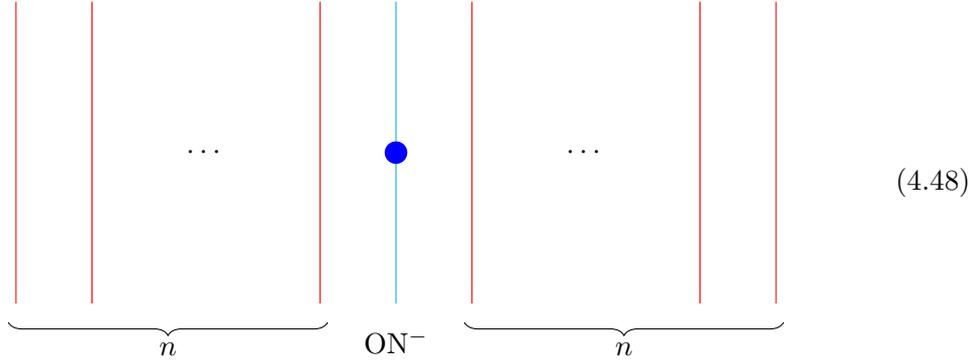
\begin{equation}
    \vcenter{\hbox{
    %
    }}
\end{equation}
Depicted at the origin of $\w{\mathrm{Spin}(2n,\mathbb{C})}{}{[0,0,\dots,0,0,1]}$, with all D3 branes and D5 branes suppressed. The quiver representing the origin is:
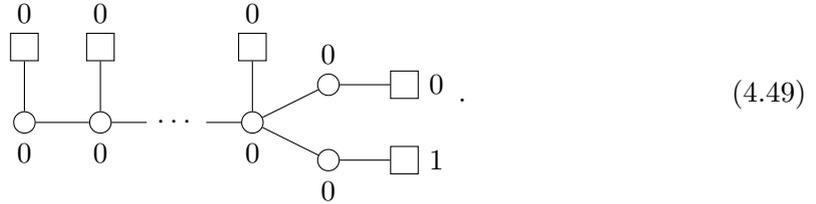
\begin{equation}
     \raisebox{-.5\height}{ %
} \, . 
\end{equation}

\paragraph{Minimal transitions.} In the $D$-type brane systems, we may perform the same transitions as in the $A$-type case. Additionally there is one more minimal transition:
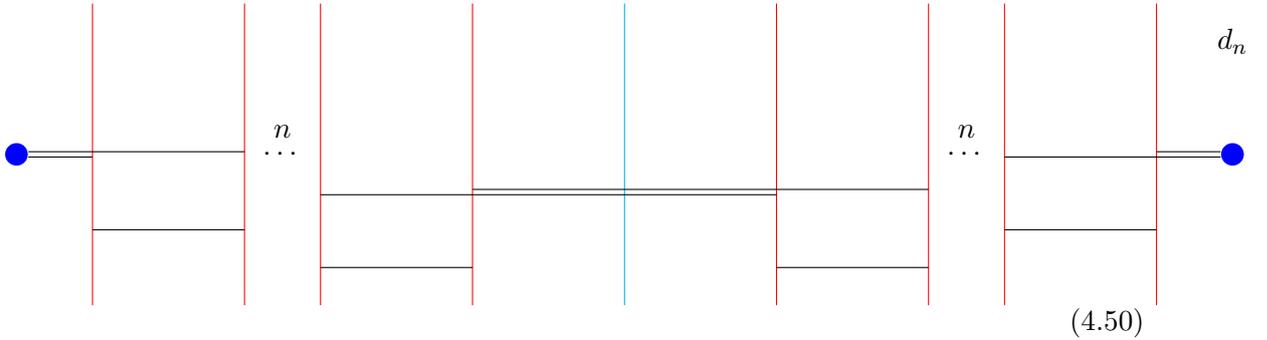
\begin{equation}
    \vcenter{\hbox{
    %
    }}
\end{equation}

\subsubsection{\texorpdfstring{The $D_2$ affine Grassmannian}{The D2 affine Grassmannian}}

There are four components:
\paragraph{Component $(0,0) \in \mathbb{Z}_2\times\mathbb{Z}_2$.} This component is studied in Figure \ref{fig:D2Branes}. The origin is
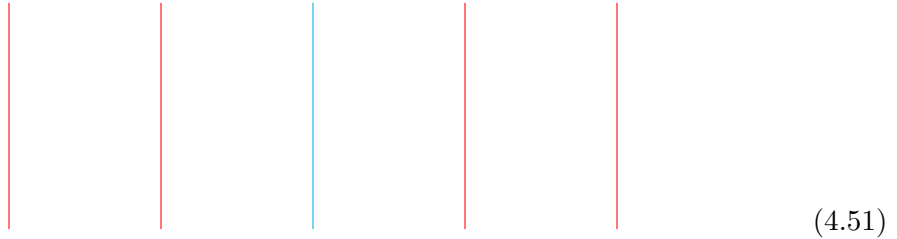
\begin{equation}
    %
\end{equation}
\paragraph{Component $(0,1) \in \mathbb{Z}_2\times\mathbb{Z}_2$.} This component is studied in Figure \ref{fig:D2Branes2}. The origin is
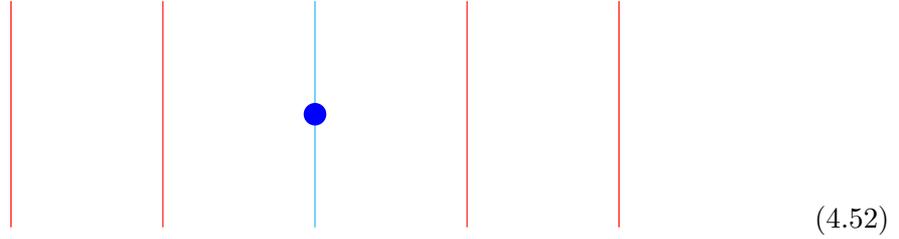
\begin{equation}
    %
\end{equation}
\paragraph{Component $(1,0) \in \mathbb{Z}_2\times\mathbb{Z}_2$.} This component is studied in Figure \ref{fig:D2Branes3}. The origin is
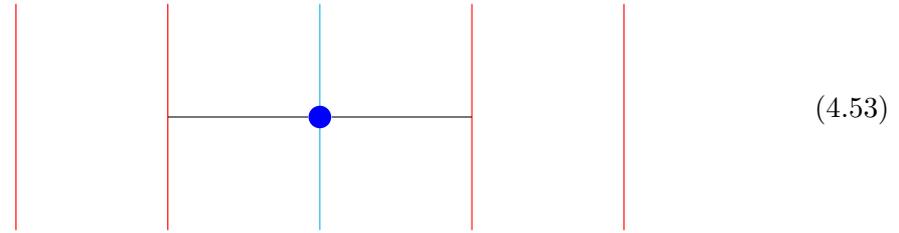
\begin{equation}
     \raisebox{-.5\height}{ %
}
\end{equation}
\paragraph{Component $(1,1) \in \mathbb{Z}_2\times\mathbb{Z}_2$.} This component is studied in Figure \ref{fig:D2Branes4}. The origin is
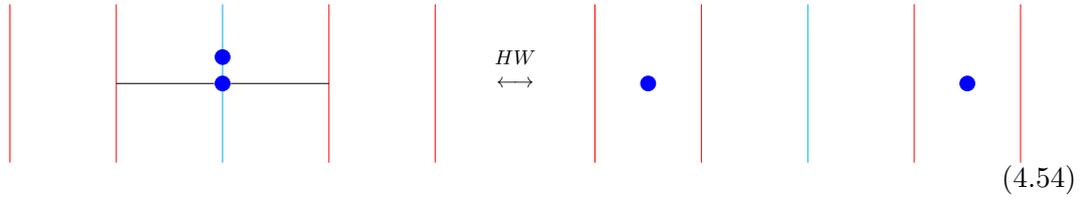
\begin{equation}
 \raisebox{-.5\height}{ \begin{adjustbox}{center}
\scalebox{0.7}{
    %
}
\end{adjustbox}}
\end{equation}

\paragraph{Transition.} While not strictly a Hanany-Witten transition, the following two brane webs are equivalent in that they depict the same leaf in the Coulomb branch:
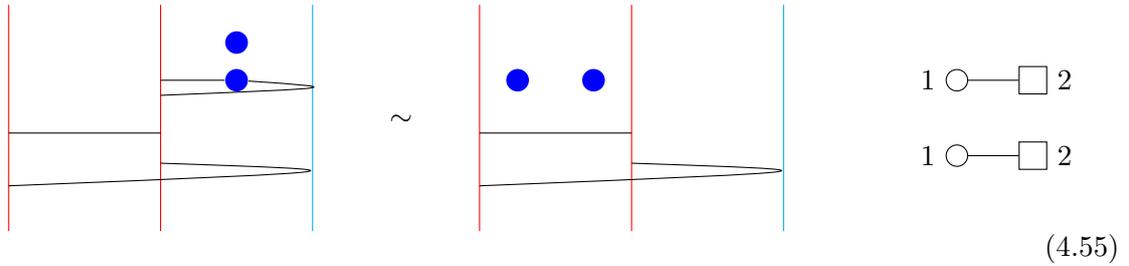
\begin{equation}
    \vcenter{\hbox{\scalebox{1}{%
}}}
        \label{eq:D2choices}
\end{equation}
In Figure \ref{fig:D2Branes} we use the representation on the left. Furthermore the following two transverse slices are equivalent and correspond to an $A_1$ transition:
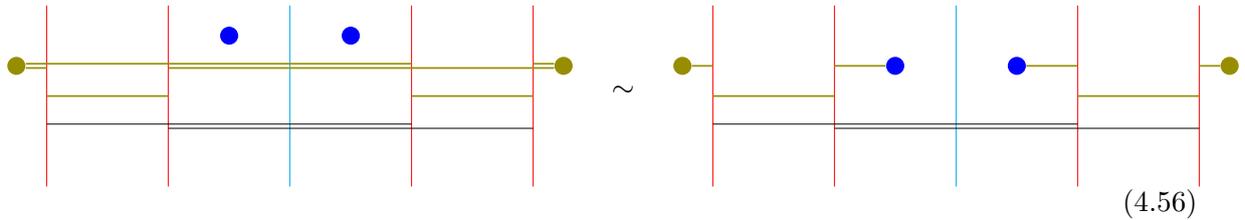
\begin{equation}
    \begin{adjustbox}{center}\raisebox{-0.5\height}{\scalebox{0.8}{%
}}}
        \end{adjustbox}
        \label{eq:D2choices2}
\end{equation}
The only difference between the two brane systems in \eqref{eq:D2choices} and \eqref{eq:D2choices2} lies in their Higgs branch.

\begin{landscape}

\begin{figure}[t]
    \centering
    \begin{adjustbox}{center}
    %
}$};
        \draw (00)--(0020)--(20)--(2040)--(40) (20)--(2022)--(22) (00)--(0002)--(02)--(0204)--(04) (02)--(0222)--(22);
        \draw (40)--(-2*\x-0.4*\x,2*\x+0.4*\x) (40)--(-2*\x+0.4*\x,2*\x+0.4*\x) (22)--(-0.4*\x,2*\x+0.4*\x) (22)--(0.4*\x,2*\x+0.4*\x) (04)--(2*\x-0.4*\x,2*\x+0.4*\x) (04)--(2*\x+0.4*\x,2*\x+0.4*\x);
    \end{tikzpicture}
    \end{adjustbox}
    \caption{Brane depiction of the low dimensional leaves [and elementary slices between them] of $\w{\mathrm{PSO}(4,\mathbb{C})}{}{[0,0]}$.}
    \label{fig:D2Branes}
\end{figure}

\begin{figure}[t]
    \centering
    \begin{adjustbox}{center}
    \begin{tikzpicture}
    \def\x{5*1cm};
        \node (00) at (0,0) {$\scalebox{0.7}{\begin{tikzpicture}
            \draw[red] (0,0)--(0,3) (2,0)--(2,3);
            \draw[cyan] (4,0)--(4,3);
            \node[D5] at (4,2.5) {};
        \end{tikzpicture}}$};
        \node (20) at (-1*\x,1*\x) {$\scalebox{0.7}{\begin{tikzpicture}
            \draw[red] (0,0)--(0,3) (2,0)--(2,3);
            \draw[cyan] (4,0)--(4,3);
            \node[D5] at (3,2.5) {};
            \draw (2,0.9) .. controls (4.9,0.8) .. (0,0.6);
            \node[D5] at (4,2.5) {};
        \end{tikzpicture}}$};
        \node (02) at (1*\x,1*\x) {$\scalebox{0.7}{\begin{tikzpicture}
            \draw[red] (0,0)--(0,3) (2,0)--(2,3);
            \draw[cyan] (4,0)--(4,3);
            \node[D5] (1) at (3,2) {};
            \draw (2,2)--(1);
            \draw (1) .. controls (4.45,1.9) .. (2,1.8);
            \draw (0,1)--(2,1);
            \node[D5] at (4,2.5) {};
        \end{tikzpicture}}$};
        \node (40) at (-2*\x,2*\x) {$\scalebox{0.7}{\begin{tikzpicture}
            \draw[red] (0,0)--(0,3) (2,0)--(2,3);
            \draw[cyan] (4,0)--(4,3);
            \node[D5] at (2.5,2.5) {};
            \node[D5] at (3.5,2.5) {};
            \draw (2,0.9) .. controls (4.9,0.8) .. (0,0.6);
            \draw (2,1.4) .. controls (4.9,1.3) .. (0,1.1);
            \node[D5] at (4,2.5) {};
        \end{tikzpicture}}$};
        \node (22) at (0*\x,2*\x) {$\scalebox{0.7}{\begin{tikzpicture}
            \draw[red] (0,0)--(0,3) (2,0)--(2,3);
            \draw[cyan] (4,0)--(4,3);
            \node[D5] at (3,2.5) {};
            \node[D5] (1) at (3,2) {};
            \draw (2,2)--(1);
            \draw (1) .. controls (4.45,1.9) .. (2,1.8);
            \draw (0,1.2)--(2,1.2);
            \draw (2,0.9) .. controls (4.9,0.8) .. (0,0.6);
            \node[D5] at (4,2.5) {};
        \end{tikzpicture}}$};
        \node (04) at (2*\x,2*\x) {$\scalebox{0.7}{\begin{tikzpicture}
            \draw[red] (0,0)--(0,3) (2,0)--(2,3);
            \draw[cyan] (4,0)--(4,3);
            \node[D5] (1) at (3,2) {};
            \draw (2,2)--(1);
            \draw (1) .. controls (4.45,1.9) .. (2,1.8);
            \node[D5] (2) at (3,1.6) {};
            \draw (2,1.6)--(2);
            \draw (2) .. controls (4.45,1.5) .. (2,1.4);
            \draw (0,1)--(2,1) (0,0.7)--(2,0.7);
            \node[D5] at (4,2.5) {};
        \end{tikzpicture}}$};
        \node[draw,label=left:{\color{olive}$A_2$}] (0020) at ($(00)!0.5!(20)$) {$\scalebox{0.4}{\begin{tikzpicture}
            \draw[red] (0,0)--(0,3) (2,0)--(2,3) (6,0)--(6,3) (8,0)--(8,3);
            \draw[cyan] (4,0)--(4,3);
            \node[D5o] (1) at (-0.5,2) {};
            \node[D5o] (2) at (8.5,2) {};
            \draw[olive,thick,transform canvas={yshift=2pt}] (1)--(2,2);
            \draw[olive,thick,transform canvas={yshift=-2pt}] (1)--(0,2);
            \draw[olive,thick,transform canvas={yshift=-2pt}] (2)--(6,2);
            \draw[olive,thick,transform canvas={yshift=2pt}] (2)--(8,2);
            \draw[olive,thick,transform canvas={yshift=2pt}] (2,1)--(8,1);
            \draw[olive,thick,transform canvas={yshift=-2pt}] (0,1)--(6,1);
            \node[D5] at (4,2.5) {};
        \end{tikzpicture}}$};
        \node[draw,label=left:{\color{olive}$A_1$}] (0002) at ($(00)!0.5!(02)$) {$\scalebox{0.4}{\begin{tikzpicture}
            \draw[red] (0,0)--(0,3) (2,0)--(2,3) (6,0)--(6,3) (8,0)--(8,3);
            \draw[cyan] (4,0)--(4,3);
            \node[D5o] (1) at (-0.5,2) {};
            \node[D5o] (2) at (8.5,2) {};
            \draw[olive,thick,transform canvas={yshift=2pt}] (1)--(6,2);
            \draw[olive,thick,transform canvas={yshift=-2pt}] (1)--(0,2);
            \draw[olive,thick,transform canvas={yshift=-2pt}] (2)--(2,2);
            \draw[olive,thick,transform canvas={yshift=2pt}] (2)--(8,2);
            \draw[olive,thick] (0,1)--(2,1) (6,1)--(8,1);
            \node[D5] at (4,2.5) {};
        \end{tikzpicture}}$};
        \node[draw,label=left:{\color{olive}$A_4$}] (2040) at ($(20)!0.5!(40)$) {$\scalebox{0.4}{\begin{tikzpicture}
            \draw[red] (0,0)--(0,3) (2,0)--(2,3) (6,0)--(6,3) (8,0)--(8,3);
            \draw[cyan] (4,0)--(4,3);
            \node[D5] at (3,2.5) {};
            \node[D5] at (5,2.5) {};
            \draw[transform canvas={yshift=2pt}] (2,0.7)--(8,0.7);
            \draw[transform canvas={yshift=-2pt}] (0,0.7)--(6,0.7);
            \node[D5o] (1) at (-0.5,2) {};
            \node[D5o] (2) at (8.5,2) {};
            \draw[olive,thick,transform canvas={yshift=2pt}] (1)--(2,2);
            \draw[olive,thick,transform canvas={yshift=-2pt}] (1)--(0,2);
            \draw[olive,thick,transform canvas={yshift=-2pt}] (2)--(6,2);
            \draw[olive,thick,transform canvas={yshift=2pt}] (2)--(8,2);
            \draw[olive,thick,transform canvas={yshift=2pt}] (2,1.2)--(8,1.2);
            \draw[olive,thick,transform canvas={yshift=-2pt}] (0,1.2)--(6,1.2);
            \node[D5] at (4,2.5) {};
        \end{tikzpicture}}$};
        \node[draw,label=left:{\color{olive}$A_1$}] (2022) at ($(20)!0.5!(22)$) {$\scalebox{0.4}{\begin{tikzpicture}
            \draw[red] (0,0)--(0,3) (2,0)--(2,3) (6,0)--(6,3) (8,0)--(8,3);
            \draw[cyan] (4,0)--(4,3);
            \node[D5] at (3,2.5) {};
            \node[D5] at (5,2.5) {};
            \draw[transform canvas={yshift=2pt}] (2,0.7)--(8,0.7);
            \draw[transform canvas={yshift=-2pt}] (0,0.7)--(6,0.7);
            \node[D5o] (1) at (-0.5,2) {};
            \node[D5o] (2) at (8.5,2) {};
            \draw[olive,thick,transform canvas={yshift=2pt}] (1)--(6,2);
            \draw[olive,thick,transform canvas={yshift=-2pt}] (1)--(0,2);
            \draw[olive,thick,transform canvas={yshift=-2pt}] (2)--(2,2);
            \draw[olive,thick,transform canvas={yshift=2pt}] (2)--(8,2);
            \draw[olive,thick,transform canvas={yshift=2pt}] (6,1.2)--(8,1.2);
            \draw[olive,thick,transform canvas={yshift=2pt}] (0,1.2)--(2,1.2);
            \node[D5] at (4,2.5) {};
        \end{tikzpicture}}$};
        \node[draw,label=left:{\color{olive}$A_2$}] (0222) at ($(02)!0.5!(22)$) {$\scalebox{0.4}{\begin{tikzpicture}
            \draw[red] (0,0)--(0,3) (2,0)--(2,3) (6,0)--(6,3) (8,0)--(8,3);
            \draw[cyan] (4,0)--(4,3);
            \node[D5] (l) at (3,1.5) {};
            \node[D5] (r) at (5,1.5) {};
            \draw (0,0.5)--(2,0.5) (6,0.5)--(8,0.5);
            \draw[transform canvas={yshift=2pt}] (2,1.5)--(l)--(6,1.5);
            \draw[transform canvas={yshift=-2pt}] (2,1.5)--(r)--(6,1.5);
            \node[D5o] (1) at (-0.5,2) {};
            \node[D5o] (2) at (8.5,2) {};
            \draw[olive,thick,transform canvas={yshift=2pt}] (1)--(2,2);
            \draw[olive,thick,transform canvas={yshift=-2pt}] (1)--(0,2);
            \draw[olive,thick,transform canvas={yshift=-2pt}] (2)--(6,2);
            \draw[olive,thick,transform canvas={yshift=2pt}] (2)--(8,2);
            \draw[olive,thick,transform canvas={yshift=2pt}] (2,1)--(8,1);
            \draw[olive,thick,transform canvas={yshift=-2pt}] (0,1)--(6,1);
            \node[D5] at (4,2.5) {};
        \end{tikzpicture}}$};
        \node[draw,label=left:{\color{olive}$A_3$}] (0204) at ($(02)!0.5!(04)$) {$\scalebox{0.4}{\begin{tikzpicture}
            \draw[red] (0,0)--(0,3) (2,0)--(2,3) (6,0)--(6,3) (8,0)--(8,3);
            \draw[cyan] (4,0)--(4,3);
            \node[D5] (l) at (3,1.5) {};
            \node[D5] (r) at (5,1.5) {};
            \draw (0,0.5)--(2,0.5) (6,0.5)--(8,0.5);
            \draw[transform canvas={yshift=2pt}] (2,1.5)--(l)--(6,1.5);
            \draw[transform canvas={yshift=-2pt}] (2,1.5)--(r)--(6,1.5);
            \node[D5o] (1) at (-0.5,2) {};
            \node[D5o] (2) at (8.5,2) {};
            \draw[olive,thick,transform canvas={yshift=2pt}] (1)--(6,2);
            \draw[olive,thick,transform canvas={yshift=-2pt}] (1)--(0,2);
            \draw[olive,thick,transform canvas={yshift=-2pt}] (2)--(2,2);
            \draw[olive,thick,transform canvas={yshift=2pt}] (2)--(8,2);
            \draw[olive,thick,transform canvas={yshift=2pt}] (6,1)--(8,1);
            \draw[olive,thick,transform canvas={yshift=2pt}] (0,1)--(2,1);
            \node[D5] at (4,2.5) {};
        \end{tikzpicture}}$};
        \draw (00)--(0020)--(20)--(2040)--(40) (20)--(2022)--(22) (00)--(0002)--(02)--(0204)--(04) (02)--(0222)--(22);
        \draw (40)--(-2*\x-0.4*\x,2*\x+0.4*\x) (40)--(-2*\x+0.4*\x,2*\x+0.4*\x) (22)--(-0.4*\x,2*\x+0.4*\x) (22)--(0.4*\x,2*\x+0.4*\x) (04)--(2*\x-0.4*\x,2*\x+0.4*\x) (04)--(2*\x+0.4*\x,2*\x+0.4*\x);
    \end{tikzpicture}
    \end{adjustbox}
    \caption{Brane depiction of the low dimensional leaves [and elementary slices between them] of $\w{\mathrm{PSO}(4,\mathbb{C})}{}{[0,1]}$.}
    \label{fig:D2Branes2}
\end{figure}

\begin{figure}[t]
    \centering
    \begin{adjustbox}{center}
    \begin{tikzpicture}
    \def\x{5*1cm};
        \node (00) at (0,0) {$\scalebox{0.7}{\begin{tikzpicture}
            \draw[red] (0,0)--(0,3) (2,0)--(2,3);
            \draw[cyan] (4,0)--(4,3);
            \node[D5] (m) at (4,2.2) {};
            \draw (2,2.2)--(m);
        \end{tikzpicture}}$};
        \node (20) at (-1*\x,1*\x) {$\scalebox{0.7}{\begin{tikzpicture}
            \draw[red] (0,0)--(0,3) (2,0)--(2,3);
            \draw[cyan] (4,0)--(4,3);
            \node[D5] at (3,2.5) {};
            \draw (2,0.9) .. controls (4.9,0.8) .. (0,0.6);
            \node[D5] (m) at (4,2.8) {};
            \draw (2,2.8)--(m);
        \end{tikzpicture}}$};
        \node (02) at (1*\x,1*\x) {$\scalebox{0.7}{\begin{tikzpicture}
            \draw[red] (0,0)--(0,3) (2,0)--(2,3);
            \draw[cyan] (4,0)--(4,3);
            \node[D5] (1) at (3,2) {};
            \draw (2,2)--(1);
            \draw (1) .. controls (4.45,1.9) .. (2,1.8);
            \draw (0,1)--(2,1);
            \node[D5] (m) at (4,2.8) {};
            \draw (2,2.8)--(m);
        \end{tikzpicture}}$};
        \node (40) at (-2*\x,2*\x) {$\scalebox{0.7}{\begin{tikzpicture}
            \draw[red] (0,0)--(0,3) (2,0)--(2,3);
            \draw[cyan] (4,0)--(4,3);
            \node[D5] at (2.5,2.5) {};
            \node[D5] at (3.5,2.5) {};
            \draw (2,0.9) .. controls (4.9,0.8) .. (0,0.6);
            \draw (2,1.4) .. controls (4.9,1.3) .. (0,1.1);
        \end{tikzpicture}}$};
        \node (22) at (0*\x,2*\x) {$\scalebox{0.7}{\begin{tikzpicture}
            \draw[red] (0,0)--(0,3) (2,0)--(2,3);
            \draw[cyan] (4,0)--(4,3);
            \node[D5] at (3,2.5) {};
            \node[D5] (1) at (3,2) {};
            \draw (2,2)--(1);
            \draw (1) .. controls (4.45,1.9) .. (2,1.8);
            \draw (0,1.2)--(2,1.2);
            \draw (2,0.9) .. controls (4.9,0.8) .. (0,0.6);
            \node[D5] (m) at (4,2.8) {};
            \draw (2,2.8)--(m);
        \end{tikzpicture}}$};
        \node (04) at (2*\x,2*\x) {$\scalebox{0.7}{\begin{tikzpicture}
            \draw[red] (0,0)--(0,3) (2,0)--(2,3);
            \draw[cyan] (4,0)--(4,3);
            \node[D5] (1) at (3,2) {};
            \draw (2,2)--(1);
            \draw (1) .. controls (4.45,1.9) .. (2,1.8);
            \node[D5] (2) at (3,1.6) {};
            \draw (2,1.6)--(2);
            \draw (2) .. controls (4.45,1.5) .. (2,1.4);
            \draw (0,1)--(2,1) (0,0.7)--(2,0.7);
            \node[D5] (m) at (4,2.8) {};
            \draw (2,2.8)--(m);
        \end{tikzpicture}}$};
        \node[draw,label=left:{\color{olive}$A_1$}] (0020) at ($(00)!0.5!(20)$) {$\scalebox{0.4}{\begin{tikzpicture}
            \draw[red] (0,0)--(0,3) (2,0)--(2,3) (6,0)--(6,3) (8,0)--(8,3);
            \draw[cyan] (4,0)--(4,3);
            \node[D5o] (1) at (-0.5,2) {};
            \node[D5o] (2) at (8.5,2) {};
            \draw[olive,thick,transform canvas={yshift=2pt}] (1)--(2,2);
            \draw[olive,thick,transform canvas={yshift=-2pt}] (1)--(0,2);
            \draw[olive,thick,transform canvas={yshift=-2pt}] (2)--(6,2);
            \draw[olive,thick,transform canvas={yshift=2pt}] (2)--(8,2);
            \draw[olive,thick,transform canvas={yshift=2pt}] (2,1)--(8,1);
            \draw[olive,thick,transform canvas={yshift=-2pt}] (0,1)--(6,1);
            \node[D5] (m) at (4,2.8) {};
            \draw (2,2.8)--(m)--(6,2.8);
        \end{tikzpicture}}$};
        \node[draw,label=left:{\color{olive}$A_2$}] (0002) at ($(00)!0.5!(02)$) {$\scalebox{0.4}{\begin{tikzpicture}
            \draw[red] (0,0)--(0,3) (2,0)--(2,3) (6,0)--(6,3) (8,0)--(8,3);
            \draw[cyan] (4,0)--(4,3);
            \node[D5o] (1) at (-0.5,2) {};
            \node[D5o] (2) at (8.5,2) {};
            \draw[olive,thick,transform canvas={yshift=2pt}] (1)--(6,2);
            \draw[olive,thick,transform canvas={yshift=-2pt}] (1)--(0,2);
            \draw[olive,thick,transform canvas={yshift=-2pt}] (2)--(2,2);
            \draw[olive,thick,transform canvas={yshift=2pt}] (2)--(8,2);
            \draw[olive,thick] (0,1)--(2,1) (6,1)--(8,1);
            \node[D5] (m) at (4,2.8) {};
            \draw (2,2.8)--(m)--(6,2.8);
        \end{tikzpicture}}$};
        \node[draw,label=left:{\color{olive}$A_3$}] (2040) at ($(20)!0.5!(40)$) {$\scalebox{0.4}{\begin{tikzpicture}
            \draw[red] (0,0)--(0,3) (2,0)--(2,3) (6,0)--(6,3) (8,0)--(8,3);
            \draw[cyan] (4,0)--(4,3);
            \node[D5] at (3,2.5) {};
            \node[D5] at (5,2.5) {};
            \draw[transform canvas={yshift=2pt}] (2,0.7)--(8,0.7);
            \draw[transform canvas={yshift=-2pt}] (0,0.7)--(6,0.7);
            \node[D5o] (1) at (-0.5,2) {};
            \node[D5o] (2) at (8.5,2) {};
            \draw[olive,thick,transform canvas={yshift=2pt}] (1)--(2,2);
            \draw[olive,thick,transform canvas={yshift=-2pt}] (1)--(0,2);
            \draw[olive,thick,transform canvas={yshift=-2pt}] (2)--(6,2);
            \draw[olive,thick,transform canvas={yshift=2pt}] (2)--(8,2);
            \draw[olive,thick,transform canvas={yshift=2pt}] (2,1.2)--(8,1.2);
            \draw[olive,thick,transform canvas={yshift=-2pt}] (0,1.2)--(6,1.2);
            \node[D5] (m) at (4,2.8) {};
            \draw (2,2.8)--(m)--(6,2.8);
        \end{tikzpicture}}$};
        \node[draw,label=left:{\color{olive}$A_2$}] (2022) at ($(20)!0.5!(22)$) {$\scalebox{0.4}{\begin{tikzpicture}
            \draw[red] (0,0)--(0,3) (2,0)--(2,3) (6,0)--(6,3) (8,0)--(8,3);
            \draw[cyan] (4,0)--(4,3);
            \node[D5] at (3,2.5) {};
            \node[D5] at (5,2.5) {};
            \draw[transform canvas={yshift=2pt}] (2,0.7)--(8,0.7);
            \draw[transform canvas={yshift=-2pt}] (0,0.7)--(6,0.7);
            \node[D5o] (1) at (-0.5,2) {};
            \node[D5o] (2) at (8.5,2) {};
            \draw[olive,thick,transform canvas={yshift=2pt}] (1)--(6,2);
            \draw[olive,thick,transform canvas={yshift=-2pt}] (1)--(0,2);
            \draw[olive,thick,transform canvas={yshift=-2pt}] (2)--(2,2);
            \draw[olive,thick,transform canvas={yshift=2pt}] (2)--(8,2);
            \draw[olive,thick,transform canvas={yshift=2pt}] (6,1.2)--(8,1.2);
            \draw[olive,thick,transform canvas={yshift=2pt}] (0,1.2)--(2,1.2);
            \node[D5] (m) at (4,2.8) {};
            \draw (2,2.8)--(m)--(6,2.8);
        \end{tikzpicture}}$};
        \node[draw,label=left:{\color{olive}$A_1$}] (0222) at ($(02)!0.5!(22)$) {$\scalebox{0.4}{\begin{tikzpicture}
            \draw[red] (0,0)--(0,3) (2,0)--(2,3) (6,0)--(6,3) (8,0)--(8,3);
            \draw[cyan] (4,0)--(4,3);
            \node[D5] (l) at (3,1.5) {};
            \node[D5] (r) at (5,1.5) {};
            \draw (0,0.5)--(2,0.5) (6,0.5)--(8,0.5);
            \draw[transform canvas={yshift=2pt}] (2,1.5)--(l)--(6,1.5);
            \draw[transform canvas={yshift=-2pt}] (2,1.5)--(r)--(6,1.5);
            \node[D5o] (1) at (-0.5,2) {};
            \node[D5o] (2) at (8.5,2) {};
            \draw[olive,thick,transform canvas={yshift=2pt}] (1)--(2,2);
            \draw[olive,thick,transform canvas={yshift=-2pt}] (1)--(0,2);
            \draw[olive,thick,transform canvas={yshift=-2pt}] (2)--(6,2);
            \draw[olive,thick,transform canvas={yshift=2pt}] (2)--(8,2);
            \draw[olive,thick,transform canvas={yshift=2pt}] (2,1)--(8,1);
            \draw[olive,thick,transform canvas={yshift=-2pt}] (0,1)--(6,1);
            \node[D5] (m) at (4,2.8) {};
            \draw (2,2.8)--(m)--(6,2.8);
        \end{tikzpicture}}$};
        \node[draw,label=left:{\color{olive}$A_4$}] (0204) at ($(02)!0.5!(04)$) {$\scalebox{0.4}{\begin{tikzpicture}
            \draw[red] (0,0)--(0,3) (2,0)--(2,3) (6,0)--(6,3) (8,0)--(8,3);
            \draw[cyan] (4,0)--(4,3);
            \node[D5] (l) at (3,1.5) {};
            \node[D5] (r) at (5,1.5) {};
            \draw (0,0.5)--(2,0.5) (6,0.5)--(8,0.5);
            \draw[transform canvas={yshift=2pt}] (2,1.5)--(l)--(6,1.5);
            \draw[transform canvas={yshift=-2pt}] (2,1.5)--(r)--(6,1.5);
            \node[D5o] (1) at (-0.5,2) {};
            \node[D5o] (2) at (8.5,2) {};
            \draw[olive,thick,transform canvas={yshift=2pt}] (1)--(6,2);
            \draw[olive,thick,transform canvas={yshift=-2pt}] (1)--(0,2);
            \draw[olive,thick,transform canvas={yshift=-2pt}] (2)--(2,2);
            \draw[olive,thick,transform canvas={yshift=2pt}] (2)--(8,2);
            \draw[olive,thick,transform canvas={yshift=2pt}] (6,1)--(8,1);
            \draw[olive,thick,transform canvas={yshift=2pt}] (0,1)--(2,1);
            \node[D5] (m) at (4,2.8) {};
            \draw (2,2.8)--(m)--(6,2.8);
        \end{tikzpicture}}$};
        \draw (00)--(0020)--(20)--(2040)--(40) (20)--(2022)--(22) (00)--(0002)--(02)--(0204)--(04) (02)--(0222)--(22);
        \draw (40)--(-2*\x-0.4*\x,2*\x+0.4*\x) (40)--(-2*\x+0.4*\x,2*\x+0.4*\x) (22)--(-0.4*\x,2*\x+0.4*\x) (22)--(0.4*\x,2*\x+0.4*\x) (04)--(2*\x-0.4*\x,2*\x+0.4*\x) (04)--(2*\x+0.4*\x,2*\x+0.4*\x);
    \end{tikzpicture}
    \end{adjustbox}
    \caption{Brane depiction of the low dimensional leaves [and elementary slices between them] of $\w{\mathrm{PSO}(4,\mathbb{C})}{}{[1,0]}$.}
    \label{fig:D2Branes3}
\end{figure}

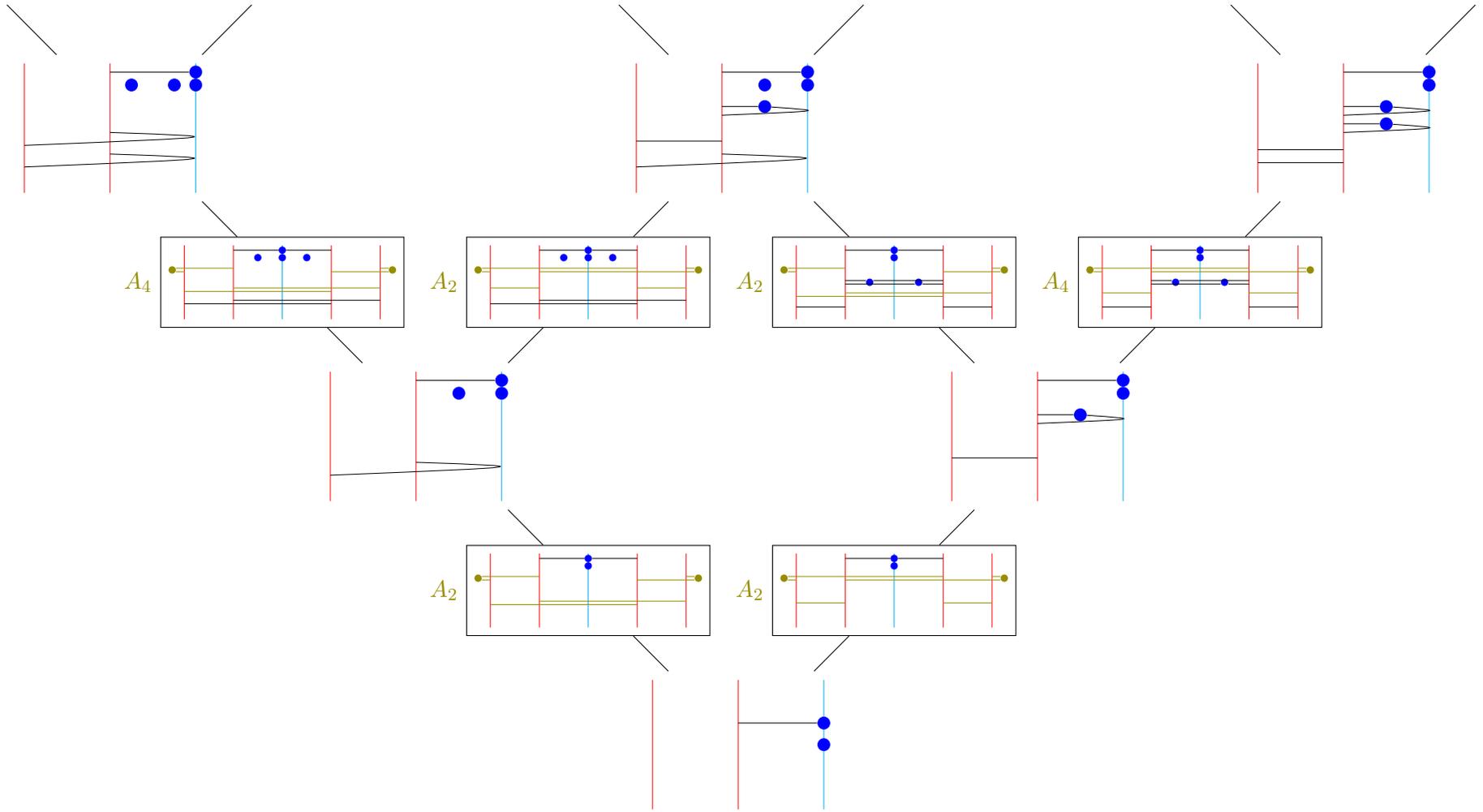
\begin{figure}[t]
    \centering
    \begin{adjustbox}{center}
    \begin{tikzpicture}
    \def\x{5*1cm};
        \node (00) at (0,0) {$\scalebox{0.7}{\begin{tikzpicture}
            \draw[red] (0,0)--(0,3) (2,0)--(2,3);
            \draw[cyan] (4,0)--(4,3);
            \node[D5] at (4,1.5) {};
            \node[D5] (m) at (4,2) {};
            \draw (2,2)--(m);
        \end{tikzpicture}}$};
        \node (20) at (-1*\x,1*\x) {$\scalebox{0.7}{\begin{tikzpicture}
            \draw[red] (0,0)--(0,3) (2,0)--(2,3);
            \draw[cyan] (4,0)--(4,3);
            \node[D5] at (3,2.5) {};
            \draw (2,0.9) .. controls (4.9,0.8) .. (0,0.6);
            \node[D5] at (4,2.5) {};
            \node[D5] (m) at (4,2.8) {};
            \draw (2,2.8)--(m);
        \end{tikzpicture}}$};
        \node (02) at (1*\x,1*\x) {$\scalebox{0.7}{\begin{tikzpicture}
            \draw[red] (0,0)--(0,3) (2,0)--(2,3);
            \draw[cyan] (4,0)--(4,3);
            \node[D5] (1) at (3,2) {};
            \draw (2,2)--(1);
            \draw (1) .. controls (4.45,1.9) .. (2,1.8);
            \draw (0,1)--(2,1);
            \node[D5] at (4,2.5) {};
            \node[D5] (m) at (4,2.8) {};
            \draw (2,2.8)--(m);
        \end{tikzpicture}}$};
        \node (40) at (-2*\x,2*\x) {$\scalebox{0.7}{\begin{tikzpicture}
            \draw[red] (0,0)--(0,3) (2,0)--(2,3);
            \draw[cyan] (4,0)--(4,3);
            \node[D5] at (2.5,2.5) {};
            \node[D5] at (3.5,2.5) {};
            \draw (2,0.9) .. controls (4.9,0.8) .. (0,0.6);
            \draw (2,1.4) .. controls (4.9,1.3) .. (0,1.1);
            \node[D5] at (4,2.5) {};
            \node[D5] (m) at (4,2.8) {};
            \draw (2,2.8)--(m);
        \end{tikzpicture}}$};
        \node (22) at (0*\x,2*\x) {$\scalebox{0.7}{\begin{tikzpicture}
            \draw[red] (0,0)--(0,3) (2,0)--(2,3);
            \draw[cyan] (4,0)--(4,3);
            \node[D5] at (3,2.5) {};
            \node[D5] (1) at (3,2) {};
            \draw (2,2)--(1);
            \draw (1) .. controls (4.45,1.9) .. (2,1.8);
            \draw (0,1.2)--(2,1.2);
            \draw (2,0.9) .. controls (4.9,0.8) .. (0,0.6);
            \node[D5] at (4,2.5) {};
            \node[D5] (m) at (4,2.8) {};
            \draw (2,2.8)--(m);
        \end{tikzpicture}}$};
        \node (04) at (2*\x,2*\x) {$\scalebox{0.7}{\begin{tikzpicture}
            \draw[red] (0,0)--(0,3) (2,0)--(2,3);
            \draw[cyan] (4,0)--(4,3);
            \node[D5] (1) at (3,2) {};
            \draw (2,2)--(1);
            \draw (1) .. controls (4.45,1.9) .. (2,1.8);
            \node[D5] (2) at (3,1.6) {};
            \draw (2,1.6)--(2);
            \draw (2) .. controls (4.45,1.5) .. (2,1.4);
            \draw (0,1)--(2,1) (0,0.7)--(2,0.7);
            \node[D5] at (4,2.5) {};
            \node[D5] (m) at (4,2.8) {};
            \draw (2,2.8)--(m);
        \end{tikzpicture}}$};
        \node[draw,label=left:{\color{olive}$A_2$}] (0020) at ($(00)!0.5!(20)$) {$\scalebox{0.4}{\begin{tikzpicture}
            \draw[red] (0,0)--(0,3) (2,0)--(2,3) (6,0)--(6,3) (8,0)--(8,3);
            \draw[cyan] (4,0)--(4,3);
            \node[D5o] (1) at (-0.5,2) {};
            \node[D5o] (2) at (8.5,2) {};
            \draw[olive,thick,transform canvas={yshift=2pt}] (1)--(2,2);
            \draw[olive,thick,transform canvas={yshift=-2pt}] (1)--(0,2);
            \draw[olive,thick,transform canvas={yshift=-2pt}] (2)--(6,2);
            \draw[olive,thick,transform canvas={yshift=2pt}] (2)--(8,2);
            \draw[olive,thick,transform canvas={yshift=2pt}] (2,1)--(8,1);
            \draw[olive,thick,transform canvas={yshift=-2pt}] (0,1)--(6,1);
            \node[D5] at (4,2.5) {};
            \node[D5] (m) at (4,2.8) {};
            \draw (2,2.8)--(m)--(6,2.8);
        \end{tikzpicture}}$};
        \node[draw,label=left:{\color{olive}$A_2$}] (0002) at ($(00)!0.5!(02)$) {$\scalebox{0.4}{\begin{tikzpicture}
            \draw[red] (0,0)--(0,3) (2,0)--(2,3) (6,0)--(6,3) (8,0)--(8,3);
            \draw[cyan] (4,0)--(4,3);
            \node[D5o] (1) at (-0.5,2) {};
            \node[D5o] (2) at (8.5,2) {};
            \draw[olive,thick,transform canvas={yshift=2pt}] (1)--(6,2);
            \draw[olive,thick,transform canvas={yshift=-2pt}] (1)--(0,2);
            \draw[olive,thick,transform canvas={yshift=-2pt}] (2)--(2,2);
            \draw[olive,thick,transform canvas={yshift=2pt}] (2)--(8,2);
            \draw[olive,thick] (0,1)--(2,1) (6,1)--(8,1);
            \node[D5] at (4,2.5) {};
            \node[D5] (m) at (4,2.8) {};
            \draw (2,2.8)--(m)--(6,2.8);
        \end{tikzpicture}}$};
        \node[draw,label=left:{\color{olive}$A_4$}] (2040) at ($(20)!0.5!(40)$) {$\scalebox{0.4}{\begin{tikzpicture}
            \draw[red] (0,0)--(0,3) (2,0)--(2,3) (6,0)--(6,3) (8,0)--(8,3);
            \draw[cyan] (4,0)--(4,3);
            \node[D5] at (3,2.5) {};
            \node[D5] at (5,2.5) {};
            \draw[transform canvas={yshift=2pt}] (2,0.7)--(8,0.7);
            \draw[transform canvas={yshift=-2pt}] (0,0.7)--(6,0.7);
            \node[D5o] (1) at (-0.5,2) {};
            \node[D5o] (2) at (8.5,2) {};
            \draw[olive,thick,transform canvas={yshift=2pt}] (1)--(2,2);
            \draw[olive,thick,transform canvas={yshift=-2pt}] (1)--(0,2);
            \draw[olive,thick,transform canvas={yshift=-2pt}] (2)--(6,2);
            \draw[olive,thick,transform canvas={yshift=2pt}] (2)--(8,2);
            \draw[olive,thick,transform canvas={yshift=2pt}] (2,1.2)--(8,1.2);
            \draw[olive,thick,transform canvas={yshift=-2pt}] (0,1.2)--(6,1.2);
            \node[D5] at (4,2.5) {};
            \node[D5] (m) at (4,2.8) {};
            \draw (2,2.8)--(m)--(6,2.8);
        \end{tikzpicture}}$};
        \node[draw,label=left:{\color{olive}$A_2$}] (2022) at ($(20)!0.5!(22)$) {$\scalebox{0.4}{\begin{tikzpicture}
            \draw[red] (0,0)--(0,3) (2,0)--(2,3) (6,0)--(6,3) (8,0)--(8,3);
            \draw[cyan] (4,0)--(4,3);
            \node[D5] at (3,2.5) {};
            \node[D5] at (5,2.5) {};
            \draw[transform canvas={yshift=2pt}] (2,0.7)--(8,0.7);
            \draw[transform canvas={yshift=-2pt}] (0,0.7)--(6,0.7);
            \node[D5o] (1) at (-0.5,2) {};
            \node[D5o] (2) at (8.5,2) {};
            \draw[olive,thick,transform canvas={yshift=2pt}] (1)--(6,2);
            \draw[olive,thick,transform canvas={yshift=-2pt}] (1)--(0,2);
            \draw[olive,thick,transform canvas={yshift=-2pt}] (2)--(2,2);
            \draw[olive,thick,transform canvas={yshift=2pt}] (2)--(8,2);
            \draw[olive,thick,transform canvas={yshift=2pt}] (6,1.2)--(8,1.2);
            \draw[olive,thick,transform canvas={yshift=2pt}] (0,1.2)--(2,1.2);
            \node[D5] at (4,2.5) {};
            \node[D5] (m) at (4,2.8) {};
            \draw (2,2.8)--(m)--(6,2.8);
        \end{tikzpicture}}$};
        \node[draw,label=left:{\color{olive}$A_2$}] (0222) at ($(02)!0.5!(22)$) {$\scalebox{0.4}{\begin{tikzpicture}
            \draw[red] (0,0)--(0,3) (2,0)--(2,3) (6,0)--(6,3) (8,0)--(8,3);
            \draw[cyan] (4,0)--(4,3);
            \node[D5] (l) at (3,1.5) {};
            \node[D5] (r) at (5,1.5) {};
            \draw (0,0.5)--(2,0.5) (6,0.5)--(8,0.5);
            \draw[transform canvas={yshift=2pt}] (2,1.5)--(l)--(6,1.5);
            \draw[transform canvas={yshift=-2pt}] (2,1.5)--(r)--(6,1.5);
            \node[D5o] (1) at (-0.5,2) {};
            \node[D5o] (2) at (8.5,2) {};
            \draw[olive,thick,transform canvas={yshift=2pt}] (1)--(2,2);
            \draw[olive,thick,transform canvas={yshift=-2pt}] (1)--(0,2);
            \draw[olive,thick,transform canvas={yshift=-2pt}] (2)--(6,2);
            \draw[olive,thick,transform canvas={yshift=2pt}] (2)--(8,2);
            \draw[olive,thick,transform canvas={yshift=2pt}] (2,1)--(8,1);
            \draw[olive,thick,transform canvas={yshift=-2pt}] (0,1)--(6,1);
            \node[D5] at (4,2.5) {};
            \node[D5] (m) at (4,2.8) {};
            \draw (2,2.8)--(m)--(6,2.8);
        \end{tikzpicture}}$};
        \node[draw,label=left:{\color{olive}$A_4$}] (0204) at ($(02)!0.5!(04)$) {$\scalebox{0.4}{\begin{tikzpicture}
            \draw[red] (0,0)--(0,3) (2,0)--(2,3) (6,0)--(6,3) (8,0)--(8,3);
            \draw[cyan] (4,0)--(4,3);
            \node[D5] (l) at (3,1.5) {};
            \node[D5] (r) at (5,1.5) {};
            \draw (0,0.5)--(2,0.5) (6,0.5)--(8,0.5);
            \draw[transform canvas={yshift=2pt}] (2,1.5)--(l)--(6,1.5);
            \draw[transform canvas={yshift=-2pt}] (2,1.5)--(r)--(6,1.5);
            \node[D5o] (1) at (-0.5,2) {};
            \node[D5o] (2) at (8.5,2) {};
            \draw[olive,thick,transform canvas={yshift=2pt}] (1)--(6,2);
            \draw[olive,thick,transform canvas={yshift=-2pt}] (1)--(0,2);
            \draw[olive,thick,transform canvas={yshift=-2pt}] (2)--(2,2);
            \draw[olive,thick,transform canvas={yshift=2pt}] (2)--(8,2);
            \draw[olive,thick,transform canvas={yshift=2pt}] (6,1)--(8,1);
            \draw[olive,thick,transform canvas={yshift=2pt}] (0,1)--(2,1);
            \node[D5] at (4,2.5) {};
            \node[D5] (m) at (4,2.8) {};
            \draw (2,2.8)--(m)--(6,2.8);
        \end{tikzpicture}}$};
        \draw (00)--(0020)--(20)--(2040)--(40) (20)--(2022)--(22) (00)--(0002)--(02)--(0204)--(04) (02)--(0222)--(22);
        \draw (40)--(-2*\x-0.4*\x,2*\x+0.4*\x) (40)--(-2*\x+0.4*\x,2*\x+0.4*\x) (22)--(-0.4*\x,2*\x+0.4*\x) (22)--(0.4*\x,2*\x+0.4*\x) (04)--(2*\x-0.4*\x,2*\x+0.4*\x) (04)--(2*\x+0.4*\x,2*\x+0.4*\x);
    \end{tikzpicture}
    \end{adjustbox}
    \caption{Brane depiction of the low dimensional leaves [and elementary slices between them] of $\w{\mathrm{PSO}(4,\mathbb{C})}{}{[1,1]}$.}
    \label{fig:D2Branes4}
\end{figure}

\end{landscape}

\section{Quiver addition}
\label{sec:quiveradd}

\subsection{Algorithm}

In recent papers \cite{Cabrera:2016vvv,Cabrera:2017njm,Cabrera:2018ann,Cabrera:2019izd,Bourget:2019aer,Cabrera:2019dob,Hanany:2019tji,Grimminger:2020dmg,Bourget:2020gzi,Bourget:2020asf,Bourget:2020mez}, the notion of quiver subtraction was developed to obtain the Hasse diagram for symplectic singularities, and has been used in various scenarios \cite{Rogers:2018dez,Rogers:2019pqe,Eckhard:2020jyr,Closset:2020scj,vanBeest:2020kou,vanBeest:2020civ,Closset:2020afy}. In this section, we consider the reverse of this process: \emph{quiver addition}. Aspects of this notion have been discussed in \cite{Rogers:2019pqe} for $D$-type Dynkin quivers. In this work, we show that the quiver addition algorithm presented below precisely reproduces the Hasse diagrams and quivers for the slices in the affine Grassmannian of any finite dimensional simple complex Lie group. For classical groups, the quiver addition algorithm can be viewed as the quiver version of the brane constructions of Section \ref{sec:branes}, expressed in a way more suitable e.g.\ to program. For exceptional groups, however, no brane systems are known, and the quiver addition algorithm provides a simple, but powerful way to obtain the Hasse diagram of any transverse slice in the affine Grassmannian. Before proceeding to quiver addition, let us review the procedure of quiver subtraction. 
 
One way to obtain the Coulomb branch Hasse diagram of a quiver $\mathsf{Q}$ is to use quiver subtraction. In quiver subtraction one identifies an elementary slice with quiver $\mathsf{D}$ from Table \ref{tab:elem} and subtracts it from $\mathsf{Q}$ following the algorithm given in \cite[Appendix A]{Bourget:2019aer} to produce a new quiver $\mathsf{Q}'=\mathsf{Q}-\mathsf{D}$. We restrict to the case where a given elementary slice cannot be subtracted more than once consecutively.\footnote{If the same slice can be subtracted more than once, we are in the realm of affine Grassmannians of affine type, rather than finite type.} In this case, the Coulomb branch of $\mathsf{Q}'$ is the closure of a minimal degeneration of the highest leaf in the Coulomb branch of $\mathsf{Q}$. 
One key point in the algorithm is that the balance of all the gauge nodes remains invariant during subtraction. If more than one quiver can be subtracted, the Hasse diagram bifurcates. As introduced above, we call the number of elementary slices which can be subtracted at any given step the disposition $\mathfrak{D}$ of the leaf.

Conversely, to a given quiver one can generically add an infinite number of elementary quivers. However, if one imposes the restriction that \emph{the Dynkin diagram formed by the gauge nodes remains invariant} then the number of possible slices allowed to be added becomes finite. We now discuss the algorithm.

\paragraph{Quiver addition algorithm.}
Let $\mathsf{Q}$ be a quiver of unitary gauge nodes forming a rank $r$ finite Dynkin diagram with Cartan matrix $(C_{\mathfrak{g}})_{i,j=1 , \dots , r}$. Let the ranks of the gauge nodes be $k_i \geq 0$, and the ranks of the flavors nodes be $N_i \geq 0$ for $i=1 , \dots , r$.\footnote{We emphasize that $k_i = 0$ is allowed, even when $N_i >0$. } 
\begin{enumerate}
\item \textbf{Identification of the elementary quivers $\mathsf{S}$ to add}. Three kinds of elementary quivers can be added to $\mathsf{Q}$. 
\begin{itemize}
    \item Define the subset of gauge nodes without any flavors 
    \begin{equation}
    J_0 = \left\{ j \in \{1 , \dots , r\} \mid N_j =0 \right\} \,.
\end{equation}
The nodes labeled by $J_0$ form a union of connected Dynkin diagrams. For each connected Dynkin diagram, one can add the quiver $\mathsf{S}$ for the closure of the minimal nilpotent orbit (see Table \ref{tab:elem}). 
\item Consider the set of gauge nodes 
\begin{equation}
    J_1 = \left\{ j \in \{1 , \dots , r\} \mid N_j \geq 1  \quad \textrm{and} \quad \forall \; j'\in \{1 , \dots , r\} ,  \; N_{j'} \geq - (C_{\mathfrak{g}})_{j'j} \right\} 
\end{equation}
such that gauge node itself has flavours and also its connected neighbours have enough flavour. For each $j \in J_1$, one can add on that node the quiver for a Kleinian singularity $A_{N_j +1}$. 
\item If there is a subset of the gauge nodes which realize the quivers on the first column of Table \ref{tab:quasiminimalAddition} with  \emph{the flavors exactly as indicated}, and with \emph{no flavorless gauge node connected to that subset}, then one can add the $\mathsf{S}$ in the second column. 
\end{itemize}
\item \textbf{Addition of the elementary quivers}. 
The result of the addition of the elementary quiver $\mathsf{S}$ to $\mathsf{Q}$ is a quiver $\mathsf{Q}'$ with gauge nodes given by the Dynkin diagram specified by $(C_{\mathfrak{g}})_{i,j=1 , \dots , r}$, where
\begin{itemize}
    \item The ranks of the gauge groups $k'_i$ are the sums of the ranks in $\mathsf{Q}$ and $\mathsf{S}$.
    \item The flavors $N'_i$ are given by the corresponding flavor in $\mathsf{S}$ if $k_i \neq k'_i$ and by rebalancing otherwise. 
\end{itemize}
\end{enumerate}

\begin{table}[]
    \centering
    \begin{tabular}{c|c}
    \toprule 
    To this section of $\mathsf{Q}$ & One can add the following $\mathsf{S}$  \\ \midrule
        \raisebox{-.5\height}{\begin{tikzpicture}
    \node[gauge] (4) at (-3,0) {};
    \node (3) at (-2,0) {$\cdots$};
    \node[gauge] (2) at (-1,0) {};
    \node[gauge] (1) at (0,0) {};
    \node[flavour,label=above:{1}] (1a) at (0,1) {};
    \draw (1a)--(1) (2)--(3)--(4) (-0.4,0.2)--(-0.6,0)--(-0.4,-0.2);
    \draw[transform canvas={yshift=-1pt}] (1)--(2);
    \draw[transform canvas={yshift=1pt}] (1)--(2);
\end{tikzpicture}} & \raisebox{-.5\height}{\begin{tikzpicture}
    \node[flavour,label=above:{1}] (10) at (4,1) {}; 
    \node[gauge,label=below:{1}] (0) at (0,0) {};
    \node[gauge,label=below:{1}] (1) at (1,0) {};
    \node (2) at (2,0) {$\cdots$};
    \node[gauge,label=below:{1}] (3) at (3,0) {};
    \node[gauge,label=below:{1}] (4) at (4,0) {};
    \node[flavour,label=above:{1}] (0a) at (0,1) {};
    \draw (0)--(0a);
    \draw (0)--(1)--(2)--(3) (3.6,0.2)--(3.4,0)--(3.6,-0.2) (4)--(10);
    \draw[transform canvas={yshift=-1pt}] (3)--(4);
    \draw[transform canvas={yshift=1pt}] (3)--(4);
\end{tikzpicture}} $=ac_n$ \\
        \raisebox{-.5\height}{\begin{tikzpicture}
    \node[gauge] (2) at (-1,0) {};
    \node[gauge] (1) at (0,0) {};
    \node[flavour,label=above:{1}] (1a) at (0,1) {};
    \draw (1a)--(1)--(2) (-0.4,0.2)--(-0.6,0)--(-0.4,-0.2);
    \draw[transform canvas={yshift=-2pt}] (1)--(2);
    \draw[transform canvas={yshift=0pt}] (1)--(2);
    \draw[transform canvas={yshift=2pt}] (1)--(2);
\end{tikzpicture}}  & \raisebox{-.5\height}{\begin{tikzpicture}
    \node[gauge,label=below:{1}] (3) at (3,0) {};
    \node[gauge,label=below:{1}] (4) at (4,0) {};
    \node[flavour,label=above:{1}] (2) at (3,1) {};
    \draw[transform canvas={yshift=-2pt}] (3)--(4);
    \draw[transform canvas={yshift=0pt}] (3)--(4);
    \draw[transform canvas={yshift=2pt}] (3)--(4);
    \draw (3.6,0.2)--(3.4,0)--(3.6,-0.2) (2)--(3);
\end{tikzpicture}} $=cg_2$ \\
\raisebox{-.5\height}{\begin{tikzpicture}
    \node[gauge] (2) at (-1,0) {};
    \node[gauge] (1) at (0,0) {};
    \node[flavour,label=above:{2}] (1a) at (0,1) {};
    \draw (1a)--(1)--(2) (-0.4,0.2)--(-0.6,0)--(-0.4,-0.2);
    \draw[transform canvas={yshift=-2pt}] (1)--(2);
    \draw[transform canvas={yshift=0pt}] (1)--(2);
    \draw[transform canvas={yshift=2pt}] (1)--(2);
\end{tikzpicture}} & \raisebox{-.5\height}{\begin{tikzpicture}
    \node[gauge,label=below:{1}] (3) at (3,0) {};
    \node[gauge,label=below:{1}] (4) at (4,0) {};
    \node[flavour,label=above:{1}] (5) at (4,1) {};
    \node[flavour,label=above:{1}] (2) at (3,1) {};
    \draw[transform canvas={yshift=-2pt}] (3)--(4);
    \draw[transform canvas={yshift=0pt}] (3)--(4);
    \draw[transform canvas={yshift=2pt}] (3)--(4);
    \draw (3.6,0.2)--(3.4,0)--(3.6,-0.2) (4)--(5) (2)--(3);
\end{tikzpicture}} $=ag_2$ \\ \bottomrule
    \end{tabular}
    \caption{Identification of the situations where the quivers for $ac_n$, $cg_2$ and $ag_2$ can be added. We insist on the fact that the flavor nodes must be exactly as indicated in the first column, and that there should be no flavorless gauge node connected to the portions of quiver drawn in the first column, so that quiver of the second column can be added. }
    \label{tab:quasiminimalAddition}
\end{table}

\vspace{1em}
Note that the total number of flavors either stays the same or increases when a slice is added.\footnote{Just like the number of D5 branes between NS5 branes stays the same or increases when moving upwards in the Hasse diagrams in Section \ref{sec:branes}} As we grow the Hasse diagram further, we reach a stage where all gauge groups have non-zero flavor nodes. Once this happens, the only slices that can be added are Kleinian $A_k$ singularities. Hence, as the Hasse diagram grows, more and more Kleinian singularities populate the Hasse diagram. 
When all the nodes have flavors, the disposition is equal to the rank of the group, and can not decrease upon quiver addition.

\subsection{Examples with simply laced quivers}
Let us exemplify how the algorithm works in the case of good linear quivers. 
The set $J_0$ of unflavored nodes decomposes into smaller $A$-type diagrams, and for each such diagram, an $a_n$-type slice can be added.\footnote{The transition in the corresponding brane system is \eqref{eq:antrans}.} For instance, consider the following scenario:
    \begin{equation}
    \raisebox{-.5\height}{
\begin{tikzpicture}[scale=0.8]
	\begin{pgfonlayer}{nodelayer}
		\node [style=gauge3] (0) at (-4.5, 0.25) {};
		\node [style=gauge3] (1) at (-3.5, 0.25) {};
		\node [style=gauge3] (2) at (-2.5, 0.25) {};
		\node [style=gauge3] (3) at (-1, 0.25) {};
		\node [style=gauge3] (4) at (0, 0.25) {};
		\node [style=gauge3] (5) at (1, 0.25) {};
		\node [style=none] (6) at (-1.75, 0.25) {$\dots$};
		\node [style=gauge3] (7) at (4, 0.25) {};
		\node [style=none] (10) at (2, 0.25) {$\dots$};
		\node [style=none] (14) at (-5.25, 0.25) {$\dots$};
		\node [style=flavour2] (16) at (4, 1.25) {};
		\node [style=flavour2] (17) at (-3.5, 1.25) {};
		\node [style=flavour2] (18) at (0, 1.25) {};
		\node [style=gauge3] (19) at (-6, 0.25) {};
		\node [style=gauge3] (20) at (3, 0.25) {};
		\node [style=none] (21) at (-6, -0.75) {};
		\node [style=none] (22) at (-4.5, -0.75) {};
		\node [style=none] (24) at (3, -0.75) {};
		\node [style=none] (25) at (-5.25, -1.5) {$m_1$};
		\node [style=none] (26) at (-1.75, -1.5) {$m_2$};
		\node [style=none] (27) at (2, -1.5) {$m_3$};
		\node [style=none] (28) at (-5.5, 2) {};
		\node [style=none] (29) at (-5.5, 1) {};
		\node [style=none] (30) at (-5.5, 6) {Can add $a_{m_1}$};
		\node [style=none] (31) at (-1.75, 2) {};
		\node [style=none] (32) at (-1.75, 1) {};
		\node [style=none] (34) at (2, 2) {};
		\node [style=none] (35) at (2, 1) {};
		\node [style=gauge3] (38) at (-4.75, 4) {};
		\node [style=gauge3] (39) at (-6.25, 4) {};
		\node [style=none] (40) at (-5.5, 4) {$\dots$};
		\node [style=flavour2] (41) at (-6.25, 5) {};
		\node [style=flavour2] (42) at (-4.75, 5) {};
		\node [style=none] (44) at (-6.25, 3.25) {};
		\node [style=none] (45) at (-4.75, 3.25) {};
		\node [style=none] (46) at (-5.5, 2.5) {$m_1$};
		\node [style=none] (47) at (-1.75, 6) {Can add $a_{m_2}$};
		\node [style=gauge3] (48) at (-1, 4) {};
		\node [style=gauge3] (49) at (-2.5, 4) {};
		\node [style=none] (50) at (-1.75, 4) {$\dots$};
		\node [style=flavour2] (51) at (-2.5, 5) {};
		\node [style=flavour2] (52) at (-1, 5) {};
		\node [style=none] (53) at (-2.5, 3.25) {};
		\node [style=none] (54) at (-1, 3.25) {};
		\node [style=none] (55) at (-1.75, 2.5) {$m_2$};
		\node [style=none] (59) at (2, 6) {Can add $a_{m_3}$};
		\node [style=gauge3] (60) at (2.75, 4) {};
		\node [style=gauge3] (61) at (1.25, 4) {};
		\node [style=none] (62) at (2, 4) {$\dots$};
		\node [style=flavour2] (63) at (1.25, 5) {};
		\node [style=flavour2] (64) at (2.75, 5) {};
		\node [style=none] (65) at (1.25, 3.25) {};
		\node [style=none] (66) at (2.75, 3.25) {};
		\node [style=none] (67) at (2, 2.5) {$m_3$};
		\node [style=none] (68) at (-6.25, 3.5) {1};
		\node [style=none] (69) at (-4.75, 3.5) {1};
		\node [style=none] (70) at (-2.5, 3.5) {1};
		\node [style=none] (71) at (-1, 3.5) {1};
		\node [style=none] (72) at (1.25, 3.5) {1};
		\node [style=none] (73) at (2.75, 3.5) {1};
		\node [style=none] (74) at (-6.25, 5.5) {1};
		\node [style=none] (75) at (-4.75, 5.5) {1};
		\node [style=none] (76) at (-2.5, 5.5) {1};
		\node [style=none] (77) at (-1, 5.5) {1};
		\node [style=none] (78) at (1.25, 5.5) {1};
		\node [style=none] (79) at (2.75, 5.5) {1};
		\node [style=none] (80) at (-2.5, -0.75) {};
		\node [style=none] (81) at (-1, -0.75) {};
		\node [style=none] (82) at (1, -0.75) {};
		\node [style=none] (83) at (-6, -0.25) {\small $k_a$};
		\node [style=none] (84) at (-4.5, -0.25) {\small $k_b$};
		\node [style=none] (85) at (-3.5, -0.25) {\small $k_c$};
		\node [style=none] (86) at (-2.5, -0.25) {\small $k_d$};
		\node [style=none] (87) at (-1, -0.25) {\small $k_f$};
		\node [style=none] (88) at (0, -0.25) {\small $k_g$};
		\node [style=none] (89) at (1, -0.25) {\small $k_h$};
		\node [style=none] (90) at (3, -0.25) {\small $k_i$};
		\node [style=none] (91) at (4, -0.25) {\small $k_j$};
		\node [style=none] (92) at (-3.5, 1.75) {$N_1$};
		\node [style=none] (93) at (0, 1.75) {$N_2$};
		\node [style=none] (94) at (4, 1.75) {$N_3$};
	\end{pgfonlayer}
	\begin{pgfonlayer}{edgelayer}
		\draw (0) to (1);
		\draw (1) to (2);
		\draw (5) to (4);
		\draw (4) to (3);
		\draw (17) to (1);
		\draw (18) to (4);
		\draw (16) to (7);
		\draw (20) to (7);
		\draw [style=->] (28.center) to (29.center);
		\draw [style=->] (31.center) to (32.center);
		\draw [style=->] (34.center) to (35.center);
		\draw (41) to (39);
		\draw (42) to (38);
		\draw (51) to (49);
		\draw (52) to (48);
		\draw (63) to (61);
		\draw (64) to (60);
		\draw [style=brace] (66.center) to (65.center);
		\draw [style=brace] (54.center) to (53.center);
		\draw [style=brace] (45.center) to (44.center);
		\draw [style=brace] (22.center) to (21.center);
		\draw [style=brace] (81.center) to (80.center);
		\draw [style=brace] (24.center) to (82.center);
	\end{pgfonlayer}
\end{tikzpicture}
}
    \end{equation}
There are three possible slices one can add $a_{m_1}$, $a_{m_2}$, and $a_{m_3}$. For example, adding the $a_{m_2}$ slice produces
\begin{equation}
\raisebox{-.5\height}{
    \begin{tikzpicture}[scale=0.8]
	\begin{pgfonlayer}{nodelayer}
		\node [style=gauge3] (0) at (-4.5, 0.25) {};
		\node [style=gauge3] (1) at (-3.5, 0.25) {};
		\node [style=gauge3] (2) at (-2.5, 0.25) {};
		\node [style=gauge3] (3) at (-1, 0.25) {};
		\node [style=gauge3] (4) at (0, 0.25) {};
		\node [style=gauge3] (5) at (1, 0.25) {};
		\node [style=none] (6) at (-1.75, 0.25) {$\dots$};
		\node [style=gauge3] (7) at (4, 0.25) {};
		\node [style=none] (10) at (2, 0.25) {$\dots$};
		\node [style=none] (14) at (-5.25, 0.25) {$\dots$};
		\node [style=flavour2] (16) at (4, 1.25) {};
		\node [style=flavour2] (17) at (-3.5, 1.25) {};
		\node [style=flavour2] (18) at (0, 1.25) {};
		\node [style=gauge3] (19) at (-6, 0.25) {};
		\node [style=gauge3] (20) at (3, 0.25) {};
		\node [style=none] (21) at (-6, -0.75) {};
		\node [style=none] (22) at (-4.5, -0.75) {};
		\node [style=none] (24) at (3, -0.75) {};
		\node [style=none] (25) at (-5.25, -1.5) {$m_1$};
		\node [style=none] (26) at (-1.75, -1.5) {$m_2$};
		\node [style=none] (27) at (2, -1.5) {$m_3$};
		\node [style=none] (80) at (-2.5, -0.75) {};
		\node [style=none] (81) at (-1, -0.75) {};
		\node [style=none] (82) at (1, -0.75) {};
		\node [style=none] (83) at (-6, -0.25) {\small $k_a$};
		\node [style=none] (84) at (-4.5, -0.25) {\small $k_b$};
		\node [style=none] (85) at (-3.5, -0.25) {\small $k_c$};
		\node [style=none] (86) at (-2.5, -0.25) {\small $k_d{+}1$};
		\node [style=none] (87) at (-1, -0.25) {\small $k_f{+}1$};
		\node [style=none] (88) at (0, -0.25) {\small $k_g$};
		\node [style=none] (89) at (1, -0.25) {\small $k_h$};
		\node [style=none] (90) at (3, -0.25) {\small $k_i$};
		\node [style=none] (91) at (4, -0.25) {\small $k_j$};
		\node [style=none] (92) at (-3.5, 1.75) {$N_1{-}1$};
		\node [style=none] (93) at (0, 1.75) {$N_2{-}1$};
		\node [style=none] (94) at (4, 1.75) {$N_3$};
		\node [style=flavour2] (95) at (-2.5, 1.25) {};
		\node [style=flavour2] (96) at (-1, 1.25) {};
		\node [style=none] (97) at (-2.5, 1.75) {1};
		\node [style=none] (98) at (-1, 1.75) {1};
	\end{pgfonlayer}
	\begin{pgfonlayer}{edgelayer}
		\draw (0) to (1);
		\draw (1) to (2);
		\draw (5) to (4);
		\draw (4) to (3);
		\draw (17) to (1);
		\draw (18) to (4);
		\draw (16) to (7);
		\draw (20) to (7);
		\draw [style=brace] (22.center) to (21.center);
		\draw [style=brace] (81.center) to (80.center);
		\draw [style=brace] (24.center) to (82.center);
		\draw (95) to (2);
		\draw (96) to (3);
	\end{pgfonlayer}
\end{tikzpicture}
}
\end{equation}
On the other hand, the $A$-type Kleinian singularities can be added in the following situation:\footnote{The transition in the corresponding brane system is \eqref{eq:Antrans}.}
    \begin{equation}
    \raisebox{-.5\height}{
 \begin{tikzpicture}[scale=0.8]
	\begin{pgfonlayer}{nodelayer}
		\node [style=gauge3] (36) at (-4, 0) {};
		\node [style=gauge3] (37) at (-3, 0) {};
		\node [style=gauge3] (38) at (-2, 0) {};
		\node [style=flavour2] (39) at (-4, 1) {};
		\node [style=flavour2] (40) at (-3, 1) {};
		\node [style=flavour2] (41) at (-2, 1) {};
		\node [style=none] (42) at (-4, 1.5) {$N_1$};
		\node [style=none] (43) at (-3, 1.5) {$N_2$};
		\node [style=none] (44) at (-2, 1.5) {$N_3$};
		\node [style=none] (45) at (-5.75, 0) {$\dots$};
		\node [style=none] (46) at (-0.25, 0) {$\dots$};
		\node [style=none] (47) at (-3, 3) {};
		\node [style=none] (48) at (-3, 2) {};
		\node [style=none] (49) at (0, 4.5) {Can add $A_{N_2{+}1}$};
		\node [style=gauge3] (50) at (-5, 0) {};
		\node [style=gauge3] (51) at (-1, 0) {};
		\node [style=gauge3] (52) at (-3, 4) {};
		\node [style=flavour2] (53) at (-3, 5) {};
		\node [style=none] (55) at (-3, 3.5) {1};
		\node [style=none] (56) at (1, 1) {};
		\node [style=none] (57) at (3, 1) {};
		\node [style=gauge3] (58) at (5.75, 0) {};
		\node [style=gauge3] (59) at (6.75, 0) {};
		\node [style=gauge3] (60) at (7.75, 0) {};
		\node [style=flavour2] (61) at (5.75, 1) {};
		\node [style=flavour2] (62) at (6.75, 1) {};
		\node [style=flavour2] (63) at (7.75, 1) {};
		\node [style=none] (64) at (5.25, 1.5) {\small $N_1{-}1$};
		\node [style=none] (65) at (6.75, 1.5) {\small $N_2{+}2$};
		\node [style=none] (66) at (8.25, 1.5) {\small $N_3{-}1$};
		\node [style=none] (67) at (4, 0) {$\dots$};
		\node [style=none] (68) at (9.5, 0) {$\dots$};
		\node [style=gauge3] (69) at (4.75, 0) {};
		\node [style=gauge3] (70) at (8.75, 0) {};
		\node [style=none] (71) at (-5, -0.5) {$k_a$};
		\node [style=none] (72) at (-4, -0.5) {$k_b$};
		\node [style=none] (73) at (-3, -0.5) {$k_c$};
		\node [style=none] (74) at (-2, -0.5) {$k_d$};
		\node [style=none] (75) at (-1, -0.5) {$k_e$};
		\node [style=none] (76) at (-3, 5.5) {$N_2{+}2$};
		\node [style=none] (77) at (4.75, -0.5) {$k_a$};
		\node [style=none] (78) at (5.75, -0.5) {$k_b$};
		\node [style=none] (79) at (6.75, -0.5) {$k_c+1$};
		\node [style=none] (80) at (7.75, -0.5) {$k_d$};
		\node [style=none] (81) at (8.75, -0.5) {$k_e$};
	\end{pgfonlayer}
	\begin{pgfonlayer}{edgelayer}
		\draw (36) to (37);
		\draw (37) to (38);
		\draw (39) to (36);
		\draw (40) to (37);
		\draw (41) to (38);
		\draw [style=->] (47.center) to (48.center);
		\draw (36) to (50);
		\draw (38) to (51);
		\draw (53) to (52);
		\draw [style=->] (56.center) to (57.center);
		\draw (58) to (59);
		\draw (59) to (60);
		\draw (61) to (58);
		\draw (62) to (59);
		\draw (63) to (60);
		\draw (58) to (69);
		\draw (60) to (70);
	\end{pgfonlayer}
\end{tikzpicture}
}
    \end{equation}
which corresponds to nodes in the set $J_1$.

\paragraph{Single unbalanced gauge group.}
As an illustration, we consider linear quivers with one unbalanced node, with imbalance $+1$. These quivers cover all cases of transverse slice to the lowest leaf in the components of $\mathrm{Gr}_{\mathrm{PSL}(N , \mathbb{C})}$ not connected to the trivial lattice. Quiver addition then reproduces the Hasse diagram and quivers for the transverse slices in these components of $\mathrm{Gr}_{\mathrm{PSL}(N , \mathbb{C})}$. Specifically, Figure \ref{unbalancedfigure} shows the first few leaves in the affine Grassmannian Hasse diagram of $\mathrm{Gr}_{\mathrm{PSL}(m_1 + m_2 + 2 , \mathbb{C})}$ for $m_1,m_2 \geq 6$ and lowest dominant coweight (see Table \ref{tab:LDC})
\begin{equation}
    \Omega_{1+m_1}=[\underbrace{0\dots0}_{m_1}1\underbrace{0\dots 0}_{m_2}] \, . 
\end{equation}
It is clear that for quivers where all gauge groups have non-zero ranks, the global symmetry is $A_{m_1}\times A_{m_2} \times \mathfrak{u}(1)$. As a feature of Hasse diagrams, the non-abelian part of the global symmetry is given by the two slices directly connected to the origin. Note, since $m_1$ and $m_2$ can be different, the Hasse diagram becomes less symmetric as one moves away from the bottom. 

We observe that in the Hasse diagram of $\mathrm{Gr}_{\mathrm{PSL}(m_1 + m_2 + 2 , \mathbb{C})}$, if we keep the leaves where only the ranks of the first $m_1$ gauge nodes are non-zero, then the resulting diagram is finite and is the Hasse diagram of the nilpotent orbits of $\mathfrak{sl}(m_1 +1, \mathbb{C})$.

\afterpage{
\begin{landscape}

\begin{figure}[t]
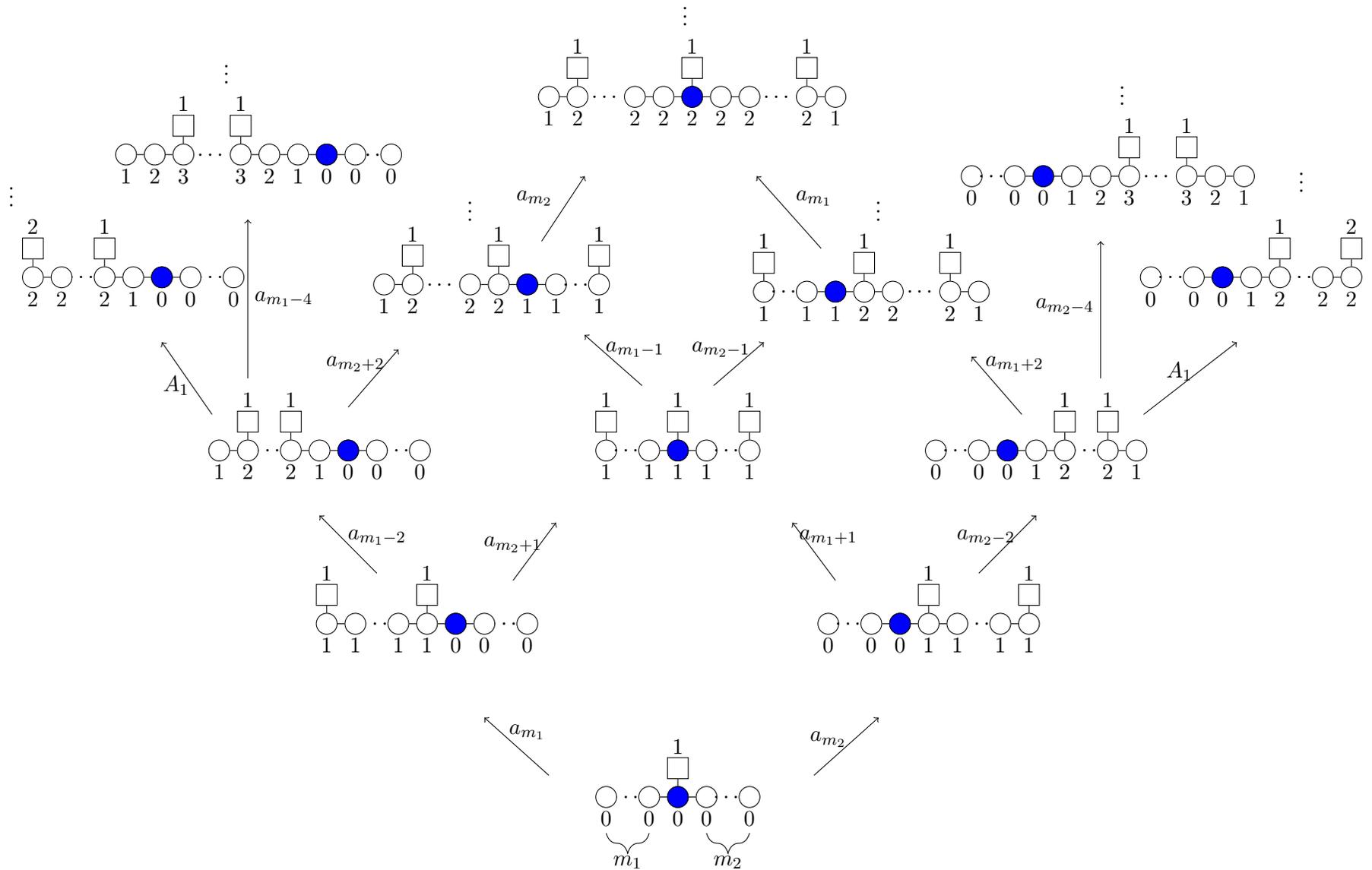

    \centering

 \caption{Hasse diagram for the first orbits in the component of $\mathrm{Gr}_{\mathrm{PSL}(m_1 + m_2 + 2 , \mathbb{C})}$ corresponding to the lowest coweight $\Omega_{1+m_1}=[\underbrace{0\dots0}_{m_1}1\underbrace{0\dots 0}_{m_2}]$ for $m_1,m_2 \geq 6$. Blue nodes represent unbalanced gauge groups. }
\label{unbalancedfigure}
\end{figure}

\end{landscape} }

\paragraph{Non-linear $D_N$, $E_{6,7,8}$ quivers.}
The treatment is almost identical to the $A$-type affine Grassmannians except for the novelty that one may add $d_n$, $e_6$, $e_7$, $e_8$ elementary slices. For $A$-type affine Grassmannian, we mentioned how $n$ unflavored gauge groups between two flavored gauge groups allow the addition of an $a_n$ slice. We can now generalize this by considering the subset of $n$ flavorless gauge nodes that are bounded by flavored gauge nodes. Then, one is allowed to add the elementary slice that is the closure of the minimal nilpotent orbit of $\mathfrak{g}_n$, where $\mathfrak{g}_n$ is the finite Dynkin diagram in the shape of the unflavored gauge nodes.  Let us demonstrate this with some slices of $\mathrm{E}_6$ affine Grassmannian:
\begin{equation}
    \raisebox{-.5\height}{

}
\end{equation}
Here, we can add a $a_5$ slice to the nodes that make up a $a_5$ finite Dynkin diagram in the red box.

\begin{equation}
    \raisebox{-.5\height}{
%
}
\end{equation}
Here, we can add a $d_5$ slice to the nodes that make up a $d_5$ finite Dynkin diagram in the red box. 

\begin{equation}
    \raisebox{-.5\height}{
%
}
\end{equation}
Here, we can add a $d_4$ slice to the nodes that make up a $d_4$ finite Dynkin diagram in the red box.

\subsection{Examples with non simply laced quivers}

\begin{figure}[t]
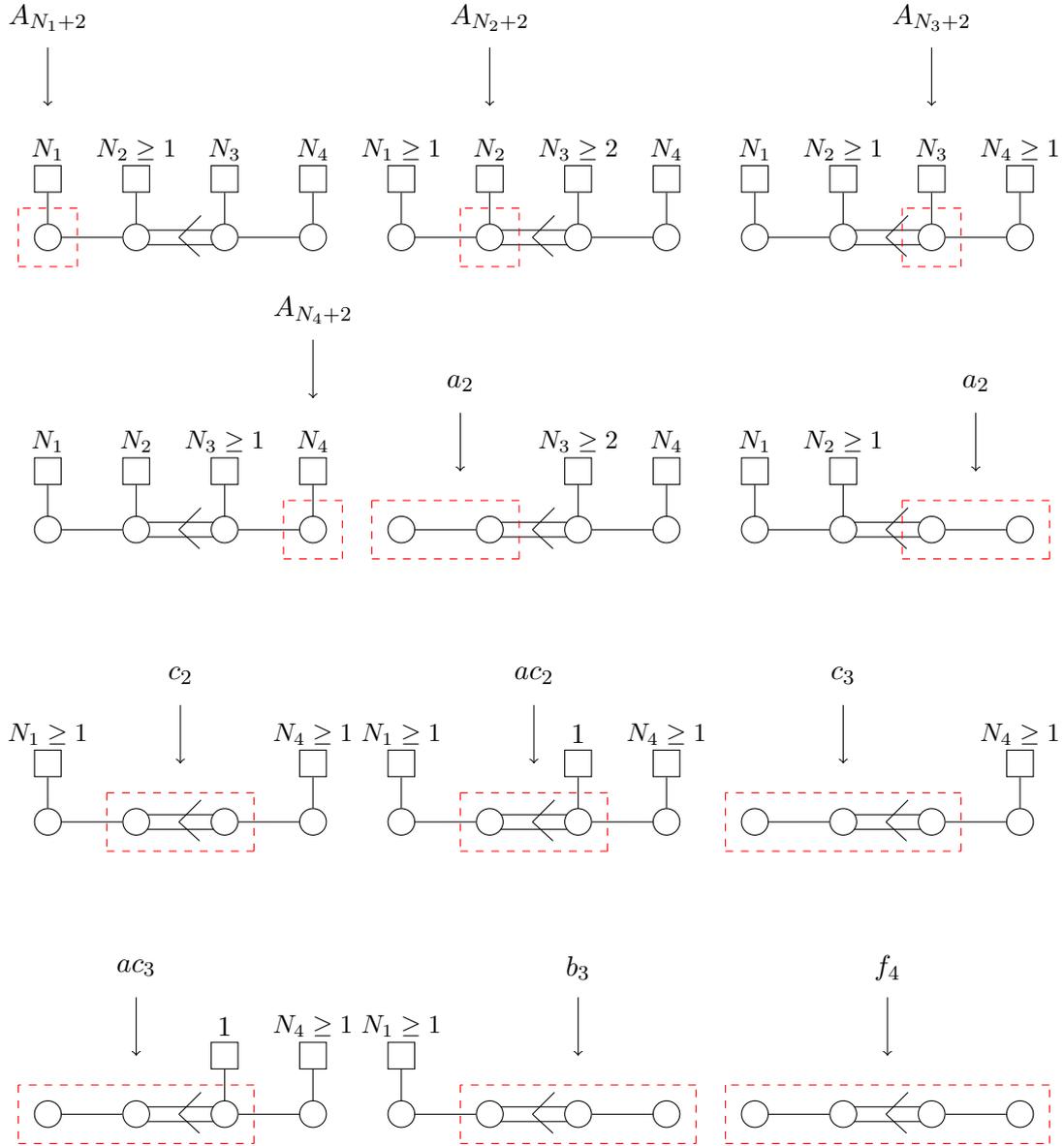

    \centering
%
    \caption{Quiver addition rules regarding slices in the affine Grassmannian of $\mathrm{F}_4$. We just illustrate step 1 of the algorithm, as once the slices to be added are identified, step 2 is straightforward. The red boxes indicate where the quivers for the slices, which can be found in Table \ref{tab:elem}, have to be added. When no condition is written on a flavor node rank $N_i$, it means that it can take any value $N_i \geq 0$. }
     \label{F4additions}
\end{figure}

We now turn to non simply laced quivers. We consider the case of $\mathrm{F}_4$, which displays all the subtleties of the algorithm. This is summarized in Figure \ref{F4additions}, where we show the 12 types of quiver additions that can be made as a function of the ranks of the flavor nodes.

\clearpage 

\section{Hilbert series of transverse slices}
\label{sec:Hilbert}

\subsection{Hilbert series for infinite dimensional transverse slices}

In this section, the Hilbert series of infinite dimensional slices, arising from the affine Grassmannian, are explored.

To every orbit $\gr{G}{\lambda}{}$, labelled by a coweight $\lambda$, in the affine Grassmannian of a certain group $G$, there is an infinite dimensional transverse slice $\w{G}{}{\lambda}$. To this transverse slice we can associate a Hilbert series $\mathrm{HS}_{G,\lambda}$, which can be obtained from an infinite rank limit of a quiver \cite{Barns-Graham:2019ydl}. $\mathrm{HS}_{G,\lambda}(t)$ can be expressed as an infinite product:
\begin{equation}
  \mathrm{HS}_{G,\lambda}(t) = \prod\limits_{j=1}^{\infty} \mathrm{PE}\left[\sum\limits_{\alpha \in \Delta(\mathrm{adj})}t^{|\langle\lambda,\alpha\rangle|+2j}\right]= \mathrm{PE}\left[\sum\limits_{\alpha \in \Delta(\mathrm{adj})}\frac{t^{2+|\langle\lambda,\alpha\rangle|}}{1-t^2}\right]  
  \,, 
\end{equation}
where PE denotes the plethystic exponential. 
This can be further refined using fundamental weight fugacities $x_i$ as follows:
\begin{equation}
\label{HSTransverseSlice}
  \mathrm{HS}_{G,\lambda}(t;x) = \mathrm{PE}\left[\sum\limits_{\alpha \in \Delta(\mathrm{adj})}\frac{x^{\alpha}t^{2+|\langle\lambda,\alpha\rangle|}}{1-t^2}\right]
\end{equation}
where 
\begin{equation}
    x^\lambda = \prod\limits_{i=1}^{r} x_i^{\langle \lambda ,\alpha_i^{\vee} \rangle} 
\end{equation}
and $r$ is the rank of $G$.

The Hilbert series $\mathrm{HS}_{G,\lambda}^{\mu}$ of a (finite dimensional) transverse slice $\wb{G}{\mu}{\lambda}=\w{G}{}{\lambda}\cap\grb{G}{\mu}{}$ approximates the Hilbert series for the infinite dimensional transverse slice \eqref{HSTransverseSlice}, and this approximation improves when $\mu$ becomes large. More precisely, 
\begin{equation}
\label{eqLimit}
    \lim\limits_{\langle \rho , \mu \rangle \rightarrow \infty } \mathrm{HS}_{G,\lambda}^{\mu} (t;x) = \mathrm{HS}_{G,\lambda} (t;x) \, ,  
\end{equation}
where $\rho$ is the Weyl vector. 
In this equation, $\mathrm{HS}_{G,\lambda}^{\mu}$ can be computed using the monopole formula on the quiver $\mathsf{Q}_{\lambda}^{\mu}$. When $\mu$ gets larger, the ranks of the gauge nodes in $\mathsf{Q}_{\lambda}^{\mu}$ increase while the imbalance stays fixed. In the remainder of this section, we elaborate on the appearance of new generators in the chiral ring when these ranks increase, focusing on two examples with $G = \mathrm{Sp}(4 , \mathbb{C})$ and $\lambda = 0$ and $\lambda = \varpi_2$ respectively.

\paragraph{The case $\lambda = 0$.}
By means of the Hilbert series, we are studying the moduli space generators and the relations between them. Let us consider slices in the affine Grassmannian of $G$ where all gauge groups are balanced (i.e.\ slices to the origin, $\lambda = 0$). If all the generators appear at order $t^2$ in the Hilbert series, the Coulomb branch is the closure of a nilpotent orbit of $G$ \cite{namikawa2018characterization}. However, by increasing the ranks of the gauge groups, whilst maintaining the balance, new generators show up at higher orders. 

For example, let us consider slices of the $G=\mathrm{Sp}(4 , \mathbb{C})$ affine Grassmannian for $\lambda=[0,0]$:
\begin{equation}   
\raisebox{-.5\height}{\begin{tikzpicture}
	\begin{pgfonlayer}{nodelayer}
		\node [style=gauge3] (0) at (0, 0) {};
		\node [style=gauge3] (1) at (1, 0) {};
		\node [style=none] (2) at (0.15, 0.1) {};
		\node [style=none] (3) at (0.15, -0.15) {};
		\node [style=none] (4) at (1.15, 0.1) {};
		\node [style=none] (5) at (1.15, -0.15) {};
		\node [style=none] (6) at (0.85, 0.35) {};
		\node [style=none] (7) at (0.5, -0.025) {};
		\node [style=none] (8) at (0.85, -0.375) {};
		\node [style=flavour2] (9) at (0, 1) {};
		\node [style=flavour2] (10) at (1, 1) {};
		\node [style=none] (11) at (0, -0.5) {1};
		\node [style=none] (12) at (1, -0.5) {2};
		\node [style=none] (13) at (1, 1.5) {$2$};
		\node [style=none] (14) at (0, 1.5) {$0$};
		\node [style=gauge3] (15) at (-0.025, 3.25) {};
		\node [style=gauge3] (16) at (0.975, 3.25) {};
		\node [style=none] (17) at (0.125, 3.35) {};
		\node [style=none] (18) at (0.125, 3.1) {};
		\node [style=none] (19) at (1.125, 3.35) {};
		\node [style=none] (20) at (1.125, 3.1) {};
		\node [style=none] (21) at (0.825, 3.6) {};
		\node [style=none] (22) at (0.475, 3.225) {};
		\node [style=none] (23) at (0.825, 2.875) {};
		\node [style=flavour2] (24) at (-0.025, 4.25) {};
		\node [style=flavour2] (25) at (0.975, 4.25) {};
		\node [style=none] (26) at (-0.025, 2.75) {1};
		\node [style=none] (27) at (0.975, 2.75) {1};
		\node [style=none] (28) at (0.975, 4.75) {0};
		\node [style=none] (29) at (-0.025, 4.75) {1};
		\node [style=gauge3] (30) at (-0.025, -3.5) {};
		\node [style=gauge3] (31) at (0.975, -3.5) {};
		\node [style=none] (32) at (0.125, -3.4) {};
		\node [style=none] (33) at (0.125, -3.65) {};
		\node [style=none] (34) at (1.125, -3.4) {};
		\node [style=none] (35) at (1.125, -3.65) {};
		\node [style=none] (36) at (0.825, -3.15) {};
		\node [style=none] (37) at (0.475, -3.525) {};
		\node [style=none] (38) at (0.825, -3.875) {};
		\node [style=flavour2] (39) at (-0.025, -2.5) {};
		\node [style=flavour2] (40) at (0.975, -2.5) {};
		\node [style=none] (41) at (-0.025, -4) {2};
		\node [style=none] (42) at (0.975, -4) {2};
		\node [style=none] (43) at (0.975, -2) {0};
		\node [style=none] (44) at (-0.025, -2) {2};
		\node [style=gauge3] (45) at (7.725, 3.5) {};
		\node [style=gauge3] (46) at (8.725, 3.5) {};
		\node [style=none] (47) at (7.875, 3.6) {};
		\node [style=none] (48) at (7.875, 3.35) {};
		\node [style=none] (49) at (8.875, 3.6) {};
		\node [style=none] (50) at (8.875, 3.35) {};
		\node [style=none] (51) at (8.575, 3.85) {};
		\node [style=none] (52) at (8.225, 3.475) {};
		\node [style=none] (53) at (8.575, 3.125) {};
		\node [style=flavour2] (54) at (7.725, 4.5) {};
		\node [style=flavour2] (55) at (8.725, 4.5) {};
		\node [style=none] (56) at (7.725, 3) {2};
		\node [style=none] (57) at (8.725, 3) {3};
		\node [style=none] (58) at (8.725, 5) {$2$};
		\node [style=none] (59) at (7.725, 5) {1};
		\node [style=gauge3] (60) at (7.725, 0) {};
		\node [style=gauge3] (61) at (8.725, 0) {};
		\node [style=none] (62) at (7.875, 0.1) {};
		\node [style=none] (63) at (7.875, -0.15) {};
		\node [style=none] (64) at (8.875, 0.1) {};
		\node [style=none] (65) at (8.875, -0.15) {};
		\node [style=none] (66) at (8.575, 0.35) {};
		\node [style=none] (67) at (8.225, -0.025) {};
		\node [style=none] (68) at (8.575, -0.375) {};
		\node [style=flavour2] (69) at (7.725, 1) {};
		\node [style=flavour2] (70) at (8.725, 1) {};
		\node [style=none] (71) at (7.725, -0.5) {2};
		\node [style=none] (72) at (8.725, -0.5) {4};
		\node [style=none] (73) at (8.725, 1.5) {4};
		\node [style=none] (74) at (7.725, 1.5) {$0$};
		\node [style=gauge3] (75) at (7.725, -3.25) {};
		\node [style=gauge3] (76) at (8.725, -3.25) {};
		\node [style=none] (77) at (7.875, -3.15) {};
		\node [style=none] (78) at (7.875, -3.4) {};
		\node [style=none] (79) at (8.875, -3.15) {};
		\node [style=none] (80) at (8.875, -3.4) {};
		\node [style=none] (81) at (8.575, -2.9) {};
		\node [style=none] (82) at (8.225, -3.275) {};
		\node [style=none] (83) at (8.575, -3.625) {};
		\node [style=flavour2] (84) at (7.725, -2.25) {};
		\node [style=flavour2] (85) at (8.725, -2.25) {};
		\node [style=none] (86) at (7.725, -3.75) {3};
		\node [style=none] (87) at (8.725, -3.75) {3};
		\node [style=none] (88) at (8.725, -1.75) {0};
		\node [style=none] (89) at (7.725, -1.75) {3};
		\node [style=none] (90) at (-0.75, 5.25) {};
		\node [style=none] (91) at (-0.75, 2.5) {};
		\node [style=none] (92) at (1.75, 2.5) {};
		\node [style=none] (93) at (1.75, 5.25) {};
		\node [style=none] (94) at (-0.75, -1.5) {};
		\node [style=none] (95) at (-0.75, -4.25) {};
		\node [style=none] (96) at (1.75, -4.25) {};
		\node [style=none] (97) at (1.75, -1.5) {};
		\node [style=none] (98) at (7, -1.5) {};
		\node [style=none] (99) at (7, -4.25) {};
		\node [style=none] (100) at (9.5, -4.25) {};
		\node [style=none] (101) at (9.5, -1.5) {};
		\node [style=none] (102) at (3.75, 6) {PL[HS]};
		\node [style=none] (103) at (3.75, 3.5) {$ [2,0]_{C_2}t^2 +\dots$};
		\node [style=none] (114) at (6.5, 6) {};
		\node [style=none] (115) at (6.5, -4) {};
		\node [style=none] (116) at (12.5, 6) {PL[HS]};
		\node [style=none] (117) at (0.5, 6) {Slices of $\gr{\mathrm{Sp}(4 , \mathbb{C})}{}{\lambda = 0}$};
		\node [style=none] (118) at (9.25, 6) {Slices of $\gr{\mathrm{Sp}(4 , \mathbb{C})}{}{\lambda = 0}$};
		\node [style=none] (119) at (1.75, 3.5) {};
		\node [style=none] (120) at (3.75, 0.5) {$ [2,0]_{C_2}t^2 +\dots$};
		\node [style=none] (121) at (4, -3) {$ [2,0]_{C_2}t^2\\ +[2,0]_{C_2}t^4 +\dots$};
		\node [style=none] (122) at (11.25, 3.5) {$ [2,0]_{C_2}t^2\\ +[2,0]_{C_2}t^4 +\dots$};
		\node [style=none] (123) at (11.25, 0.5) {$ [2,0]_{C_2}t^2\\ +[2,0]_{C_2}t^4 +\dots$};
		\node [style=none] (124) at (15.25, -3) {\parbox{10cm}{$ [2,0]_{C_2}t^2 +[2,0]_{C_2}t^4\\ +[2,0]_{C_2}t^6 +\dots$}};
	\end{pgfonlayer}
	\begin{pgfonlayer}{edgelayer}
		\draw (2.center) to (4.center);
		\draw (3.center) to (5.center);
		\draw (6.center) to (7.center);
		\draw (7.center) to (8.center);
		\draw (9) to (0);
		\draw (10) to (1);
		\draw (17.center) to (19.center);
		\draw (18.center) to (20.center);
		\draw (21.center) to (22.center);
		\draw (22.center) to (23.center);
		\draw (24) to (15);
		\draw (25) to (16);
		\draw (32.center) to (34.center);
		\draw (33.center) to (35.center);
		\draw (36.center) to (37.center);
		\draw (37.center) to (38.center);
		\draw (39) to (30);
		\draw (40) to (31);
		\draw (47.center) to (49.center);
		\draw (48.center) to (50.center);
		\draw (51.center) to (52.center);
		\draw (52.center) to (53.center);
		\draw (54) to (45);
		\draw (55) to (46);
		\draw (62.center) to (64.center);
		\draw (63.center) to (65.center);
		\draw (66.center) to (67.center);
		\draw (67.center) to (68.center);
		\draw (69) to (60);
		\draw (70) to (61);
		\draw (77.center) to (79.center);
		\draw (78.center) to (80.center);
		\draw (81.center) to (82.center);
		\draw (82.center) to (83.center);
		\draw (84) to (75);
		\draw (85) to (76);
		\draw [style=Bluedotted] (90.center) to (91.center);
		\draw [style=Bluedotted] (91.center) to (92.center);
		\draw [style=Bluedotted] (92.center) to (93.center);
		\draw [style=Bluedotted] (93.center) to (90.center);
		\draw [style=Bluedotted] (94.center) to (95.center);
		\draw [style=Bluedotted] (95.center) to (96.center);
		\draw [style=Bluedotted] (96.center) to (97.center);
		\draw [style=Bluedotted] (97.center) to (94.center);
		\draw [style=Bluedotted] (98.center) to (99.center);
		\draw [style=Bluedotted] (99.center) to (100.center);
		\draw [style=Bluedotted] (100.center) to (101.center);
		\draw [style=Bluedotted] (101.center) to (98.center);
		\draw (114.center) to (115.center);
	\end{pgfonlayer}
\end{tikzpicture}
}
\label{c2generatorscompare}
\end{equation}
where the first few terms of the PL of the Hilbert series are written, giving the quantum numbers of the generators of the Coulomb branch. In general, one starts with the affine Dynkin quiver of $C_2$ which contributes $[2,0]t^2$ for the generators. Once all gauge nodes increase their rank by 1, a generator transforming in $[2,0]t^4$ is added. This pattern persists for any addition of 1 to all gauge nodes, and we see the quivers in the blue boxes are precisely those where new generators in the adjoint representation appear. In general, for the quiver 
\begin{equation}
    \raisebox{-.5\height}{
\begin{tikzpicture}
	\begin{pgfonlayer}{nodelayer}
		\node [style=gauge3] (0) at (-0.5, 0) {};
		\node [style=gauge3] (1) at (0.5, 0) {};
		\node [style=none] (2) at (-0.35, 0.1) {};
		\node [style=none] (3) at (-0.35, -0.15) {};
		\node [style=none] (4) at (0.65, 0.1) {};
		\node [style=none] (5) at (0.65, -0.15) {};
		\node [style=none] (6) at (0.35, 0.35) {};
		\node [style=none] (7) at (0, -0.025) {};
		\node [style=none] (8) at (0.35, -0.375) {};
		\node [style=flavour2] (9) at (-0.5, 1) {};
		\node [style=flavour2] (10) at (0.5, 1) {};
		\node [style=none] (11) at (-0.5, -0.5) {$1+x$};
		\node [style=none] (12) at (0.5, -0.5) {$1+x$};
		\node [style=none] (13) at (0.5, 1.5) {0};
		\node [style=none] (14) at (-0.5, 1.5) {$1+x$};
	\end{pgfonlayer}
	\begin{pgfonlayer}{edgelayer}
		\draw (2.center) to (4.center);
		\draw (3.center) to (5.center);
		\draw (6.center) to (7.center);
		\draw (7.center) to (8.center);
		\draw (9) to (0);
		\draw (10) to (1);
	\end{pgfonlayer}
\end{tikzpicture}
}
\label{generalC2}
\end{equation}
one finds 
\begin{eqnarray}
    \mathrm{PL} \left[ \mathrm{HS}_{\mathrm{Sp}(4 ,\mathbb{C}),\lambda = [0,0]}^{\mu = [1+x,0]} \right] &=& \sum\limits_{i=0}^{x}[2,0]_{C_2}t^{2(1+i)}+ O(t^{2(2+x)})  \nonumber \\ &=& [2,0]_{C_2} t^2 \frac{1- t^{2+2x}}{1-t^2}+ O(t^{2(2+x)}) \, . 
\end{eqnarray}
As $x \rightarrow \infty$, we see the Coulomb branch Hilbert series takes the following form:
\begin{equation}
    \mathrm{HS}_{\mathrm{Sp}(4 ,\mathbb{C}),[0,0]}  \mathrm{PE}\left[ [2,0]_{C_2}  \frac{t^2}{1-t^2}\right] \,.
    \label{sp2infty}
\end{equation}
In this equation, it is understood that $[2,0]_{C_2}$ stands for a character whose variables participate in the plethystic exponential. Explicitly, using fundamental weight fugacities $x_i$ of $C_2$, this is 
\begin{equation}
    \mathrm{HS}_{\mathrm{Sp}(4,\mathbb{C}),[0,0]}(t;x_i)= \mathrm{PE}\left[
    \left(
     x_1^2
     +x_2
     +\frac{x_2^2}{x_1^2}
     +\frac{x_1^2}{x_2}
     +2
    +\frac{x_2}{x_1^2}
    +\frac{x_1^2}{x_2^2}
    +\frac{1}{x_2}
    +\frac{1}{x_1^2}
    \right)
    \frac{t^2}{1-t^2}\right]
\end{equation}
This reproduces the limit (\ref{eqLimit}).

The reasoning behind this is as follows: if one increases the rank of \emph{all} the gauge groups, then new generators appear. This is most easily seen in the cases of algebras of type $A$ and $C$ where the comarks are all equal to 1. If we increase all the ranks by 1, then the number of new generators is the dimension of the adjoint representation. For type $A$ and $C$ and all nodes balanced, we are able to simultaneously increase the rank of all gauge groups by 1 and at the same time maintaining a fully balanced quiver. Since the balance is maintained, the new generators transform always in the adjoint representation. Each time this is done, a new set of generators in the adjoint appears at higher and higher order.
One can consider repeating this counting process for other Lie algebras, but since some comarks are $>1$, the situations is more involved and is left for future work.

\paragraph{The cases $\lambda \neq  0$. }
If $\lambda \neq  0$, the quiver is unbalanced. However, the total number of generators is still an integer multiple of the dimension of the adjoint representation of $G$, but with generators spread over different orders of $t$. 
Consider $\lambda=[0,1]$ for $G=\mathrm{Sp}(4,\mathbb{C})$:
\begin{equation}
    \raisebox{-.5\height}{
\begin{tikzpicture}
	\begin{pgfonlayer}{nodelayer}
		\node [style=gauge3] (0) at (0, 0) {};
		\node [style=gauge3] (1) at (1, 0) {};
		\node [style=none] (2) at (0.15, 0.1) {};
		\node [style=none] (3) at (0.15, -0.15) {};
		\node [style=none] (4) at (1.15, 0.1) {};
		\node [style=none] (5) at (1.15, -0.15) {};
		\node [style=none] (6) at (0.85, 0.35) {};
		\node [style=none] (7) at (0.5, -0.025) {};
		\node [style=none] (8) at (0.85, -0.375) {};
		\node [style=flavour2] (9) at (0, 1) {};
		\node [style=flavour2] (10) at (1, 1) {};
		\node [style=none] (11) at (0, -0.5) {1};
		\node [style=none] (12) at (1, -0.5) {2};
		\node [style=none] (13) at (1, 1.5) {3};
		\node [style=none] (14) at (0, 1.5) {$0$};
		\node [style=gauge3] (15) at (-0.025, 3.25) {};
		\node [style=gauge3] (16) at (0.975, 3.25) {};
		\node [style=none] (17) at (0.125, 3.35) {};
		\node [style=none] (18) at (0.125, 3.1) {};
		\node [style=none] (19) at (1.125, 3.35) {};
		\node [style=none] (20) at (1.125, 3.1) {};
		\node [style=none] (21) at (0.825, 3.6) {};
		\node [style=none] (22) at (0.475, 3.225) {};
		\node [style=none] (23) at (0.825, 2.875) {};
		\node [style=flavour2] (24) at (-0.025, 4.25) {};
		\node [style=flavour2] (25) at (0.975, 4.25) {};
		\node [style=none] (26) at (-0.025, 2.75) {1};
		\node [style=none] (27) at (0.975, 2.75) {1};
		\node [style=none] (28) at (0.975, 4.75) {1};
		\node [style=none] (29) at (-0.025, 4.75) {1};
		\node [style=gauge3] (30) at (-0.025, -3.5) {};
		\node [style=gauge3] (31) at (0.975, -3.5) {};
		\node [style=none] (32) at (0.125, -3.4) {};
		\node [style=none] (33) at (0.125, -3.65) {};
		\node [style=none] (34) at (1.125, -3.4) {};
		\node [style=none] (35) at (1.125, -3.65) {};
		\node [style=none] (36) at (0.825, -3.15) {};
		\node [style=none] (37) at (0.475, -3.525) {};
		\node [style=none] (38) at (0.825, -3.875) {};
		\node [style=flavour2] (39) at (-0.025, -2.5) {};
		\node [style=flavour2] (40) at (0.975, -2.5) {};
		\node [style=none] (41) at (-0.025, -4) {2};
		\node [style=none] (42) at (0.975, -4) {2};
		\node [style=none] (43) at (0.975, -2) {1};
		\node [style=none] (44) at (-0.025, -2) {2};
		\node [style=gauge3] (45) at (6.975, 3.25) {};
		\node [style=gauge3] (46) at (7.975, 3.25) {};
		\node [style=none] (47) at (7.125, 3.35) {};
		\node [style=none] (48) at (7.125, 3.1) {};
		\node [style=none] (49) at (8.125, 3.35) {};
		\node [style=none] (50) at (8.125, 3.1) {};
		\node [style=none] (51) at (7.825, 3.6) {};
		\node [style=none] (52) at (7.475, 3.225) {};
		\node [style=none] (53) at (7.825, 2.875) {};
		\node [style=flavour2] (54) at (6.975, 4.25) {};
		\node [style=flavour2] (55) at (7.975, 4.25) {};
		\node [style=none] (56) at (6.975, 2.75) {2};
		\node [style=none] (57) at (7.975, 2.75) {3};
		\node [style=none] (58) at (7.975, 4.75) {3};
		\node [style=none] (59) at (6.975, 4.75) {1};
		\node [style=gauge3] (60) at (6.975, -0.25) {};
		\node [style=gauge3] (61) at (7.975, -0.25) {};
		\node [style=none] (62) at (7.125, -0.15) {};
		\node [style=none] (63) at (7.125, -0.4) {};
		\node [style=none] (64) at (8.125, -0.15) {};
		\node [style=none] (65) at (8.125, -0.4) {};
		\node [style=none] (66) at (7.825, 0.1) {};
		\node [style=none] (67) at (7.475, -0.275) {};
		\node [style=none] (68) at (7.825, -0.625) {};
		\node [style=flavour2] (69) at (6.975, 0.75) {};
		\node [style=flavour2] (70) at (7.975, 0.75) {};
		\node [style=none] (71) at (6.975, -0.75) {2};
		\node [style=none] (72) at (7.975, -0.75) {4};
		\node [style=none] (73) at (7.975, 1.25) {5};
		\node [style=none] (74) at (6.975, 1.25) {$0$};
		\node [style=gauge3] (75) at (6.975, -3.5) {};
		\node [style=gauge3] (76) at (7.975, -3.5) {};
		\node [style=none] (77) at (7.125, -3.4) {};
		\node [style=none] (78) at (7.125, -3.65) {};
		\node [style=none] (79) at (8.125, -3.4) {};
		\node [style=none] (80) at (8.125, -3.65) {};
		\node [style=none] (81) at (7.825, -3.15) {};
		\node [style=none] (82) at (7.475, -3.525) {};
		\node [style=none] (83) at (7.825, -3.875) {};
		\node [style=flavour2] (84) at (6.975, -2.5) {};
		\node [style=flavour2] (85) at (7.975, -2.5) {};
		\node [style=none] (86) at (6.975, -4) {3};
		\node [style=none] (87) at (7.975, -4) {3};
		\node [style=none] (88) at (7.975, -2) {1};
		\node [style=none] (89) at (6.975, -2) {3};
		\node [style=none] (90) at (-1, 5.25) {};
		\node [style=none] (91) at (-1, 2.5) {};
		\node [style=none] (92) at (2, 2.5) {};
		\node [style=none] (93) at (2, 5.25) {};
		\node [style=none] (94) at (-1, -1.5) {};
		\node [style=none] (95) at (-1, -4.25) {};
		\node [style=none] (96) at (2, -4.25) {};
		\node [style=none] (97) at (2, -1.5) {};
		\node [style=none] (115) at (3.75, 0.5) {$([2]_{A_1}+[0]_{A_1})t^2$};
		\node [style=none] (117) at (3.75, -1.75) {$([2]_{A_1}+[0]_{A_1})t^2$};
		\node [style=none] (119) at (3.75, -2.75) {$+([2]_{A_1}+[0]_{A_1})t^4$};
		\node [style=none] (135) at (3.75, 0) {$+(q+q^{-1})[2]_{A_1}t^3+\dots$};
		\node [style=none] (138) at (3.75, -2.25) {$+(q+q^{-1})[2]_{A_1}t^3$};
		\node [style=none] (139) at (3.75, -3.25) {$ +(q+q^{-1})[2]_{A_1}t^5+\dots$};
		\node [style=none] (148) at (11.25, -4) {$+([2]_{A_1}+[0]_{A_1})t^6$};
		\node [style=none] (149) at (11.25, -4.5) {$+(q+q^{-1})[2]_{A_1}t^7+\dots$};
		\node [style=none] (150) at (6, 6) {};
		\node [style=none] (151) at (6, -5) {};
		\node [style=none] (152) at (6.25, -1.75) {};
		\node [style=none] (153) at (6.25, -4.5) {};
		\node [style=none] (154) at (9.25, -4.5) {};
		\node [style=none] (155) at (9.25, -1.75) {};
		\node [style=none] (158) at (4, 6.5) {PL[HS]};
		\node [style=none] (159) at (11.75, 6.5) {PL[HS]};
		\node [style=none] (160) at (0.75, 6.5) {Slices of $\gr{\mathrm{Sp}(4 , \mathbb{C})}{}{\lambda = [0,1]}$};
		\node [style=none] (161) at (8.5, 6.5) {Slices of $\gr{\mathrm{Sp}(4 , \mathbb{C})}{}{\lambda = [0,1]}$};
		\node [style=none] (162) at (3.75, 3.5) {$([2]_{A_1}+[0]_{A_1})t^2$};
		\node [style=none] (163) at (3.75, 3) {$+(q+q^{-1})[2]_{A_1}t^3+\dots$};
		\node [style=none] (164) at (11.5, 4.5) {$([2]_{A_1}+[0]_{A_1})t^2$};
		\node [style=none] (165) at (11.5, 3.5) {$+([2]_{A_1}+[0]_{A_1})t^4$};
		\node [style=none] (166) at (11.5, 4) {$+(q+q^{-1})[2]_{A_1}t^3$};
		\node [style=none] (167) at (11.5, 3) {$ +(q+q^{-1})[2]_{A_1}t^5+\dots$};
		\node [style=none] (168) at (11.5, 1.5) {$([2]_{A_1}+[0]_{A_1})t^2$};
		\node [style=none] (169) at (11.5, 0.5) {$+([2]_{A_1}+[0]_{A_1})t^4$};
		\node [style=none] (170) at (11.5, 1) {$+(q+q^{-1})[2]_{A_1}t^3$};
		\node [style=none] (171) at (11.5, 0) {$+ (q+q^{-1})[2]_{A_1}t^5+\dots$};
		\node [style=none] (172) at (11.25, -2) {$([2]_{A_1}+[0]_{A_1})t^2$};
		\node [style=none] (173) at (11.25, -3) {$+([2]_{A_1}+[0]_{A_1})t^4$};
		\node [style=none] (174) at (11.25, -2.5) {$+(q+q^{-1})[2]_{A_1}t^3$};
		\node [style=none] (175) at (11.25, -3.5) {$ +(q+q^{-1})[2]_{A_1}t^5$};
	\end{pgfonlayer}
	\begin{pgfonlayer}{edgelayer}
		\draw (2.center) to (4.center);
		\draw (3.center) to (5.center);
		\draw (6.center) to (7.center);
		\draw (7.center) to (8.center);
		\draw (9) to (0);
		\draw (10) to (1);
		\draw (17.center) to (19.center);
		\draw (18.center) to (20.center);
		\draw (21.center) to (22.center);
		\draw (22.center) to (23.center);
		\draw (24) to (15);
		\draw (25) to (16);
		\draw (32.center) to (34.center);
		\draw (33.center) to (35.center);
		\draw (36.center) to (37.center);
		\draw (37.center) to (38.center);
		\draw (39) to (30);
		\draw (40) to (31);
		\draw (47.center) to (49.center);
		\draw (48.center) to (50.center);
		\draw (51.center) to (52.center);
		\draw (52.center) to (53.center);
		\draw (54) to (45);
		\draw (55) to (46);
		\draw (62.center) to (64.center);
		\draw (63.center) to (65.center);
		\draw (66.center) to (67.center);
		\draw (67.center) to (68.center);
		\draw (69) to (60);
		\draw (70) to (61);
		\draw (77.center) to (79.center);
		\draw (78.center) to (80.center);
		\draw (81.center) to (82.center);
		\draw (82.center) to (83.center);
		\draw (84) to (75);
		\draw (85) to (76);
		\draw [style=Bluedotted] (90.center) to (91.center);
		\draw [style=Bluedotted] (91.center) to (92.center);
		\draw [style=Bluedotted] (92.center) to (93.center);
		\draw [style=Bluedotted] (93.center) to (90.center);
		\draw [style=Bluedotted] (94.center) to (95.center);
		\draw [style=Bluedotted] (95.center) to (96.center);
		\draw [style=Bluedotted] (96.center) to (97.center);
		\draw [style=Bluedotted] (97.center) to (94.center);
		\draw (150.center) to (151.center);
		\draw [style=Bluedotted] (152.center) to (153.center);
		\draw [style=Bluedotted] (153.center) to (154.center);
		\draw [style=Bluedotted] (154.center) to (155.center);
		\draw [style=Bluedotted] (155.center) to (152.center);
	\end{pgfonlayer}
\end{tikzpicture}
}
\end{equation}
Due to the overbalanced gauge node, the global symmetry is now $\mathfrak{su}(2)\times \mathfrak{u}(1)$ with $[\;\; ]_{A_1}$ and $q$ Dynkin labels respectively. Compared with \eqref{c2generatorscompare}, we see that the total number of generators (at all orders of $t$ combined) remains the same. However, due to the unbalanced gauge node, some generators appear at higher orders and acquire non-trivial $\mathfrak{u}(1)$ charge. The precise irreducible representations for these generators depend on which node of the balanced quiver it is connected to \cite{Cabrera:2018uvz}. This easily generalizes to $\lambda=[0,\lambda_2]$ such that in the large rank limit, we can once again derive the freely generated Hilbert series:
\begin{equation}
\begin{aligned}
    \mathrm{HS}_{\mathrm{Sp}(2),[0,\lambda_2]}(t) &=\mathrm{PE}\left[\sum\limits_{i=1}^{\infty}(\mathrm{dim}([2]_{A_1})+1)t^{2i}+2\mathrm{dim}([2]_{A_1}))t^{2i+\lambda_2}\right] \\ 
    &=\mathrm{PE}\left[\sum\limits_{i=1}^{\infty}4t^{2i}+6t^{2i+\lambda_2}\right] 
    \end{aligned}
    \label{sp2infty[01]}
\end{equation}
and refining with $x_i$ and $q$, the fundamental weight fugacities of $\mathfrak{su}(2)\times \mathfrak{u}(1)$, yields: 
\begin{equation}
     \mathrm{HS}_{\mathrm{Sp}(2),[0,\lambda_2]}(t;x_i,q) = \mathrm{PE}\left[\frac{(2+x^2+\frac{1}{x^2})t^2+(q+\frac{1}{q})(1+x^2+\frac{1}{x^2})t^{2+\lambda_2}}{1-t^2}\right] \,.
\end{equation}
This agrees with (\ref{HSTransverseSlice}). 
A similar analysis can be repeated for any coweight $\lambda$ and any group $G$. 

\paragraph{Observation about number of generators.} We have the following generalization of the observations above. Take a slice $\wb{G}{\lambda}{\mu}$ in the affine Grassmannian of a group $G$, with global symmetry $G'\subset G$. The adjoint representation, identified with the Lie algebra, is decomposed as
\begin{equation}
    \mathfrak{g}\mapsto \underbrace{\mathfrak{g}'}_{\textnormal{appears at }t^2}+\underbrace{R'}_{\textnormal{appears at higher orders}}
\end{equation}
where $R'$ is a (possibly reducible) representation of $G'$. In the Hilbert series of $\wb{G}{\lambda}{\mu}$, $\mathfrak{g}'$ appears at $t^2$ while irreducible representations in $R'$ appear at higher orders of $t$. The only further sets of generators (if there are any) are in the same representations $\mathfrak{g}'+R'$, but appear at higher orders in $t$, shifted by a fixed integer. The total number of generators is hence an integer multiple of the dimension of $G$. One could say that the adjoint of $G$ is ``stretched'' over higher orders.

\subsection{A note on infinity}

The affine Grassmannian is infinite dimensional, and this aspect pervades through all parts of this paper, in which we met infinite Hasse diagrams, configurations with infinitely many branes, Hilbert series with infinitely many generators. However in all those cases, these infinite quantities do not represent a problem, because they can be approached in a controlled way by finite quantities. In our language, this translates into equalities of the type 
\begin{equation}
\label{limitCoulombBranch}
    \lim\limits_{N \rightarrow \infty} \mathcal{C}^{3d} \left( \raisebox{-.5\height}{\scalebox{.8}{\begin{tikzpicture}[xscale=.7,yscale=.7]
    \node[flavour,label=right:{$2N$}] (1) at (0,1) {}; 
    \node[gauge,label=right:{$N$}] (0) at (0,0) {};
    \draw (0)--(1);
\end{tikzpicture}}} \right) =  \mathrm{Gr}_{\mathrm{SL}(2 \mathbb{C})} \, . 
\end{equation}
The affine Grassmannian appears here as a controlled limit of finite dimensional Coulomb branches. The reason why the limit (\ref{limitCoulombBranch}) makes sense is because each term in the sequence is a subset of the next term: 
\begin{equation}
    \mathcal{C}^{3d} \left( \raisebox{-.5\height}{\scalebox{.8}{\begin{tikzpicture}[xscale=.7,yscale=.7]
    \node[flavour,label=right:{$2N$}] (1) at (0,1) {}; 
    \node[gauge,label=right:{$N$}] (0) at (0,0) {};
    \draw (0)--(1);
\end{tikzpicture}}} \right) \subset \mathcal{C}^{3d} \left( \raisebox{-.5\height}{\scalebox{.8}{\begin{tikzpicture}[xscale=.7,yscale=.7]
    \node[flavour,label=right:{$2(N+1)$}] (1) at (0,1) {}; 
    \node[gauge,label=right:{$N+1$}] (0) at (0,0) {};
    \draw (0)--(1);
\end{tikzpicture}}} \right) \, . 
\end{equation}
This property is the reason why we can build the bottom part of the Hasse diagrams of the infinite dimensional affine Grassmannian using only data from the finite dimensional Coulomb branches corresponding to transverse slices. In algebraic geometry, this is the main idea behind the construction of so-called \emph{ind-schemes} \cite{2016arXiv160305593Z}, of which the affine Grassmannian is a prime example. 

All the infinite quantities that appear in this paper can be regularized that way. For instance, the brane setup (\ref{braneInfinity}) should be seen as the limit
\begin{equation}
\label{limitBraneSystem}
    \lim\limits_{N \rightarrow \infty} \left( \raisebox{-.5\height}{\scalebox{.8}{
	\begin{tikzpicture}
		\draw[red] (0,0)--(0,4) (2,0)--(2,4) (6,0)--(6,4);
		\node at (4,2) {$\cdots$};
		\node (1) at (-3,2) {$\cdots$};
		\node[D5] (2) at (-2,2) {};
		\node[D5] (3) at (-1,2) {};
		\node[D5] (4) at (7,2) {};
		\node[D5] (5) at (8,2) {};
		\node (6) at (9,2) {$\cdots$};
		\draw[thick] (1)--(2)--(3)--(3,2) (5,2)--(4)--(5)--(6);
		\node[rotate=-45] at (-1.9,2.7) {\small $N-1$ D3};
		\node at (1,2.3) {\small $N$ D3};
		\node[rotate=-45] at (7.1,2.7) {\small $N-1$ D3};
		\draw [decorate,decoration={brace,amplitude=5pt}] (6.2,-0.2)--(-0.2,-0.2);
		\node at (3,-1) {$n$ NS5};
		\begin{scope}
		\clip (-3,1) rectangle(3,2);
		\draw [decorate,decoration={brace,amplitude=5pt}] (-0.8,1.8)--(-3.5,1.8);
		\end{scope}
		\node at (-2,1) {$N$ D5};
		\begin{scope}
		\clip (6,1) rectangle(9,2);
		\draw [decorate,decoration={brace,amplitude=5pt}] (9.5,1.8)--(6.8,1.8);
		\end{scope}
		\node at (8,1) {$N$ D5};
	\end{tikzpicture}}} \right)
\end{equation}
where the fundamental property that the system for a given $N$ is included into the system for $N+1$ is again satisfied. On that diagram, we can now compute the linking number of the D5 branes at finite $N$ to be $N - (N-1) -N + (N-1) = 0$, independent of $N$. Therefore in the infinite limit, the D5 branes also have linking number 0. 
On the other hand, if one adds a D5 brane between the two NS5 branes in (\ref{limitBraneSystem}), its linking number is equal to $N - N -(N-1) + N = 1$ for all $N$, and therefore is 1 in the limit, as claimed below the brane system  (\ref{eq:ASecondCompGeneral}).

Finally, the same logic is behind the formulas for the Hilbert series as infinite products (\ref{HSTransverseSlice}) obtained as a limit (\ref{eqLimit}), and underlies the observations about the generators made in the previous subsection. This can be shown explicitly in the case of the $\mathrm{PSL}(2,\mathbb{C})$ affine Grassmannian, using the Hilbert series given in \cite[Sec 5.1]{Cremonesi:2013lqa}. The Hilbert series for $\wb{\mathrm{PSL}(2,\mathbb{C})}{\mu}{\lambda}$, with $\mu , \lambda \in \mathbb{N}$ and $\mu - \lambda \in 2 \mathbb{N}$, is 
\begin{eqnarray}
    \mathrm{HS}_{\mathrm{PSL}(2,\mathbb{C}) , \lambda}^{\mu} (t;x) 
    &=& \prod\limits_{j=1}^{\frac{1}{2} (\mu - \lambda)} \frac{1-t^{2j + \mu  + \lambda}}{(1-t^{2j})(1-x^2 t^{2j   + \lambda}) (1- x^{-2} t^{2j    + \lambda})} \nonumber \\ 
     &=& \mathrm{PE} \left[ \left( t^2   + (x^2 + x^{-2})   t^{2+\lambda}  - t^{2+\lambda + \mu} \right)  \frac{1 -t^{\mu - \lambda}}{1-t^{2}}  \right] \, . 
\end{eqnarray}
From this we can take the formal limit $\mu \rightarrow \infty$ using $\lim\limits_{\mu \rightarrow \infty} t^{\mu} = 0$: 
\begin{eqnarray}
    \mathrm{HS}_{\mathrm{PSL}(2,\mathbb{C}) , \lambda} (t;x)
     &=&  \lim\limits_{\mu \rightarrow \infty} \mathrm{HS}_{\mathrm{PSL}(2,\mathbb{C}) , \lambda}^{\mu} (t;x) \nonumber  \\
     &=& \mathrm{PE} \left[  \frac{t^2 +  (x^2 + x^{-2}) t^{2+ \lambda}}{1-t^2} \right] \, . 
\end{eqnarray}
This reproduces the general result (\ref{HSTransverseSlice}). For $\lambda = 0$, the character $1+x^2 + x^{-2}$ appears in front of $t^2$, showing the symmetry enhancement from $\mathfrak{u}(1)$ for $\lambda >0$ to $\mathfrak{su}(2)$ for $\lambda = 0$.

\section{Outlook}

In this paper we explored the Hasse diagrams and transverse slices of various affine Grassmannians, using both quivers and branes. Every slice in the affine Grassmannian, specified by two dominant coweights $\lambda\leq\mu$, of a finite dimensional simple group $G$ is the Coulomb branch of a good framed quiver in the shape of the Dynkin diagram of $G$. The coweight $\mu$ (in coweight basis) is the framing, while $\lambda$ is the imbalance. For classical groups Kraft-Procesi transitions in brane systems naturally reproduce the Hasse diagram of the affine Grassmannian and all transverse slices. Quiver subtraction proves effective in the realm of the affine Grassmannian. A new notion called \emph{quiver addition} is introduced. Requiring the reverse of quiver subtraction in the sense that for a quiver $\mathsf{Q}$ we take an `addable' quiver $\mathsf{S}$ and produce a new quiver $\mathsf{Q}'$, such that $\mathsf{Q}'=\mathsf{Q}-\mathsf{S}$ leads to an infinite number of possibilities. When we limit to slices in the affine Grassmannian however, we have a finite number of possibilities. Starting with a coweight, quiver addition produces the entire Hasse diagram of slices to this coweight. This agrees perfectly with the brane construction.

\paragraph{`Ugly' and `bad' quivers.} Of course quivers, as well as brane systems, are not limited to `good' theories. Theories which have a negative imbalance have Coulomb branches dubbed \emph{generalized affine Grassmannian slices} \cite{Braverman:2016pwk}, which are also specified by two coweights, but the lower coweight, $\lambda$, is not required to be dominant, i.e.\ the imbalance is allowed to be negative. Quivers and brane systems for these generalized slices are easily produced, and the Hasse diagram may be obtained using quiver subtraction or brane moves. It would be nice to explore these theories in the future.

\paragraph{Other brane systems.} We expect that for many brane systems consisting of D$p$, NS5, and D$(p+2)$ branes (with $p\le6$), and more general systems, there exists a corresponding slice in the affine Grassmannian (provided that every gauge group in the effective theory has enough flavors). We hope to explore brane systems, other than the D3, NS5, and D5 systems in this paper, in the future.

\paragraph{Relation to nilpotent orbits and $T^{\sigma}_{\rho}[G]$ theories.} Quivers for slices in the affine Grassmannian of $A$-type and $T^{\sigma}_{\rho}[SU]$ theories coincide. They are the good framed linear quivers. For other groups this is not the case. While the construction for $T^{\sigma}_{\rho}$ theories for other classical groups uses O3 orientifold planes and linear orthosymplectic quivers, the construction for affine Grassmannian slices uses ON planes and Dynkin shape unitary quivers. The moduli spaces only coincide in few cases, and are related through quotients in few more. Exploring the relationship between $T^{\sigma}_{\rho}$ theories and slices in the affine Grassmannian is left for future work.

\paragraph{Symplectic duality.} The quivers associated to slices in the affine Grassmannian for simply laced groups also possess a Higgs branch. Their Hasse diagrams are simply obtained from \emph{inversion} \cite{Grimminger:2020dmg}. It could be interesting to ask for some notion of the symplectic dual of the entire affine Grassmannian.

\paragraph{Relation to other work.} We expect there to be many other interesting lines of research involving the affine Grassmannian in physics, such as in the context of little strings \cite{Haouzi:2016ohr,Haouzi:2016yyg}, domain walls \cite{Bachas:2000dx}, or topological quantum field theories \cite{Gukov:2020lqm}.

\section*{Acknowledgements}
AH would like to extend special thanks to Jacob Matherne for explaining basic concepts about the affine Grassmannian, and for email exchange together with Santiago Cabrera, where early versions of the main topic of this paper were conceived. Also to Hiraku Nakajima for explaining many concepts related to the affine Grassmanian. We are grateful to Alex Weekes for helping us understand the mathematical literature and for vital insights. Furthermore we would like to thank Travis Schedler for many helpful discussions throughout the years. AB would like to thank Daniel Juteau for many profitable exchanges. The work of AB, JFG, AH and ZZ is supported by STFC grant ST/P000762/1 and ST/T000791/1. The work of MS is supported by the National Thousand-Young-Talents Program of China, the National Natural Science Foundation of China (grant no.\ 11950410497), and the China Postdoctoral Science Foundation (grant no.\ 2019M650616). MS and ZZ are grateful to Fudan University, Department of Physics for hospitality during part of this work.

\appendix

\section{Branes, ON planes and quivers}
\label{app:orientifolds}
In this appendix we review allowed configurations of branes in the presence of ON planes. We construct brane systems of NS5 branes, D5 branes and D3 branes in Type IIB String Theory as first developed in \cite{Hanany:1996ie}, with the addition of ON planes. The space-time extension of the branes are collected in Table \ref{tab:setup}.
\begin{table}[ht]
\begin{center}
\begin{tabular}{c|c|c|c|c|c|c|c|c|c|c}
& $x^0$ & $x^1$ & $x^2$ & $x^3$ & $x^4$ & $x^5$ & $x^6$ & $x^7$ & $x^8$ & $x^9$\\
\hline
{\color{red}NS5} & x & x & x & x & x & x & & & &  \\
\hline
D3 & x & x & x & & & & x & & & \\
\hline
{\color{blue}D5} & x & x & x & & & & & x & x & x \\
\hline
{\color{gray}ON} & x & x & x & x & x & x & & & &  
\end{tabular}
\caption{Type IIB set-up. The 'x' mark the spacetime directions spanned by the various branes and, if present, the ON plane. All NS5 branes are localized at the same value of $(x^7,x^8,x^9)$. All D5 branes are localized at the same value of $(x^3,x^4,x^5)$.}
\label{tab:setup}
\end{center}
\end{table}

The low energy theory on the world-volume of D3 branes suspended between NS5 branes is a $3d$ $\mathcal{N}=4$ quiver gauge theory, where D5 branes provide flavour nodes. For example, the following set-up:
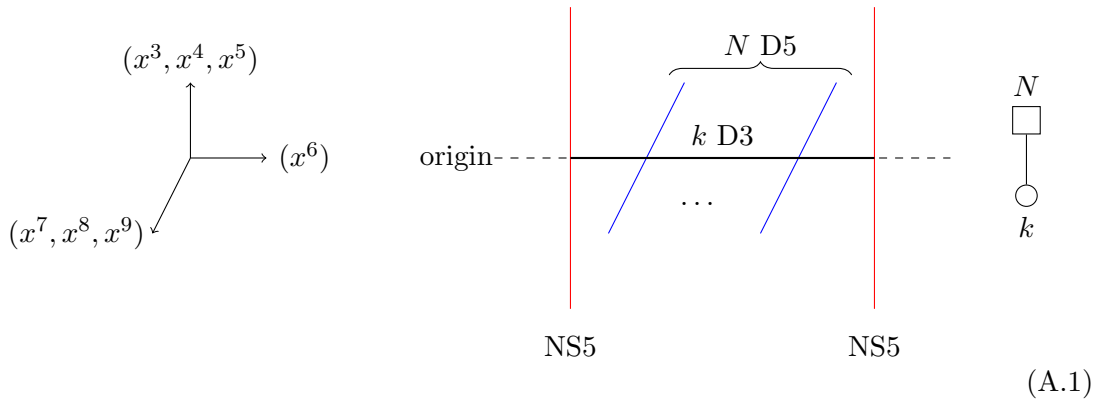
\begin{equation}
	\vcenter{\hbox{\makebox[\textwidth][c]{
    \begin{tikzpicture}
        \begin{scope}[shift={(-2,1)}]
        \draw[->] (-3,1)--(-2,1);
        \node at (-1.5,1) {$(x^6)$};
        \draw[->] (-3,1)--(-3,2);
        \node at (-3,2.3) {$(x^3,x^4,x^5)$};
        \draw[->] (-3,1)--(-3.5,0);
        \node at (-4.5,0) {$(x^7,x^8,x^9)$};
        \end{scope}
        \draw[red] (0,0)--(0,4) (4,0)--(4,4);
        \node at (0,-0.5) {NS5};
        \node at (4,-0.5) {NS5};
        \draw[blue] (0.5,1)--(1.5,3) (2.5,1)--(3.5,3);
        \draw[dashed] (-1,2)--(5,2);
        \node at (-1.5,2) {origin};
        \draw[thick] (0,2)--(4,2);
        \node at (2,2.3) {$k$ D3};
        \draw [decorate,decoration={brace,amplitude=5pt}] (1.3,3.1)--(3.7,3.1);
        \node at (2.5,3.5) {$N$ D5};
        \node at (1.7,1.4) {\dots};
        \begin{scope}[shift={(6,1.5)}]
        \node[gauge,label=below:{$k$}] (1) at (0,0) {};
        \node[flavour,label=above:{$N$}] (2) at (0,1) {};
        \draw (1)--(2);
        \end{scope}
    \end{tikzpicture}}}}
    \label{eq:BranesNode}
\end{equation}
We will supress the $(x^7,x^8,x^9)$ direction from now on and draw D5 branes as $\vcenter{\hbox{\scalebox{0.5}{\begin{tikzpicture}\node[D5]{};\end{tikzpicture}}}}$. The position of D3 branes along the NS5 branes, as depicted in \eqref{eq:Coulomb}, correspond to Coulomb branch moduli:
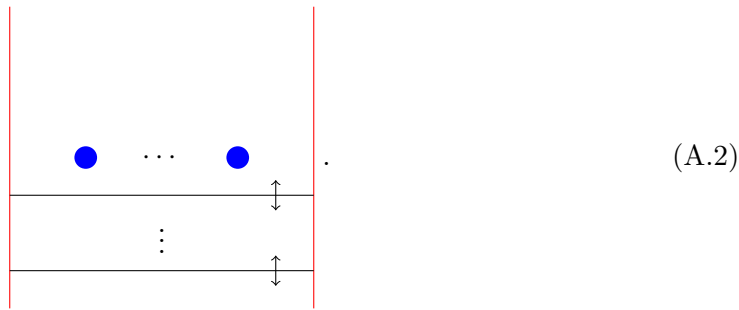
\begin{equation}
	\vcenter{\hbox{\scalebox{1}{
	\centering
    \begin{tikzpicture}
        \draw[red] (0,0)--(0,4) (4,0)--(4,4);
        \node[D5] (1) at (1,2) {};
        \node[D5] (2) at (3,2) {};
        \node at (2,2) {$\cdots$};
        \draw (0,1.5)--(4,1.5) (0,0.5)--(4,0.5);
        \node at (2,1) {$\vdots$};
        \draw[<->] (3.5,0.3)--(3.5,0.7);
        \draw[<->] (3.5,1.3)--(3.5,1.7);
    \end{tikzpicture}}}}\;.
    \label{eq:Coulomb}
\end{equation}

\paragraph{Including ON planes.} The gauge group on the world-volume of a stack of NS5 branes on top of an ON plane is given by the allowed D1 branes (and their mirror images) ending on NS5 branes and their mirror images. This is analogous to stacks of D3 branes on top of O3 planes with fundamental strings stretched between them, studied in \cite{Hanany:2001iy}. The allowed D1 states and the corresponding gauge/electric algebra on the world-volume of the NS5 branes are:

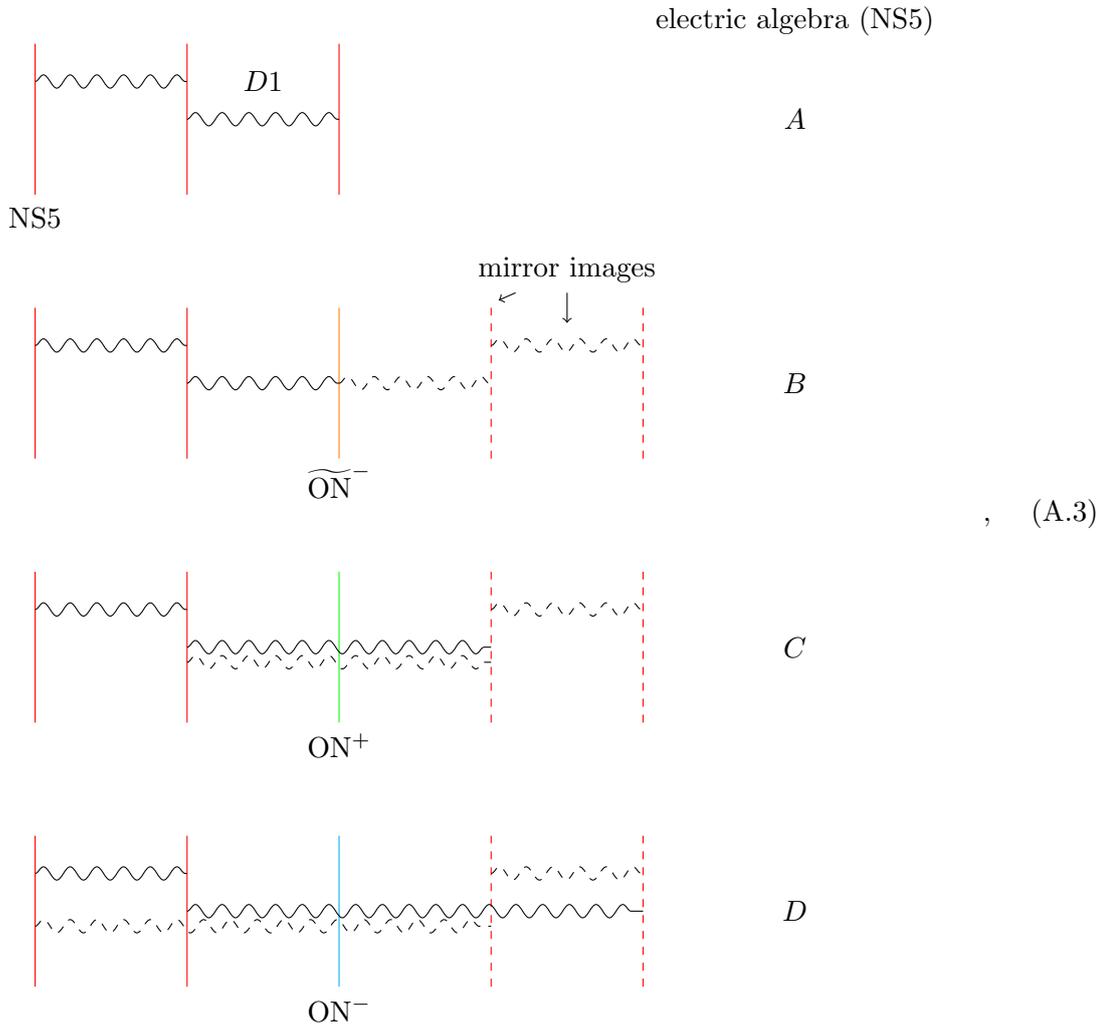
\begin{equation}
	\vcenter{\hbox{
	\begin{tikzpicture}
	    \begin{scope}[shift={(0,0)}]
		\draw[red] (0,0)--(0,2) (2,0)--(2,2) (4,0)--(4,2);
		\draw[wiggler] (0,1.5)--(2,1.5);
		\draw[wiggler] (2,1)--(4,1);
	    \end{scope}
		\begin{scope}[shift={(0,-3.5)}]
		\draw[red] (0,0)--(0,2) (2,0)--(2,2);
		\draw[red, dashed] (6,0)--(6,2) (8,0)--(8,2);
		\draw[orange] (4,0)--(4,2);
		\draw[wiggler] (0,1.5)--(2,1.5);
		\draw[wiggler,dashed] (6,1.5)--(8,1.5);
		\draw[wiggler] (2,1)--(4,1);
		\draw[wiggler,dashed] (4,1)--(6,1);
		\end{scope}
	    \begin{scope}[shift={(0,-7)}]
		\draw[red] (0,0)--(0,2) (2,0)--(2,2);
		\draw[red, dashed] (6,0)--(6,2) (8,0)--(8,2);
		\draw[green] (4,0)--(4,2);
		\draw[wiggler] (0,1.5)--(2,1.5);
		\draw[wiggler,dashed] (6,1.5)--(8,1.5);
		\draw[wiggler] (2,1)--(6,1);
		\draw[wiggler,dashed] (2,0.8)--(6,0.8);
		\end{scope}
		\begin{scope}[shift={(0,-10.5)}]
		\draw[red] (0,0)--(0,2) (2,0)--(2,2);
		\draw[red, dashed] (6,0)--(6,2) (8,0)--(8,2);
		\draw[cyan] (4,0)--(4,2);
		\draw[wiggler] (0,1.5)--(2,1.5);
		\draw[wiggler,dashed] (6,1.5)--(8,1.5);
		\draw[wiggler] (2,1)--(8,1);
		\draw[wiggler,dashed] (0,0.8)--(6,0.8);
		\end{scope}
		\node at (10,2.3) {electric algebra (NS5)};
		\node at (10,1) {$A$};
		\node at (10,-2.5) {$B$};
		\node at (10,-6) {$C$};
		\node at (10,-9.5) {$D$};
		\node at (3,1.5) {$D1$};
		\node at (0,-0.3) {NS5};
		\node at (4,-3.8) {$\widetilde{\textnormal{ON}}^-$};
		\node at (4,-7.3) {$\textnormal{ON}^+$};
		\node at (4,-10.8) {$\textnormal{ON}^-$};
		\node (mi) at (7,-1) {mirror images};
		\draw[->] (mi)--(6.1,-1.4);
		\draw[->] (mi)--(7,-1.7);
	\end{tikzpicture}
	}}\quad,
\end{equation}
where dashed lines correspond to mirror images (in the main text we use solid lines). 

The endpoints of D3 branes play the role of magnetic monopoles on the world-volume of the NS5 branes, and the D3 branes play the role of roots of the GNO dual/magnetic algebra \cite{GNO}. The allowed D3 states and corresponding magnetic algebras are:

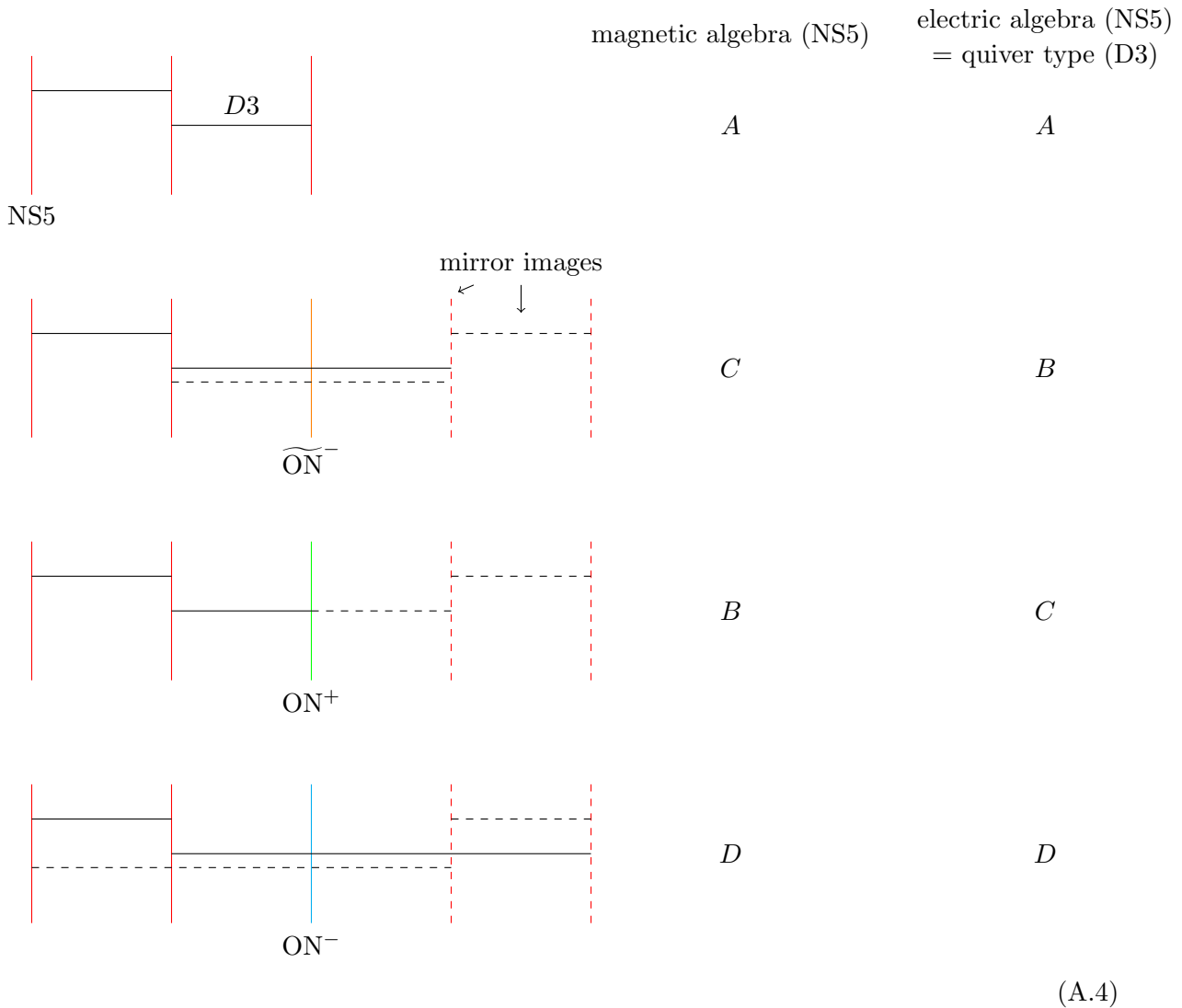
\begin{equation}
	\begin{adjustbox}{center}
		\begin{tikzpicture}
		\begin{scope}[shift={(0,0)}]
		\draw[red] (0,0)--(0,2) (2,0)--(2,2) (4,0)--(4,2);
		\draw (0,1.5)--(2,1.5) (2,1)--(4,1);
		\end{scope}
		\begin{scope}[shift={(0,-3.5)}]
		\draw[red] (0,0)--(0,2) (2,0)--(2,2);
		\draw[red, dashed] (6,0)--(6,2) (8,0)--(8,2);
		\draw[orange] (4,0)--(4,2);
		\draw (0,1.5)--(2,1.5);
		\draw[dashed] (6,1.5)--(8,1.5);
		\draw (2,1)--(6,1);
		\draw[dashed] (2,0.8)--(6,0.8);
		\end{scope}
		\begin{scope}[shift={(0,-7)}]
		\draw[red] (0,0)--(0,2) (2,0)--(2,2);
		\draw[red, dashed] (6,0)--(6,2) (8,0)--(8,2);
		\draw[green] (4,0)--(4,2);
		\draw (0,1.5)--(2,1.5);
		\draw[dashed] (6,1.5)--(8,1.5);
		\draw (2,1)--(4,1);
		\draw[dashed] (4,1)--(6,1);
		\end{scope}
		\begin{scope}[shift={(0,-10.5)}]
		\draw[red] (0,0)--(0,2) (2,0)--(2,2);
		\draw[red, dashed] (6,0)--(6,2) (8,0)--(8,2);
		\draw[cyan] (4,0)--(4,2);
		\draw (0,1.5)--(2,1.5);
		\draw[dashed] (6,1.5)--(8,1.5);
		\draw (2,1)--(8,1);
		\draw[dashed] (0,0.8)--(6,0.8);
		\end{scope}
		\node at (10,2.3) {magnetic algebra (NS5)};
		\node at (10,1) {$A$};
		\node at (10,-2.5) {$C$};
		\node at (10,-6) {$B$};
		\node at (10,-9.5) {$D$};
		\node at (3,1.3) {$D3$};
		\node at (0,-0.3) {NS5};
		\node at (4,-3.8) {$\widetilde{\textnormal{ON}}^-$};
		\node at (4,-7.3) {$\textnormal{ON}^+$};
		\node at (4,-10.8) {$\textnormal{ON}^-$};
		\node at (14.5,2.5) {electric algebra (NS5)};
		\node at (14.5,2) {= quiver type (D3)};
		\node at (14.5,1) {$A$};
		\node at (14.5,-2.5) {$B$};
		\node at (14.5,-6) {$C$};
		\node at (14.5,-9.5) {$D$};
		\node (mi) at (7,-1) {mirror images};
		\draw[->] (mi)--(6.1,-1.4);
		\draw[->] (mi)--(7,-1.7);
	\end{tikzpicture}
	\end{adjustbox}
	\label{eq:D3_suspension}
\end{equation}
However, the quiver representation of the low energy theory on the world-volume of D3 branes suspended between NS5 branes in presence of an ON plane is not in the form of the Dynkin diagram associated to the magnetic algebra of the ON plane given in \eqref{eq:D3_suspension}. Rather, D3 branes corresponding to a short root of the magnetic algebra will produce a gauge node which is a long node in the quiver and vice versa, as pointed out in \cite{Cremonesi:2014xha}. The quiver type is also indicated in \eqref{eq:D3_suspension}. 

\paragraph{Reading the quiver.} In the following, and in the main text, we denote both a brane and its mirror image with a solid line. A stack of $k$ D3 branes between two NS5 branes away from an ON leads to a $\mathrm{U}(k)$ gauge node in the corresponding quiver, as depicted in \eqref{eq:BranesNode}. The presence of an ON plane does not have an effect on the type of gauge groups in the quiver (they remain unitary), but one has to use the following rules to read the shape of the quiver:

\subsection{\texorpdfstring{$\widetilde{\textnormal{ON}}^-$}{ON-t}}
In the presence of an $\widetilde{\textnormal{ON}}^-$ we can study the following brane system.
    \begin{equation}
    	\vcenter
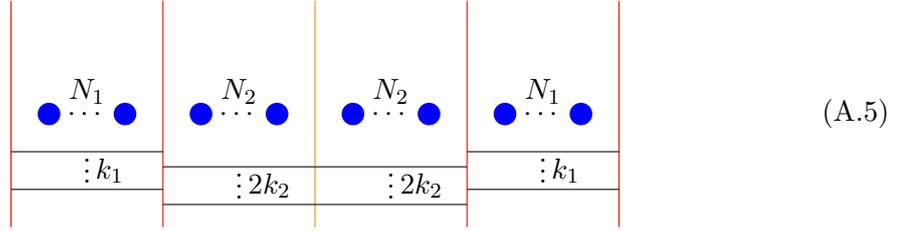
{\hbox{
        \begin{tikzpicture}
            \draw[red] (0,0)--(0,3) (2,0)--(2,3) (6,0)--(6,3) (8,0)--(8,3);
            \draw[orange] (4,0)--(4,3);
            \node[D5] at (0.5,1.5) {};
            \node at (1,1.5) {$\cdots$};
            \node at (1,1.8) {$N_1$};
            \node[D5] at (1.5,1.5) {};
            \node[D5] at (2.5,1.5) {};
            \node at (3,1.5) {$\cdots$};
            \node at (3,1.8) {$N_2$};
            \node[D5] at (3.5,1.5) {};
            \node[D5] at (4.5,1.5) {};
            \node at (5,1.5) {$\cdots$};
            \node at (5,1.8) {$N_2$};
            \node[D5] at (5.5,1.5) {};
            \node[D5] at (6.5,1.5) {};
            \node at (7,1.5) {$\cdots$};
            \node at (7,1.8) {$N_1$};
            \node[D5] at (7.5,1.5) {};
            \draw (0,0.5)--(2,0.5) (0,1)--(2,1);
            \node at (1,0.85) {$\vdots$};
            \node at (1.3,0.75) {$k_1$};
            \draw (2,0.3)--(4,0.3) (2,0.8)--(4,0.8);
            \node at (3,0.65) {$\vdots$};
            \node at (3.4,0.55) {$2k_2$};
            \draw (4,0.3)--(6,0.3) (4,0.8)--(6,0.8);
            \node at (5,0.65) {$\vdots$};
            \node at (5.4,0.55) {$2k_2$};
            \draw (6,0.5)--(8,0.5) (6,1)--(8,1);
            \node at (7,0.85) {$\vdots$};
            \node at (7.3,0.75) {$k_1$};
        \end{tikzpicture}
        }}
    \end{equation}
    We can restrict to the physical system and read the corresponding quiver:
    \begin{equation}
    	\vcenter
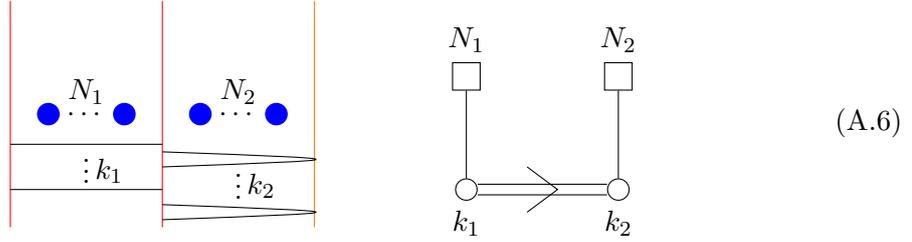
{\hbox{
        \begin{tikzpicture}
            \draw[red] (0,0)--(0,3) (2,0)--(2,3);
            \draw[orange] (4,0)--(4,3);
            \node[D5] at (0.5,1.5) {};
            \node at (1,1.5) {$\cdots$};
            \node at (1,1.8) {$N_1$};
            \node[D5] at (1.5,1.5) {};
            \node[D5] at (2.5,1.5) {};
            \node at (3,1.5) {$\cdots$};
            \node at (3,1.8) {$N_2$};
            \node[D5] at (3.5,1.5) {};
            \draw (0,0.5)--(2,0.5) (0,1.1)--(2,1.1);
            \node at (1,0.85) {$\vdots$};
            \node at (1.3,0.75) {$k_1$};
            \draw (2,0.3) .. controls (4.7,0.2) .. (2,0.1);
            \draw (2,0.8).. controls (4.7,0.9) .. (2,1);
            \node at (3,0.65) {$\vdots$};
            \node at (3.3,0.55) {$k_2$};
            \begin{scope}[shift={(6,0.5)}]
            \node[gauge,label=below:{$k_1$}] (g1) at (0,0) {};
            \node[gauge,label=below:{$k_2$}] (g2) at (2,0) {};
            \node[flavour,label=above:{$N_1$}] (f1) at (0,1.5) {};
            \node[flavour,label=above:{$N_2$}] (f2) at (2,1.5) {};
            \draw (g1)--(f1) (g2)--(f2);
            \draw[transform canvas={yshift=-2pt}] (g1)--(g2);
            \draw[transform canvas={yshift=2pt}] (g1)--(g2);
            \draw (0.8,0.3)--(1.2,0)--(0.8,-0.3);
            \end{scope}
        \end{tikzpicture}
        }}
    \end{equation}

\subsection{\texorpdfstring{$\textnormal{ON}^+$}{ON+}}
In the presence of an $\textnormal{ON}^+$ we can study the following brane system. (D3 branes are drawn away from the origin for clarity)
    \begin{equation}
    	\vcenter
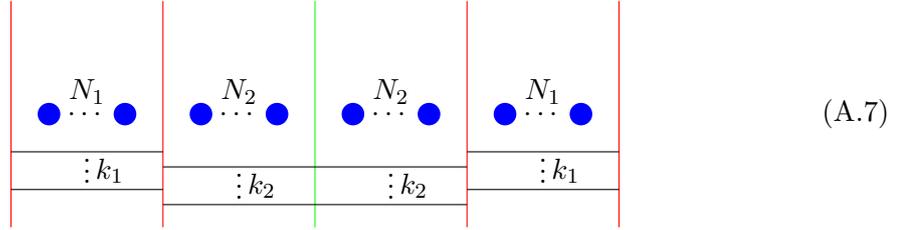
{\hbox{
        \begin{tikzpicture}
            \draw[red] (0,0)--(0,3) (2,0)--(2,3) (6,0)--(6,3) (8,0)--(8,3);
            \draw[green] (4,0)--(4,3);
            \node[D5] at (0.5,1.5) {};
            \node at (1,1.5) {$\cdots$};
            \node at (1,1.8) {$N_1$};
            \node[D5] at (1.5,1.5) {};
            \node[D5] at (2.5,1.5) {};
            \node at (3,1.5) {$\cdots$};
            \node at (3,1.8) {$N_2$};
            \node[D5] at (3.5,1.5) {};
            \node[D5] at (4.5,1.5) {};
            \node at (5,1.5) {$\cdots$};
            \node at (5,1.8) {$N_2$};
            \node[D5] at (5.5,1.5) {};
            \node[D5] at (6.5,1.5) {};
            \node at (7,1.5) {$\cdots$};
            \node at (7,1.8) {$N_1$};
            \node[D5] at (7.5,1.5) {};
            \draw (0,0.5)--(2,0.5) (0,1)--(2,1);
            \node at (1,0.85) {$\vdots$};
            \node at (1.3,0.75) {$k_1$};
            \draw (2,0.3)--(4,0.3) (2,0.8)--(4,0.8);
            \node at (3,0.65) {$\vdots$};
            \node at (3.3,0.55) {$k_2$};
            \draw (4,0.3)--(6,0.3) (4,0.8)--(6,0.8);
            \node at (5,0.65) {$\vdots$};
            \node at (5.3,0.55) {$k_2$};
            \draw (6,0.5)--(8,0.5) (6,1)--(8,1);
            \node at (7,0.85) {$\vdots$};
            \node at (7.3,0.75) {$k_1$};
        \end{tikzpicture}
        }}
    \end{equation}
    We can restrict to the physical system, move onto the Coulomb branch, and read the corresponding quiver:
    \begin{equation}
    	\vcenter
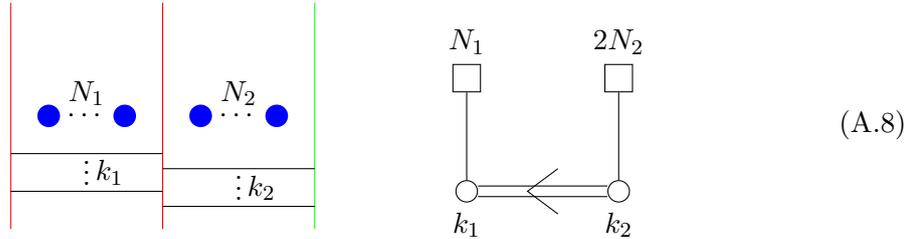
{\hbox{
        \begin{tikzpicture}
            \draw[red] (0,0)--(0,3) (2,0)--(2,3);
            \draw[green] (4,0)--(4,3);
            \node[D5] at (0.5,1.5) {};
            \node at (1,1.5) {$\cdots$};
            \node at (1,1.8) {$N_1$};
            \node[D5] at (1.5,1.5) {};
            \node[D5] at (2.5,1.5) {};
            \node at (3,1.5) {$\cdots$};
            \node at (3,1.8) {$N_2$};
            \node[D5] at (3.5,1.5) {};
            \draw (0,0.5)--(2,0.5) (0,1)--(2,1);
            \node at (1,0.85) {$\vdots$};
            \node at (1.3,0.75) {$k_1$};
            \draw (2,0.3)--(4,0.3) (2,0.8)--(4,0.8);
            \node at (3,0.65) {$\vdots$};
            \node at (3.3,0.55) {$k_2$};
            \begin{scope}[shift={(6,0.5)}]
            \node[gauge,label=below:{$k_1$}] (g1) at (0,0) {};
            \node[gauge,label=below:{$k_2$}] (g2) at (2,0) {};
            \node[flavour,label=above:{$N_1$}] (f1) at (0,1.5) {};
            \node[flavour,label=above:{$2N_2$}] (f2) at (2,1.5) {};
            \draw (g1)--(f1) (g2)--(f2);
            \draw[transform canvas={yshift=-2pt}] (g1)--(g2);
            \draw[transform canvas={yshift=2pt}] (g1)--(g2);
            \draw (1.2,0.3)--(0.8,0)--(1.2,-0.3);
            \end{scope}
        \end{tikzpicture}
        }}
    \end{equation}

\subsection{\texorpdfstring{$\textnormal{ON}^-$}{ON-}}
In the presence of an $\textnormal{ON}^-$ we can study the following brane system. 
    \begin{equation}
    	\vcenter
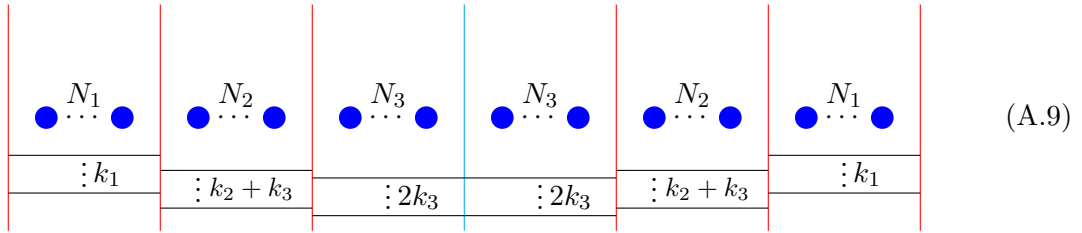
{\hbox{
        \begin{tikzpicture}
            \draw[red] (-2,0)--(-2,3) (0,0)--(0,3) (2,0)--(2,3) (6,0)--(6,3) (8,0)--(8,3) (10,0)--(10,3);
            \draw[cyan] (4,0)--(4,3);
            \node[D5] at (-1.5,1.5) {};
            \node at (-1,1.5) {$\cdots$};
            \node at (-1,1.8) {$N_1$};
            \node[D5] at (-0.5,1.5) {};
            \node[D5] at (0.5,1.5) {};
            \node at (1,1.5) {$\cdots$};
            \node at (1,1.8) {$N_2$};
            \node[D5] at (1.5,1.5) {};
            \node[D5] at (2.5,1.5) {};
            \node at (3,1.5) {$\cdots$};
            \node at (3,1.8) {$N_3$};
            \node[D5] at (3.5,1.5) {};
            \node[D5] at (4.5,1.5) {};
            \node at (5,1.5) {$\cdots$};
            \node at (5,1.8) {$N_3$};
            \node[D5] at (5.5,1.5) {};
            \node[D5] at (6.5,1.5) {};
            \node at (7,1.5) {$\cdots$};
            \node at (7,1.8) {$N_2$};
            \node[D5] at (7.5,1.5) {};
            \node[D5] at (8.5,1.5) {};
            \node at (9,1.5) {$\cdots$};
            \node at (9,1.8) {$N_1$};
            \node[D5] at (9.5,1.5) {};
            \draw (-2,0.5)--(0,0.5) (-2,1)--(0,1);
            \node at (-1,0.85) {$\vdots$};
            \node at (-0.7,0.75) {$k_1$};
            \draw (0,0.3)--(2,0.3) (0,0.8)--(2,0.8);
            \node at (0.5,0.65) {$\vdots$};
            \node at (1.2,0.55) {\small$k_2+k_3$};
            \draw (2,0.2)--(4,0.2) (2,0.7)--(4,0.7);
            \node at (3,0.55) {$\vdots$};
            \node at (3.4,0.45) {$2k_3$};
            \draw (4,0.2)--(6,0.2) (4,0.7)--(6,0.7);
            \node at (5,0.55) {$\vdots$};
            \node at (5.4,0.45) {$2k_3$};
            \draw (6,0.3)--(8,0.3) (6,0.8)--(8,0.8);
            \node at (6.5,0.65) {$\vdots$};
            \node at (7.2,0.55) {\small$k_2+k_3$};
            \draw (8,0.5)--(10,0.5) (8,1)--(10,1);
            \node at (9,0.85) {$\vdots$};
            \node at (9.3,0.75) {$k_1$};
        \end{tikzpicture}
        }}
    \end{equation}
    In order to move onto the Coulomb branch and read a quiver from the brane system, one has to be careful. For a general system, where $N_3>0$, there will be several quivers one can read, depending on boundary conditions. This is investigated in detail in \cite{Ncones} and we only state the possibilities here. The first option is the most natural one, identified in the literature before \cite{Ferlito:2016grh}:
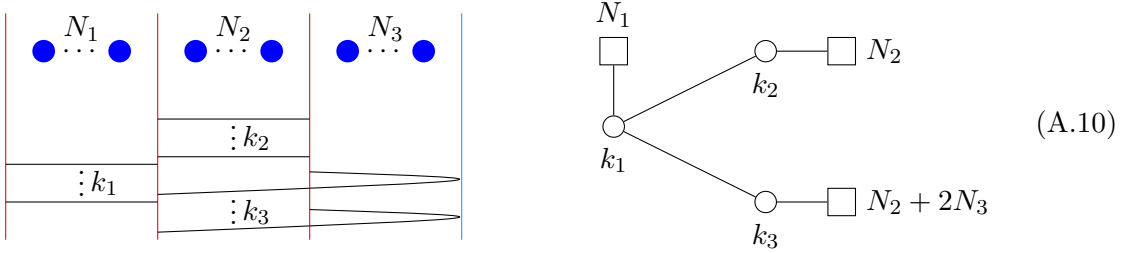
    \begin{equation}
    	\vcenter{\hbox{
        \begin{tikzpicture}
            \draw[red] (-2,0)--(-2,3) (0,0)--(0,3) (2,0)--(2,3);
            \draw[cyan] (4,0)--(4,3);
            \node[D5] at (-1.5,2.5) {};
            \node at (-1,2.5) {$\cdots$};
            \node at (-1,2.8) {$N_1$};
            \node[D5] at (-0.5,2.5) {};
            \node[D5] at (0.5,2.5) {};
            \node at (1,2.5) {$\cdots$};
            \node at (1,2.8) {$N_2$};
            \node[D5] at (1.5,2.5) {};
            \node[D5] at (2.5,2.5) {};
            \node at (3,2.5) {$\cdots$};
            \node at (3,2.8) {$N_3$};
            \node[D5] at (3.5,2.5) {};
            \draw (-2,0.5)--(0,0.5) (-2,1)--(0,1);
            \node at (-1,0.85) {$\vdots$};
            \node at (-0.7,0.75) {$k_1$};
            \draw (0,1.1)--(2,1.1) (0,1.6)--(2,1.6);
            \node at (1,1.45) {$\vdots$};
            \node at (1.3,1.35) {$k_2$};
            \draw (2,0.9) .. controls (4.9,0.8) .. (0,0.6);
            \draw (2,0.4) .. controls (4.9,0.3) .. (0,0.1);
            \node at (1,0.47) {$\vdots$};
            \node at (1.3,0.37) {$k_3$};
            \begin{scope}[shift={(6,1.5)}]
            \node[gauge,label=below:{$k_1$}] (g1) at (0,0) {};
            \node[gauge,label=below:{$k_2$}] (g2) at (2,1) {};
            \node[gauge,label=below:{$k_3$}] (g3) at (2,-1) {};
            \node[flavour,label=above:{$N_1$}] (f1) at (0,1) {};
            \node[flavour,label=right:{$N_2$}] (f2) at (3,1) {};
            \node[flavour,label=right:{$N_2+2N_3$}] (f3) at (3,-1) {};
            \draw (f3)--(g3)--(g1)--(g2)--(f2) (g1)--(f1);
            \end{scope}
        \end{tikzpicture}
        }}
    \end{equation}
    The second option is obtained, when one of the $2k_3$ D3 branes crossing the $\textnormal{ON}^-$ ends on one of the $N_3$ D5 branes instead of NS5 branes. This leads to a different possibility of breaking the D3 branes along the NS5 branes, and provides a different electric quiver (The horizontal separation of D5 branes is only for ease of reading):
    
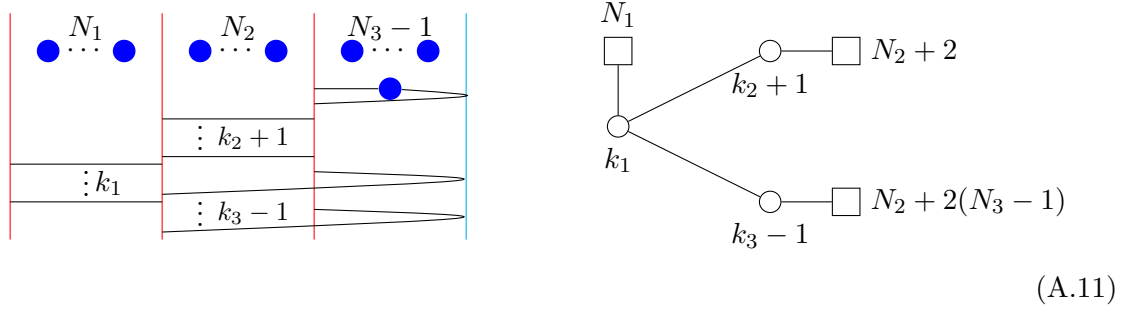
\begin{equation}
        \begin{tikzpicture}
            \draw[red] (-2,0)--(-2,3) (0,0)--(0,3) (2,0)--(2,3);
            \draw[cyan] (4,0)--(4,3);
            \node[D5] at (-1.5,2.5) {};
            \node at (-1,2.5) {$\cdots$};
            \node at (-1,2.8) {$N_1$};
            \node[D5] at (-0.5,2.5) {};
            \node[D5] at (0.5,2.5) {};
            \node at (1,2.5) {$\cdots$};
            \node at (1,2.8) {$N_2$};
            \node[D5] at (1.5,2.5) {};
            \node[D5] at (2.5,2.5) {};
            \node at (3,2.5) {$\cdots$};
            \node at (3,2.8) {$N_3-1$};
            \node[D5] at (3.5,2.5) {};
            \draw (-2,0.5)--(0,0.5) (-2,1)--(0,1);
            \node at (-1,0.85) {$\vdots$};
            \node at (-0.7,0.75) {$k_1$};
            \draw (0,1.1)--(2,1.1) (0,1.6)--(2,1.6);
            \node at (0.5,1.45) {$\vdots$};
            \node at (1.2,1.35) {\small$k_2+1$};
            \draw (2,0.9) .. controls (4.9,0.8) .. (0,0.6);
            \draw (2,0.4) .. controls (4.9,0.3) .. (0,0.1);
            \node at (0.5,0.45) {$\vdots$};
            \node at (1.2,0.36) {\small$k_3-1$};
            \node[D5] (1) at (3,2) {};
            \draw (2,2)--(1);
            \draw (1) .. controls (4.45,1.9) .. (2,1.8);
            \begin{scope}[shift={(6,1.5)}]
            \node[gauge,label=below:{$k_1$}] (g1) at (0,0) {};
            \node[gauge,label=below:{$k_2+1$}] (g2) at (2,1) {};
            \node[gauge,label=below:{$k_3-1$}] (g3) at (2,-1) {};
            \node[flavour,label=above:{$N_1$}] (f1) at (0,1) {};
            \node[flavour,label=right:{$N_2+2$}] (f2) at (3,1) {};
            \node[flavour,label=right:{$N_2+2(N_3-1)$}] (f3) at (3,-1) {};
            \draw (f3)--(g3)--(g1)--(g2)--(f2) (g1)--(f1);
            \end{scope}
        \end{tikzpicture}
    \end{equation}
    One can keep going in this way. Overall there are $\textnormal{min}\{N_3+1,k_3+1\}$ possibilities, labelled by $l\in\{0,\dots,\textnormal{min}\{N_3,k_3\}\}$:
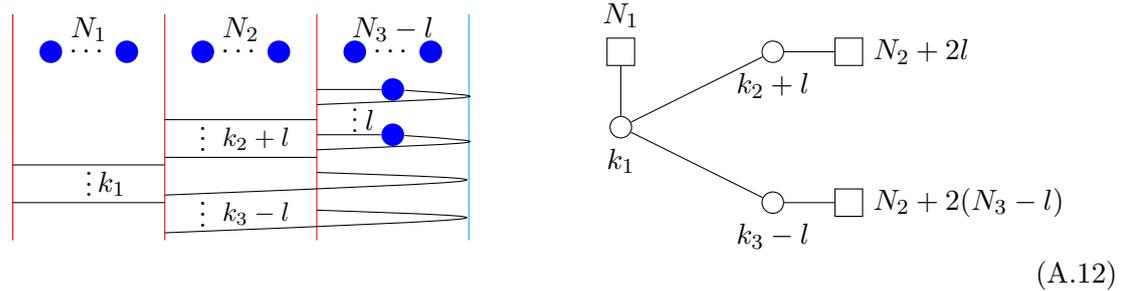
    \begin{equation}
    	\vcenter{\hbox{
        \begin{tikzpicture}
            \draw[red] (-2,0)--(-2,3) (0,0)--(0,3) (2,0)--(2,3);
            \draw[cyan] (4,0)--(4,3);
            \node[D5] at (-1.5,2.5) {};
            \node at (-1,2.5) {$\cdots$};
            \node at (-1,2.8) {$N_1$};
            \node[D5] at (-0.5,2.5) {};
            \node[D5] at (0.5,2.5) {};
            \node at (1,2.5) {$\cdots$};
            \node at (1,2.8) {$N_2$};
            \node[D5] at (1.5,2.5) {};
            \node[D5] at (2.5,2.5) {};
            \node at (3,2.5) {$\cdots$};
            \node at (3,2.8) {$N_3-l$};
            \node[D5] at (3.5,2.5) {};
            \draw (-2,0.5)--(0,0.5) (-2,1)--(0,1);
            \node at (-1,0.85) {$\vdots$};
            \node at (-0.7,0.75) {$k_1$};
            \draw (0,1.1)--(2,1.1) (0,1.6)--(2,1.6);
            \node at (0.5,1.45) {$\vdots$};
            \node at (1.2,1.35) {\small$k_2+l$};
            \draw (2,0.9) .. controls (4.9,0.8) .. (0,0.6);
            \draw (2,0.4) .. controls (4.9,0.3) .. (0,0.1);
            \node at (0.5,0.45) {$\vdots$};
            \node at (1.2,0.36) {\small$k_3-l$};
            \node[D5] (1) at (3,2) {};
            \draw (2,2)--(1);
            \draw (1) .. controls (4.45,1.9) .. (2,1.8);
            \node[D5] (2) at (3,1.4) {};
            \draw (2,1.4)--(2);
            \draw (2) .. controls (4.45,1.3) .. (2,1.2);
            \node at (2.5,1.7) {\small$\vdots$};
            \node at (2.7,1.6) {$l$};
            \begin{scope}[shift={(6,1.5)}]
            \node[gauge,label=below:{$k_1$}] (g1) at (0,0) {};
            \node[gauge,label=below:{$k_2+l$}] (g2) at (2,1) {};
            \node[gauge,label=below:{$k_3-l$}] (g3) at (2,-1) {};
            \node[flavour,label=above:{$N_1$}] (f1) at (0,1) {};
            \node[flavour,label=right:{$N_2+2l$}] (f2) at (3,1) {};
            \node[flavour,label=right:{$N_2+2(N_3-l)$}] (f3) at (3,-1) {};
            \draw (f3)--(g3)--(g1)--(g2)--(f2) (g1)--(f1);
            \end{scope}
        \end{tikzpicture}
        }}
        \label{eq:Dgeneral}
    \end{equation}
Note that in order to transition between two leaves, whose closure is described by two quivers with different $l$ in \eqref{eq:Dgeneral}, one needs to move onto a lower leaf which appears in the stratification of both spaces.

\subsection{Leaf closures and transverse slices in brane systems}
In this section we investigate how to not only to identify the low energy theory living on D3 branes in a brane system, but how to analyse the stratification of its moduli space using branes. For a given brane system there are various distinct phases, which correspond to the symplectic leaves that make up its moduli space. In the following we will only consider Coulomb branch phases. For an analysis of the full moduli space, including a brane perspective, see \cite{Grimminger:2020dmg}; we repeat what is needed for this paper in the following. Let $\mathcal{C}$ be the Coulomb branch of the low energy theory living on the brane system. For every symplectic leaf $\mathcal{L}\subset\mathcal{C}$ there is a Coulomb phase $\mathcal{P}$ in the brane system. From the phase $\mathcal{P}$ we can read a quiver $\mathsf{Q}$, whose Coulomb branch $\mathcal{C}(\mathsf{Q})=\overline{\mathcal{L}}$ is the closure of the symplectic leaf $\mathcal{L}$ corresponding to the phase $\mathcal{P}$. The phases of the brane system are distinguished, by how many branes coincide at the origin, i.e.\ how many massless states are present in the field theory when on the leaf $\mathcal{L}$ on its Coulomb branch. Let us look at the example of $\mathrm{U}(2)$ with $4$ fundamental hypers. The Hasse diagram of its Coulomb branch is straight forwardly computed, e.g.\ through quiver subtraction:
\begin{equation}
    \begin{tikzpicture}
        \node[bd,label=right:{$\mathcal{L}_a$}] (1) at (0,0) {};
        \node[bd,label=right:{$\mathcal{L}_b$}] (2) at (0,1) {};
        \node[bd,label=right:{$\mathcal{L}_c$}] (3) at (0,2) {};
        \draw (1)--(2)--(3);
        \node at (-0.3,0.5) {$A_1$};
        \node at (-0.3,1.5) {$A_3$};
        \node at (2,1) {for};
        \node at (4,1) {$\mathcal{C}\left(\vcenter{\hbox{\begin{tikzpicture}
        \node[gauge,label=below:{$2$}] (g) at (0,0.5) {};
        \node[flavour,label=above:{$4$}] (f) at (0,1.5) {};
        \draw (g)--(f);
        \end{tikzpicture}}}\right)$};
    \end{tikzpicture}
\end{equation}
Where we labelled the leaves by $\mathcal{L}_i$, $i=a,b,c$. The three distinct Coulomb phases of the brane system are:
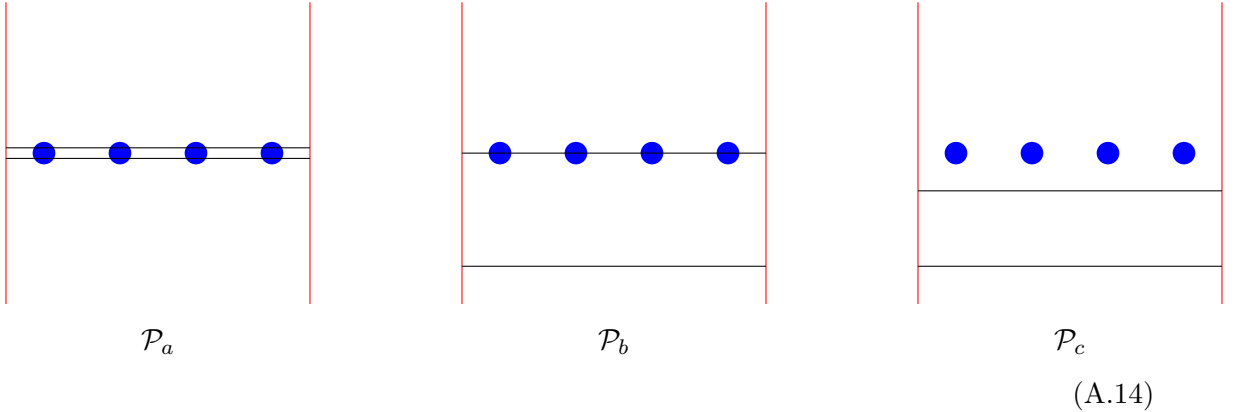
\begin{equation}
    \begin{tikzpicture}
        \draw[red] (0,0)--(0,4) (4,0)--(4,4);
        \node[D5] at (0.5,2) {};
        \node[D5] at (1.5,2) {};
        \node[D5] at (2.5,2) {};
        \node[D5] at (3.5,2) {};
        \draw[transform canvas={yshift=-2pt}] (0,2)--(4,2);
        \draw[transform canvas={yshift=2pt}] (0,2)--(4,2);
        \begin{scope}[shift={(6,0)}]
        \draw[red] (0,0)--(0,4) (4,0)--(4,4);
        \node[D5] at (0.5,2) {};
        \node[D5] at (1.5,2) {};
        \node[D5] at (2.5,2) {};
        \node[D5] at (3.5,2) {};
        \draw (0,2)--(4,2);
        \draw (0,0.5)--(4,0.5);
        \end{scope}
        \begin{scope}[shift={(12,0)}]
        \draw[red] (0,0)--(0,4) (4,0)--(4,4);
        \node[D5] at (0.5,2) {};
        \node[D5] at (1.5,2) {};
        \node[D5] at (2.5,2) {};
        \node[D5] at (3.5,2) {};
        \draw (0,1.5)--(4,1.5);
        \draw (0,0.5)--(4,0.5);
        \end{scope}
        \node at (2,-0.5) {$\mathcal{P}_a$};
        \node at (8,-0.5) {$\mathcal{P}_b$};
        \node at (14,-0.5) {$\mathcal{P}_c$};
    \end{tikzpicture}
\end{equation}
The quiver for the most general Coulomb phase, $\mathcal{P}_c$, is easily read off. It is the theory itself:
\begin{equation}
    \begin{tikzpicture}
        \node at (-2,0.5) {$\mathsf{Q}_c=$};
        \node[gauge,label=below:{$2$}] (1) at (0,0) {};
        \node[flavour,label=above:{$4$}] (2) at (0,1) {};
        \draw (1)--(2);
    \end{tikzpicture}
\end{equation}
The quiver not only describes the phase $\mathcal{P}_c$, but also the possibility of the D3 branes moving to the origin. It is a quantum field theory in its own right with a moduli space consisting of several leaves. Hence its Coulomb branch is not the leaf $\mathcal{L}_c$ itself, but the closure of the leaf $\overline{\mathcal{L}}_c$, or in other words, the transverse slice from the origin, $\mathcal{L}_a$, to the leaf $\mathcal{L}_c$.\\

The quiver for the phase $\mathcal{P}_b$ is more tricky to work out. There is one D3 brane resting at the origin. We first have to make this D3 brane end on D5 branes rather than NS5 branes. This can be done through a Hanany-Witten transition. The resulting brane system is:
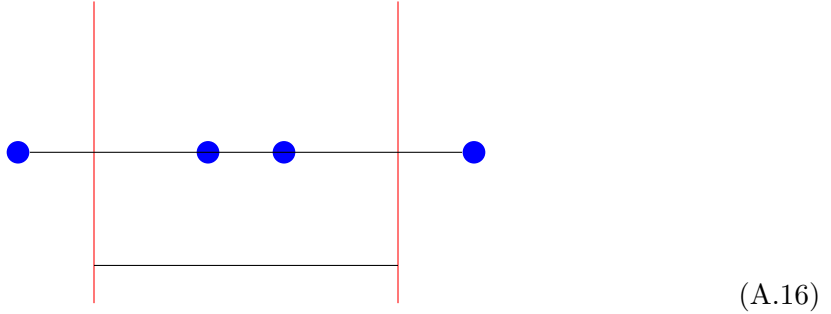
\begin{equation}
    \begin{tikzpicture}
        \draw[red] (0,0)--(0,4) (4,0)--(4,4);
        \node[D5] (1) at (-1,2) {};
        \node[D5] at (1.5,2) {};
        \node[D5] at (2.5,2) {};
        \node[D5] (2) at (5,2) {};
        \draw (1)--(2);
        \draw (0,0.5)--(4,0.5);
    \end{tikzpicture}
\end{equation}
Now we have to ignore the possibility of the D3 brane, which is supposed to rest at the origin, to move along the NS5 branes. We do this by simply not drawing it:
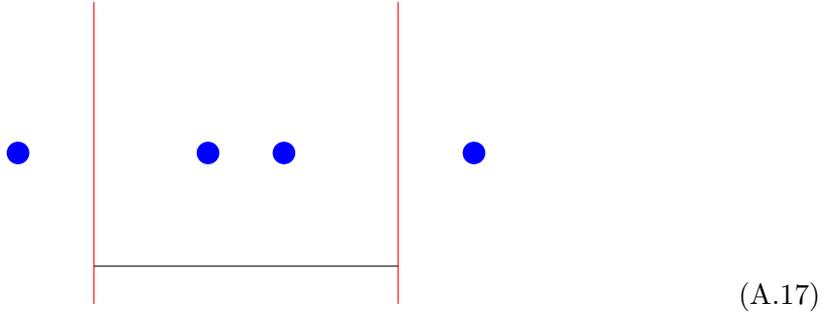
\begin{equation}
    \begin{tikzpicture}
        \draw[red] (0,0)--(0,4) (4,0)--(4,4);
        \node[D5] (1) at (-1,2) {};
        \node[D5] at (1.5,2) {};
        \node[D5] at (2.5,2) {};
        \node[D5] (2) at (5,2) {};
        \draw (0,0.5)--(4,0.5);
    \end{tikzpicture}
\end{equation}
Now we can read a quiver from this brane system:
\begin{equation}
    \begin{tikzpicture}
        \node at (-2,0.5) {$\mathsf{Q}_b=$};
        \node[gauge,label=below:{$1$}] (1) at (0,0) {};
        \node[flavour,label=above:{$2$}] (2) at (0,1) {};
        \draw (1)--(2);
    \end{tikzpicture}
\end{equation}
The Coulomb branch of $\mathsf{Q}_b$ is the closure $\overline{\mathcal{L}}_b$.\\

The quiver associated to $\mathcal{P}_a$, the origin, is trivial:
\begin{equation}
    \begin{tikzpicture}
        \node at (-2,0.5) {$\mathsf{Q}_a=$};
        \node[gauge,label=below:{$0$}] (1) at (0,0) {};
        \node[flavour,label=above:{$0$}] (2) at (0,1) {};
        \draw (1)--(2);
    \end{tikzpicture}
\end{equation}
The quivers we have computed represent closures of the leaves. This corresponds to transverse slices from the origin $\mathcal{L}_a$ to the leaf $\mathcal{L}_i$ in question. We can also obtain a quiver for any other transverse slice. The remaining non-trivial slice for our theory is the one from $\mathcal{L}_b$ to $\mathcal{L}_c$, we refer to the corresponding quiver as $\mathsf{Q}_{b \rightarrow c}$. We can obtain the quiver from the brane system in the following way: Comparing the two phases $\mathcal{P}_b$ and $\mathcal{P}_c$ we have to ignore the modulus which is turned on in both phases. We achieve this by simply not drawing the D3 brane in question; one can think of it as sending the D3 brane off to infinity \footnote{We have two instances of ignoring a D3 brane by not drawing it: Either fixing it to remain at the origin, or sending it to infinity. When computing the quiver associated to the transition between two non-trivial phases of the brane system, one in general has to use both.}. The resulting brane system is:
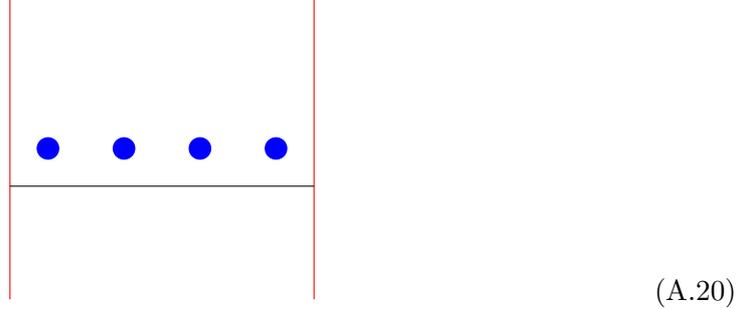
\begin{equation}
    \begin{tikzpicture}
        \draw[red] (0,0)--(0,4) (4,0)--(4,4);
        \node[D5] at (0.5,2) {};
        \node[D5] at (1.5,2) {};
        \node[D5] at (2.5,2) {};
        \node[D5] at (3.5,2) {};
        \draw (0,1.5)--(4,1.5);
    \end{tikzpicture}
\end{equation}
From which a quiver is easily read off:
\begin{equation}
    \begin{tikzpicture}
        \node at (-2,0.5) {$\mathsf{Q}_{b \rightarrow c}=$};
        \node[gauge,label=below:{$1$}] (1) at (0,0) {};
        \node[flavour,label=above:{$4$}] (2) at (0,1) {};
        \draw (1)--(2);
    \end{tikzpicture}
\end{equation}
The Coulomb branch of this quiver is the transverse slice from $\mathcal{L}_b$ to $\mathcal{L}_c$.\\

Of course none of these quivers are new, as they are exactly the quivers involved in quiver subtraction. The point of this exercise was to show, how all transverse slices of the Coulomb branch show up as transitions in the brane system.\\

In Section \ref{sec:branes} we propose brane systems whose Coulomb branch is the transverse slice to the lowest leaf in a connected component of the affine Grassmannian of a classical group. This space is infinite dimensional reflected in the fact that we have an infinite amount of D3 branes present in the brane system. We fix all but a finite amount of D3 branes to be at the origin, in order to study the symplectic leaves the space is made up of from bottom up. For every such leaf we look at the corresponding phase in the brane system and obtain the quiver whose Coulomb branch is the closure of the leaf. Furthermore one can associate a quiver to a transition between any two phases in the brane system. Every good quiver in shape of a classical Dynkin diagram appears this way in one of the brane systems.

\bibliographystyle{JHEP}
\bibliography{bibli.bib}

\providecommand{\href}[2]{#2}\begingroup\raggedright\begin{thebibliography}{10}

\bibitem{beauville1994conformal}
A.~Beauville and Y.~Laszlo, \emph{Conformal blocks and generalized theta
  functions}, {\emph{Communications in mathematical physics} {\bfseries 164}
  (1994) 385--419}.

\bibitem{kumar1994infinite}
S.~Kumar, M.~Narasimhan and A.~Ramanathan, \emph{Infinite grassmannians and
  moduli spaces}, {\emph{Math. Ann} {\bfseries 300} (1994) 41--75}.

\bibitem{beauville1998picard}
A.~Beauville, Y.~Laszio and C.~Sorger, \emph{The picard group of the moduli
  of-bundles on a curve}, {\emph{Compositio Mathematica} {\bfseries 112} (1998)
  183--216}.

\bibitem{pappas2008twisted}
G.~Pappas and M.~Rapoport, \emph{Twisted loop groups and their affine flag
  varieties}, {\emph{Advances in Mathematics} {\bfseries 219} (2008) 118--198}.

\bibitem{sorgerlectures}
C.~Sorger, \emph{Lectures on moduli of principal g-bundles over algebraic
  curves}, {\emph{ICTP Lecture Notes} 3}.

\bibitem{beilinson1991quantization}
A.~Beilinson and V.~Drinfeld, ``Quantization of hitchin’s integrable system
  and hecke eigensheaves.''

\bibitem{lusztig1983singularities}
G.~Lusztig, \emph{Singularities, character formulas, and a q-analog of weight
  multiplicities}, {\emph{Ast{\'e}risque} {\bfseries 101} (1983) 208--229}.

\bibitem{ginzburg1995perverse}
V.~Ginzburg, \emph{Perverse sheaves on a loop group and langlands' duality},
  {\emph{arXiv preprint alg-geom/9511007} (1995) }.

\bibitem{mirkovic2007geometric}
I.~Mirkovi{\'c} and K.~Vilonen, \emph{Geometric langlands duality and
  representations of algebraic groups over commutative rings}, {\emph{Annals of
  mathematics} (2007) 95--143}.

\bibitem{Hanany:1996ie}
A.~Hanany and E.~Witten, \emph{{Type IIB superstrings, BPS monopoles, and
  three-dimensional gauge dynamics}},
  \href{http://dx.doi.org/10.1016/S0550-3213(97)00157-0,
  10.1016/S0550-3213(97)80030-2}{\emph{Nucl. Phys.} {\bfseries B492} (1997)
  152--190}, [\href{https://arxiv.org/abs/hep-th/9611230}{{\ttfamily
  hep-th/9611230}}].

\bibitem{2003math:5095M}
A.~{Malkin}, V.~{Ostrik} and M.~{Vybornov}, \emph{{The minimal degeneration
  singularities in the affine Grassmannians}}, {\emph{arXiv Mathematics
  e-prints} (May, 2003) math/0305095},
  [\href{https://arxiv.org/abs/math/0305095}{{\ttfamily math/0305095}}].

\bibitem{kamnitzer2014yangians}
J.~Kamnitzer, B.~Webster, A.~Weekes and O.~Yacobi, \emph{Yangians and
  quantizations of slices in the affine grassmannian}, {\emph{Algebra \& Number
  Theory} {\bfseries 8} (2014) 857--893}.

\bibitem{Braverman:2016pwk}
A.~Braverman, M.~Finkelberg and H.~Nakajima, \emph{{Coulomb branches of $3d$
  $\mathcal{N}=4$ quiver gauge theories and slices in the affine
  Grassmannian}},
  \href{http://dx.doi.org/10.4310/ATMP.2019.v23.n1.a3}{\emph{Adv. Theor. Math.
  Phys.} {\bfseries 23} (2019) 75--166},
  [\href{https://arxiv.org/abs/1604.03625}{{\ttfamily 1604.03625}}].

\bibitem{kraft1980minimal}
H.~Kraft and C.~Procesi, \emph{Minimal singularities in ${GL}_n$},
  {\emph{Inventiones mathematicae} {\bfseries 62} (1980) 503--515}.

\bibitem{Kraft1982}
H.~Kraft and C.~Procesi, \emph{On the geometry of conjugacy classes in
  classical groups},
  \href{http://dx.doi.org/10.1007/BF02565876}{\emph{Commentarii Mathematici
  Helvetici} {\bfseries 57} (Dec, 1982) 539--602}.

\bibitem{Cabrera:2016vvv}
S.~Cabrera and A.~Hanany, \emph{{Branes and the Kraft-Procesi Transition}},
  \href{http://dx.doi.org/10.1007/JHEP11(2016)175}{\emph{JHEP} {\bfseries 11}
  (2016) 175}, [\href{https://arxiv.org/abs/1609.07798}{{\ttfamily
  1609.07798}}].

\bibitem{Cabrera:2017njm}
S.~Cabrera and A.~Hanany, \emph{{Branes and the Kraft-Procesi transition:
  classical case}},
  \href{http://dx.doi.org/10.1007/JHEP04(2018)127}{\emph{JHEP} {\bfseries 04}
  (2018) 127}, [\href{https://arxiv.org/abs/1711.02378}{{\ttfamily
  1711.02378}}].

\bibitem{Chalmers:1996xh}
G.~Chalmers and A.~Hanany, \emph{{Three-dimensional gauge theories and
  monopoles}},
  \href{http://dx.doi.org/10.1016/S0550-3213(97)00036-9}{\emph{Nucl. Phys. B}
  {\bfseries 489} (1997) 223--244},
  [\href{https://arxiv.org/abs/hep-th/9608105}{{\ttfamily hep-th/9608105}}].

\bibitem{deBoer:1996mp}
J.~de~Boer, K.~Hori, H.~Ooguri and Y.~Oz, \emph{{Mirror symmetry in
  three-dimensional gauge theories, quivers and D-branes}},
  \href{http://dx.doi.org/10.1016/S0550-3213(97)00125-9}{\emph{Nucl. Phys. B}
  {\bfseries 493} (1997) 101--147},
  [\href{https://arxiv.org/abs/hep-th/9611063}{{\ttfamily hep-th/9611063}}].

\bibitem{deBoer:1996ck}
J.~de~Boer, K.~Hori, H.~Ooguri, Y.~Oz and Z.~Yin, \emph{{Mirror symmetry in
  three-dimensional theories, SL(2,Z) and D-brane moduli spaces}},
  \href{http://dx.doi.org/10.1016/S0550-3213(97)00115-6}{\emph{Nucl. Phys. B}
  {\bfseries 493} (1997) 148--176},
  [\href{https://arxiv.org/abs/hep-th/9612131}{{\ttfamily hep-th/9612131}}].

\bibitem{Cherkis:1997aa}
S.~A. Cherkis and A.~Kapustin, \emph{{Singular monopoles and supersymmetric
  gauge theories in three-dimensions}},
  \href{http://dx.doi.org/10.1016/S0550-3213(98)00341-1}{\emph{Nucl. Phys. B}
  {\bfseries 525} (1998) 215--234},
  [\href{https://arxiv.org/abs/hep-th/9711145}{{\ttfamily hep-th/9711145}}].

\bibitem{Tong:1998fa}
D.~Tong, \emph{{Three-dimensional gauge theories and ADE monopoles}},
  \href{http://dx.doi.org/10.1016/S0370-2693(98)01583-4}{\emph{Phys. Lett. B}
  {\bfseries 448} (1999) 33--36},
  [\href{https://arxiv.org/abs/hep-th/9803148}{{\ttfamily hep-th/9803148}}].

\bibitem{Bullimore:2015lsa}
M.~Bullimore, T.~Dimofte and D.~Gaiotto, \emph{{The Coulomb Branch of 3d
  ${\mathcal{N}= 4}$ Theories}},
  \href{http://dx.doi.org/10.1007/s00220-017-2903-0}{\emph{Commun. Math. Phys.}
  {\bfseries 354} (2017) 671--751},
  [\href{https://arxiv.org/abs/1503.04817}{{\ttfamily 1503.04817}}].

\bibitem{finkelberg2017double}
M.~Finkelberg, \emph{Double affine grassmannians and coulomb branches of 3d n=
  4 quiver gauge theories},  in \emph{Proceedings of the International Congress
  of Mathematicians}, World Scientific, 2017.
\newblock \href{http://dx.doi.org/10.1142/9789813272880_0097}{DOI}.

\bibitem{Nakajima:2019olw}
H.~Nakajima and A.~Weekes, \emph{{Coulomb branches of quiver gauge theories
  with symmetrizers}},  \href{https://arxiv.org/abs/1907.06552}{{\ttfamily
  1907.06552}}.

\bibitem{braverman2010pursuing}
A.~Braverman, M.~Finkelberg et~al., \emph{Pursuing the double affine
  grassmannian, i: Transversal slices via instantons on $ a\_k
  $-singularities},
  \href{http://dx.doi.org/10.1215/00127094-2010-011}{\emph{Duke Mathematical
  Journal} {\bfseries 152} (2010) 175--206}.

\bibitem{Cremonesi:2013lqa}
S.~Cremonesi, A.~Hanany and A.~Zaffaroni, \emph{{Monopole operators and Hilbert
  series of Coulomb branches of $3d$ $\mathcal{N} = 4$ gauge theories}},
  \href{http://dx.doi.org/10.1007/JHEP01(2014)005}{\emph{JHEP} {\bfseries 01}
  (2014) 005}, [\href{https://arxiv.org/abs/1309.2657}{{\ttfamily 1309.2657}}].

\bibitem{Nakajima:2015txa}
H.~Nakajima, \emph{{Towards a mathematical definition of Coulomb branches of
  $3$-dimensional $\mathcal{N}=4$ gauge theories, I}},
  \href{http://dx.doi.org/10.4310/ATMP.2016.v20.n3.a4}{\emph{Adv. Theor. Math.
  Phys.} {\bfseries 20} (2016) 595--669},
  [\href{https://arxiv.org/abs/1503.03676}{{\ttfamily 1503.03676}}].

\bibitem{Nakajima2015}
H.~{Nakajima}, \emph{{Questions on provisional Coulomb branches of
  $3$-dimensional $\mathcal N=4$ gauge theories}}, {\emph{arXiv e-prints} (Oct,
  2015) arXiv:1510.03908}, [\href{https://arxiv.org/abs/1510.03908}{{\ttfamily
  1510.03908}}].

\bibitem{Braverman:2016wma}
A.~Braverman, M.~Finkelberg and H.~Nakajima, \emph{{Towards a mathematical
  definition of Coulomb branches of $3$-dimensional $\mathcal{N} = 4$ gauge
  theories, II}},
  \href{http://dx.doi.org/10.4310/ATMP.2018.v22.n5.a1}{\emph{Adv. Theor. Math.
  Phys.} {\bfseries 22} (2018) 1071--1147},
  [\href{https://arxiv.org/abs/1601.03586}{{\ttfamily 1601.03586}}].

\bibitem{Nakajima:2017bdt}
H.~Nakajima, \emph{{Introduction to a provisional mathematical definition of
  Coulomb branches of $3$-dimensional $\mathcal N=4$ gauge theories}},
  \href{https://arxiv.org/abs/1706.05154}{{\ttfamily 1706.05154}}.

\bibitem{kamnitzer2018reducedness}
J.~Kamnitzer, D.~Muthiah, A.~Weekes and O.~Yacobi, \emph{Reducedness of affine
  grassmannian slices in type a}, {\emph{Proceedings of the American
  Mathematical Society} {\bfseries 146} (2018) 861--874}.

\bibitem{muthiah2019symplectic}
D.~Muthiah and A.~Weekes, \emph{Symplectic leaves for generalized affine
  grassmannian slices}, {\emph{arXiv preprint arXiv:1902.09771} (2019) }.

\bibitem{weekes2019generators}
A.~Weekes, \emph{Generators for coulomb branches of quiver gauge theories},
  {\emph{arXiv preprint arXiv:1903.07734} (2019) }.

\bibitem{kamnitzer2019category}
J.~Kamnitzer, P.~Tingley, B.~Webster, A.~Weekes and O.~Yacobi, \emph{On
  category o for affine grassmannian slices and categorified tensor products},
  {\emph{Proceedings of the London Mathematical Society} {\bfseries 119} (2019)
  1179--1233}.

\bibitem{Bourget:2019aer}
A.~Bourget, S.~Cabrera, J.~F. Grimminger, A.~Hanany, M.~Sperling, A.~Zajac
  et~al., \emph{{The Higgs mechanism \textemdash{} Hasse diagrams for
  symplectic singularities}},
  \href{http://dx.doi.org/10.1007/JHEP01(2020)157}{\emph{JHEP} {\bfseries 01}
  (2020) 157}, [\href{https://arxiv.org/abs/1908.04245}{{\ttfamily
  1908.04245}}].

\bibitem{beauville2000symplectic}
A.~Beauville, \emph{Symplectic singularities}, {\emph{Inventiones mathematicae}
  {\bfseries 139} (2000) 541--549}.

\bibitem{stembridge1998partial}
J.~R. Stembridge, \emph{The partial order of dominant weights}, {\emph{Advances
  in Mathematics} {\bfseries 136} (1998) 340--364}.

\bibitem{Ito:2011ea}
Y.~Ito, T.~Okuda and M.~Taki, \emph{{Line operators on $S^1 \times \mathbb
  {R}^3$ and quantization of the Hitchin moduli space}},
  \href{http://dx.doi.org/10.1007/JHEP03(2016)085}{\emph{JHEP} {\bfseries 04}
  (2012) 010}, [\href{https://arxiv.org/abs/1111.4221}{{\ttfamily 1111.4221}}].

\bibitem{Gomis:2011pf}
J.~Gomis, T.~Okuda and V.~Pestun, \emph{{Exact Results for 't Hooft Loops in
  Gauge Theories on $S^4$}},
  \href{http://dx.doi.org/10.1007/JHEP05(2012)141}{\emph{JHEP} {\bfseries 05}
  (2012) 141}, [\href{https://arxiv.org/abs/1105.2568}{{\ttfamily 1105.2568}}].

\bibitem{Dedushenko:2017avn}
M.~Dedushenko, Y.~Fan, S.~S. Pufu and R.~Yacoby, \emph{{Coulomb Branch
  Operators and Mirror Symmetry in Three Dimensions}},
  \href{http://dx.doi.org/10.1007/JHEP04(2018)037}{\emph{JHEP} {\bfseries 04}
  (2018) 037}, [\href{https://arxiv.org/abs/1712.09384}{{\ttfamily
  1712.09384}}].

\bibitem{Dedushenko:2018icp}
M.~Dedushenko, Y.~Fan, S.~S. Pufu and R.~Yacoby, \emph{{Coulomb Branch
  Quantization and Abelianized Monopole Bubbling}},
  \href{http://dx.doi.org/10.1007/JHEP10(2019)179}{\emph{JHEP} {\bfseries 10}
  (2019) 179}, [\href{https://arxiv.org/abs/1812.08788}{{\ttfamily
  1812.08788}}].

\bibitem{Assel:2019iae}
B.~Assel and A.~Sciarappa, \emph{{On monopole bubbling contributions to
  \textquoteright{}t Hooft loops}},
  \href{http://dx.doi.org/10.1007/JHEP05(2019)180}{\emph{JHEP} {\bfseries 05}
  (2019) 180}, [\href{https://arxiv.org/abs/1903.00376}{{\ttfamily
  1903.00376}}].

\bibitem{Assel:2019yzd}
B.~Assel, S.~Cremonesi and M.~Renwick, \emph{{Quantized Coulomb branches,
  monopole bubbling and wall-crossing phenomena in 3d $ \mathcal{N} $ = 4
  theories}}, \href{http://dx.doi.org/10.1007/JHEP04(2020)213}{\emph{JHEP}
  {\bfseries 04} (2020) 213},
  [\href{https://arxiv.org/abs/1910.01650}{{\ttfamily 1910.01650}}].

\bibitem{segal1985loop}
G.~Segal, \emph{Loop groups},  in \emph{Arbeitstagung Bonn 1984}, pp.~155--168.
\newblock Springer, 1985.

\bibitem{gortz2010affine}
U.~G{\"o}rtz, \emph{Affine springer fibers and affine deligne-lusztig
  varieties},  in \emph{Affine flag manifolds and principal bundles},
  pp.~1--50.
\newblock Springer, 2010.

\bibitem{2016arXiv160305593Z}
X.~{Zhu}, \emph{{An introduction to affine Grassmannians and the geometric
  Satake equivalence}}, {\emph{arXiv e-prints} (Mar., 2016) arXiv:1603.05593},
  [\href{https://arxiv.org/abs/1603.05593}{{\ttfamily 1603.05593}}].

\bibitem{baumann2018notes}
P.~Baumann and S.~Riche, \emph{Notes on the geometric satake equivalence},  in
  \emph{Relative aspects in representation theory, Langlands functoriality and
  automorphic forms}, pp.~1--134.
\newblock Springer, 2018.

\bibitem{RicharzLec}
T.~Richarz, \emph{Notes: Basics on affine grassmannians},
  \url{https://timo-richarz.com/wp-content/uploads/2020/02/BoAG_02.pdf},
  accessed November 2020.

\bibitem{AcharLec}
P.~Achar, \emph{Lecture: Introduction to affine grassmannians and the geometric
  satake equivalence},  Notes taken by Jose Simental,
  \url{https://web.northeastern.edu/iloseu/Achar_lectures.html}, accessed
  November 2020.

\bibitem{brunson2014matrix}
J.~C. Brunson, \emph{Matrix Schubert varieties for the affine Grassmannian}.
\newblock PhD thesis, Virginia Tech, 2014.

\bibitem{Kapustin:2006pk}
A.~Kapustin and E.~Witten, \emph{{Electric-Magnetic Duality And The Geometric
  Langlands Program}},
  \href{http://dx.doi.org/10.4310/CNTP.2007.v1.n1.a1}{\emph{Commun. Num. Theor.
  Phys.} {\bfseries 1} (2007) 1--236},
  [\href{https://arxiv.org/abs/hep-th/0604151}{{\ttfamily hep-th/0604151}}].

\bibitem{Witten:2009mh}
E.~Witten, \emph{{Geometric Langlands And The Equations Of Nahm And
  Bogomolny}},  \href{https://arxiv.org/abs/0905.4795}{{\ttfamily 0905.4795}}.

\bibitem{Frenkel:2005pa}
E.~Frenkel, \emph{{Lectures on the Langlands program and conformal field
  theory}},  in \emph{{Les Houches School of Physics: Frontiers in Number
  Theory, Physics and Geometry}}, pp.~387--533, 2007.
\newblock \href{https://arxiv.org/abs/hep-th/0512172}{{\ttfamily
  hep-th/0512172}}.
\newblock \href{http://dx.doi.org/10.1007/978-3-540-30308-4_11}{DOI}.

\bibitem{juteau2008modular}
D.~Juteau, \emph{Modular representations of reductive groups and geometry of
  affine grassmannians}, {\emph{arXiv preprint arXiv:0804.2041} (2008) }.

\bibitem{Gaiotto:2008ak}
D.~Gaiotto and E.~Witten, \emph{{S-Duality of Boundary Conditions In N=4 Super
  Yang-Mills Theory}},
  \href{http://dx.doi.org/10.4310/ATMP.2009.v13.n3.a5}{\emph{Adv. Theor. Math.
  Phys.} {\bfseries 13} (2009) 721--896},
  [\href{https://arxiv.org/abs/0807.3720}{{\ttfamily 0807.3720}}].

\bibitem{Zhou:2020bwa}
Y.~Zhou, \emph{{Note on some properties of generalized affine Grassmannian
  slices}},  \href{https://arxiv.org/abs/2011.04109}{{\ttfamily 2011.04109}}.

\bibitem{gorbatsevich1994lie}
V.~Gorbatsevich, \emph{Lie groups and Lie algebras III: Structure of Lie groups
  and Lie algebras}, vol.~41.
\newblock Springer Science \& Business Media, 1994.

\bibitem{Hanany:2014dia}
A.~Hanany and R.~Kalveks, \emph{{Highest Weight Generating Functions for
  Hilbert Series}},
  \href{http://dx.doi.org/10.1007/JHEP10(2014)152}{\emph{JHEP} {\bfseries 10}
  (2014) 152}, [\href{https://arxiv.org/abs/1408.4690}{{\ttfamily 1408.4690}}].

\bibitem{Bourget:2020asf}
A.~Bourget, J.~F. Grimminger, A.~Hanany, M.~Sperling, G.~Zafrir and Z.~Zhong,
  \emph{{Magnetic quivers for rank 1 theories}},
  \href{http://dx.doi.org/10.1007/JHEP09(2020)189}{\emph{JHEP} {\bfseries 09}
  (2020) 189}, [\href{https://arxiv.org/abs/2006.16994}{{\ttfamily
  2006.16994}}].

\bibitem{Bourget:2020mez}
A.~Bourget, S.~Giacomelli, J.~F. Grimminger, A.~Hanany, M.~Sperling and
  Z.~Zhong, \emph{{S-fold magnetic quivers}},
  \href{http://dx.doi.org/10.1007/JHEP02(2021)054}{\emph{JHEP} {\bfseries 02}
  (2021) 054}, [\href{https://arxiv.org/abs/2010.05889}{{\ttfamily
  2010.05889}}].

\bibitem{Cabrera:2018ann}
S.~Cabrera and A.~Hanany, \emph{{Quiver Subtractions}},
  \href{http://dx.doi.org/10.1007/JHEP09(2018)008}{\emph{JHEP} {\bfseries 09}
  (2018) 008}, [\href{https://arxiv.org/abs/1803.11205}{{\ttfamily
  1803.11205}}].

\bibitem{Cabrera:2019izd}
S.~Cabrera, A.~Hanany and M.~Sperling, \emph{{Magnetic quivers, Higgs branches,
  and 6d $N$=(1,0) theories}},
  \href{http://dx.doi.org/10.1007/JHEP06(2019)071}{\emph{JHEP} {\bfseries 06}
  (2019) 071}, [\href{https://arxiv.org/abs/1904.12293}{{\ttfamily
  1904.12293}}].

\bibitem{Cabrera:2019dob}
S.~Cabrera, A.~Hanany and M.~Sperling, \emph{{Magnetic quivers, Higgs branches,
  and 6d $N$=(1,0) theories — orthogonal and symplectic gauge groups}},
  \href{http://dx.doi.org/10.1007/JHEP02(2020)184}{\emph{JHEP} {\bfseries 02}
  (2020) 184}, [\href{https://arxiv.org/abs/1912.02773}{{\ttfamily
  1912.02773}}].

\bibitem{Hanany:2019tji}
A.~Hanany and R.~Kalveks, \emph{{Quiver Theories and Hilbert Series of
  Classical Slodowy Intersections}},
  \href{http://dx.doi.org/10.1016/j.nuclphysb.2020.114939}{\emph{Nucl. Phys. B}
  {\bfseries 952} (2020) 114939},
  [\href{https://arxiv.org/abs/1909.12793}{{\ttfamily 1909.12793}}].

\bibitem{Grimminger:2020dmg}
J.~F. Grimminger and A.~Hanany, \emph{{Hasse diagrams for 3d $ \mathcal{N} $ =
  4 quiver gauge theories \textemdash{} Inversion and the full moduli space}},
  \href{http://dx.doi.org/10.1007/JHEP09(2020)159}{\emph{JHEP} {\bfseries 09}
  (2020) 159}, [\href{https://arxiv.org/abs/2004.01675}{{\ttfamily
  2004.01675}}].

\bibitem{Bourget:2020gzi}
A.~Bourget, J.~F. Grimminger, A.~Hanany, M.~Sperling and Z.~Zhong,
  \emph{{Magnetic Quivers from Brane Webs with O5 Planes}},
  \href{http://dx.doi.org/10.1007/JHEP07(2020)204}{\emph{JHEP} {\bfseries 07}
  (2020) 204}, [\href{https://arxiv.org/abs/2004.04082}{{\ttfamily
  2004.04082}}].

\bibitem{Rogers:2018dez}
J.~Rogers and R.~Tatar, \emph{{Moduli space singularities for 3d $
  \mathcal{N}=4 $ circular quiver gauge theories}},
  \href{http://dx.doi.org/10.1007/JHEP11(2018)022}{\emph{JHEP} {\bfseries 11}
  (2018) 022}, [\href{https://arxiv.org/abs/1807.01754}{{\ttfamily
  1807.01754}}].

\bibitem{Rogers:2019pqe}
J.~Rogers and R.~Tatar, \emph{{$D_n$ Dynkin quiver moduli spaces}},
  \href{http://dx.doi.org/10.1088/1751-8121/ab4344}{\emph{J. Phys. A}
  {\bfseries 52} (2019) 425401},
  [\href{https://arxiv.org/abs/1902.10019}{{\ttfamily 1902.10019}}].

\bibitem{Eckhard:2020jyr}
J.~Eckhard, S.~Sch\"afer-Nameki and Y.-N. Wang, \emph{{Trifectas for T$_{N}$ in
  5d}}, \href{http://dx.doi.org/10.1007/JHEP07(2020)199}{\emph{JHEP} {\bfseries
  07} (2020) 199}, [\href{https://arxiv.org/abs/2004.15007}{{\ttfamily
  2004.15007}}].

\bibitem{Closset:2020scj}
C.~Closset, S.~Schafer-Nameki and Y.-N. Wang, \emph{{Coulomb and Higgs Branches
  from Canonical Singularities: Part 0}},
  \href{http://dx.doi.org/10.1007/JHEP02(2021)003}{\emph{JHEP} {\bfseries 02}
  (2021) 003}, [\href{https://arxiv.org/abs/2007.15600}{{\ttfamily
  2007.15600}}].

\bibitem{vanBeest:2020kou}
M.~van Beest, A.~Bourget, J.~Eckhard and S.~Schafer-Nameki, \emph{{(Symplectic)
  Leaves and (5d Higgs) Branches in the Poly(go)nesian Tropical Rain Forest}},
  \href{http://dx.doi.org/10.1007/JHEP11(2020)124}{\emph{JHEP} {\bfseries 11}
  (2020) 124}, [\href{https://arxiv.org/abs/2008.05577}{{\ttfamily
  2008.05577}}].

\bibitem{vanBeest:2020civ}
M.~van Beest, A.~Bourget, J.~Eckhard and S.~Schafer-Nameki, \emph{{(5d RG-flow)
  Trees in the Tropical Rain Forest}},
  \href{https://arxiv.org/abs/2011.07033}{{\ttfamily 2011.07033}}.

\bibitem{Closset:2020afy}
C.~Closset, S.~Giacomelli, S.~Sch\"afer-Nameki and Y.-N. Wang, \emph{{5d and 4d
  SCFTs: Canonical Singularities, Trinions and S-Dualities}},
  \href{https://arxiv.org/abs/2012.12827}{{\ttfamily 2012.12827}}.

\bibitem{Barns-Graham:2019ydl}
A.~E. Barns-Graham, \emph{{Much ado about nothing: The superconformal index and
  Hilbert series of three dimensional $\mathcal N=4$ vacua}}.
\newblock PhD thesis, Cambridge U., DAMTP, 2018.
\newblock 10.17863/CAM.35266.

\bibitem{namikawa2018characterization}
Y.~Namikawa, \emph{A characterization of nilpotent orbit closures among
  symplectic singularities}, {\emph{Mathematische Annalen} {\bfseries 370}
  (2018) 811--818}.

\bibitem{Cabrera:2018uvz}
S.~Cabrera, A.~Hanany and A.~Zajac, \emph{{Minimally Unbalanced Quivers}},
  \href{http://dx.doi.org/10.1007/JHEP02(2019)180}{\emph{JHEP} {\bfseries 02}
  (2019) 180}, [\href{https://arxiv.org/abs/1810.01495}{{\ttfamily
  1810.01495}}].

\bibitem{Haouzi:2016ohr}
N.~Haouzi and C.~Schmid, \emph{{Little String Origin of Surface Defects}},
  \href{http://dx.doi.org/10.1007/JHEP05(2017)082}{\emph{JHEP} {\bfseries 05}
  (2017) 082}, [\href{https://arxiv.org/abs/1608.07279}{{\ttfamily
  1608.07279}}].

\bibitem{Haouzi:2016yyg}
N.~Haouzi and C.~Schmid, \emph{{Little String Defects and Bala-Carter Theory}},
   \href{https://arxiv.org/abs/1612.02008}{{\ttfamily 1612.02008}}.

\bibitem{Bachas:2000dx}
C.~Bachas, J.~Hoppe and B.~Pioline, \emph{{Nahm equations, N=1* domain walls,
  and D strings in AdS(5) x S(5)}},
  \href{http://dx.doi.org/10.1088/1126-6708/2001/07/041}{\emph{JHEP} {\bfseries
  07} (2001) 041}, [\href{https://arxiv.org/abs/hep-th/0007067}{{\ttfamily
  hep-th/0007067}}].

\bibitem{Gukov:2020lqm}
S.~Gukov, P.-S. Hsin, H.~Nakajima, S.~Park, D.~Pei and N.~Sopenko,
  \emph{{Rozansky-Witten geometry of Coulomb branches and logarithmic knot
  invariants}},  \href{https://arxiv.org/abs/2005.05347}{{\ttfamily
  2005.05347}}.

\bibitem{Hanany:2001iy}
A.~Hanany and J.~Troost, \emph{{Orientifold planes, affine algebras and
  magnetic monopoles}},
  \href{http://dx.doi.org/10.1088/1126-6708/2001/08/021}{\emph{JHEP} {\bfseries
  08} (2001) 021}, [\href{https://arxiv.org/abs/hep-th/0107153}{{\ttfamily
  hep-th/0107153}}].

\bibitem{GNO}
P.~Goddard, J.~Nuyts and D.~I. Olive, \emph{{Gauge Theories and Magnetic
  Charge}}, \href{http://dx.doi.org/10.1016/0550-3213(77)90221-8}{\emph{Nucl.
  Phys.} {\bfseries B125} (1977) 1--28}.

\bibitem{Cremonesi:2014xha}
S.~Cremonesi, G.~Ferlito, A.~Hanany and N.~Mekareeya, \emph{{Coulomb Branch and
  The Moduli Space of Instantons}},
  \href{http://dx.doi.org/10.1007/JHEP12(2014)103}{\emph{JHEP} {\bfseries 12}
  (2014) 103}, [\href{https://arxiv.org/abs/1408.6835}{{\ttfamily 1408.6835}}].

\bibitem{Ncones}
A.~Bourget, J.~F. Grimminger, A.~Hanany, R.~Kalveks, M.~Sperling and Z.~Zhong,
  \emph{{A Tale of N Cones}}, {\emph{to appear} }.

\bibitem{Ferlito:2016grh}
G.~Ferlito and A.~Hanany, \emph{{A tale of two cones: the Higgs Branch of Sp(n)
  theories with 2n flavours}},
  \href{https://arxiv.org/abs/1609.06724}{{\ttfamily 1609.06724}}.

\end{thebibliography}\endgroup

\end{document}